\documentclass[twocolumn]{aastex63}
\usepackage{amsmath}
\usepackage{amssymb}
\usepackage{chngcntr}
\usepackage{hyperref}
\counterwithout{figure}{section}

\received{}
\revised{}
\accepted{}
\submitjournal{ApJ}

\shorttitle{Multiwavelength variability PSDs}
\shortauthors{Goyal et al.}

\begin{document}

\title{Multiwavelength variability power spectrum analysis of the blazars 3C\,279 and PKS\,1510$-$089 on multiple timescales}

\correspondingauthor{Arti~Goyal}
\email{arti.goyal@uj.edu.pl}

\author[0000-0002-2224-6664]{Arti~Goyal}
\affiliation{Astronomical Observatory of the Jagiellonian University, Orla 171, 30-244 Krakow, Poland}

\author{Marian~Soida}
\affiliation{Astronomical Observatory of the Jagiellonian University, Orla 171, 30-244 Krakow, Poland}

\author{{\L}ukasz~Stawarz}
\affiliation{Astronomical Observatory of the Jagiellonian University, Orla 171, 30-244 Krakow, Poland}

\author{Paul~J.~Wiita}
\affiliation{Department of Physics, The College of New Jersey, 2000 Pennington Rd., Ewing, NJ 08628-0718, USA}

\author{Kari~Nilsson}
\affiliation{Tuorla Observatory, Department of Physics and Astronomy, University of Turku, {Turku} {FI-21500}, Finland}

\author{Svetlana~Jorstad}
\affiliation{Institute for Astrophysical Research, Boston University, 725 Commonwealth Avenue, Boston, MA 02215, USA}

\author{Alan~P.~Marscher}
\affiliation{Institute for Astrophysical Research, Boston University, 725 Commonwealth Avenue, Boston, MA 02215, USA}

\author{Margo~F.~Aller}
\affiliation{University of Michigan, Department of Astronomy, 1085 South University Avenue, Ann Arbor MI 48109, USA}

\author{Hugh~D.~Aller}
\affiliation{University of Michigan, Department of Astronomy, 1085 South University Avenue, Ann Arbor MI 48109, USA}

\author{Anne~L\"ahteenm\"aki}
\affiliation{Aalto University Mets\"ahovi Radio Observatory, Mets\"ahovintie 114, FI-02540 Kylm\"al\"a, Finland}
\affiliation{Aalto University Department of Electronics and Nanoengineering, P.O. BOX 15500, FI-00076 Aalto, Finland}

\author{Talvikki~Hovatta}
\affiliation{Finnish Centre for Astronomy with ESO (FINCA), University of Turku, FI-20014, Turku, Finland}
\affiliation{Aalto University Mets\"ahovi Radio Observatory, Mets\"ahovintie 114, 02540 Kylm\"al\"a, Finland}

\author[0000-0003-3609-382X]{Staszek Zo{\l}a}
\affiliation{Astronomical Observatory of the Jagiellonian University, Orla 171, 30-244 Krakow, Poland}

\author{Krzysztof~Nalewajko}
\affiliation{Nicolaus Copernicus Astronomical Center, Polish Academy of Sciences, Bartycka 18, PL-00-716 Warszawa, Poland}

\author{Merja~Tornikoski}
\affiliation{Aalto University Mets\"ahovi Radio Observatory, Mets\"ahovintie 114, 02540 Kylm\"al\"a, Finland}

\author{Joni~Tammi}
\affiliation{Aalto University Mets\"ahovi Radio Observatory, Mets\"ahovintie 114, 02540 Kylm\"al\"a, Finland}

\author{Mark~Hodges}
\affiliation{Owens Valley Radio Observatory, California Institute of Technology, Pasadena, CA 91125, USA}

\author{Sebastian~Kiehlmann}
\affiliation{Institute of Astrophysics, Foundation for Research and Technology-Hellas, GR-71110 Heraklion,Greece}
\affiliation{Department of Physics, University of Crete, GR-70013 Heraklion, Greece}

\author{Anthony~C.~S.~Readhead}
\affiliation{Owens Valley Radio Observatory, California Institute of Technology, Pasadena, CA 91125, USA}

\author{Walter~Max-Moerbeck}
\affiliation{Departamento de Astronom\'ia, Universidad de Chile, Camino El Observatorio 1515, Las Condes, Santiago, Chile}

\author{Elina~Lindfors}
\affiliation{Finnish Centre for Astronomy with ESO (FINCA), University of Turku, FI-20014, Turku, Finland}

\author{Vandad~Fallah~Ramazani}
\affiliation{Finnish Centre for Astronomy with ESO (FINCA), University of Turku, FI-20014, Turku, Finland}

\author{D.~E.~Reichart}
\affiliation{University of North Carolina at Chapel Hill, Chapel Hill, NC 27599, USA}

\author{D.~B.~Caton}
\affiliation{Dark Sky Observatory, Dept. of Physics and Astronomy, Appalachian State University, Boone, NC 28608, USA}

\author{Janeth~Valverde}
\affiliation{Center for Space Sciences and Technology, University of Maryland Baltimore County, Baltimore, MD 21250, USA}
\affiliation{NASA Goddard Space Flight Center, Greenbelt, MD 20771, USA}

\author{Deirdre~Horan}
\affiliation{Laboratoire Leprince-Ringuet, \'Ecole polytechnique, CNRS/IN2P3, F-91128 Palaiseau, France}

\author{Roopesh~Ojha}
\affiliation{NASA Goddard Space Flight Center, Greenbelt, MD 20771, USA}

\author{Pfesesani van Zyl}
\affiliation{The South African Radio Astronomy Observatory (SARAO), Farm 502 JQ, Hartebeesthoek, Broederstroom Road, Hartebeesthoek, 1740, South Africa}

\begin{abstract}
We present the results of variability power spectral density (PSD) analysis using multiwavelength radio to GeV\,$\gamma$-ray light curves covering decades/years to days/minutes timescales for the blazars 3C\,279 and PKS\,1510$-$089. The PSDs are modeled as single power-laws, and the best-fit spectral shape is derived using the `power spectral response' method. With more than ten years of data obtained with weekly/daily sampling intervals, most of the PSDs cover $\sim$2-4 decades in temporal frequency; moreover, in the optical band, the  PSDs cover $\sim$6 decades for 3C\,279 due to the availability of intranight light curves. Our main results are the following: (1) on timescales ranging from decades to days, the synchrotron and the inverse Compton spectral components, in general, exhibit red-noise (slope $\sim$2) and flicker-noise (slope $\sim$1) type variability, respectively; (2) the slopes of $\gamma$-ray variability PSDs obtained using a 3-hr integration bin and a 3-weeks total duration exhibit a range between $\sim$1.4 and $\sim$2.0 (mean slope = 1.60$\pm$0.70), consistent within errors with the slope on longer timescales;  (3) comparisons of fractional variability indicate more power on timescales $\leq$100\, days at $\gamma$-ray frequencies as compared to longer wavelengths, in general (except between $\gamma$-ray and optical frequencies for PKS 1510$-$089); (4) the normalization of intranight optical PSDs for 3C\,279 appears to be a simple extrapolation from longer timescales, indicating a continuous (single) process driving the variability at optical wavelengths; (5) the emission at optical/infrared wavelengths may involve a combination of disk and jet processes for PKS\,1510$-$089.
\end{abstract}


\section{Introduction}\label{sec:intro}

Characterized by flux and polarization variability, blazars---including the BL Lacertae objects and flat-spectrum radio quasars (FSRQs)---constitute a prominent subset of active galactic nuclei (AGNs) whose radiative output is dominated by non-thermal processes occurring inside the relativistic, non-stationary, and magnetized jets \citep[see][for a recent review]{Blandford19}. Their broadband spectral energy distribution (SED) is typically composed of two peaks: (1) a low-energy segment ranging from radio to optical frequencies (sometimes extending up to X-rays in case of BL Lac objects) which is unequivocally attributed to synchrotron radiation of charged particles accelerated up to TeV energies, and (2) a high-energy segment ranging from UV/X-rays up to GeV/TeV\,$\gamma$-ray frequencies which are attributed to inverse-Compton (IC) radiation of the seed photons produced locally (synchrotron self-Compton; SSC) or externally (external Compton; EC) to the jet plasma within the leptonic scenario of emission \citep[e.g.,][and references therein]{Madejski16a}. Alternatively, in the `hadronic' scenario for emission, the higher-frequency emission peak originates from protons accelerated to $\simeq$\,PeV-EeV energies which could produce $\gamma$-rays via either a direct synchrotron process or meson decay and synchrotron emission by the secondaries produced in proton-photon interactions \citep[e.g.,][]{Bottcher13}.

The random, aperiodic flux variability exhibited by these sources over a wide range of emission frequencies (radio to TeV gamma rays) and variability timescales (decades to minutes) is classified broadly into long-term (decades/years to weeks), short-term (weeks to $\sim$day), and intranight ($\leq$day) variability \citep[e.g.][]{Ulrich97, Hovatta19}. The variability power spectral densities (PSDs) are known to typically exhibit a single power-law form, defined as $P(\nu_k) \propto$ $\nu_k^{-\beta}$, where $\beta$ is the slope and $\nu_k$ is the temporal frequency ($\equiv$timescale$^{-1}$), which indicate that variability is a colored noise type stochastic process \citep[e.g.,][]{Finke14, Goyal17}. Specifically, $\beta \sim$1 and $\sim$2 are known as flicker/pink-noise and damped random-walk/red-noise type {\it correlated} stochastic processes, respectively, while $\beta \sim$0 is called an {\it uncorrelated} white-noise type stochastic process \citep[e.g.,][]{Press78}. Detection of breaks in PSDs are of importance as they provide information on the `characteristic/relaxation' timescales in the system generating the variability, such as the size of the emission zone or the particle cooling or escape timescales \citep[e.g.,][]{Kastendieck11, Sobolewska14, Finke14, Chen16, Kushwaha17, Chatterjee18, Ryan19, Bhattacharyya20}.

\begin{figure*}

\includegraphics[width=0.93\textwidth]{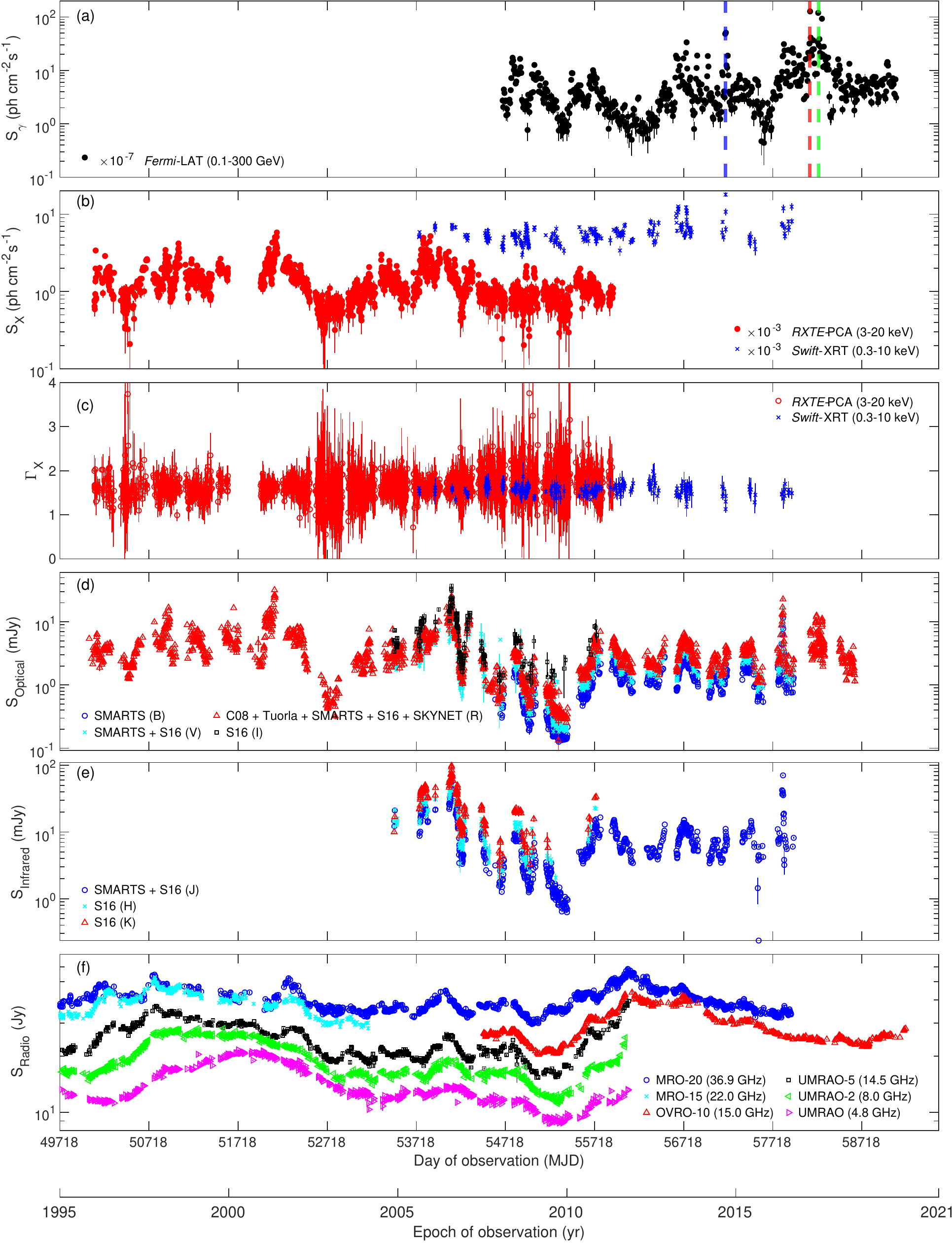}
\caption{Multiwavelength light curves of the blazar 3C\,279. Panel (a) gives the 7-day integration bin $\gamma$-ray fluxes from the {\it Fermi-}LAT satellite (0.1--300\,GeV). The blue, red and green dashed vertical lines indicate three epochs for which 3-hr integration bin light curves are generated (panels a, b, and c of Figure~\ref{fig:fermi3hr}). Panel (b) shows the X-ray fluxes from the {\it RXTE-}PCA (3--20\,keV) and {\it Swift-}XRT (0.3--10\,keV) satellites, respectively, while panel (c) gives the photon index derived by fitting a single power-law model to these datasets. Panels (d) and (e) give the $B$, $V$, $R$, $I$, $J$, $H$, and $K$-band fluxes from  various, labeled, observing programs. Panel (f) gives radio fluxes, with noted offsets, from the Mets{\"a}hovi (36.9\,GHz and 22.0 GHz), OVRO (15\,GHz) and UMRAO (14.5\,GHz, 8.0\,GHz, and 4.8\,GHz) monitoring programs, respectively. Radio light curves for the entire duration of monitoring are given in Figure~\ref{fig:mwrad}(a). }
\label{fig:lc3c279}
\end{figure*}

\begin{figure*}
\includegraphics[width=1\textwidth]{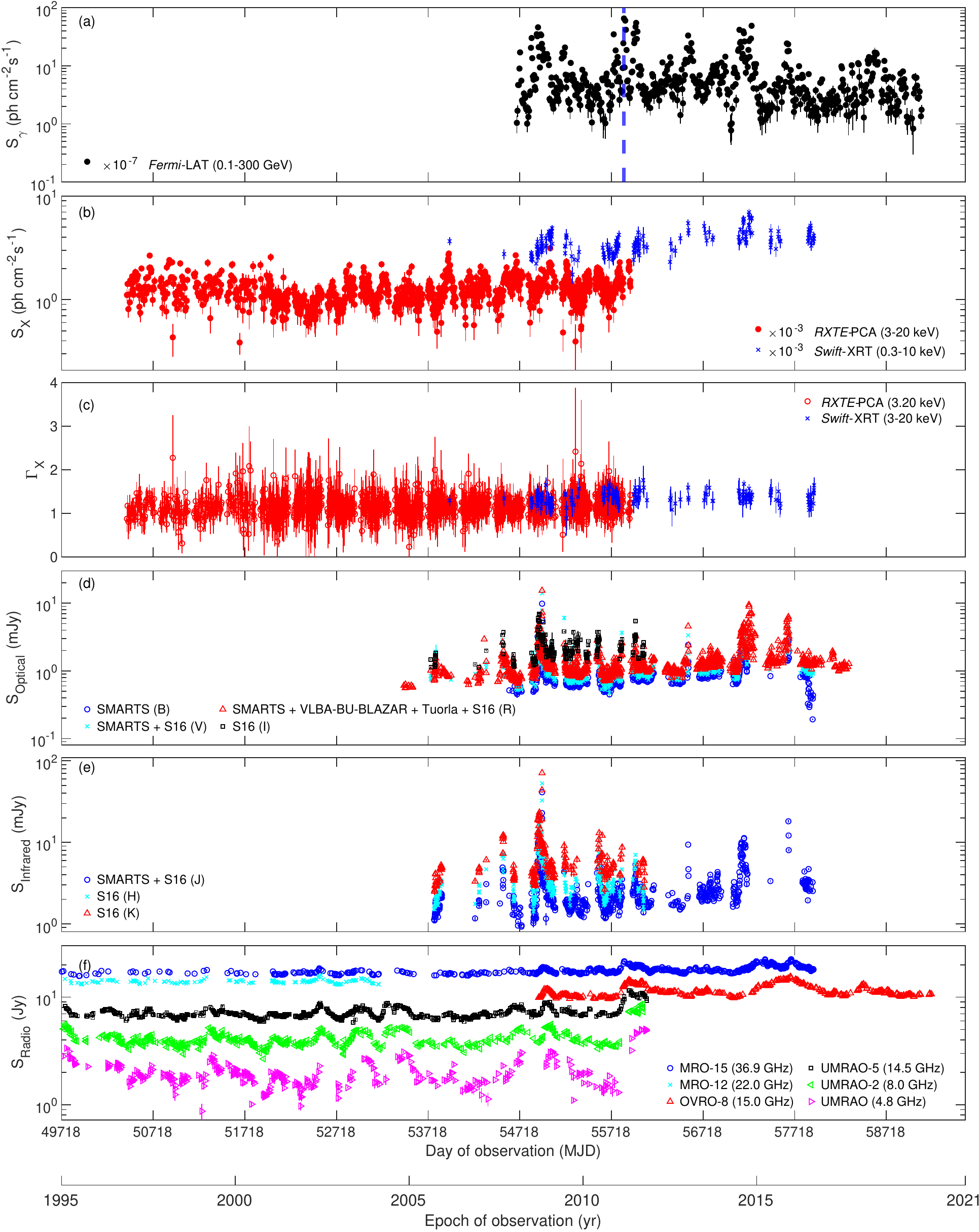}
\caption{ As in Figure~\ref{fig:lc3c279}, for the blazar PKS\,1510$-$089.  In panel (a) the  blue dashed vertical line indicates an epoch for which 3-hr integration bin light curve is generated (panel (d) of Figure~\ref{fig:fermi3hr}).}
\label{fig:lcpks1510}
\end{figure*}

\begin{deluxetable}{ccccc}
\tablenum{1}
\tablecaption{Basic properties of the blazars 3C\,279 and PKS\,1510$-$089 studied here\label{tab:sample}}
\tablewidth{0pt}
\tabletypesize{\small}
\tablehead{
\colhead{Name} & \colhead{R.A.(J2000)} & \colhead{Dec.(J2000)} &  $z$ & $M_{BH}$\\
\colhead{} &  \colhead{(h m s)} & \colhead{($^\circ$ $\prime$ ${\prime\prime}$)} &  \colhead{} & \colhead{(10$^8$M$_\odot$)} 
}
\decimalcolnumbers
\startdata
3C\,279    & 12 56 11.166 & $-$05 47 21.52 &   0.536$^a$ & 3.0-8.0$^c$ \\
PKS\,1510$-$089 & 15 12 50.532 & $-$09 05 59.82 &   0.360$^b$ & 0.4-0.7$^d$ \\ 
\enddata
\tablecomments{
(1) the name of the blazar, 
	(2-3) right ascension and declination from the \citet[][]{ned},
(4) spectroscopic redshift. $^a$\citet[][]{Marziani96}; $^b$\citet{Thompson90},
(5) mass of the SMBH. $^c$ \citet{Hayashida15}; $^d$ \citet{Rakshit20}. 
}
\end{deluxetable}

The multiwavelength ``snapshots'' of blazar activity, i.e., the SEDs, can provide important clues to the emission processes and the physical parameters related to them within the relativistic jet, at least within a particular model set. This is usually attempted through solving the kinetic equation of the population of electrons injected with a certain energy distribution and evolving under radiative and adiabatic energy losses in, and escaping from, a single emission zone \citep{Chiaberge99, Lindfors05, Sikora09, Abdo11a, Abdo11b, Paliya15, Hayashida15, Aleksic15b, Saito15, Dutka17, Hu21}. Such modeling efforts usually include an additional accretion disk component to fully describe the SED, although this component is more often necessary in the case of FSRQs \citep[e.g.,][]{Castignani13, Sbarrato16} than for BL Lac objects \citep[e.g.,][]{Kushwaha18}. Interestingly, the broad-band SEDs of the blazar PKS\,1510$-$089 often reveal a prominent accretion disk component peaking at optical/$IR$ emission frequencies \citep[][]{Nalewajko12} in contrast to the SEDs of the blazar 3C\,279 which rarely show such a component \citep[e.g.,][ except during a low $\gamma$-ray activity state when an upturn at optical-UV wavelengths was attributed to an accretion disk component by \citealt{Paliya15}]{Hayashida12, Hayashida15}. Studies of light curve variability and their PSDs provide information complementary to this as well as shedding some light on the jet dynamics and it is on this aspect that we focus in this paper. \citet{Finke14} analyzed the continuity equation in the {\it Fourier-}domain for the basic models mentioned above and predicted that PSD slopes at EC emission frequencies will be the same as that of the synchrotron emission frequencies (which is presumably more relevant for FSRQs). However, they found that the PSD slopes for the SSC emission frequencies varied with the slopes of the synchrotron emission frequencies (more relevant for BL Lac objects). \citet{Chen16} included the effects of the decline of particle acceleration in this model setup and obtained PSDs at synchrotron and SSC emission frequencies. Breaks in PSDs on timescales of hours-days have been either identified with light-crossing, electron cooling, or escape timescales \citep{Finke14}, or with relaxation timescales of the system \citep{Chen16}. 

Due to the availability of large datasets from various monitoring programs started some decades ago, although restricted to, at best, daily/weekly sampling of flux measurements, several groups have attempted to characterize the shape(s) of variability PSDs covering $\sim$3-3000 day timescales. In particular, using $\sim$30-40 years of radio data at 4.8, 8.0 and 14.5\,GHz frequencies from the University of Michigan Radio Observatory (UMRAO) program, \citet{Park17} reported $\beta$$\sim$1-3 for a sample of 43 blazars/AGNs. \citet{Max-Moerbeck14a} reported $\beta$$\sim$1.6-2.5 for a sample of 41 blazar sources using the 15 GHz radio light curves of 4-year duration from the Owens Valley Radio Observatory (OVRO) monitoring program. At optical frequencies, using the decade-long $R$-band data from the Tuorla Observatory monitoring program, \citet{Nilsson18} reported $\beta$ $\sim$1-1.5 for a sample of 31 TeV detected blazars. At X-ray energies, using $\sim$2-3 day-long Advanced Satellite for Cosmology and Astrophysics (ASCA) and {\it Rossi X-ray Timing Explorer (RXTE)} datasets, \citet{Kataoka01} reported $\beta$$\sim$1-1.5 for the blazars Mrk\,421, Mrk\,501 and PKS\,2155$-$304 with a break in the PSD at a frequency of $\sim$10$^{-5}$\,Hz \citep[see also][for confirmation of this break in the case of Mrk\,421 using longer datasets]{Isobe15, Chatterjee18}. At high energy (HE)\,$\gamma$-ray energies using 11 months of the {\it Fermi-}Large Area Telescope (LAT) data, \citet{Abdo10a} reported $\beta$$\sim$1-2 for a sample of 28 blazars; these slopes remain essentially unchanged when the PSD analyses were extended to longer timescales using the decade-long {\it Fermi-}LAT light curves \citep{Meyer19, Bhatta20, Tarnopolski20}. At VHE\,$\gamma$-ray energies, using $\geq$decade-long TeV light curves from the Very Energetic Radiation Imaging Telescope Array System (13.4 years) and High Energy Stereoscopic System (8.4 years) monitoring datasets, \citet{Goyal20} reported $\beta$$\sim$1 for the blazars Mrk\,421 and PKS\,2155$-$304, respectively.

Even though  blazar emissions have been detected for over 18 decades of  wavelengths and are variable over a wide range of timescales, very few efforts have been made to characterize the variability PSDs over different emission frequencies covering similar variability timescales for the {\it same} set of blazar sources; this is due to the lack of good quality datasets at a wide range of emission and variability frequencies. Notably, \citet{Chatterjee08} performed the PSD analysis for the blazar\,3C 279 at X-ray, optical, and radio frequencies using the decade-long {\it RXTE}-PCA data, $R$-band, and 14.5\,GHz UMRAO datasets for the period 1996-2007. We revisit these results in the discussion. \citet{Chatterjee12} reported similar PSDs slopes for a sample of 6 blazars using 2-year long $\gamma$-ray and optical/infrared light curves. Studies by \citet{Sandrinelli16} and \citet{Sandrinelli17} focussed on searching for  quasi-periodic oscillations (QPOs) using $>$6-year long $\gamma$-ray and optical light curves of the blazar sources. Keeping this in mind, we started a program in 2015 to uniformly characterize the PSDs for many objects by uniquely selecting blazar sources for which long-duration ($>$decade), densely sampled (daily/weekly binned), high photometric accuracy ($\sim$1-15\%) multiwavelength datasets exist for a large number of emission frequencies, either from public archives or from our own monitoring programs (particularly relevant for covering intranight timescales at optical frequencies). Our previous attempts focussed on characterizing the multiwavelength PSDs of the BL Lac objects PKS\,0735+178 \citep[radio, optical, and HE\,$\gamma$-ray energies;][]{Goyal17}, OJ\,287 \citep[radio, X-ray, optical, HE\,$\gamma$-ray energies where the PSDs were characterized by modeling the light curves as a continuous-time auto regressive moving average process (CARMA);][]{Goyal18}, Mrk\,421 and PKS\,2155$-$304 \citep[radio, X-ray, optical, infrared, HE\,$\gamma$-ray, and VHE\,$\gamma$-ray energies;][]{Goyal20}. Our main results were the following: (1) on timescales ranging from decades to months/days, the variability PSDs exhibited $\beta \sim$1 for X-rays, HE\,and VHE\,$\gamma$-rays and $\beta\sim$2 for the radio and optical/infrared frequencies; (2) the normalization of optical intranight PSDs turned out to be a smooth extrapolation from longer timescales; however, the PSD slopes showed a wider range with $\beta$$\sim$1-4, and; (3) more variability power was found on timescales $<$100\,day in the X-ray and $\gamma$-ray bands as compared to lower energies. We discuss these results in Section~\ref{sec:discussion}.

In order to understand the underlying phyical processes responsible for the blazar emission and its variability, here we expand our efforts to study the shape of the multiwavelength variability PSDs of two {\it bona-fide} FSRQs, namely, 3C\,279 ($z$=0.536) and PKS\,1510$-$089 ($z$=0.360). Table~\ref{tab:sample} gives the basic properties of these sources. These FSRQs have been detected at VHE\,$\gamma$-ray energies \citep[][]{MAGIC08, HESS13} and show rapid (minute-like) variability at {\it Fermi-}LAT energies \citep[][]{Foschini11, Ackermann16, Shukla20}. These FSRQs are well studied: 3C\,279 \citep{Hartman92, Larionov08, Chatterjee08, Abdo10b, Hayashida15};  PKS\,1510$-$089 \citep{Abdo10d, DAmmando11, Chen12, Castignani17, Ahnen17}. We selected these two sources as they have been a target of extensive multiwavelength campaigns including long-term monitoring with the {\it RXTE-}PCA instrument, thus allowing us to gather good-quality, long-duration light curves in the X-rays as well as at several GHz\,band radio frequencies, multiple infrared and optical bands, along with GeV\,$\gamma$-rays from the {\it Fermi} satellite. These light curves cover $\sim$6--30\,yr durations with typical sampling intervals ranging from days to weeks depending on the monitoring program. The luminosity distances for 3C\,279 and PKS\,1510$-$089 are 3.1\,Gpc and 1.9\,Gpc, respectively, computed using the cosmological calculator\footnote{\url{http://www.astro.ucla.edu/~wright/CosmoCalc.html}} \citep[][]{Wright06}  and assuming a concordant cosmology with the Hubble constant $ H_0=$ 69.6\,km\,s$^{-1}$\,Mpc$^{-1}$, $\Omega_{\rm m} =$ 0.286, and $\Omega_\Lambda =$ 0.714 \citep{Bennett14}.\\  

This paper is organized as follows. Section~\ref{sec:data} gives details on the data acquisition and the generation of multiwavelength light curves. Section~\ref{sec:analysis} describes the analysis methods, in particular, the derivation of PSDs and the estimation of the best-fit PSD shape using numerical simulations of the light curves. Section~\ref{sec:results} provides the main results while the discussion and summary are given in Section~\ref{sec:discussion}.

\begin{figure*}
\hbox{
\includegraphics[width=0.45\textwidth]{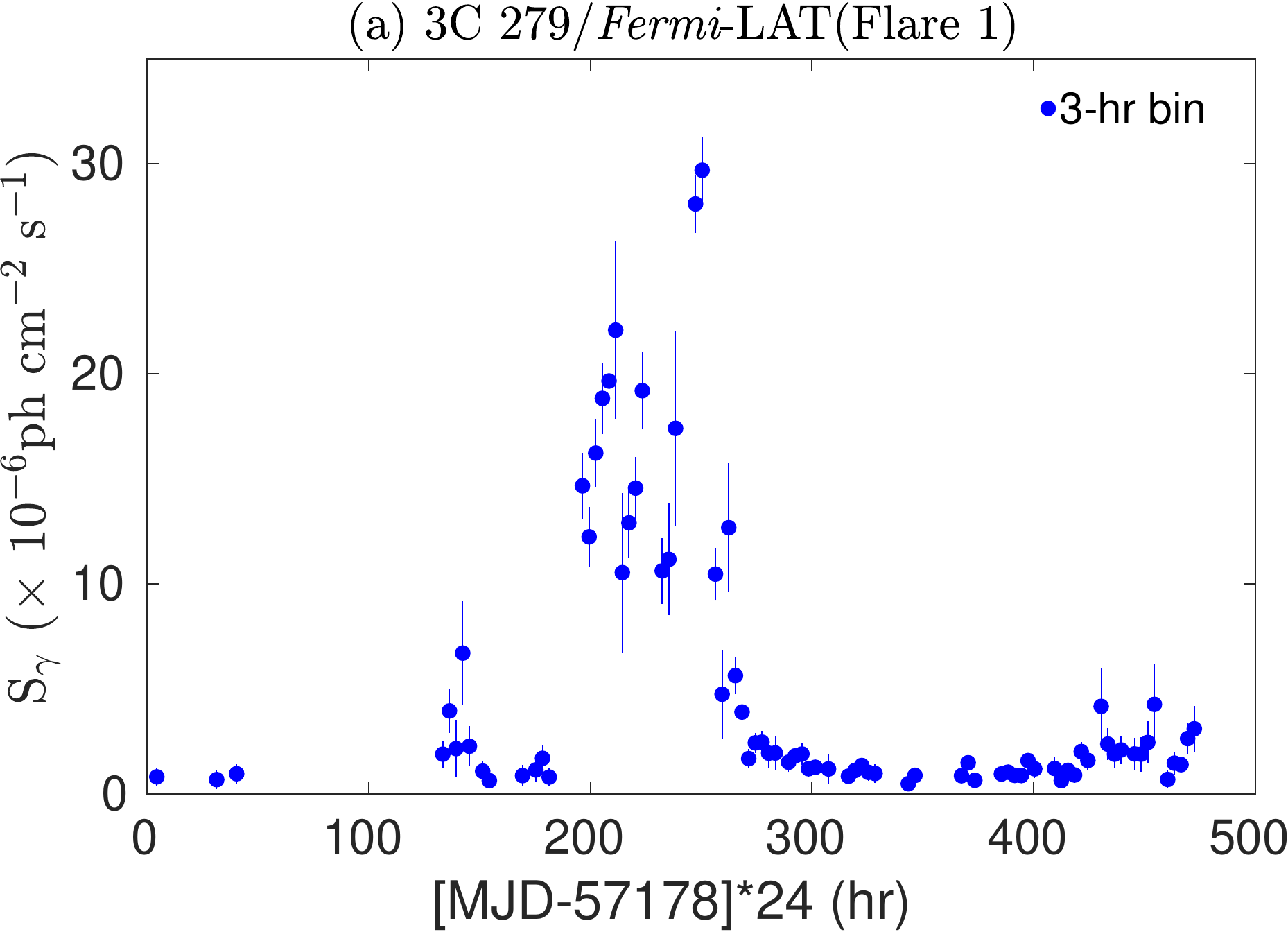}
\includegraphics[width=0.45\textwidth]{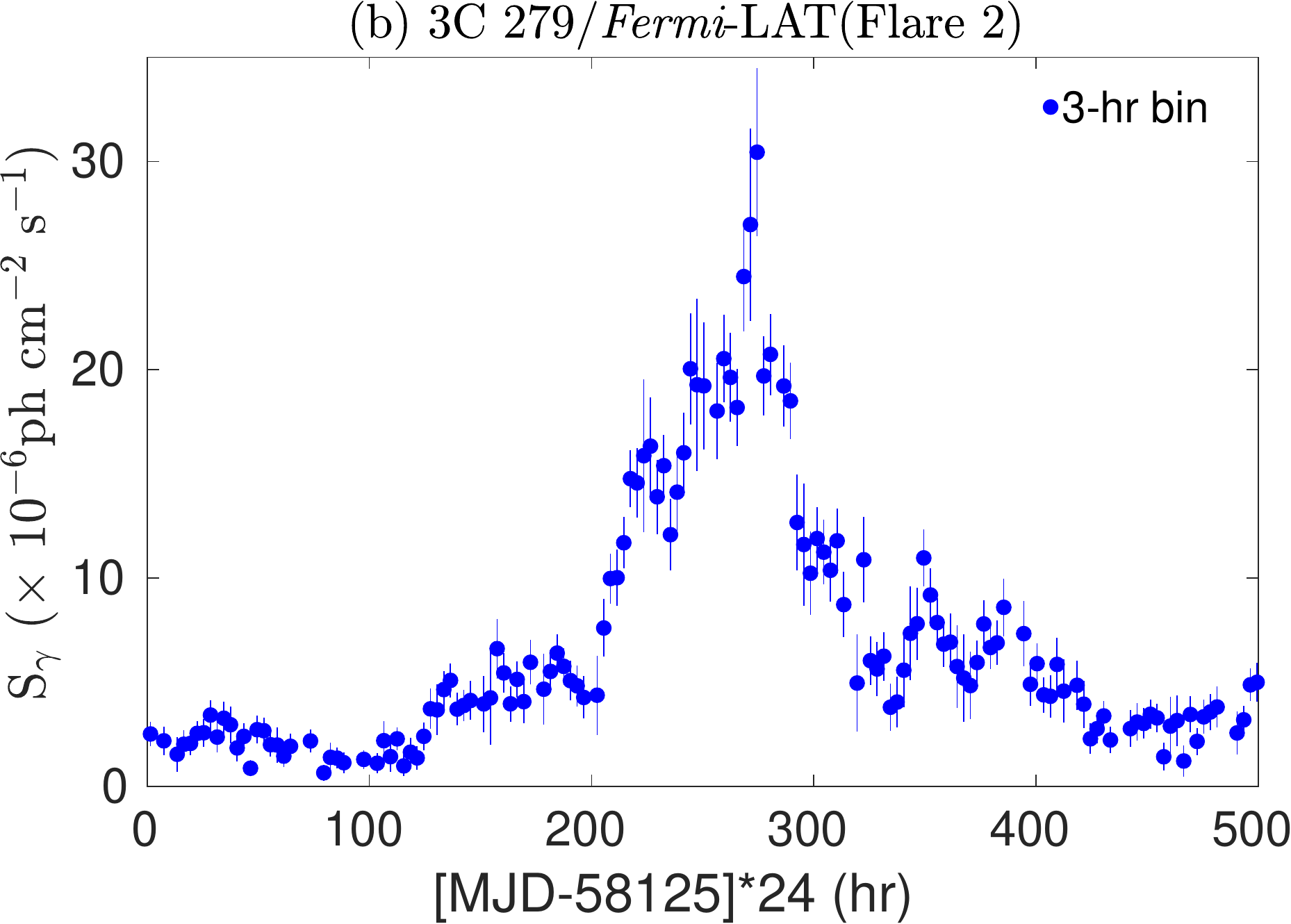}
}
\hbox{
\includegraphics[width=0.45\textwidth]{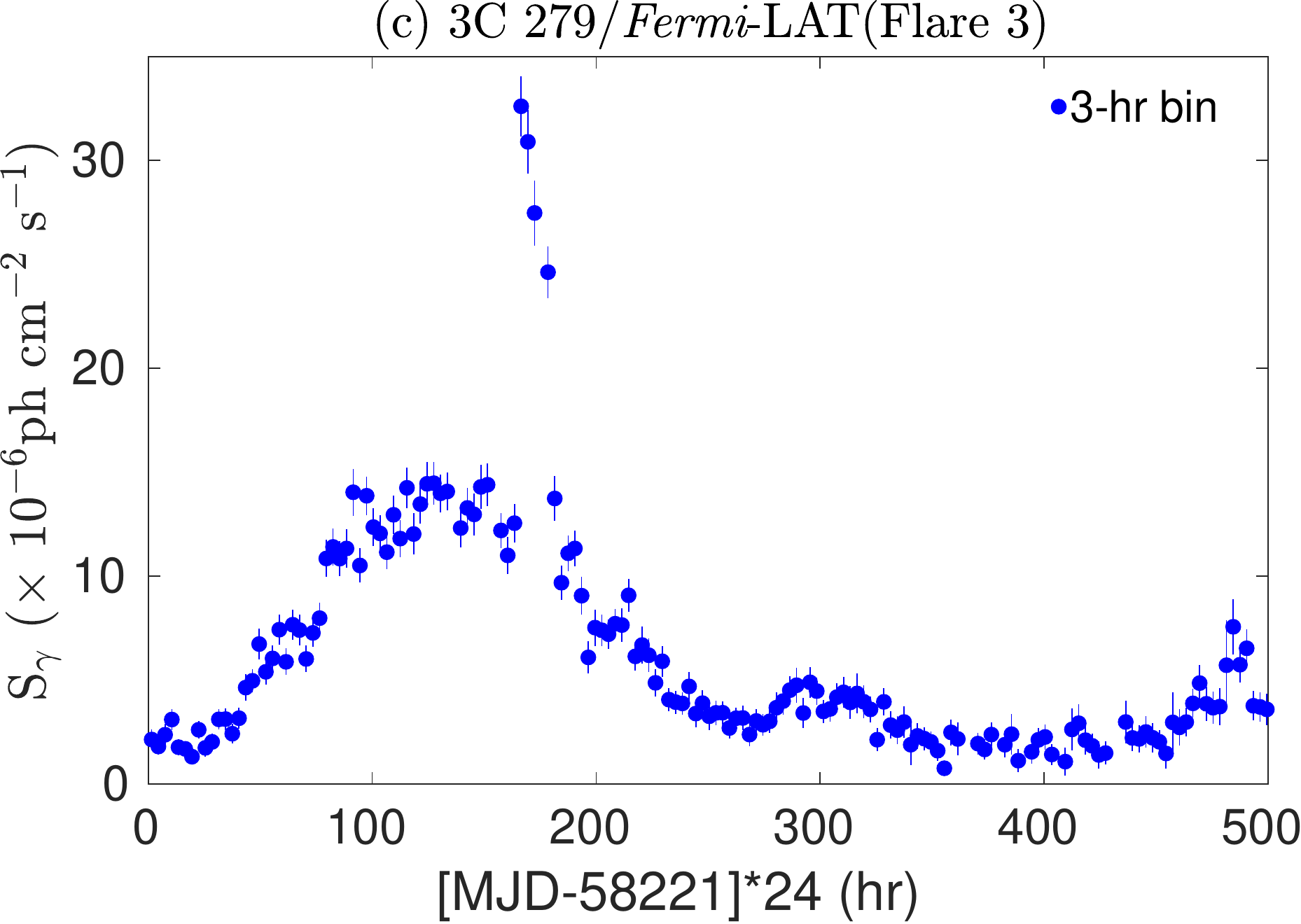}
\includegraphics[width=0.45\textwidth]{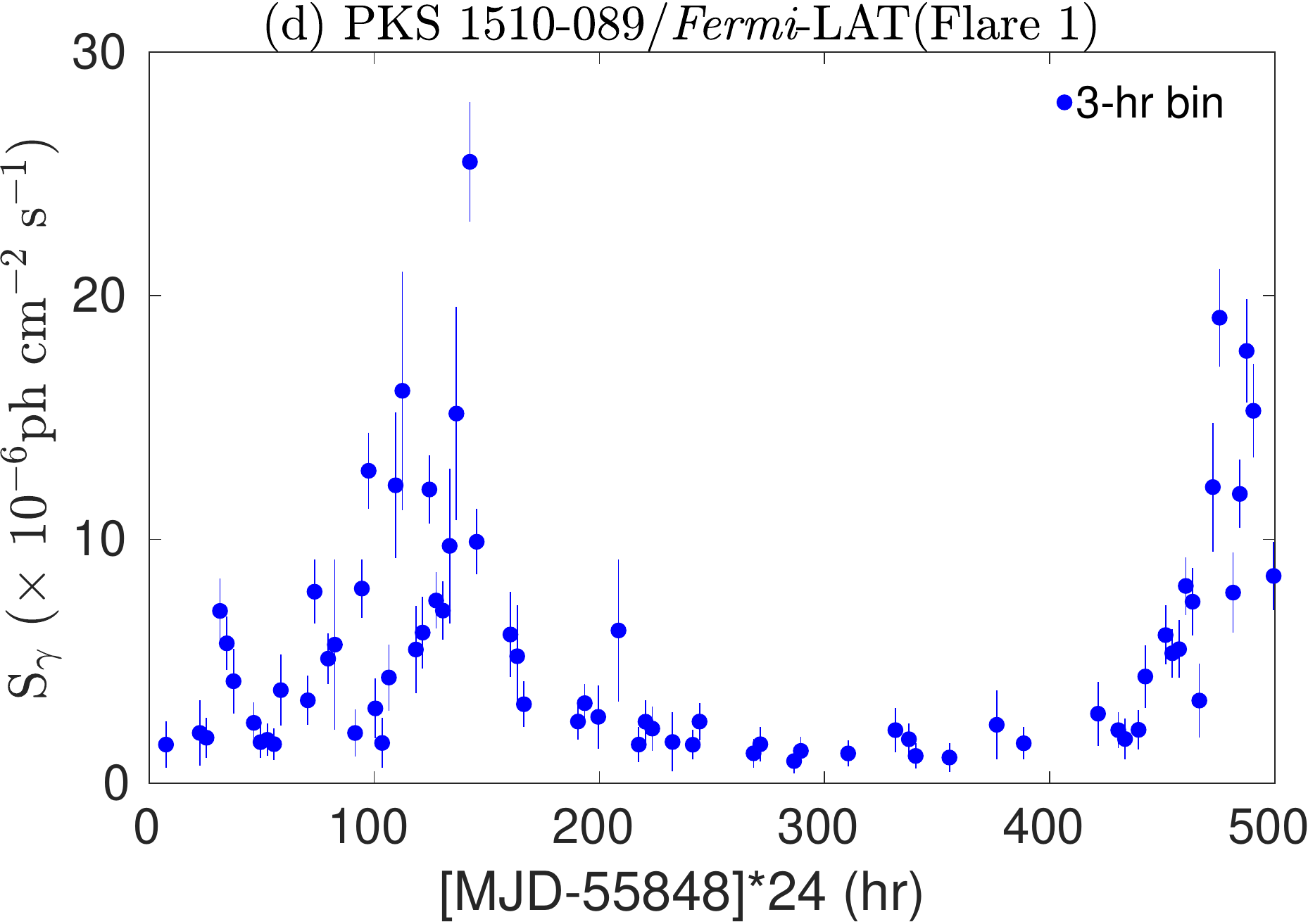}
}
\caption{  {\it Fermi-}LAT light curves obtained using 3-hr integration binning interval for selected time ranges for 3C\,279 (panels a, b, and c) and PKS\,1510$-$089 (panel d). \label{fig:fermi3hr}}
\end{figure*}

\begin{figure*}
\hbox{
\includegraphics[width=0.50\textwidth]{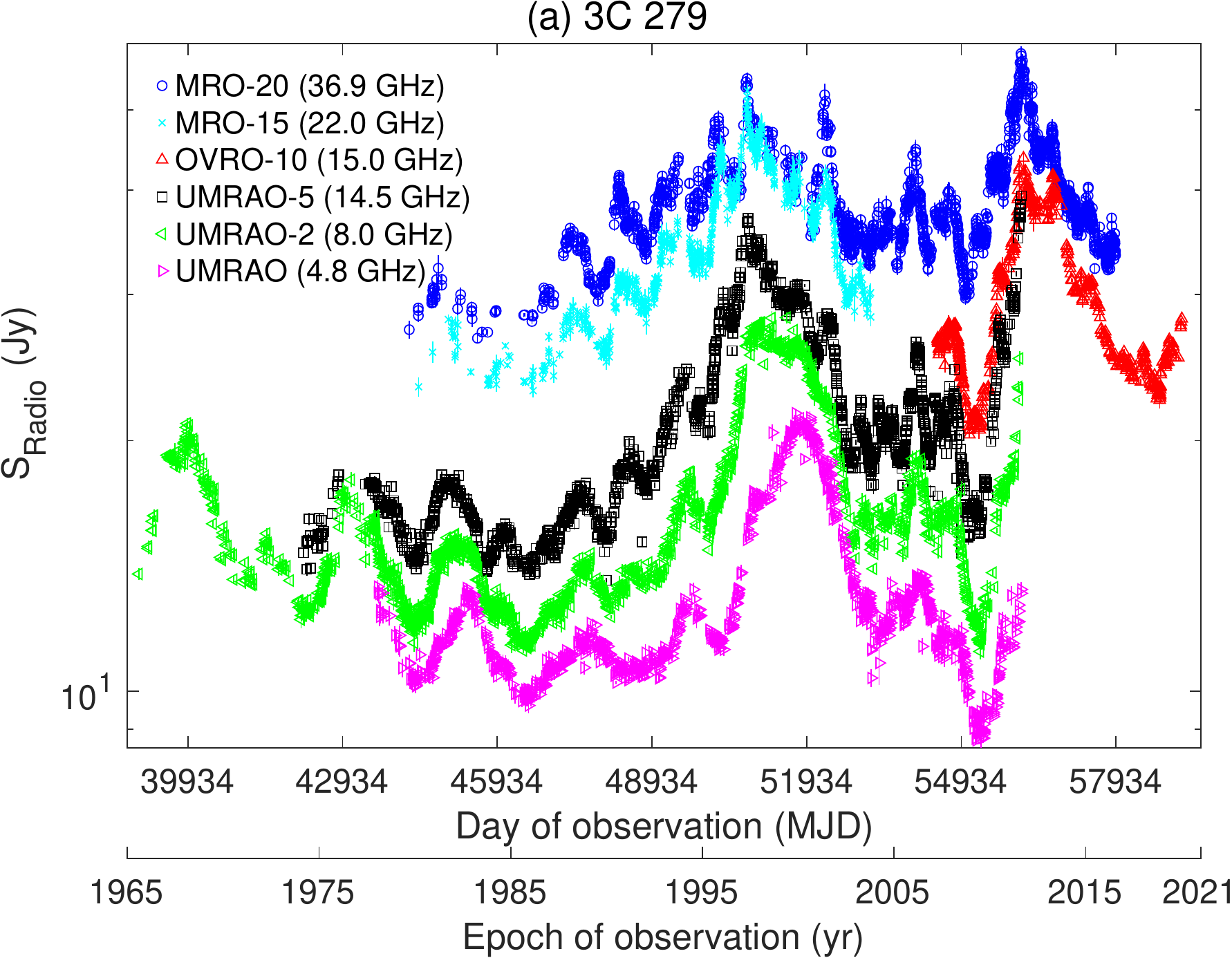}
\includegraphics[width=0.50\textwidth]{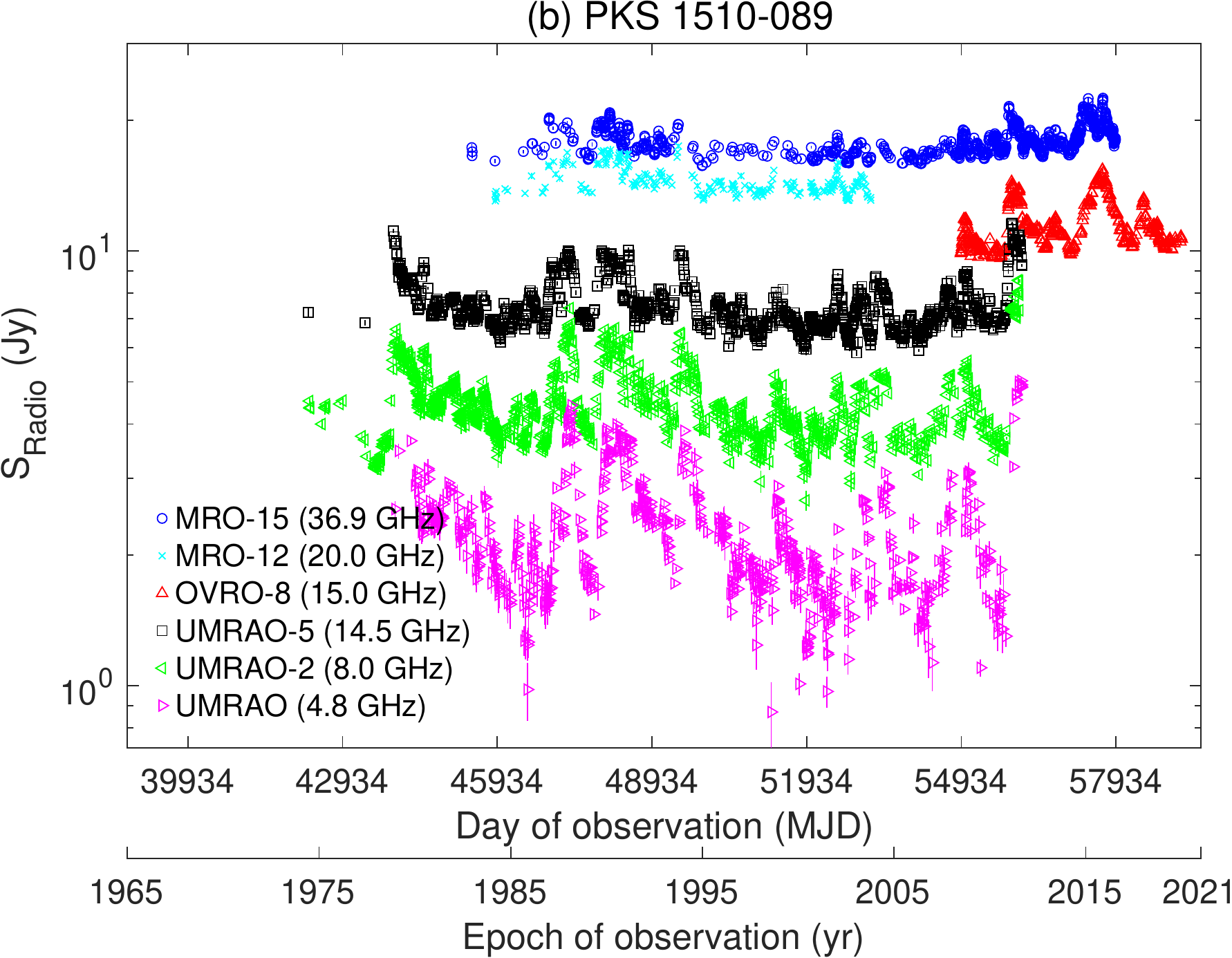}
}
\caption{Multifrequency radio light curves for the entire duration of monitoring for the FSRQs 3C\,279 (a) and PKS\,1510$-$089 (b). The GHz band radio fluxes, with noted offsets, are from the Mets{\"a}hovi (36.9\,GHz and 22.0 GHz), OVRO (15\,GHz), and UMRAO (14.5\,GHz, 8.0\,GHz, and 4.8\,GHz) monitoring programs, respectively. \label{fig:mwrad}}
\end{figure*}

\section{Multiwavelength light curves}\label{sec:data}
\subsection{HE\,\,$\gamma$-rays: {\it Fermi}-LAT }
We analyzed the {\it Fermi}-LAT \citep{Atwood09, Atwood13, Bruel18} data for 3C\,279 and PKS\,1510$-$089 for the 0.1--300\,GeV band, for the period 2008 August 8 -- 2020 September 16 with an integration time of seven days using the {\it Fermi} ScienceTools package Fermitools-conda (version 2.0.18) available from {\it Fermi} Science Support Center\footnote{\url{https://fermi.gsfc.nasa.gov/ssc/data/analysis/}} (FSSC) and the latest instrument response function {\sc p8r3\_source\_v3}. The standard unbinned likelihood analysis was performed\footnote{\url{http://fermi.gsfc.nasa.gov/ssc/data/analysis/scitools/}}. The procedure started with selecting the events for a time interval (1-week) with standard quality cuts (DATA\_QUAL$>$0)\&\&(LAT\_CONFIG == 1) and zenith angle cut of $\theta$ $<$ 90$^\circ$ to avoid contamination from the Earth's limb using the tasks {\sc gtselect} and {\sc gtmktime}. For the target, we selected a 20$^\circ$$\times$20$^\circ$ region of interest (ROI) centered at the source catalog position. Next, the exposure map was created from the livetime cube ({\sc gtltcube}) using the task {\sc gtexpmap} with a total of 35 logarithmically spaced energy intervals bounded within 0.1-300\,GeV. The choice of energy resolution is governed by the FSSC recommendation, as 10 bins per decade was found to be sufficient to accommodate the change in effective area below $\sim$1\,GeV. Finally, in the likelihood analysis ({\sc gtlike}), all the sources from data release 2 (DR2) of the 10-yr {\it Fermi}-LAT catalog (4FGL-DR2; \citealt[][]{Abdollahi20, Ballet20}), within a 30$^\circ$ radius of the target were included, as were the standard templates for diffuse emission from our Galaxy ({\sc gll\_iem\_v07.fits}) and the isotropic $\gamma$-ray background ({\sc iso\_p8r3\_source\_V3\_v1.txt}). In the fitting ({\sc gtlike}), the spectral normalization of the target and sources within 5$^\circ$ of the ROI center were allowed to be free while for the remaining sources the spectral shapes were fixed at their catalog-derived values. The normalization of the Galactic diffuse background was kept frozen while that of the isotropic $\gamma$-ray background component was allowed to vary. Specifically, the $\gamma$-ray spectra of the targets were modeled with a log-parabolic function given by the 4FGL-DR2 catalog. The light curves were generated including 0.1-300 GeV fluxes for which the test statistic (TS; defined as twice the difference between the log-likelihood which is maximized by adjusting all the parameters of the model computed with and without including the source; \citealt[][]{Mattox96}) was found to be $>$10. This gives us preliminary light curves for 3C\,279 and PKS\,1510$-$089. 

Next, we checked for the contamination of the targets' fluxes due to proximity to, or occultation by, the Sun and the Moon as it has been demonstrated that the Sun and the Moon are sources of $\gamma$-ray emission \citep[][]{Abdo11c, Abdo12}. We extracted the equatorial positions (R.A. and Dec.) of the Sun at mission elapsed time (MET) from the spacecraft files and computed the angular distances from the targets in the sky plane. The Sun is known to occult 3C\,279 on October 8 each year \citep[][]{Barbiellini14}. The closest distance between the Sun and PKS\,1510$-$089 is $\approx$8.5$^\circ$, occurring on November 11. Similarly, the positions (R.A. and Dec.) of the Moon were computed at the METs given in the spacecraft files in the Geocentric Celestial Reference System (GCRS) using the {\sc get\_moon} task from the astropy.coordinates library\footnote{\url{https://docs.astropy.org/en/stable/api/astropy.coordinates.get_moon.html}} and the angular distances between the targets and Moon were computed. The Moon occults 3C\,279 while the closest distance between the Moon and PKS\,1510$-$089 is $\sim$3.2$^\circ$. For the blazar 3C\,279, we extracted all the weekly time bins for which the distances between the targets and the Sun/Moon were found to be smaller than 10$^\circ$ and analyzed them separately. As the Moon moves much faster than the Sun in the sky plane ($\sim$13$^\circ$\,day$^{-1}$ and $\sim$1$^\circ$\,day$^{-1}$, respectively) and therefore, is expected to occult 3C\,279 more often than the Sun, we proceeded to account for the contamination due to Sun/Moon to the target fluxes in two steps.
 
First, we extracted all the weekly time bins for which the angular distances between the Moon and 3C\,279 were found to be $<$10$^\circ$ without having the Sun nearby (i.e., the angular distance between the Sun and 3C\,279 was $>$20$^\circ$). Of these, we randomly selected a few time bins during which the Moon was passing the target's ROI (separation $<$10$^\circ$). We generated the lunar templates for these bins following the FSSC guidance\footnote{\url{https://fermi.gsfc.nasa.gov/ssc/data/analysis/scitools/solar_template.html}}. The description of the Solar System Tools which models the quiescent emission from the Sun and Moon is given in \citet[][]{Johannesson13}. An all-sky livetime cube was generated for the lunar emission using the task {\sc gtltcube2} with 35 logarithmically spaced energy intervals with livetime cube and counts map option from the output of the task {\sc gtbin}. Next, we generated the livetime cube for the Moon using the task {\sc gtltcubesun} with body=Moon and instrument angle=1$^\circ$. As a final step, the template for the Moon emission was generated using the task {\sc gtsuntemp} using the lunar profile template\footnote{\url{https://fermi.gsfc.nasa.gov/ssc/data/access/lat/8yr\_catalog/}} and the all-sky livetime cube generated from the task {\sc gtltcube2}. Finally, the Moon template was included in the model file for the likelihood analysis. The Moon's SED was modeled as a power-law for which the normalization was allowed to vary while the index was fixed during the fitting. We compared the target fluxes with and without including the Moon template in the likelihood analysis and found that the change in flux (before Moon - after Moon/before Moon) was comparable to the measurement uncertainties in the weekly time bins with TS$>$10. Thus, we conclude that the Moon does not have a significant impact on the weekly binned fluxes of 3C\,279.
 
Next, we proceeded to compute the contamination due to the Sun's proximity to 3C\,279. We generated the solar template following the same steps needed to generate the lunar template with appropriate changes in input parameters (such as including body=Sun and instrument angle=180$^\circ$ in the task {\sc gtltcubesun}) and the likelihood analysis was performed. The normalization of the power-law SED of the Sun was allowed to vary while the index was fixed during the fitting. We compared the change in flux (before Sun-after Sun/before Sun) and found that for a few time bins, the change in target flux was up to 50\% of the non-corrected value when the solar template was included in the model fitting. In two time bins, 3C\,279 was detected with TS$<$10 (as compared to TS$>$10 without including the solar template). In general though we found that the change in flux is lower ($<$10\%) and therefore comparable to measurement uncertainties when the Sun was $>$2$^\circ$ away from the target. In a few time bins analyzed, the target and the Sun's fluxes were found to be highly correlated (correlation coefficient $>$0.5), so the flux measurements in these time bins were discarded. The final light curve of the blazar 3C\,279 (Figure~\ref{fig:lc3c279}) includes all flux measurements corrected for solar emission and TS$>$10. 

Since the analysis performed for weekly integrated time bins for 3C\,279 indicated negligible change when the Sun is $>$2$^\circ$ away from the target, we did not perform the checks for the Sun's contamination for PKS\,1510$-$089 (closest distance$>$8$^\circ$). We also did not correct for the Moon's contamination as weekly integrated fluxes did not show a significant change in flux after the inclusion of the lunar template. Figure~\ref{fig:lcpks1510} shows the final light curve including measurements with TS$>$10.             

 In order to extend the gamma-ray PSD temporal range down to hourly timescales we needed to examine periods of high-fluxes.  So we generated short-duration (3-weeks) LAT light curves using a 3-hr integration binning interval for time ranges when the daily LAT flux from the sources exceeded 1$\times$10$^{-5}$ ph\,cm$^{-2}$\,s$^{-1}$. We selected three time ranges for the blazar 3C\,279, denoted as `Flare 1' \citep[][]{Cutini15, Ackermann16}, `Flare 2' \citep[][]{vanZyl18}, and `Flare 3' \citep[][]{Ojha18}, respectively. For the blazar PKS\,1510$-$089, we selected one time range, denoted as `Flare 1' \citep[][]{Hungwe11, Orienti13}. In each case, the selected time range was centered around the day of the flare. The analysis was performed in a similar fashion as that of 1-week integration time bins. The final light curves were checked for the goodness of {\sc gtlike} results by plotting Flux/$\Delta$Flux vs. $\sqrt{TS}$. For a good fit, one expects a linear trend between these quantities.  We removed one flux point from the `Flare 3' and `Flare 1' light curves of 3C\,279 and PKS\,1510$-$089, respectively, which did not follow a linear trend. For 3C\,279, we checked for the contamination due to the Moon alone as the selected time ranges did not coincide with the passage of the Sun. We note that the change in flux due to inclusion of lunar templates was $<$30\%, i.e., within the measurement uncertainties provided by these measurements. Therefore, we conclude that the occultation of the Moon does not have a significant effect on 3-hr integration bin flux measurements. For PKS\,1510$-$089, the selected time range did not coincide with the passage of the Moon in the ROI ($>$10$^\circ$). Therefore, we do not correct for the Moon's effect in the 3-hr integration bin light curves for either of the targets. Figure~\ref{fig:fermi3hr} gives the 3-hr integration bin light curves for `Flare 1', `Flare 2', and `Flare 3' for the blazar 3C\,279 (panels a, b, and c) and `Flare 1' for the blazar PKS\,1510$-$089 (panel d).

\subsection{X-rays: {\it RXTE-}PCA and {\it Swift}-XRT} 
\label{sec:rxte}

We downloaded the 1998 and 1334 Stdprods (Standard2 data product) for 3C\,279 and PKS\,1510$-$089, respectively, from the {\it RXTE quick look analysis} from the HEASARC website\footnote{\url{https://heasarc.gsfc.nasa.gov/db-perl/W3Browse/w3browse.pl}}. These source and background Standard2 spectra were reduced by a pipeline using a standard set of FTOOLS and PCA backgrounds with a general script and employed one single set of filtering criteria for all observations\footnote{\url{https://heasarc.gsfc.nasa.gov/docs/xte/recipes/stdprod_guide.html}}. The Standard2 mode of PCA has 129 energy channels, covering the entire range 2--25\,keV of the PCA detector. The spectra were fitted within the energy range 3.0--20.0\,keV with a power-law of the form $F(E) = F_0 E^{-\Gamma_X}$, where $F_0$ is the normalization and $\Gamma_X$ is the photon index, found via {\it xspec} with  $\chi^2$ statistics. In the spectral analyses, the Galactic absorption corresponding to neutral hydrogen column density in the direction of the source was fixed \citep{Kalberla05}. The unabsorbed photon fluxes were obtained within the 3--20\,keV energy range by fitting {\it wabs*cpflux*powerlaw}. We averaged the fluxes using a binning interval of 1\,day. Figures~\ref{fig:lc3c279}(b) and ~\ref{fig:lcpks1510}(b) show these X-ray photon fluxes for 3C\,279 and PKS\,1510$-$089, respectively. A few earlier studies report part of the {\it RXTE}-PCA datasets examined by us for 3C\,279 \citep[][]{Hartman01, Larionov06, Bottcher07, Chatterjee08, Abdo10b} and PKS\,1510$-$089 \citep[][]{Marscher10, Castignani17}.

The archival data from the {\it Swift}-XRT \citep{Gehrels04}, consisting of 415 and 262 pointed observations made between 2005 September 17 and 2016 July 17, for the blazars 3C\,279 and PKS\,1510$-$089, respectively, were also analyzed by us. We used the latest version of the calibration database ({\sc CALDB}) and version 6.19 of the {\sc heasoft} package\footnote{\url{http://heasarc.gsfc.nasa.gov/docs/software/lheasoft/}}. For each dataset, we used the level 2 cleaned event files of the `photon counting' (PC) and `window timing' (WT) data acquisition modes generated using the standard {\sc xrtpipeline} tool. The details of the XRT data analysis are given in \citet{Goyal18} and \citet{Goyal20} and we briefly outline the procedure here. For the PC mode data, the source and background spectra were generated using a circular aperture with appropriate region sizes and grade filtering using the {\sc xselect} tool. An aperture radius of $47^{\prime\prime}$ around the source position and four source-free regions of $118^{\prime\prime}$ radius were used to estimate the source and background spectra, respectively. For the WT mode data, the source region of $47^{\prime\prime}$ circular aperture and an annular region with inner and outer radii of $187^{\prime\prime}$ and  $281^{\prime\prime}$, respectively, for background estimation were selected to estimate the source and background spectra. Following the XRT tutorial\footnote{\url{http://www.swift.ac.uk/analysis/xrt/backscal.php}}, appropriate corrections were made in the `BACKSCAL' keyword in the extracted source and background spectra to account for the different sizes of the source and background estimation regions. We also checked for the recommended pile-up limit for the PC ($\sim$0.5 count\,s$^{-1}$) and WT ($\sim$100 count\,s$^{-1}$) modes. The  ancillary response matrix was generated using the task {\sc xrtmkarf} for the exposure map generated by {\sc xrtexpomap}. All the source spectra were then binned over 20 points and corrected for the background using the task {\sc grppha}. For each exposure, the spectral analysis between 0.3 and 10 keV was performed in an identical manner to that of the {\it RXTE-}PCA data. The unabsorbed 0.3--10\,keV photon fluxes were averaged using a 1-day binning interval and these light curves are shown in Figures~\ref{fig:lc3c279}(c) and Figure~\ref{fig:lcpks1510}(c) for the targets 3C\,279 and PKS\,1510$-$089, respectively. In addition, we show the fitted ${\Gamma_X}$ for the {\it RXTE-}PCA and {\it Swift-}XRT datasets analyzed in Figures~\ref{fig:lc3c279}(c) and ~\ref{fig:lcpks1510}(c) for the targets 3C\,279 and PKS\,1510$-$089, respectively. A few earlier studies report part of the {\it Swift}-XRT datasets analyzed by us for 3C\,279 \citep[][]{Abdo10b, Hayashida15, Larionov20} and PKS\,1510$-$089 \citep[][]{Abdo10d, DAmmando11, Nalewajko12}.   

\begin{deluxetable*}{ccccccccccc}

\tablenum{2}
\tablecaption{Summary of the observation and PSD analysis\label{tab:psd}}
\tablewidth{0pt}
\tabletypesize{\footnotesize}
\tablehead{
	\colhead{Light Curve (Energy Range)} & \colhead{Epoch of Monitoring} &  $N_{obs}$ &  \colhead{$T_{\rm total}$} &  \colhead{$T_{\rm mean}$} &  \colhead{$T_{\rm in}$} & \colhead{Var$_{sp}$}  & \colhead{$\rm \log_{10} (P_{\rm stat})$} & \colhead{$\log_{10}(\nu_k) $ Range} & \colhead{{$\beta \pm $err}} & \colhead{$p_\beta^b$}\\
	\colhead{} &  \colhead{(Start-End)}  & \colhead{} &   \colhead{(yr)} &    \colhead{(day)} & \colhead{(day)} &  &  \colhead{($\frac{\mathrm rms}{\mathrm mean})^2$day} & \colhead{(day$^{-1}$)} &   \colhead{} 
}
\decimalcolnumbers
\startdata
\multicolumn{11}{c}{\bf 3C\,279}\\
	{\it Fermi-}LAT (0.1--300\,GeV)&  2008 Aug 8 - 2020 Sep 16       &601   &  12.1     &  7.3   &    1.0  &   0.25 &  $-$0.84   &  $-$3.64 to $-$1.40   &1.0$\pm$0.7 & 0.595 \\
	{\it RXTE-}PCA (3--20\,keV)   &  1996 Jan 22 - 2011 Dec 30      &1751  &  15.9     &  3.3   &    0.5  &    3.16 &  $-$0.96   &  $-$3.76 to $-$1.11   &1.2$\pm$0.3 & 0.496 \\
	{\it Swift-}XRT (0.3--10\,keV) &  2006 Jan 13 - 2017 Jun 28      &230   &  11.5     &  18.2  &    0.5  &   2.57 &  $-$0.43   &  $-$3.62 to $-$1.78   &1.4$\pm$0.7 & 0.999 \\
	Optical ({\it B})              &  2008 May 17 - 2017 Jul 19      &624   &  9.2      &  5.4   &    0.5  &   3.05 &  $-$1.41   &  $-$3.52 to $-$1.28   &1.8$\pm$1.3 & 0.567 \\
	Optical ({\it V})              &  2005 Apr 7 - 2017 Jul 19       &775   &  12.3     &  5.8   &    0.5  &   3.26 &  $-$0.21   &  $-$3.65 to $-$1.20   &2.1$\pm$1.4 & 0.651 \\
	Optical ({\it R})              &  1995 Nov 25 - 2019 Jun 4       &1829  &  23.5     &  4.7   &    0.5  &   3.17 &  $-$1.24   &  $-$3.93 to $-$1.07   &1.4$\pm$0.5 & 0.635 \\
	Optical ({\it I})              &  2005 Apr 2 - 2011 Jan 30       &183   &  6.2      &  12.5  &    0.5  &   2.82 &  $-$0.15   &  $-$3.35 to $-$1.53   &1.6$\pm$1.8 & 0.924 \\
	Infrared ({\it J})             &  2005 Apr 9 - 2017 Jul 18       &727   &  12.3     &  6.2   &    0.5  &   3.34 &  $-$1.37   &  $-$3.65 to $-$1.20   &2.1$\pm$1.5 & 0.832 \\
	Infrared ({\it H})             &  2005 Apr 9 - 2011 Jun 23       &197   &  6.2      &  11.5  &    0.5  &   3.37 &  $-$0.83   &  $-$3.35 to $-$1.53   &1.8$\pm$1.8 & 0.994 \\
	Infrared ({\it K})             &  2005 Apr 9 - 2011 Jun 23       &170   &  6.2      &  13.3  &    0.5  &   3.58 &  $-$0.31   &  $-$3.35 to $-$1.53   &2.9$\pm$1.6 & 0.992 \\
	MRO (36.9\,GHz)                 &  1979 Dec 21 - 2017 Jun 19      &1740  &  37.5     &  7.9   &    0.5  &  2.97 &  $-$1.68   &  $-$4.13 to $-$1.48   &1.7$\pm$0.3 & 0.968 \\
	MRO (22.0\,GHz)                 &  1980 Jun 19 - 2004 Jun 21      &751   &  24.0     &  11.7  &    0.5  &  2.55 &  $-$1.38   &  $-$3.94 to $-$1.49   &2.9$\pm$1.1 & 0.225 \\
	OVRO (15\,GHz)                  &  2008 Jan 8 - 2020 Dec 26       &535   &  13.0     &  8.9   &    0.5  &  1.35 &  $-$2.11   &  $-$3.67 to $-$1.43   &2.5$\pm$1.1 & 0.813 \\
	UMRAO (14.5\,GHz)               &  1974 May 8 - 2012 Jun 23       &1567  &  38.1     &  8.9   &    0.5  &  2.00 &  $-$2.39   &  $-$4.14 to $-$1.49   &2.6$\pm$1.0 & 0.977 \\
	UMRAO (8.0\,GHz)                &  1965 Aug 13 - 2012 May 16      &1653  &  46.8     &  10.3  &    0.5  &  1.59 &  $-$2.40   &  $-$4.23 to $-$1.52   &2.4$\pm$0.9 & 0.871 \\
	UMRAO (4.8\,GHz)               &  1978 Apr 15 - 2012 Jun 14      &1101  &  34.2     &  11.3  &    0.5  &   1.64 &  $-$2.46   &  $-$4.09 to $-$1.64   &2.1$\pm$0.6 & 0.309 \\
	Flare 1$^a$ (0.1--300\,GeV)    &  2015 Jun 5 - 2015 Jun 26       &80    &  0.057    & 0.24   & 0.042   &   1.84 &  $-$1.45   &  $-$1.29 to $+$0.11   &1.4$\pm$3.9 & 0.500 \\
	Flare 2$^a$ (0.1--300\,GeV)    &  2018 Jan 7 - 2018 Jan 28       &146   &  0.057    & 0.14   & 0.042   &   0.38 &  $-$1.88   &  $-$1.31 to $+$0.30   &2.0$\pm$2.8 & 0.808 \\
	Flare 3$^a$ (0.1--300\,GeV)    &  2018 Apr 13 - 2018 May 4       &157   &  0.057    & 0.14   & 0.042   &   0.27 &  $-$2.35   &  $-$1.31 to $+$0.30   &1.5$\pm$2.9 & 0.680 \\
\hline                                                                                                     
\multicolumn{11}{c}{ \bf PKS\,1510$-$089}\\                                                                
	{\it Fermi-}LAT (0.1--300\,GeV)&  2008 Aug 8 - 2020 Sep 16       &603   &  12.1     &  7.3   &    1.0   &   0.30 &  $-$0.76   &  $-$3.64 to $-$1.40   &0.9$\pm$0.6 & 0.300 \\
	{\it RXTE-}PCA (3--20\,keV)   &  1996 Dec 13 - 2011 Dec 30      &1252  &  15.1     &  4.4   &    0.5   &    1.47 &  $-$0.79   &  $-$3.73 to $-$1.08   &1.4$\pm$0.3 & 0.482 \\
	{\it Swift-}XRT (0.3--10\,keV) &  2006 Aug 4 - 2017 Jun 26       &181   &  10.9     &  22.0  &    0.5   &   3.04 &  $-$0.25   &  $-$3.59 to $-$1.76   &1.5$\pm$1.0 & 0.994 \\
	Optical ({\it B})             &  2008 May 17 - 2017 Jun 14      &542   &  9.1      &  6.1   &    0.5   &    3.29 &  $-$2.26   &  $-$3.52 to $-$1.28   &1.5$\pm$0.5 & 0.185 \\
	Optical ({\it V})             &  2006 Jan 10 - 2017 Jun 14      &611   &  11.4     &  6.8   &    0.5   &    3.48 &  $-$1.27   &  $-$3.62 to $-$1.38   &1.2$\pm$0.3 & 0.003 \\
	Optical ({\it R})             &  2005 Mar 25 - 2018 Jul 10      &1207  &  13.3     &  4.0   &    0.5   &    3.12 &  $-$2.15   &  $-$3.68 to $-$1.03   &0.7$\pm$0.4 & 0.532 \\
	Optical ({\it I})             &  2006 Jan 11 - 2012 May 28      &225   &  6.4      &  10.3  &    0.5   &    3.75 &  $-$1.09   &  $-$3.36 to $-$1.12   &0.7$\pm$0.8 & 0.245 \\
	Infrared ({\it J})            &  2006 Feb 17 - 2017 Jun 11      &675   &  11.3     &  6.1   &    0.5   &    3.86 &  $-$1.75   &  $-$3.61 to $-$1.37   &0.2$\pm$1.9 & 0.317 \\
	Infrared ({\it H})            &  2006 Feb 17 - 2012 Jun 1       &304   &  6.3      &  7.5   &    0.5   &    4.06 &  $-$1.44   &  $-$3.36 to $-$1.32   &1.3$\pm$4.4 & 0.003 \\
	Infrared ({\it K})            &  2006 Feb 22 - 2012 Jun 1       &223   &  6.3      &  10.3  &    0.5   &    3.59 &  $-$1.27   &  $-$3.36 to $-$1.52   &0.2$\pm$2.1 & 0.011 \\
	MRO (36.9\,GHz)                &  1983 Apr 16 - 2017 Jun 19      &885   &  34.2     &  14.1  &    0.5   &   2.51 &  $-$0.97   &  $-$4.09 to $-$1.65   &1.8$\pm$0.7 & 0.559\\
	MRO (22.0\,GHz)                &  1984 Jul 12 - 2004 Jun 21      &239   &  19.9     &  30.5  &    0.5   &   1.44 &  $-$0.52   &  $-$3.86 to $-$2.03   &1.4$\pm$0.5 & 0.386 \\
	OVRO (15.0\,GHz)               &  2009 Apr 2 - 2020 Dec 27       &427   &  11.7     &  10.0  &    1.0   &   1.18 &  $-$2.17   &  $-$3.63 to $-$1.39   &2.1$\pm$1.0 & 0.505 \\
	UMRAO (14.5\,GHz)              &  1974 Aug 14 - 2012 Jun 27      &1345  &  37.9     &  10.3  &    0.5   &   3.41 &  $-$1.70   &  $-$4.14 to $-$1.48   &1.6$\pm$0.4 & 0.145 \\
	UMRAO (8.0\,GHz)               &  1974 Sep 4 - 2012 May 18       &1147  &  37.7     &  12.0  &    0.5   &   1.68 &  $-$1.55   &  $-$4.13 to $-$1.48   &1.6$\pm$0.3 & 0.164 \\
	UMRAO (4.8\,GHz)               &  1979 Mar 23 - 2012 Jun 16      &672   &  33.2     &  18.1  &    0.5   &   1.56 &  $-$1.19   &  $-$4.08 to $-$1.64   &1.7$\pm$0.3 & 0.199 \\
	Flare 1$^a$ (0.1--300\,GeV)   &  2011 Oct 14 - 2011 Nov 4       &74    & 0.057     &  0.27  &  0.042   &    0.95 &  $-$1.34   &  $-$1.31 to $+$0.09   &1.5$\pm$3.6 & 0.978 \\
\enddata
\tablecomments{
(1) the instrument/band (energy range/frequency band used for observations in parentheses); $^a$ refers to {\it Fermi-}LAT light curves made with 3-hr integration bins (Figure~\ref{fig:fermi3hr}),
(2) the beginning and ending dates of the light curve used,
(3) the number of data points in the light curve,
(4) the total duration of the observed light curve,
(5) the mean sampling interval of the light curve (= Col.\ 4/Col.\ 3),
(6) the interpolation interval used in the PSD analysis,
(7) square root of the variance of the sampling pattern for the data normalized by $T_{\rm mean}$,
(8) the noise level in the PSD due to the measurement uncertainty,
(9) the temporal frequency range covered by the binned logarithmic power spectrum,
(10) the best-fit power-law slope of the PSD derived using the PSRESP method along with the corresponding errors representing the 98\% confidence limit (see Section~\ref{sec:psresp})
(11) the corresponding $p_\beta$: $^b$ a power-law model is considered a poor fit for $p_\beta\,<\,0.1$ as the corresponding rejection confidence for the model is $\geq$90\% (Section~\ref{sec:psresp}).
}
\end{deluxetable*}

\subsection{Optical and Infrared: SMARTS, REM, Tuorla, VLBA-BU-BLAZAR, and SKYNET monitoring} 
\label{sec:swift}

The blazars 3C\,279 and PKS\,1510$-$089 have been the targets of regular monitoring on daily timescales at multiple optical ($B$, $V$, $R$, and $I$) and Infrared ($J$, $H$, and $K$) bands and here we combine many of these observations. We begin by using publicly available $B$, $V$, $R$, $J$, and $K$-band light curves from the Small and Moderate Aperture Research Telescope System (SMARTS; \citealt{Bonning12}) program coordinated by Yale University\footnote{\url{http://www.astro.yale.edu/smarts/glast/home.php}} for the period $\sim$2008 until 2017. We also take $V$, $R$, $I$, $J$, $H$, and $K$-band light curves, covering the period 2005--2012, from the Rapid Eye Mounting telescope (REM; \citealt{Zerbi01}) which were published in \citet[][S16 hereafter]{Sandrinelli16}. Many $R$-band optical photometric data were also obtained from the Tuorla Blazar monitoring program for the period 2004--2016\footnote{\url{http://users.utu.fi/kani/1m/}} \citep{Takalo08} and those published in \citet[][C08 hereafter]{Chatterjee08} for the period 1996-2004. Additionally, for the blazar PKS\,1510$-$089 $R$-band photometric data  were obtained from the VLBA-BU-BLAZAR blazar monitoring program for the period 2005--2018\footnote{\url{https://www.bu.edu/blazars/VLBAproject.html}} \citep{Jorstad16}. For the blazar 3C\,279, additional $R$-band photometric data for the period 2018--2019 were obtained from the Skynet Robotic Telescope Network program (SKYNET)\footnote{\url{https://skynet.unc.edu}}. 

There is a great deal  of overlap between the SMARTS and the REM programs' data in the $V$, $R$, $J$, and $K$-bands and they complement each other well at $V$, $R$, and $J$-bands, with the pairs of magnitudes within 0.1 mag on the same nights, consistent with the calibration uncertainties provided by these programs. Moreover, the $R$-band photometric data obtained from the VLBA-BU and the Tuorla monitoring programs are comparable to other datasets within the measurement uncertainties ($<$0.1 mag) on the same nights. As a majority of these programs have mean sampling intervals $>$1-day, we have combined the datasets at $V$, $R$, and $J$-bands and average magnitudes were computed with 1-day binning intervals. This is done to avoid obtaining poor sampling on intranight timescales as well as to obtain homogeneous sampling on daily timescales. On the other hand, the $K$-band magnitudes from the SMARTS program has a magnitude offset $\geq$0.1 mag from that provided by S16 on the same nights. On visual inspection, the $K$-band SMARTS light curve appears more variable than the S16 light curve on similar timescales. Therefore, in this study, we use only the $K$-band data provided by S16. On a few occasions, fluxes at $B$, $I$, $H$, and $K$-bands were also obtained multiple times on a given night and we have  averaged those observations over a 1-day binning interval. 

Next, the optical and infrared magnitudes, $m$, were converted to flux densities using $m_{zero}$\,$\times 10^{-0.4\times\,m}$, where $m_{zero}$ (=4063\,Jy, 3636\,Jy, 3064\,Jy, 2635\,Jy, 1590\,Jy, 1020\,Jy, and 600\,Jy for the $B$, $V$, $R$, $I$, $J$, $H$, and $K$-bands, respectively) refer to zero-point magnitude fluxes of the photometric system \citep{Glass99}. The errors in fluxes were derived using standard error propagation \citep{Bevington03}. The resulting optical and infrared-band light curves for the targets 3C\,279 and PKS\,1510$-$089 are presented in Figures~\ref{fig:lc3c279}(d, e) and ~\ref{fig:lcpks1510}(d, e), respectively.

\subsection{Radio: MRO, UMRAO, and OVRO} 
\label{sec:radio}
GHz-band radio light curves of the targets were gathered from the Aalto University Mets{\"a}hovi Radio Observatory (MRO) at 36.9 and 22.0\,GHz \citep{Teraesranta98}, the UMRAO at 14.5, 8.0, and 4.8\,GHz \citep{Aller99}, and the OVRO monitoring programs at 15\,GHz \citep{Richards11}. All the light curves gathered were binned using 1-day binning intervals as the typical sampling intervals of these programs are much larger ($>$3\,days to 2 weeks). The radio light curves of the blazars 3C\,279 and PKS\,1510$-$089 for periods (1995--2017) that overlap monitoring  with other frequencies are presented in Figures~\ref{fig:lc3c279}(f) and ~\ref{fig:lcpks1510}(f), respectively. The radio light curves for the full durations of the monitoring  are shown in panels (a) and (b) of Figure~\ref{fig:mwrad}. 
\\

\section{PSD Analysis}\label{sec:analysis}  

The derivation of PSDs using the discrete Fourier transformation (DFT) and the estimation of spectral shapes using the `power spectral response (PSRESP)' method are given in detail in \citet[][]{Goyal20, Goyal21}. Here we provide the main features.

\subsection{Derivation of PSDs: discrete Fourier transform}\label{sec:dft}
The rms-normalized periodogram is given as the squared modulus of its DFT for the evenly sampled light curve $f(t_i)$, observed at discrete times $t_i$ and consisting of $N$ data points,
\begin{multline} 
P(\nu_k)  =   \frac{2 \, T}{\mu^2 \, N^2} \, \Bigg\{ \Bigg[ \sum_{i=1}^{N} f(t_i) \, \cos(2\pi\nu_k t_i)  \Bigg]^2  + \\ \Bigg[ \sum_{i=1}^{N} f(t_i) \, \sin(2\pi\nu_k t_i)  \Bigg]^2 \, \Bigg \} ,\label{eq:psd}
\end{multline}
 where $\mu$ is the mean of the light curve and is subtracted from the flux values, $f(t_i)$. The DFT is computed for evenly spaced frequencies ranging from the total duration of the light curve, $T$, down to the Nyquist sampling frequency ($\nu_{\rm Nyq}$). Specifically, in our analysis, the frequencies corresponding to $\nu_{k} = k/T$ with $k=1, ..., N/2$, $\nu_{\rm Nyq}= N/2T$, and $T = t_{N}-t_1$ are considered. The constant noise floor level from measurement uncertainties is computed using  \citep[][]{Isobe15}:
\begin{equation}
P_{\rm stat} = \frac{2 \, T}{\mu^2 \, N} \, \sigma_{\rm stat}^2 \, ,\label{eq:poi_psd}
\end{equation}
where, $\sigma_{stat}^2= \sum_{j=1}^{j=N} \Delta f(t_j)^2 / N$ is the mean variance of the measurement uncertainties for the flux values, $\Delta f\!(t_j)$, in the observed light curve at times $t_j$, with $N$ denoting the number of data points in the observed light curve. As expected, the noise floor level due to measurement uncertainties has a white noise character as the variability power contributed by measurement uncertainties alone is equal at all variability frequencies. 

The unevenly sampled time series is rendered evenly sampled through linear interpolation between two consecutive data points with the interpolation interval, $T_{in}$, roughly 5--10 times smaller than the mean observed sampling interval; otherwise, the `spectral window function' corresponding to the sampling times gives a non-zero response in the {\it Fourier} domain, resulting in spurious powers in the periodograms \citep[see, Appendix A of][]{Goyal20}. In this work, the PSDs have been generated for the actual duration of the light curves down to the mean (observed) sampling intervals, $T_{\rm mean}$. We note that interpolation is a smoothing operation, so it alters the high-frequency PSD of a data set, and this would be very obvious if we reported the spectrum up to 1/(2*$T_{in}$). The effects are reduced by cutting it off at 1/(2*$T_{\rm mean}$), but it does not necessarily solve the problem. It depends on how much the actual sampling varies, or equivalently what the typical distance between the actual data point and the interpolate point used in the spectrum at  1/(2*$T_{\rm mean}$) is. Since the light curves are relatively densely sampled, the PSDs have adequate coverage of many different frequencies. We expect a little extra noise because we do not have as many data pairs per frequency as would be the case for even sampling. These ramifications are not problematic in our case as we have plenty of coverage of the frequencies, and gridding the data is simply expedient. We demonstrate this qualitatively by calculating the typical separation of each sampling point from the even (i.e., $T_{\rm mean}$) grid and computing the variance of it. For our light curves, we computed the square root of the variance of the sampling pattern normalized by $T_{\rm mean}$, Var$_{sp}$= variance$^{1/2}$/$T_{\rm mean}$ (Column 7 of Table~\ref{tab:psd}). This value should be zero for precisely evenly sampled data. As can be seen from Table~\ref{tab:psd}, these values are close to zero for weekly-binned and $<$2 for 3-hr binned {\it Fermi-}LAT light curves for both targets, as they are better sampled. For the optical and infrared light curves, the values range between $\sim$3.0 and $\sim$4.0 because of the relatively large (4--6 months) systematic gaps in monitoring due to the target's proximity to the Sun. For the X-ray and GHz band radio light curves, these values range between $\sim$1.4 and $\sim$3.0 for the majority of the gathered light curves, except for the 14.5\,GHz UMRAO light curve for the target PKS\,1510$-$089 for which it is 3.4. As only relatively small (weeks) systematic gaps are expected in the {\it RXTE}, {\it Swift}, and the radio data because of pointing constraints due to the target's proximity to the Sun, these light curves can be considered better sampled than the optical ones. Since for the great majority of our data, these Var$_{sp}$ values are relatively small numbers, our cutting the spectrum down to 1/(2*$T_{\rm mean}$) does not alter its shape. 

The sampling distribution and the corresponding spectral window functions of the observed light curves are shown in Appendices~\ref{appfig:swf3c} and ~\ref{appfig:swfpks} for the blazars 3C\,279 and PKS\,1510$-$089, respectively. In our analysis, to avoid the distortions introduced due to the finite duration of the light curve, known as `red-noise leakage',
 the PSDs are generated using the `Hanning' window function \citep[e.g.,][]{Press92, Max-Moerbeck14a}. The distortions in the PSD due to discrete sampling of the time series, i.e., aliasing, are not serious for red-noise dominated time series and usually contribute a small amount of power at all frequencies \citep[][]{Uttley02}. 
The `raw' periodograms, obtained using  Eq.~(\ref{eq:psd}), provide a noisy estimate of the spectral power \citep[as it consists of independently distributed $\chi^2$ variables with two degrees of freedom (DOF);][]{TK95, Papadakis93, Vaughan03}; therefore, we average a number of them to obtain a reliable estimate. A binning factor of 1.6 is used, with the representative frequency taken as the geometric mean of each bin in our analysis (except for R-band and 8.0\,GHz PSDs of 3C\,279 for which a binning factor of 1.4 is used). Next, the `true' power spectrum in the log-log space is obtained by offsetting the observed power spectrum by the expectation value of a $\chi^2$ distribution with 2 DOF, which is equal to $-$0.25068 \citep[][]{Vaughan05}.

\begin{deluxetable*}{ccc}
\tablenum{3}
\tablecaption{Mean PSD slope estimates \label{tab:psdmean}}
\tablewidth{0pt}
\tabletypesize{\small}
\tablehead{
\colhead{Frequency band} & \colhead{$\beta_{\rm mean}\pm$err} & \colhead{Frequencies used for averaging$^a$}
}
\decimalcolnumbers
\startdata
\multicolumn{3}{c}{\bf 3C\,279}\\
radio-synchrotron    &  2.34$\pm$0.16      & MRO (36.9 and 22.0\,GHz), UMRAO (14.5, 8.0, and 4.8 GHz)  \\
optical-synchrotron  &  1.95$\pm$0.22      & $B$, $V$, $R$, $I$, $J$, $H$, and $K$   \\
 IC   &  1.20$\pm$0.15      & {\it RXTE}, {\it Swift}, and {\it Fermi-}LAT \\
\hline                                                                              
\multicolumn{3}{c}{ \bf PKS\,1510$-$089}\\                                         
radio-synchrotron    &     1.60$\pm$0.08   & MRO (36.9 and 22.0\,GHz), UMRAO (14.5, 8.0, and 4.8 GHz)  \\ 
optical-synchrotron  &  0.77$\pm$0.23      & $B$, $R$, $I$, and $J$ \\
IC  &  1.30$\pm$0.16      & {\it RXTE}, {\it Swift}, and {\it Fermi-}LAT \\
\enddata
\tablecomments{(1) name of the frequency band, (2) mean PSD slope and the corresponding 1$\sigma$ error, (3) frequencies used for averaging to obtain the mean PSD slope; $^a$Only PSD slopes with $p_\beta>$0.1 are used to estimate the $\beta_{\rm mean}$ in the representative frequency band (see Table~\ref{tab:psd}).
}
\end{deluxetable*}

\subsection{Estimation of the spectral shape: the PSRESP method}\label{sec:psresp}
As the PSDs are subjected to various distortions due to Fourier transformation, the PSRESP method is an efficient approach to obtain a best-fit spectral model \citep[e.g.,][]{Uttley02, Chatterjee08, Max-Moerbeck14a, Meyer19}. We have previously employed this method in \citet[][]{Goyal20} and \citet[][]{Goyal21}; therefore, we only briefly discuss its main features here. In this method, an (input) PSD model is tested against the observed PSD and the estimation of best-fit model parameters and their uncertainties is performed by varying the model parameters. A large number of mock light curves are generated with a known underlying power-spectral shape via Monte Carlo (MC) simulations. These simulated light curves are rebinned to mimic the observed data and subjected to the Fourier transformation in exactly the same manner as that of the observed data. The averaging of a large number of such PSDs gives the mean of the distorted model (input) power spectrum while the standard deviation around the mean periodogram gives the errors on the modeled (input) power spectrum. The goodness of fit of the model is estimated by computing two functions, similar to $\chi^2$, defined as

\begin{equation}
\chi^2_{\rm obs} = \sum_{\nu_{k}=\nu_{min}}^{\nu_{k}=\nu_{max}} \frac{[\overline{ \log_{10}P_{\rm sim}}(\nu_k)-\log_{10}P_{\rm obs}(\nu_k)]^2}{\Delta \overline{\log_{10}P_{\rm sim}}(\nu_k)^2},
\label{chiobs}
\end{equation}
and 
\begin{equation}
\chi^2_{\rm dist, i} = \sum_{\nu_{k}=\nu_{min}}^{\nu_{k}=\nu_{max}} \frac{[\overline{ \log_{10}P_{\rm sim}}(\nu_k)-\log_{10}P_{\rm sim,i}(\nu_k)]^2}{\Delta \overline{\log_{10}P_{\rm sim}}(\nu_k)^2}.
\label{chidist}
\end{equation}
Here, $\log_{10}P_{\rm obs}$ and $\log_{10}P_{\rm {sim, i}}$ are the observed and the simulated log-binned periodograms, respectively; $\overline{ \log_{10}P_{\rm sim}}$ and $\Delta \overline{\log_{10}P_{\rm sim}}$ are the mean and the standard deviation obtained by averaging large number of PSDs; $k$ represents the number of frequencies in the Log-binned power spectrum (ranging from $\nu_{k,min}$ to $\nu_{k,max}$), while $i$ runs over the number of simulated light curves for a given $\beta$. The $\chi^2_{\rm obs}$ determines the minimum $\chi^2$ for the model compared to the data and the $\chi^2_{\rm dist}$ values determine the goodness of the fit corresponding to the $\chi^2_{\rm obs}$, as neither $\chi^2_{\rm dist,i}$ nor $\chi^2_{\rm obs}$ follow a standard $\chi^2$ distribution. For this, the $\chi^2_{\rm dist}$ are sorted in an ascending order. The probability, or $p_{\beta}$, that a given model can be rejected is then given by the percentile of $\chi^2_{\rm dist}$ distribution above which $\chi^2_{\rm dist}$ is found to be greater than $\chi^2_{\rm obs}$ for a given $\beta$ \citep[][see also \citet{Chatterjee08} who call $p_\beta$ as a success fraction]{Uttley02}. A large value of $p_{\beta}$ or the success fraction represents a good-fit in the sense that large number of random realizations represents the shape and slope of the intrinsic (input) PSD. Subsequently, the PSRESP method allows for an estimation of best-fit model as well as the goodness of fit. It essentially uses the MC approach toward a frequentist estimation of the quality of the model compared to the data and so is a good approach when the fit-statistics are not well-understood \citep{Press92}.

We use the method of \citet{Emmanoulopoulos13} for light curve simulations as it preserves both the probability density function (PDF) of the flux distribution as well as the underlying power spectral shape and not just the power spectral shape as is the case with the method of \citet[][]{TK95}. In our light curve simulations, we assume a single power-law spectral shape with a given $\beta$ and a log-normal flux distribution. For this purpose, the mean and the standard deviation were computed by fitting a Gaussian function to the logarithmically transformed flux distributions and the measurement errors were incorporated by adding a Gaussian random variable with mean 0 and standard deviation equal to the mean error of the measurement uncertainties on the observed flux values. In such a manner, 1,000 light curves are simulated in the $\beta$ range 0.1 to 3.0, with a step of 0.1 (and in a range 0.1 to 4.0 for the 3-hr integration bin LAT light curves). The best-fit PSD slope for the observed PSD is given by the one with the highest $p_{\beta}$ value. The error on the best-fit PSD slope is obtained by fitting a Gaussian function to the $p_{\beta}$ distribution curve \citep[e.g.,][]{Chatterjee08, Bhatta16b, Bhattacharyya20} and is given as 2.354$\sigma$ where $\sigma$ is the standard deviation of the Gaussian function. Therefore, the errors provide a 98\% confidence interval on the best-fit PSD slope \citep[][]{Goyal21}. 

We note that a common approach to derive PSDs of an unevenly sampled time series is the Lomb-Scargle periodogram \citep[LSP;][]{Lomb76, Scargle82}, which is also frequently used in the AGN variability literature \citep[][]{Rani09, Rani10, Bhatta16a, Sandrinelli16, Castignani17, Sandrinelli17, Gaur18, Gupta19, Zywucka20, Bhatta20, Tarnopolski20, Li21}. However, with numerical simulations, we demonstrate in Appendix~\ref{app:C} that the LSP method does not usually reproduce the shape of the underlying power spectrum correctly up to the highest (Nyquist) frequency, whereas the DFT method does. Therefore, the DFT method (with linear interpolation) is a more appropriate choice as far as the characterization of the PSD shape is concerned when the available sampling is not very close to uniform.

\begin{figure*}
\hbox{
\includegraphics[width=0.25\textwidth]{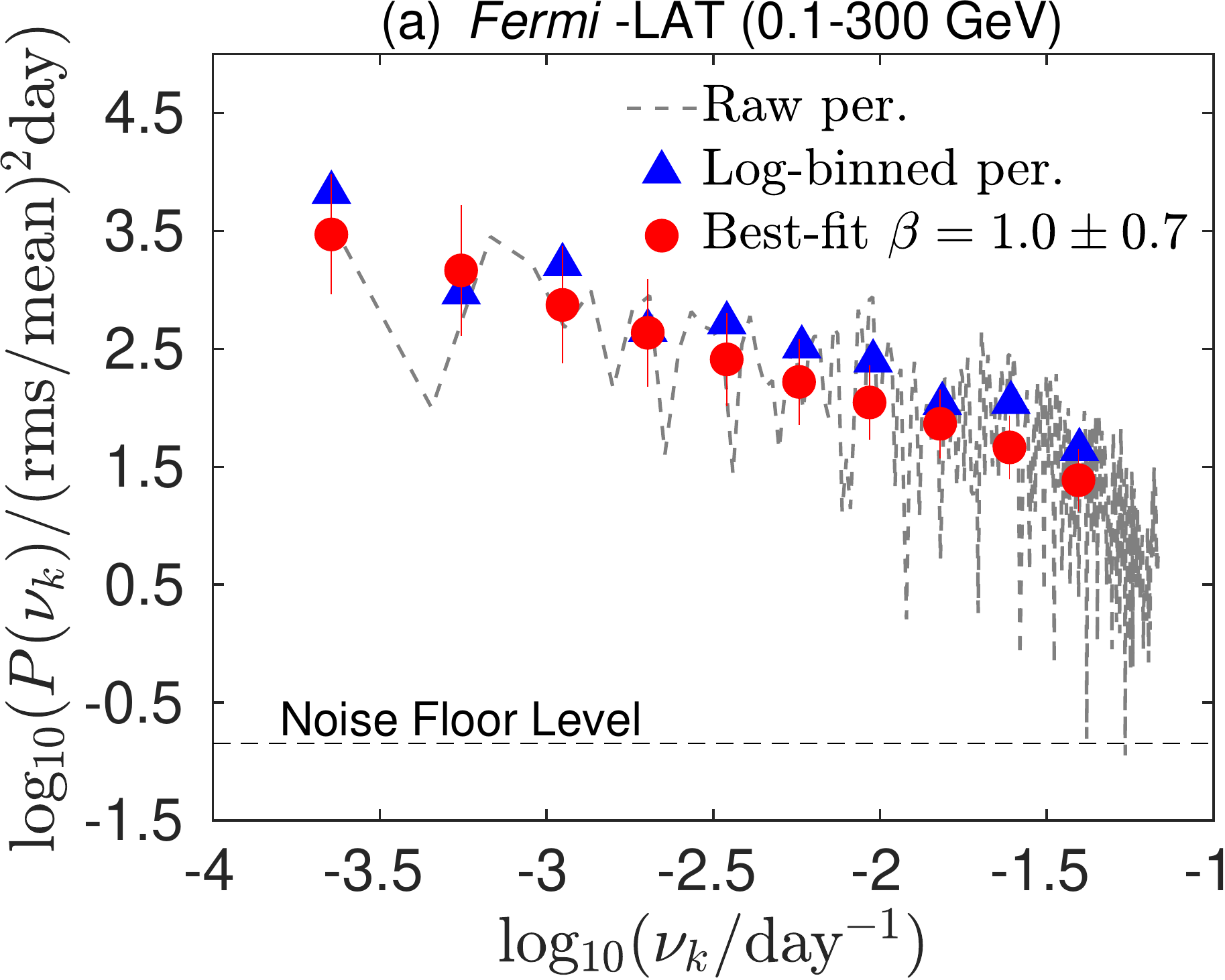}
\includegraphics[width=0.25\textwidth]{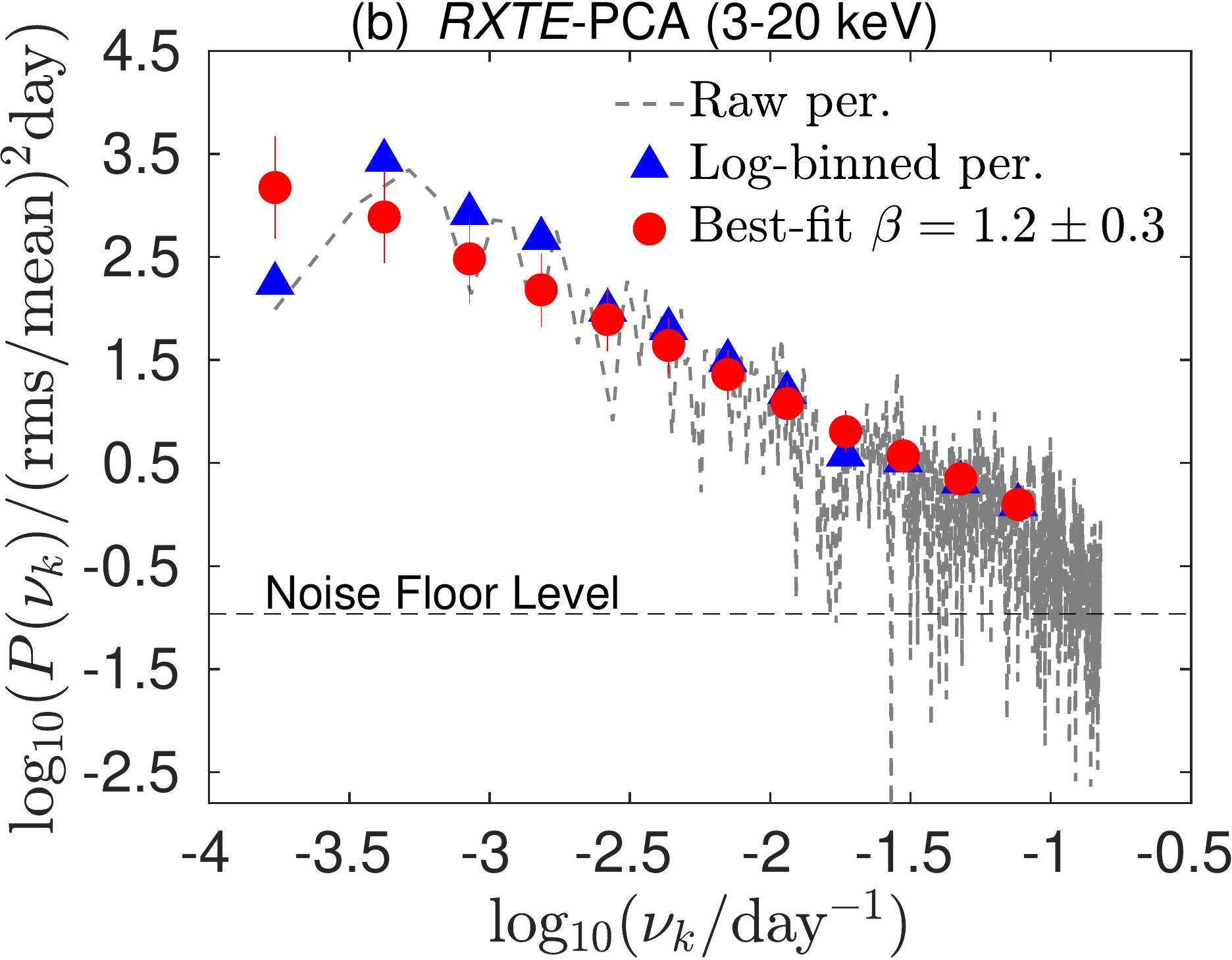}
\includegraphics[width=0.25\textwidth]{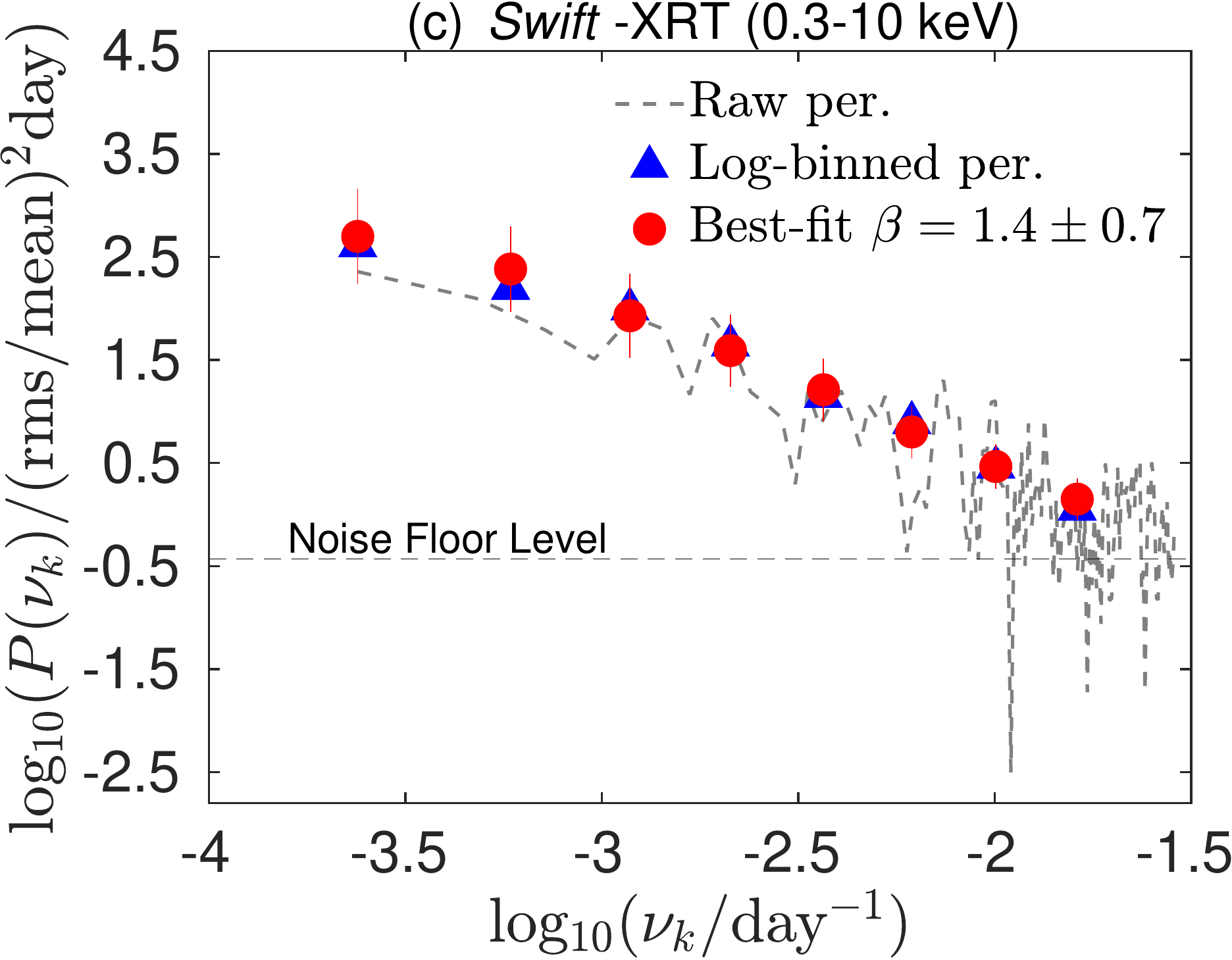}
\includegraphics[width=0.25\textwidth]{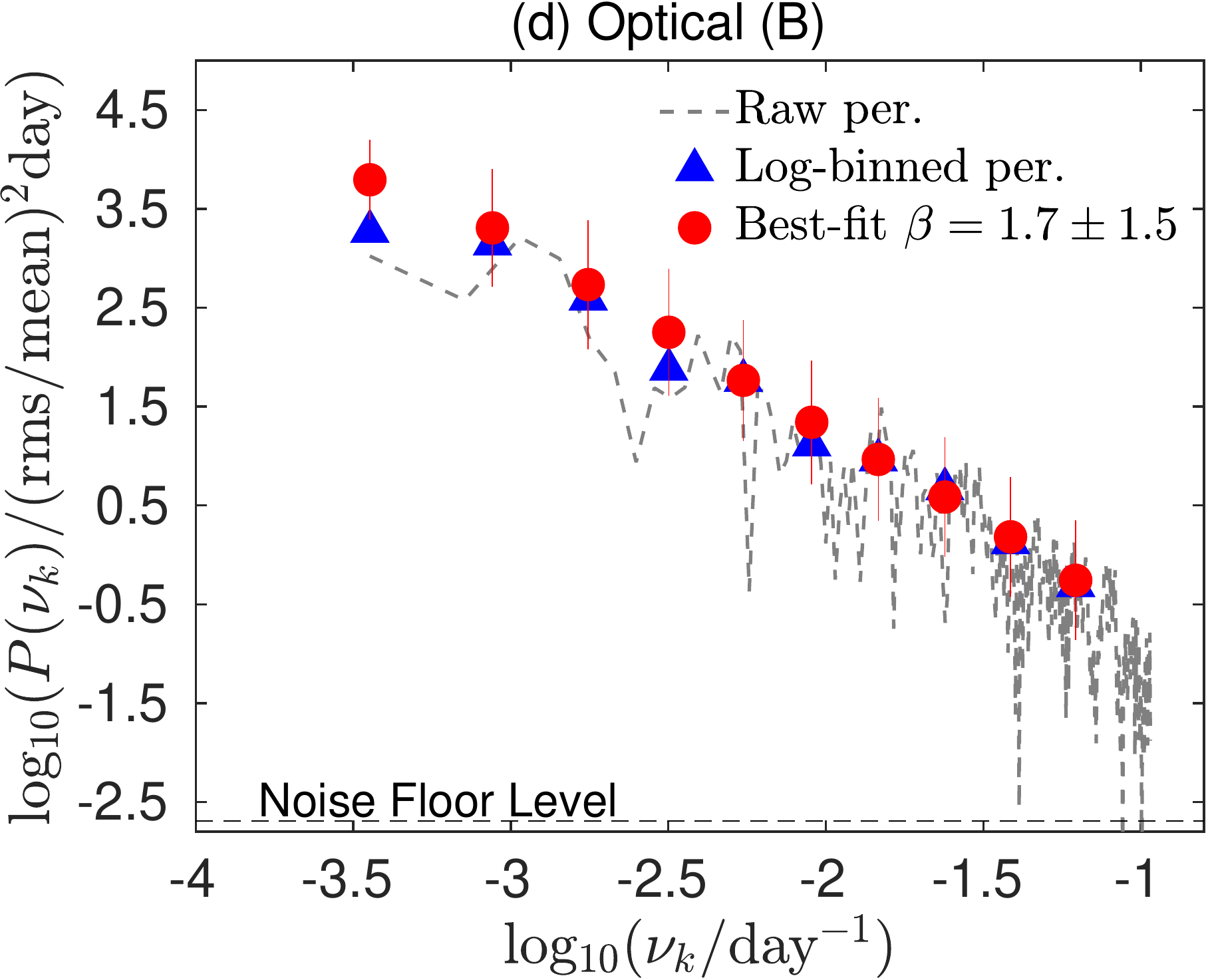}
}
\hbox{
\includegraphics[width=0.25\textwidth]{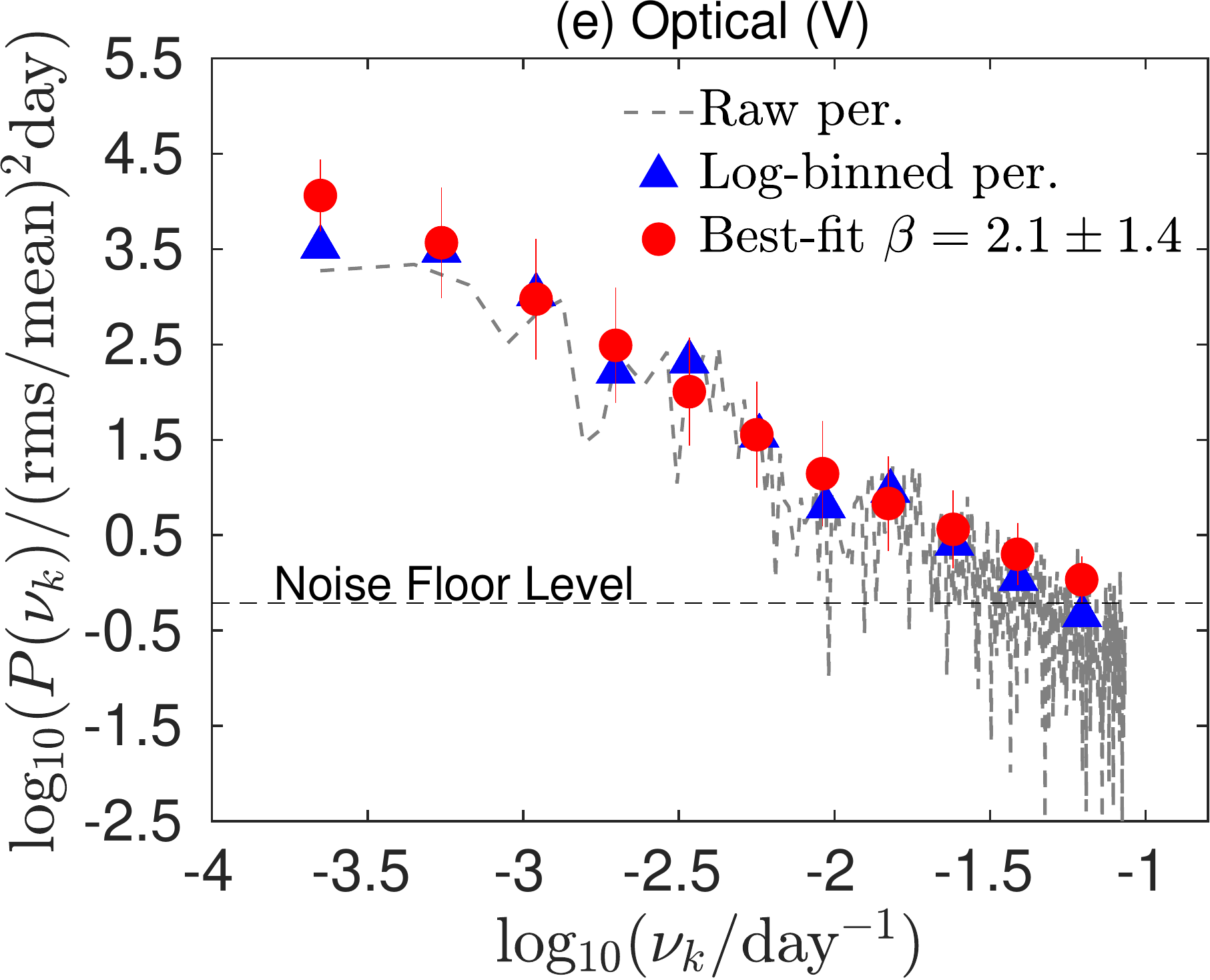}
\includegraphics[width=0.25\textwidth]{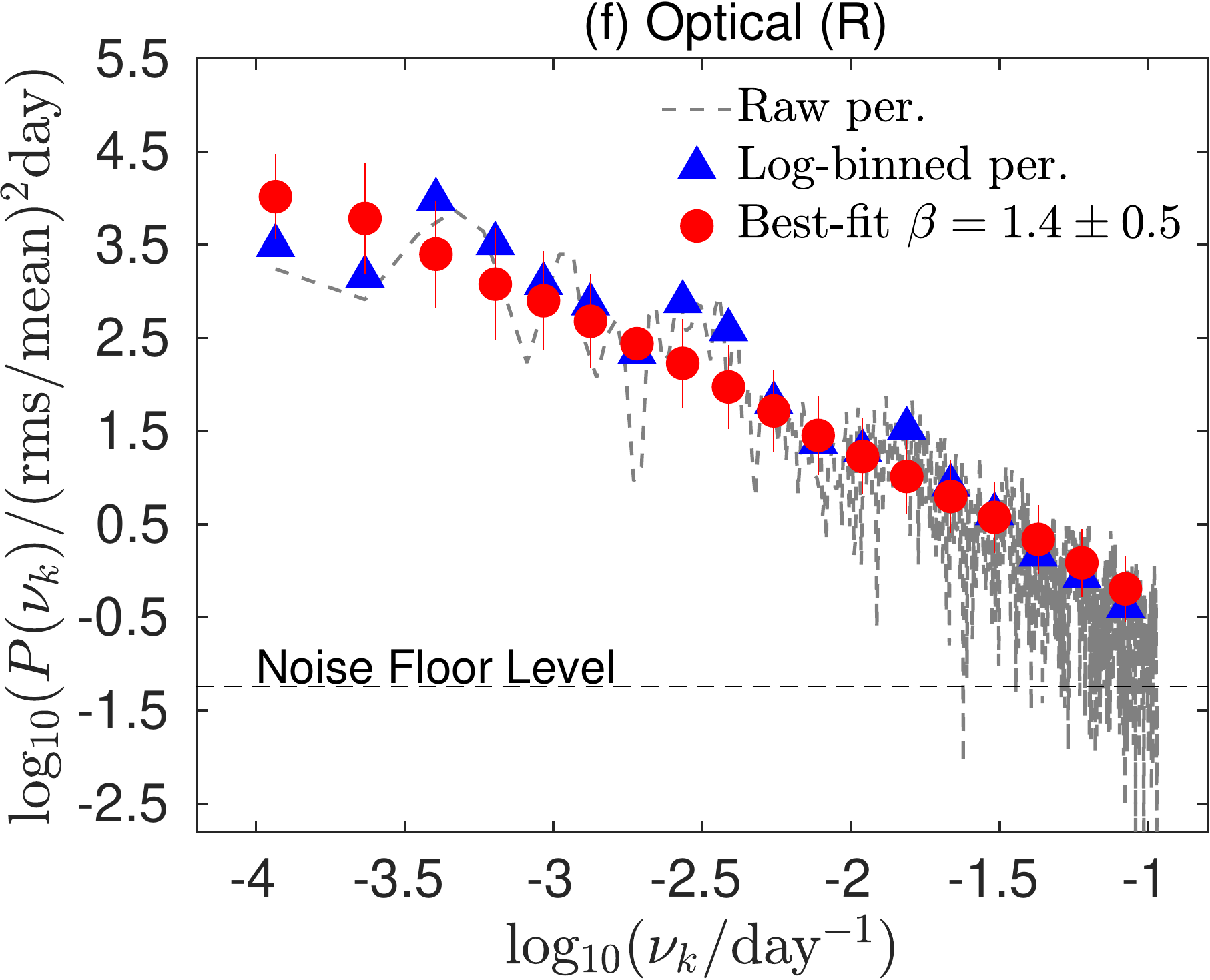}
\includegraphics[width=0.25\textwidth]{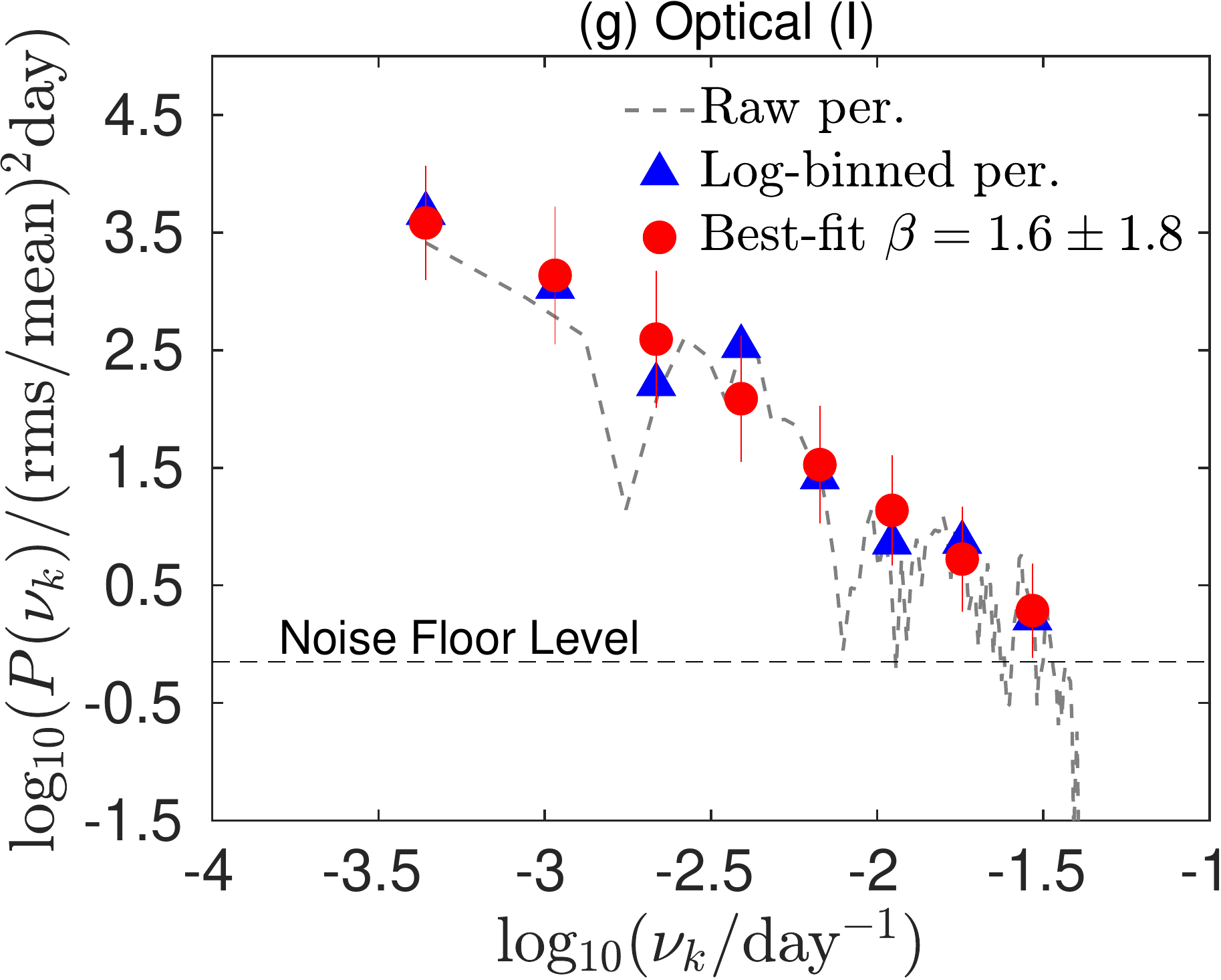}
\includegraphics[width=0.25\textwidth]{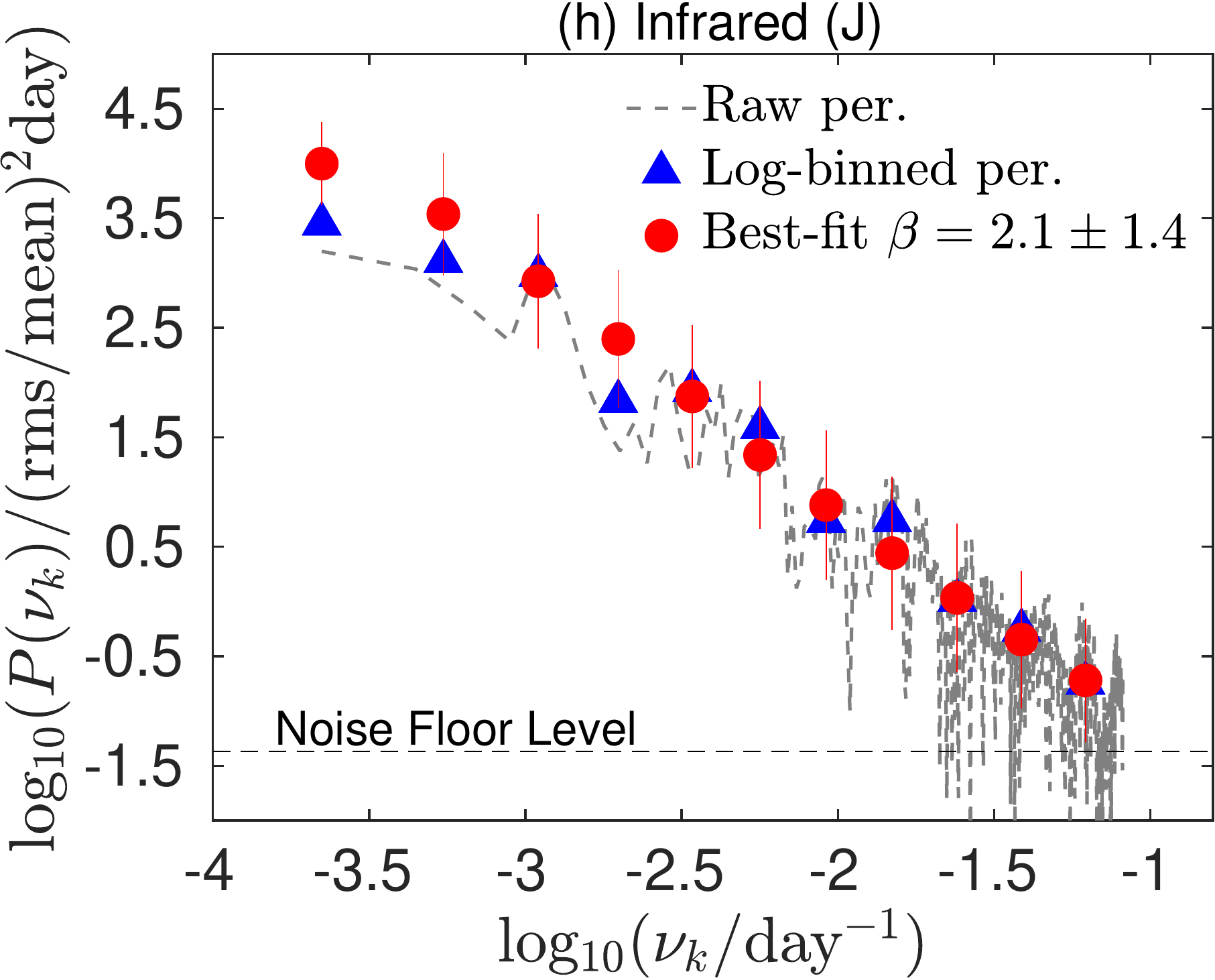}
}
\hbox{
\includegraphics[width=0.25\textwidth]{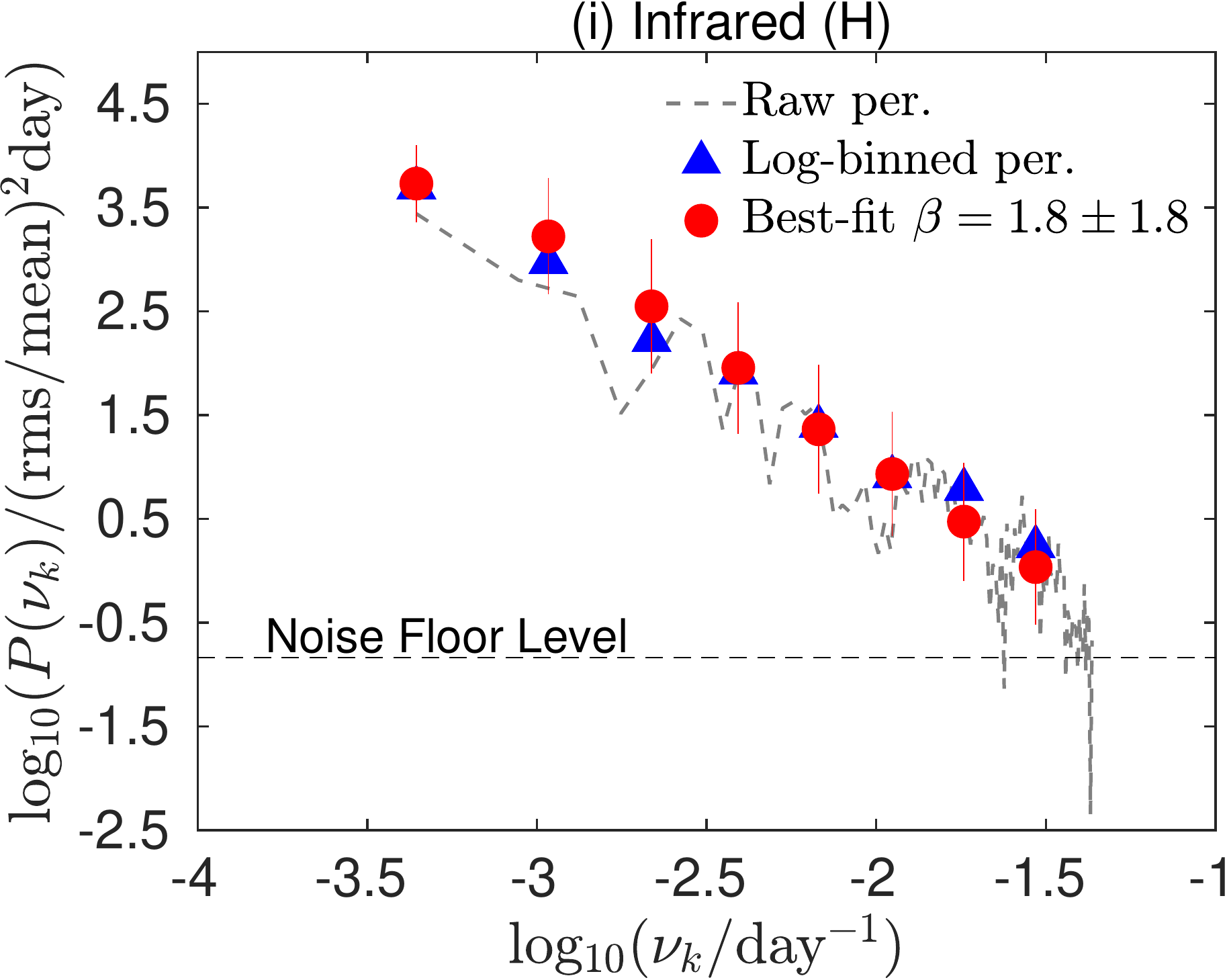}
\includegraphics[width=0.25\textwidth]{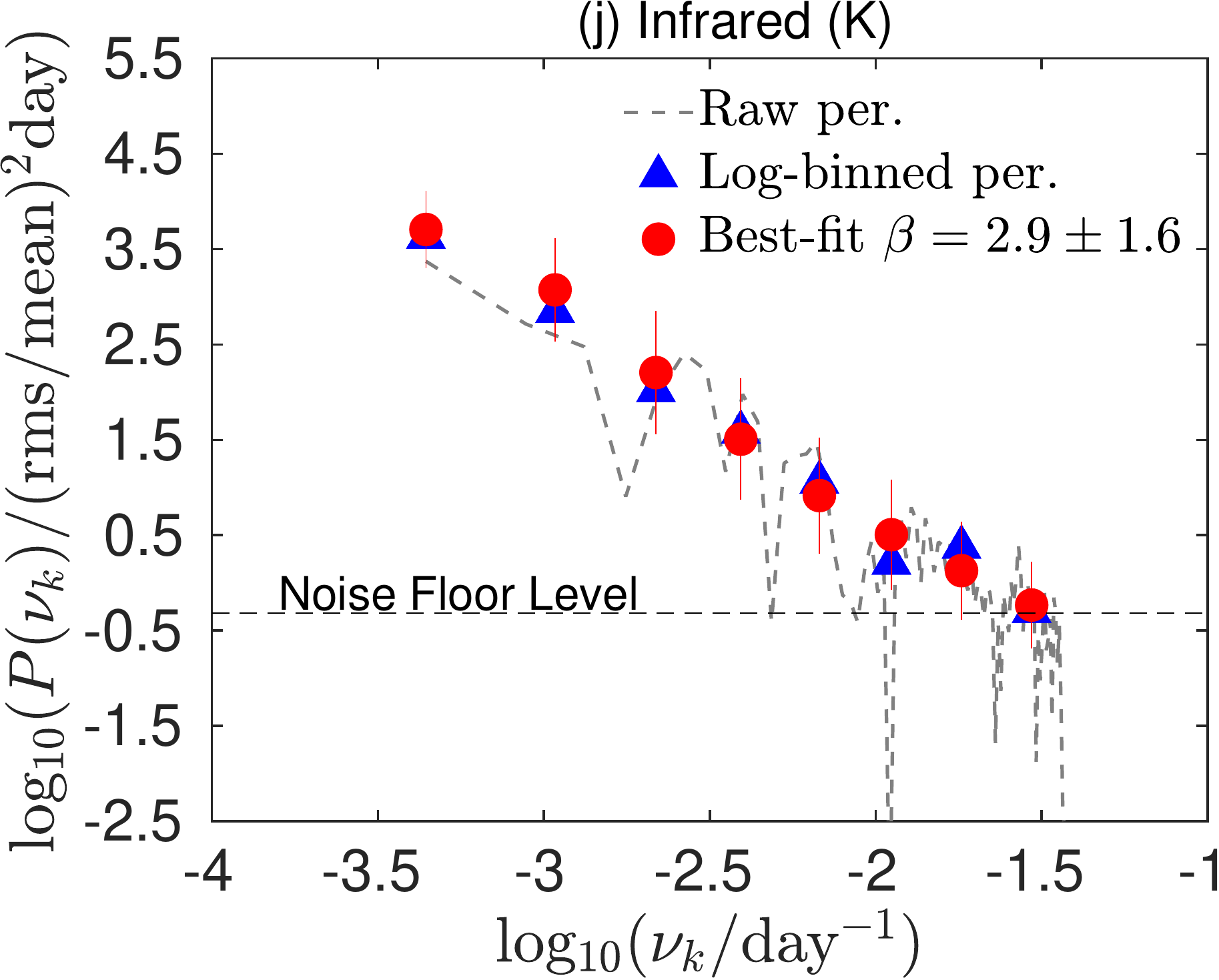}
\includegraphics[width=0.25\textwidth]{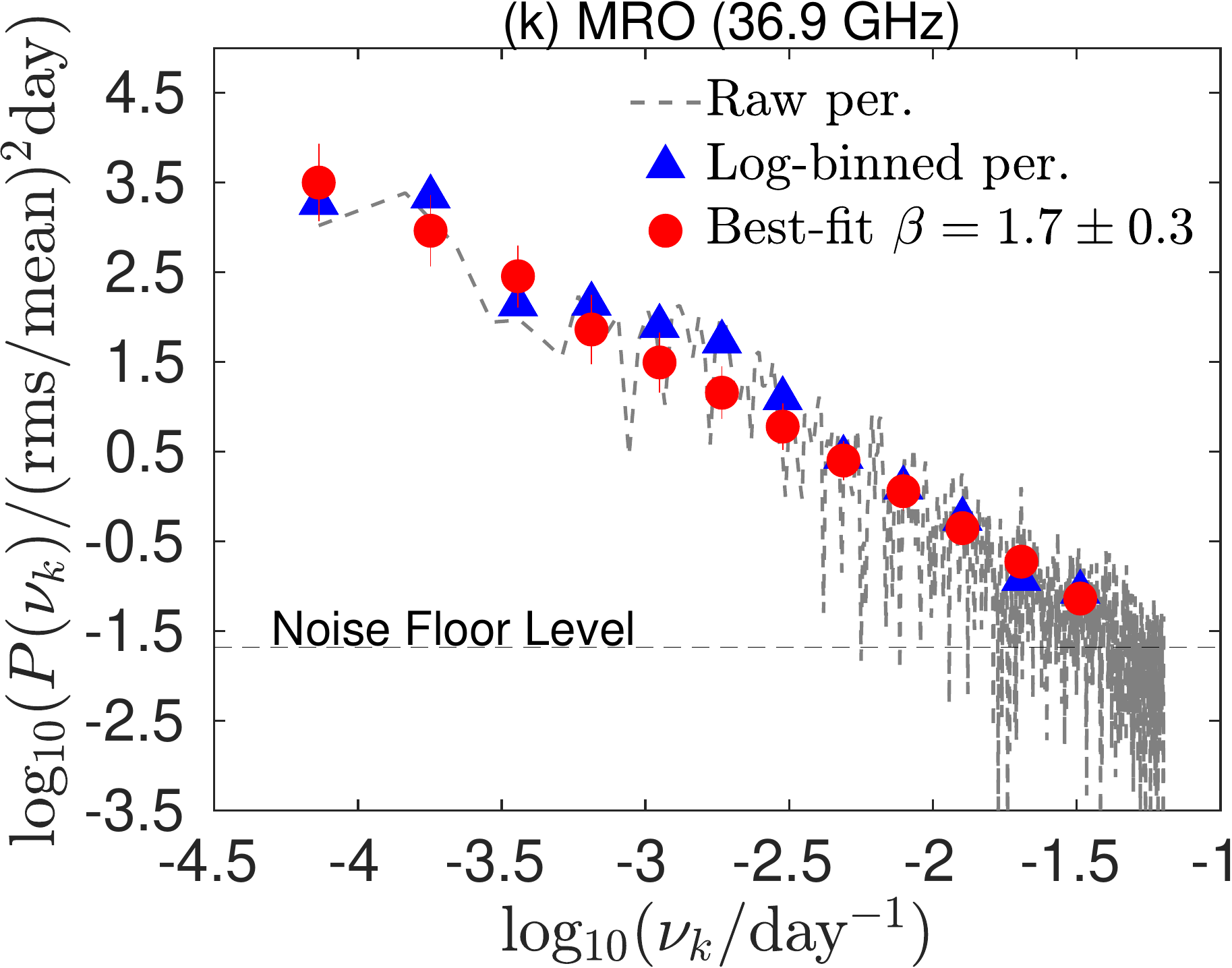}
\includegraphics[width=0.25\textwidth]{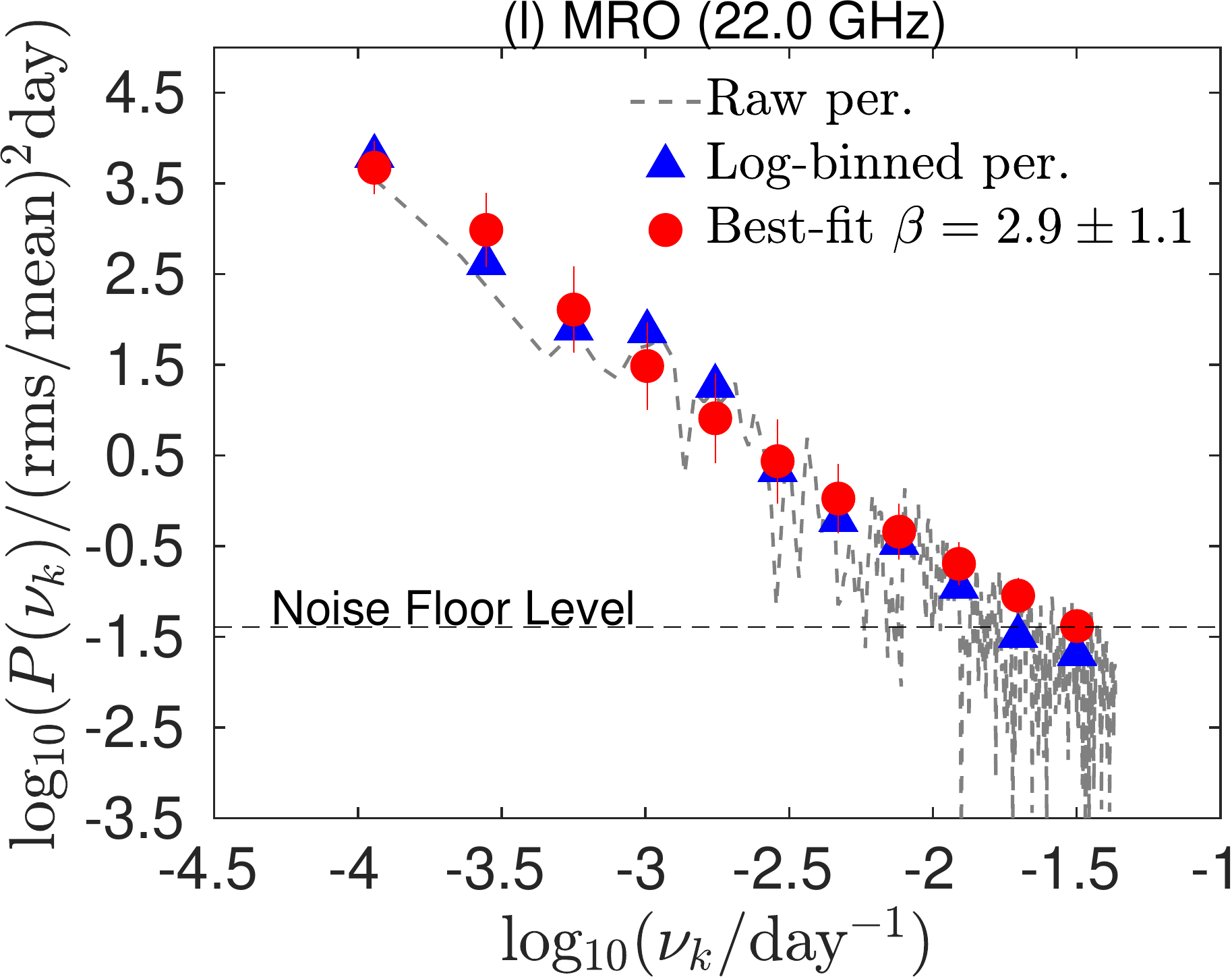}
}
\hbox{
\includegraphics[width=0.25\textwidth]{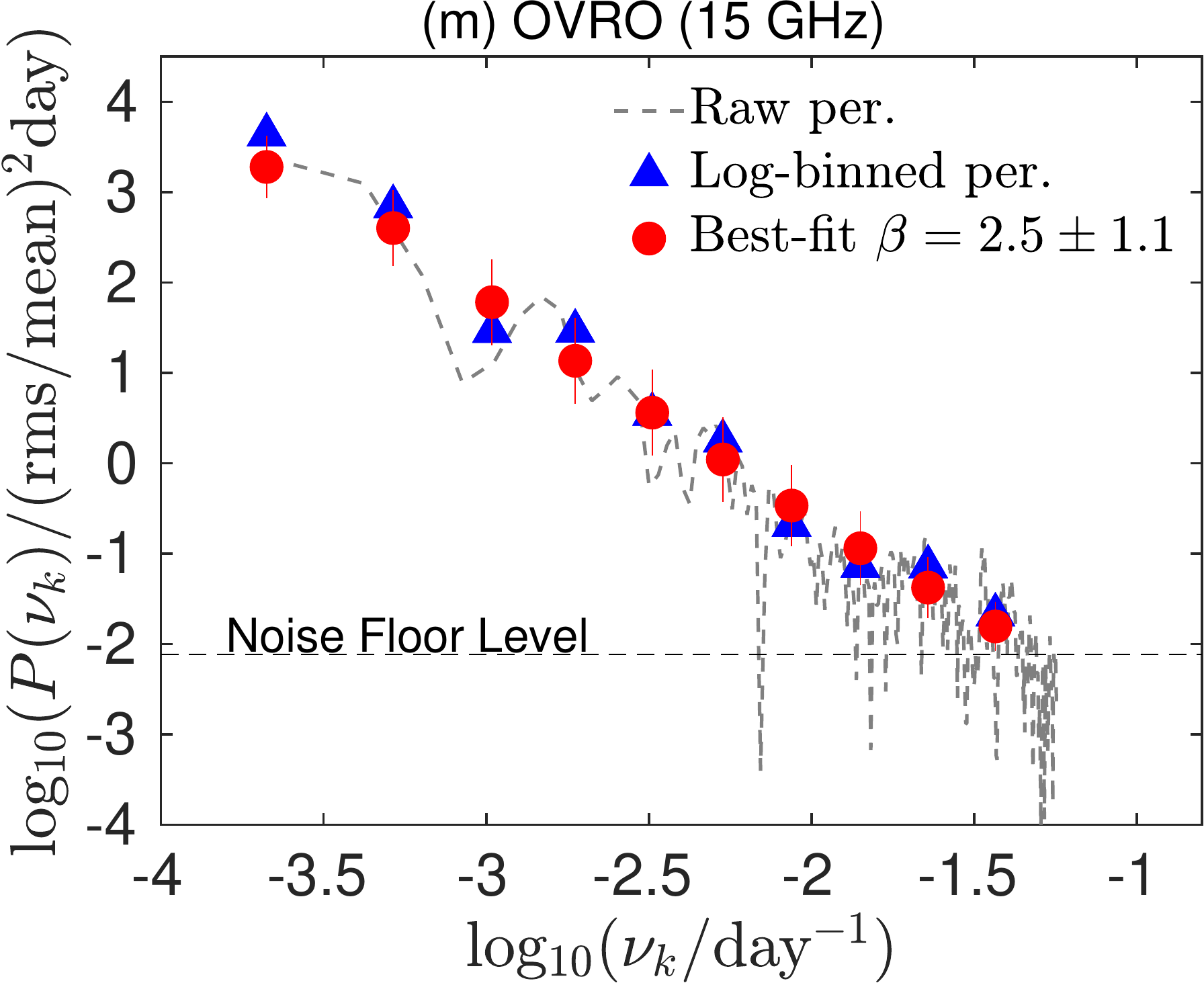}
\includegraphics[width=0.25\textwidth]{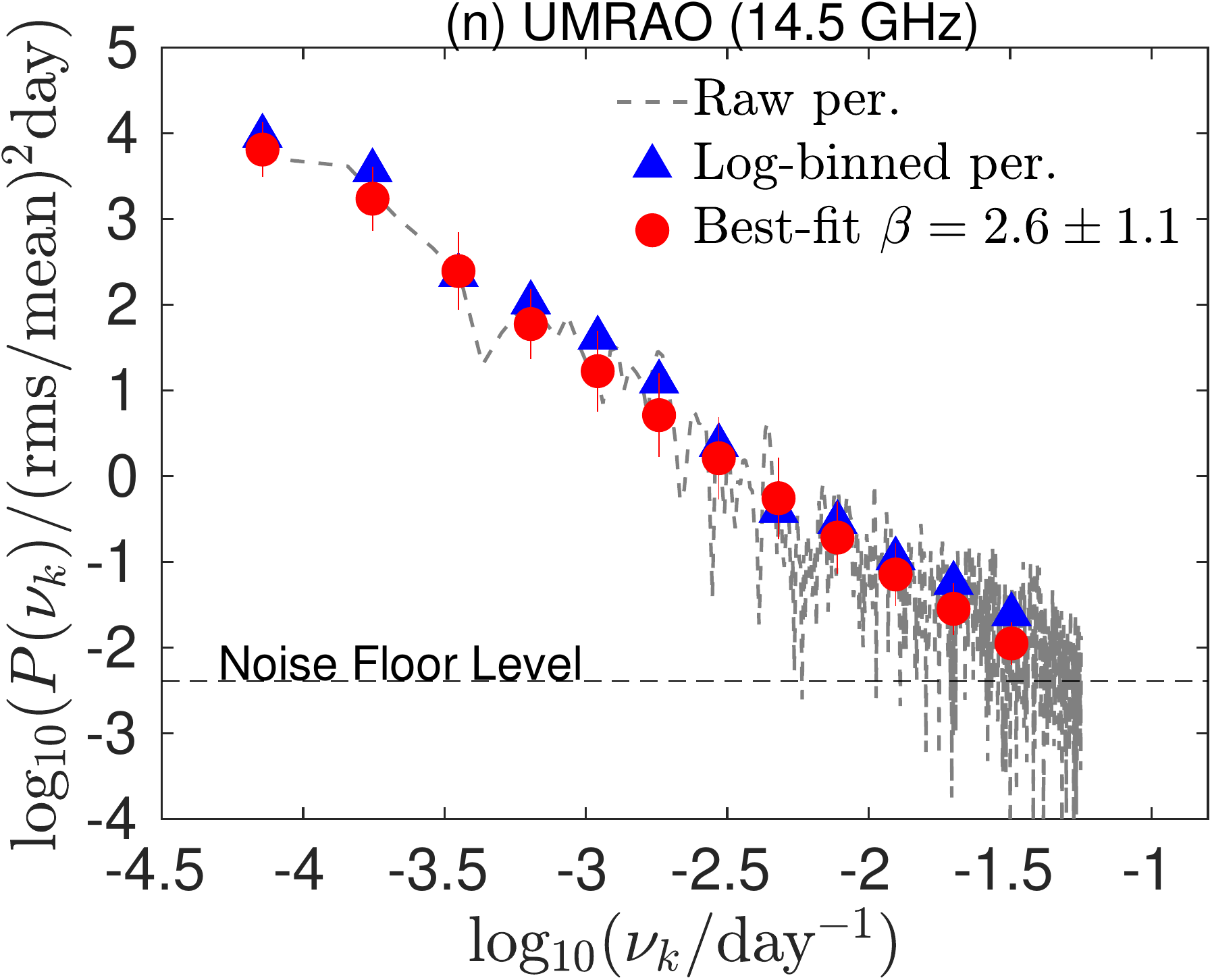}
\includegraphics[width=0.25\textwidth]{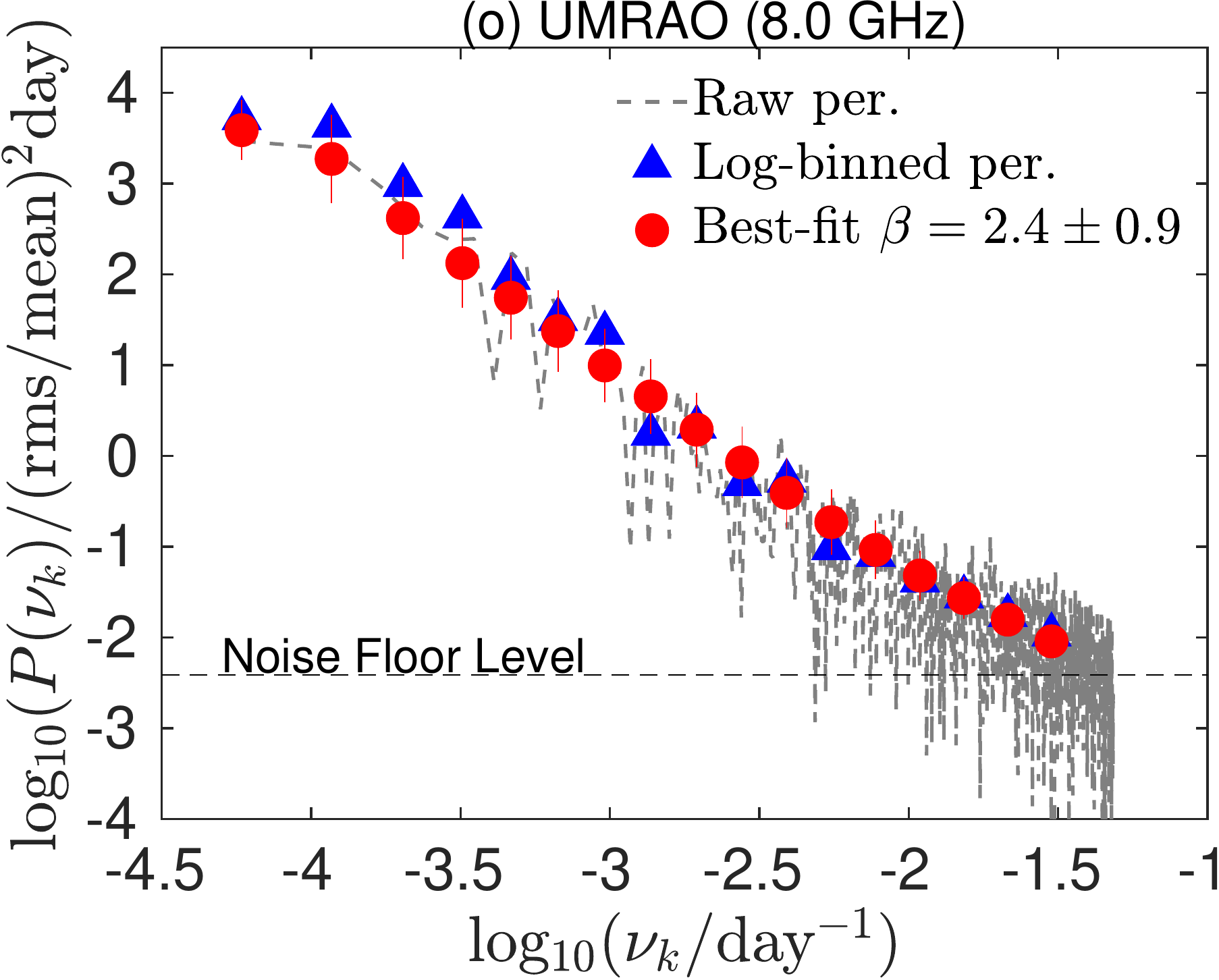}
\includegraphics[width=0.25\textwidth]{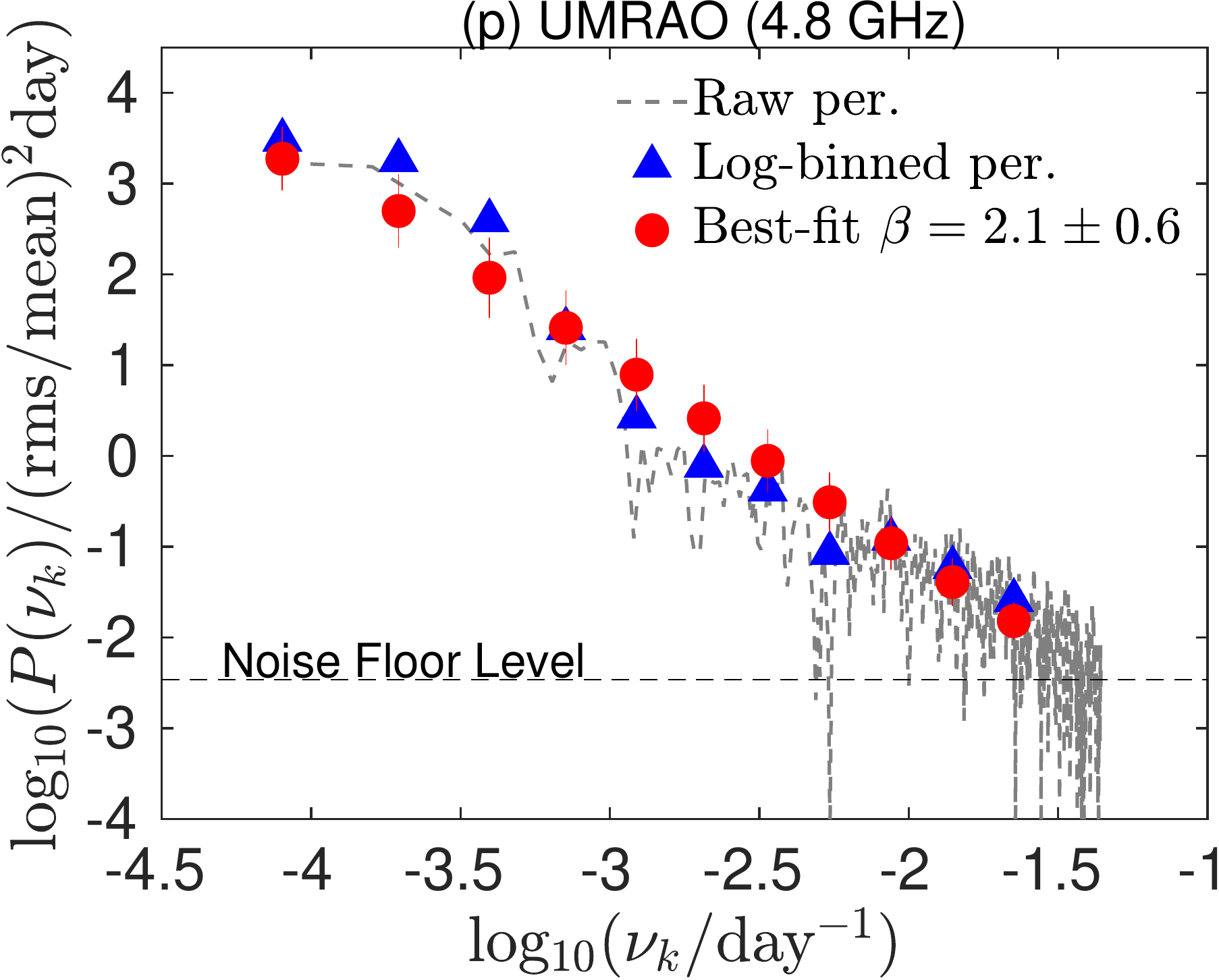}
}
\hbox{
\includegraphics[width=0.25\textwidth]{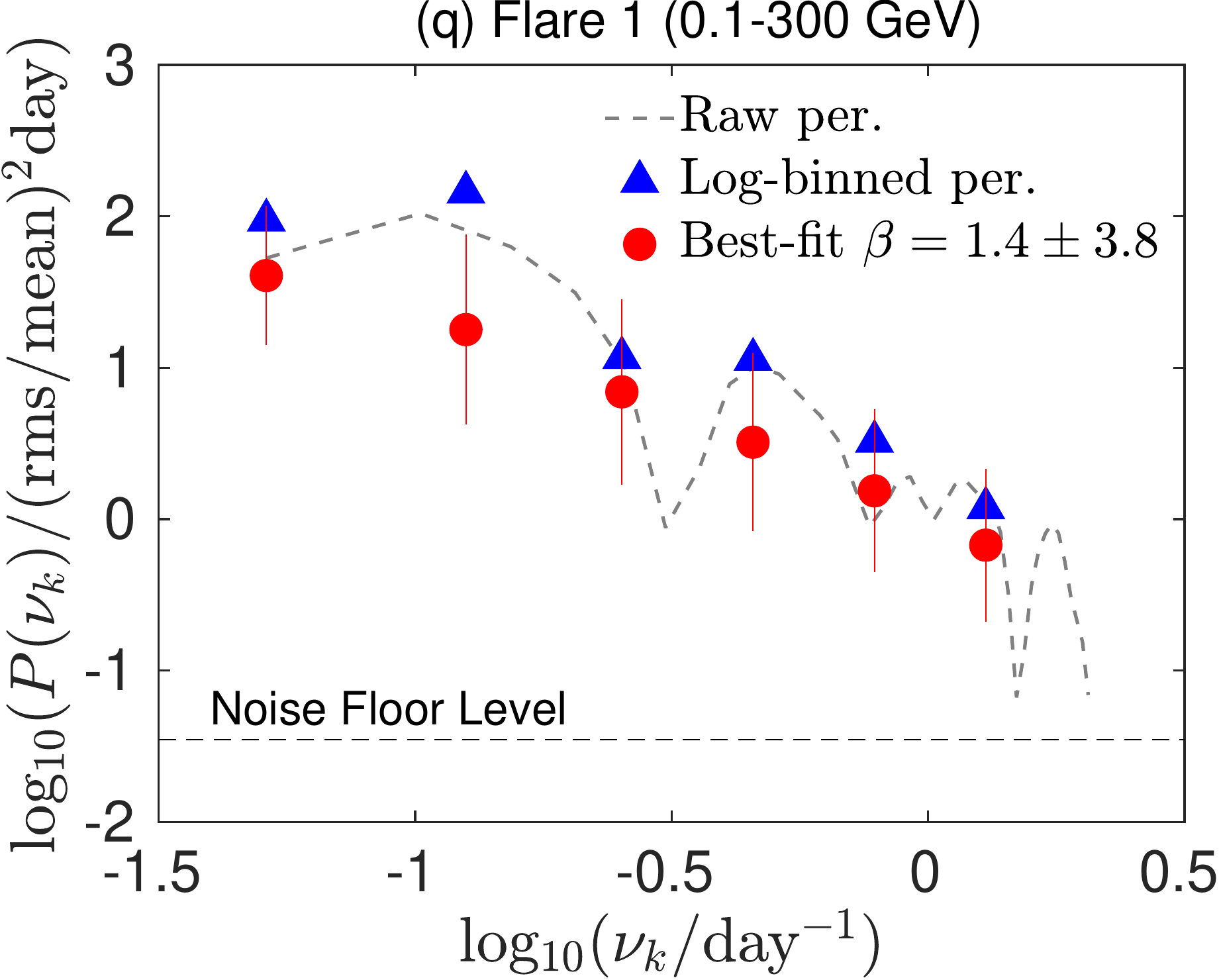}
\includegraphics[width=0.25\textwidth]{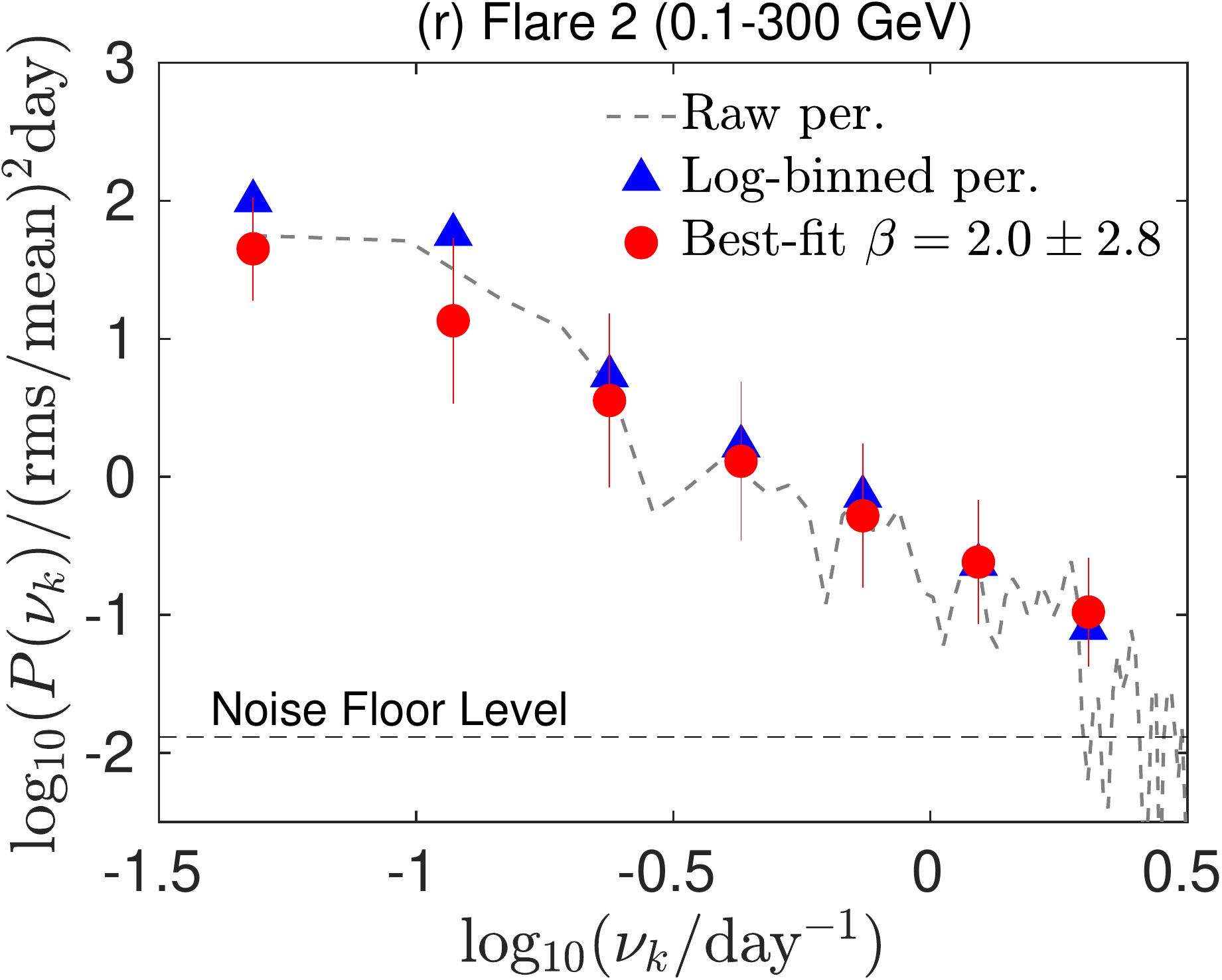}
\includegraphics[width=0.25\textwidth]{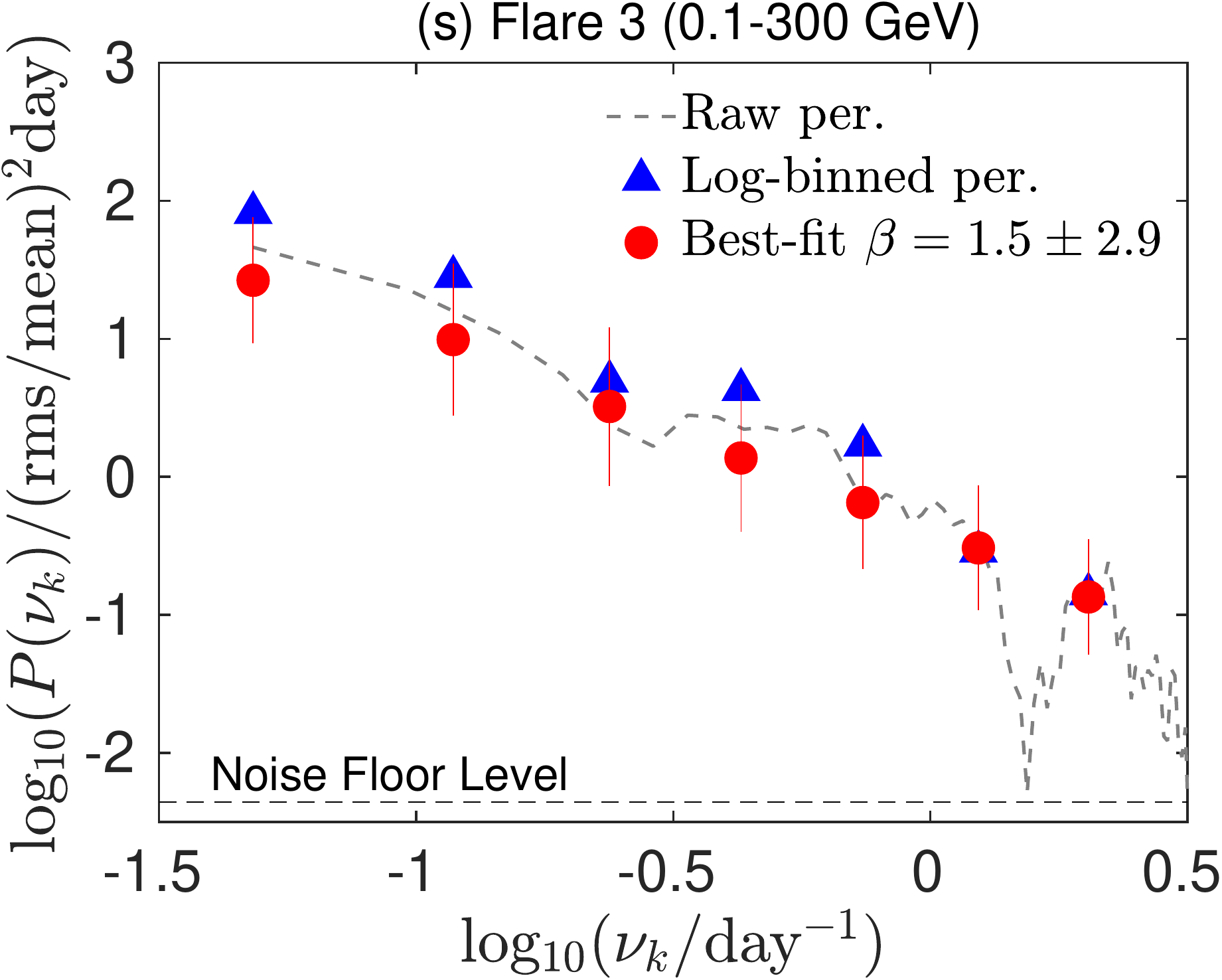}
}

\caption{Best-fit PSDs of the multiwavelength light curves shown in Figures~\ref{fig:lc3c279}, ~\ref{fig:fermi3hr}(a, b, and c), and ~\ref{fig:mwrad}(a) for the blazar 3C\,279, derived using the PSRESP method. Variability power spectrum down to the  Nyquist sampling frequency of the (mean) observed data is derived. The dashed line shows the `raw' periodogram while the blue triangles and red circles give the `logarithmically binned power spectrum' and the best-fit power spectrum, respectively. The error on the best-fit PSD slope corresponds to a 98\% confidence limit. The dashed horizontal line corresponds to the statistical noise floor level due to measurement noise. }
\label{fig:psd3c279}
\end{figure*}

\begin{figure*}
\hbox{
\includegraphics[width=0.25\textwidth]{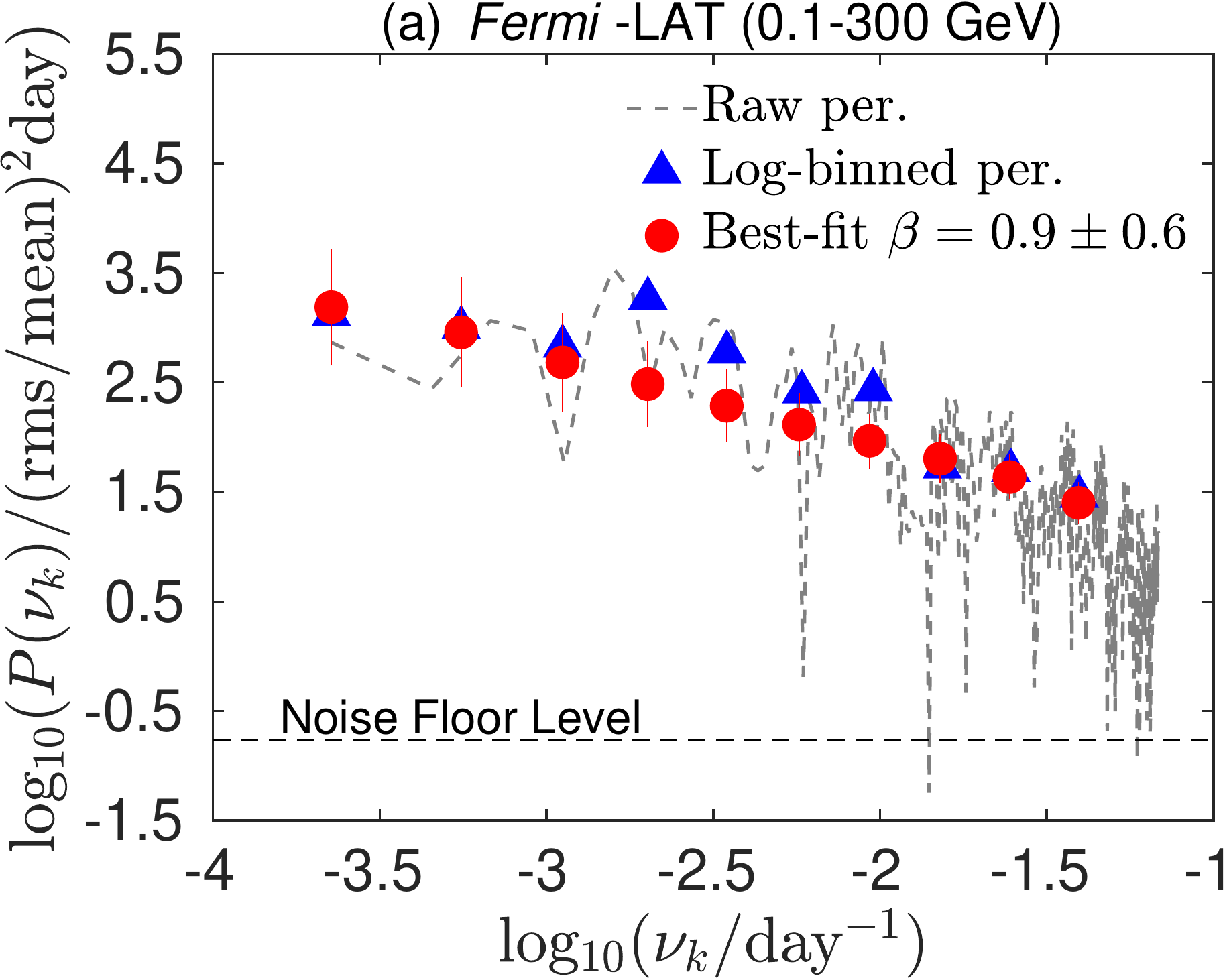}
\includegraphics[width=0.25\textwidth]{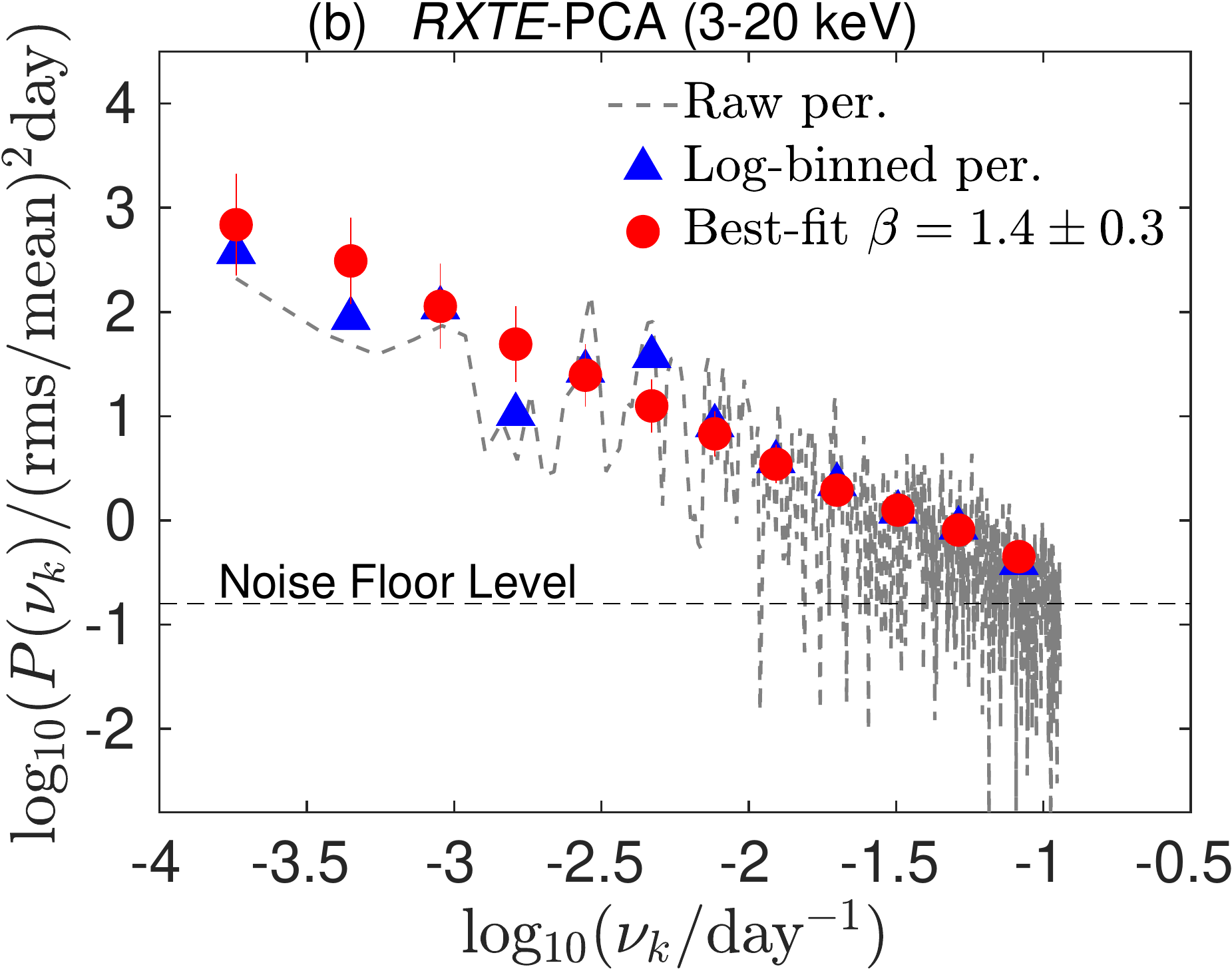}
\includegraphics[width=0.25\textwidth]{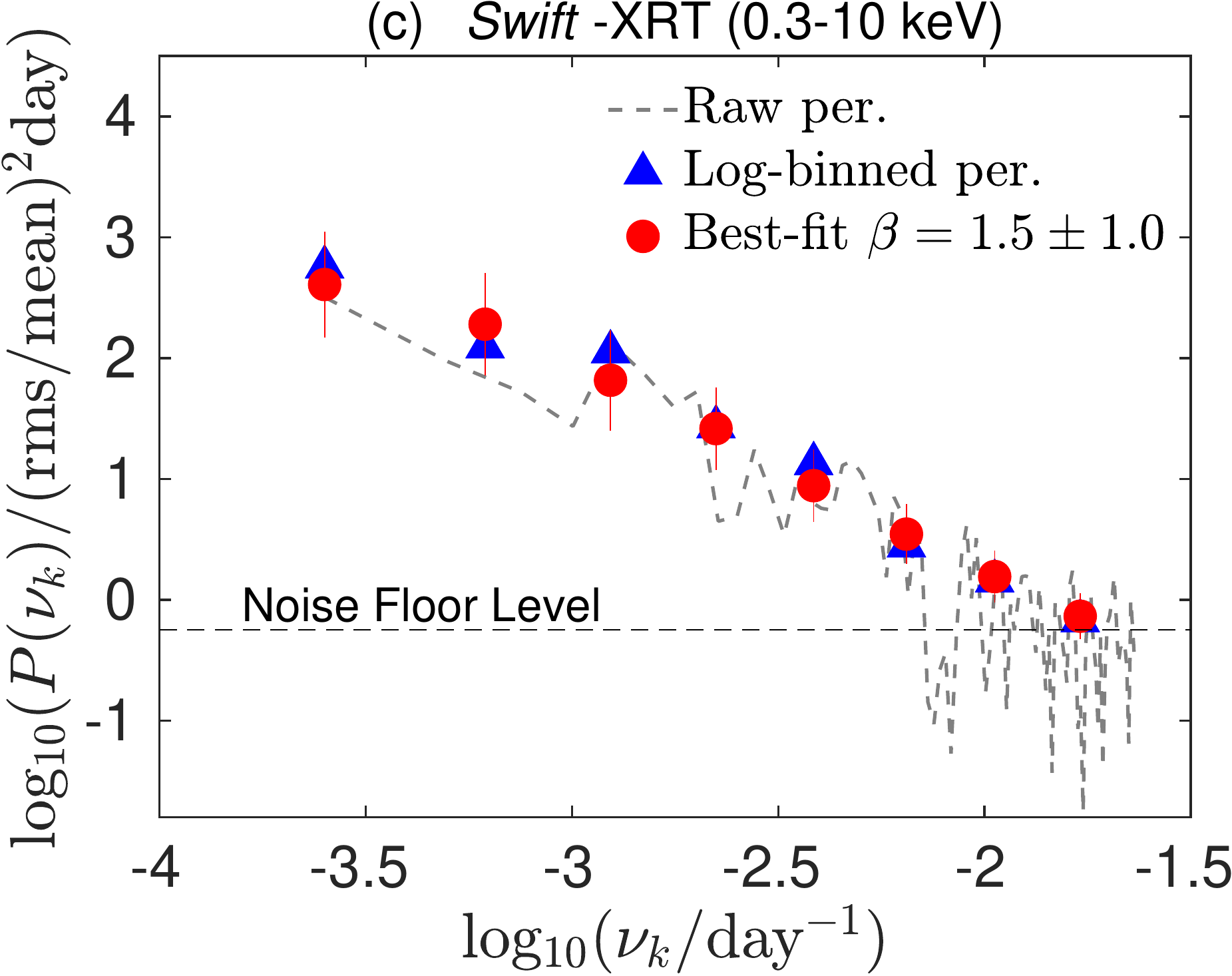}
\includegraphics[width=0.25\textwidth]{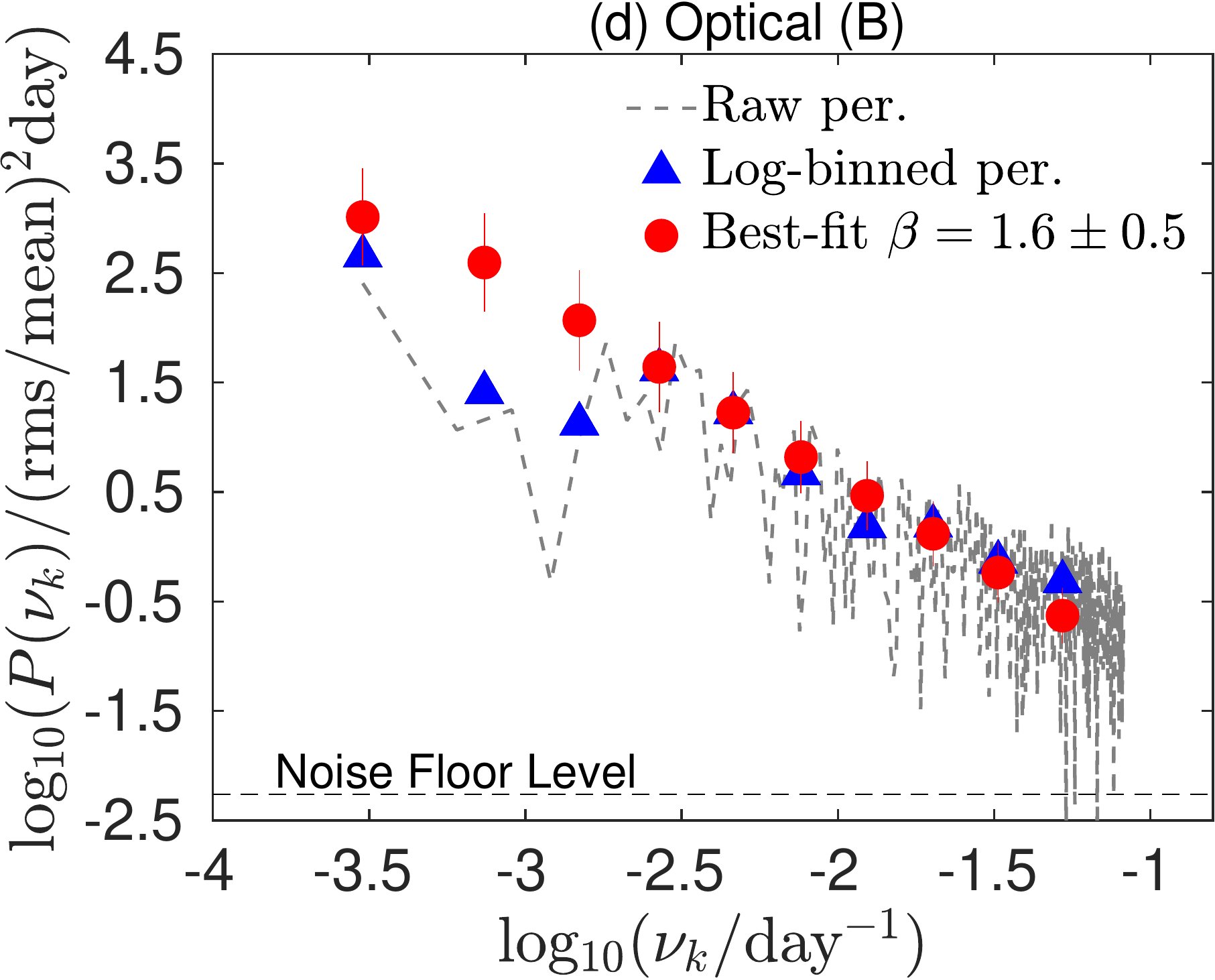}
}
\hbox{
\includegraphics[width=0.25\textwidth]{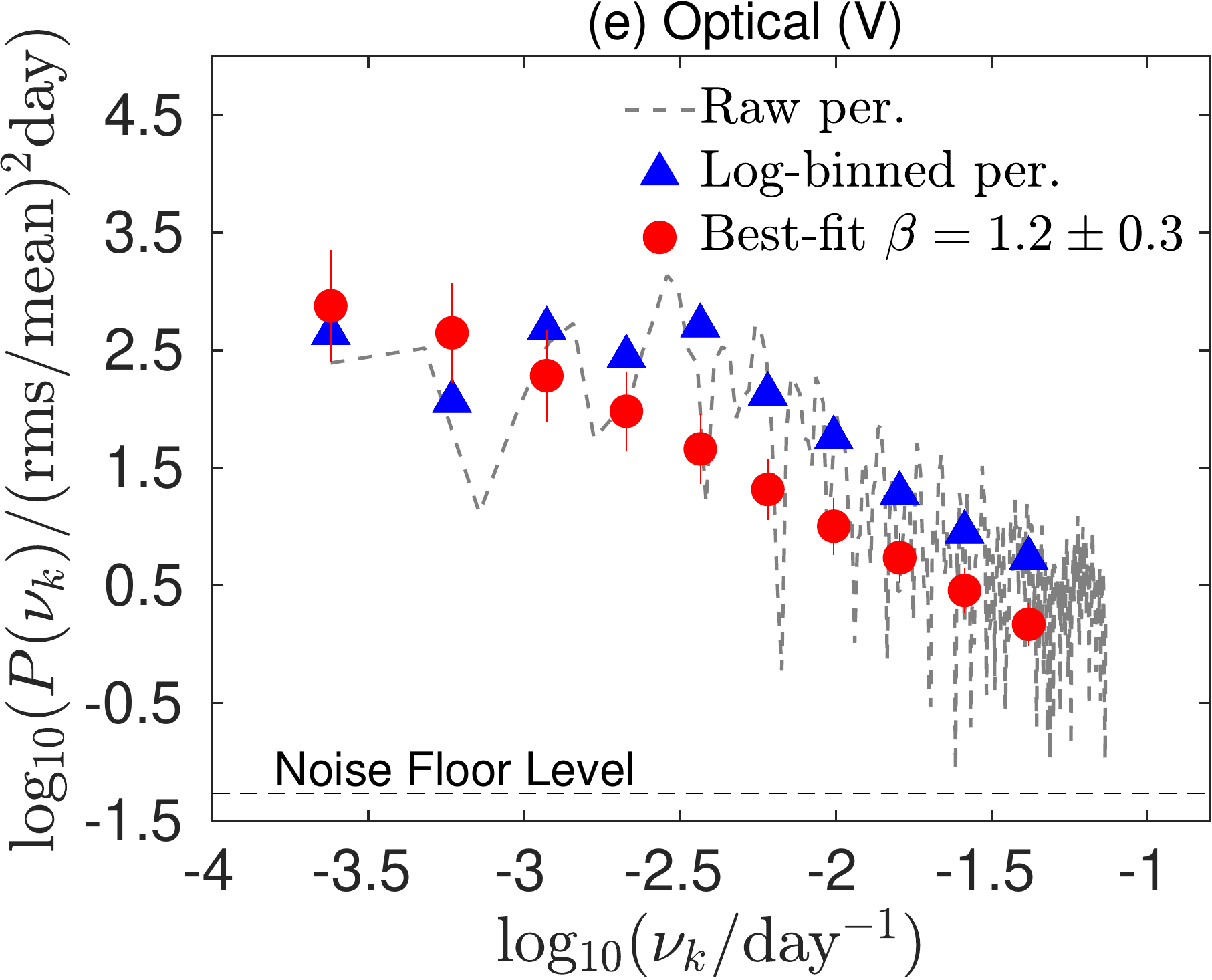}
\includegraphics[width=0.25\textwidth]{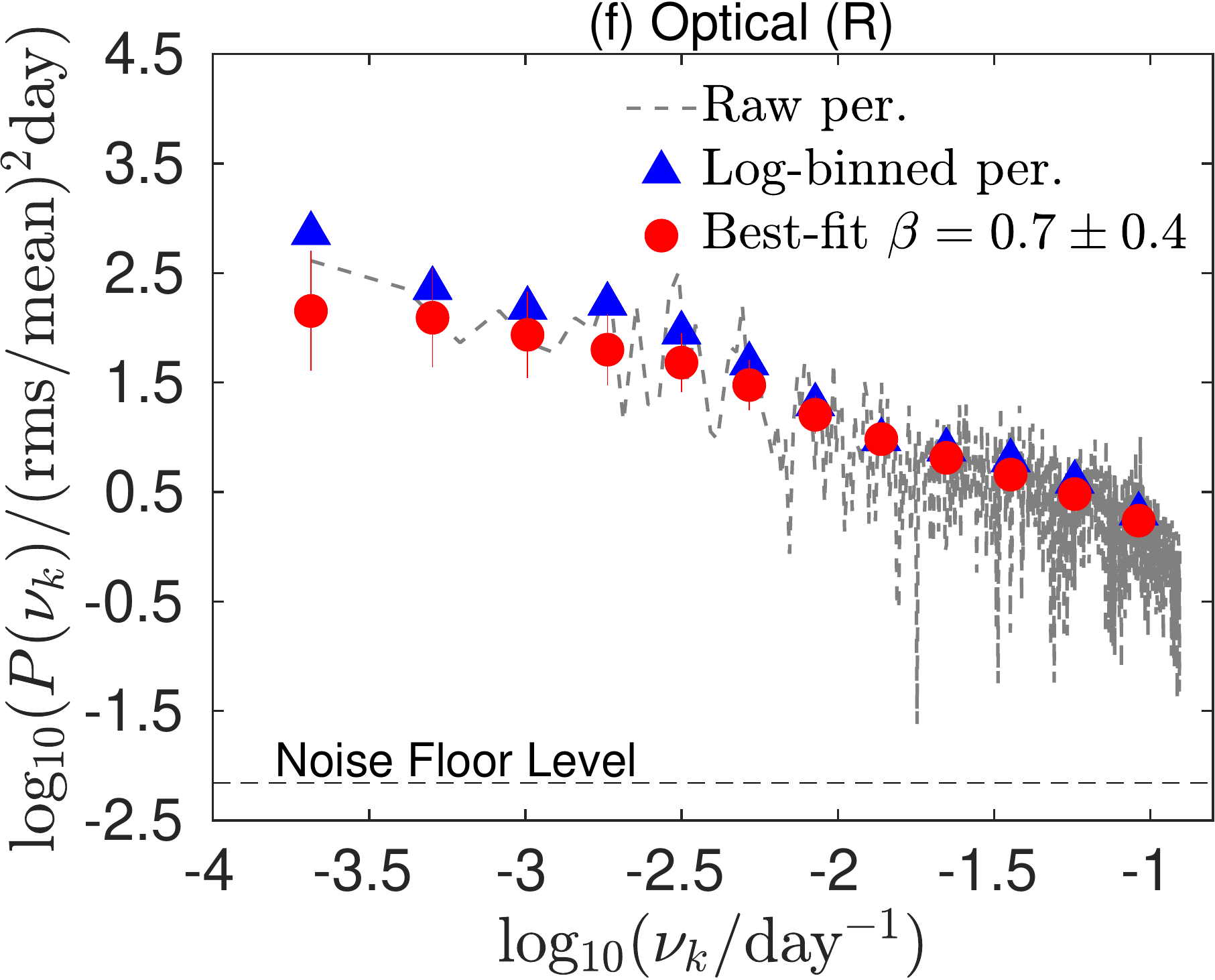}
\includegraphics[width=0.25\textwidth]{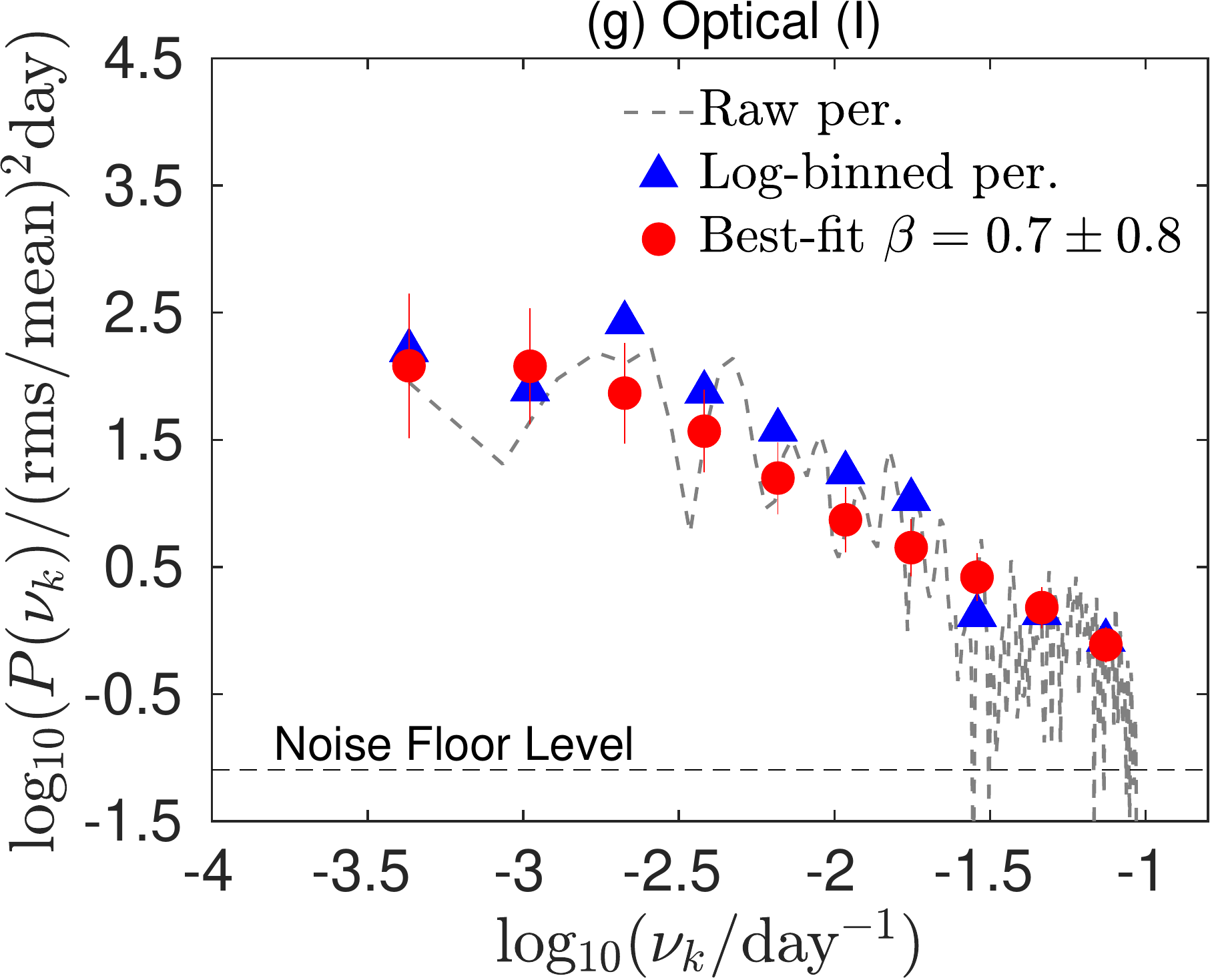}
\includegraphics[width=0.25\textwidth]{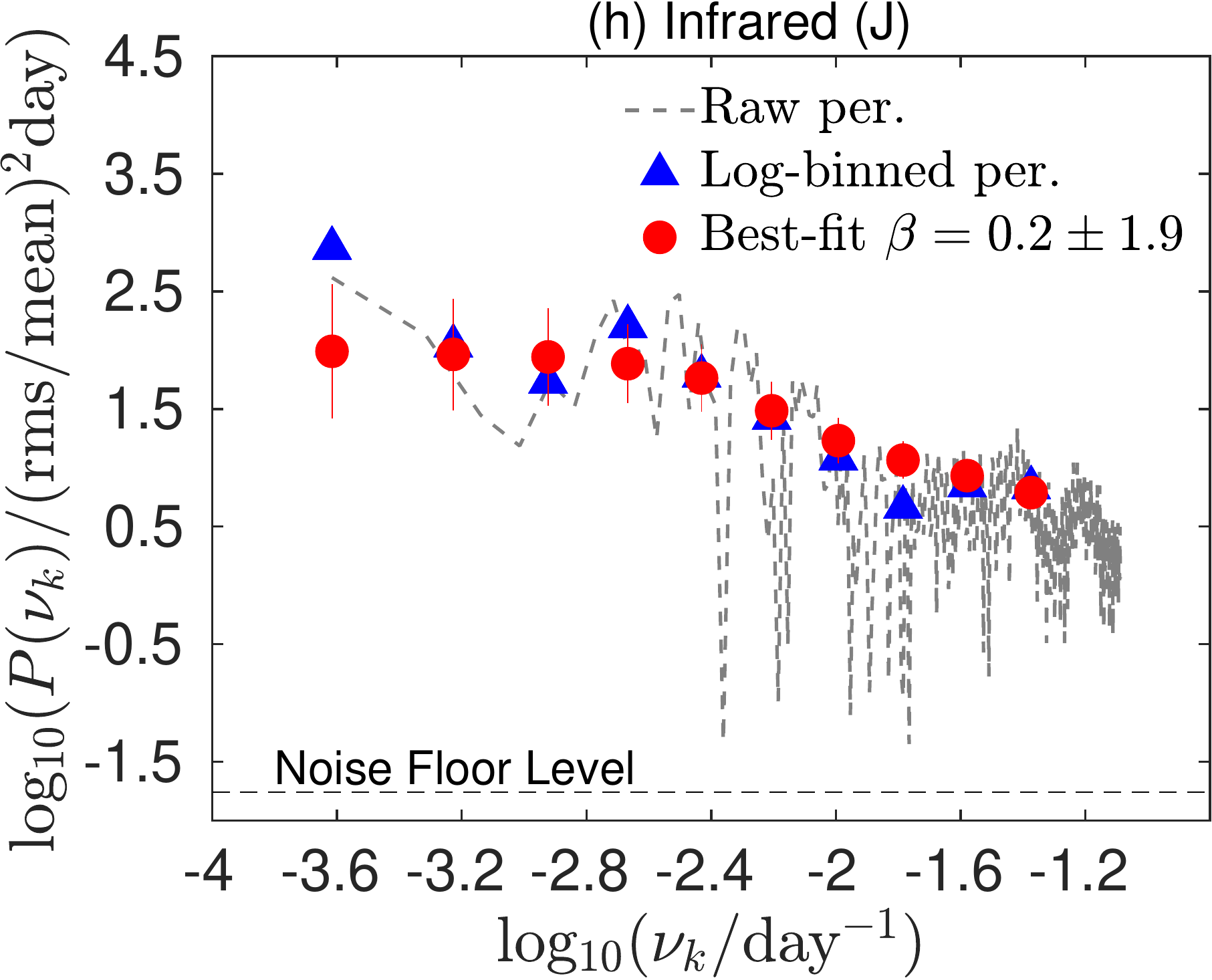}
}
\hbox{
\includegraphics[width=0.25\textwidth]{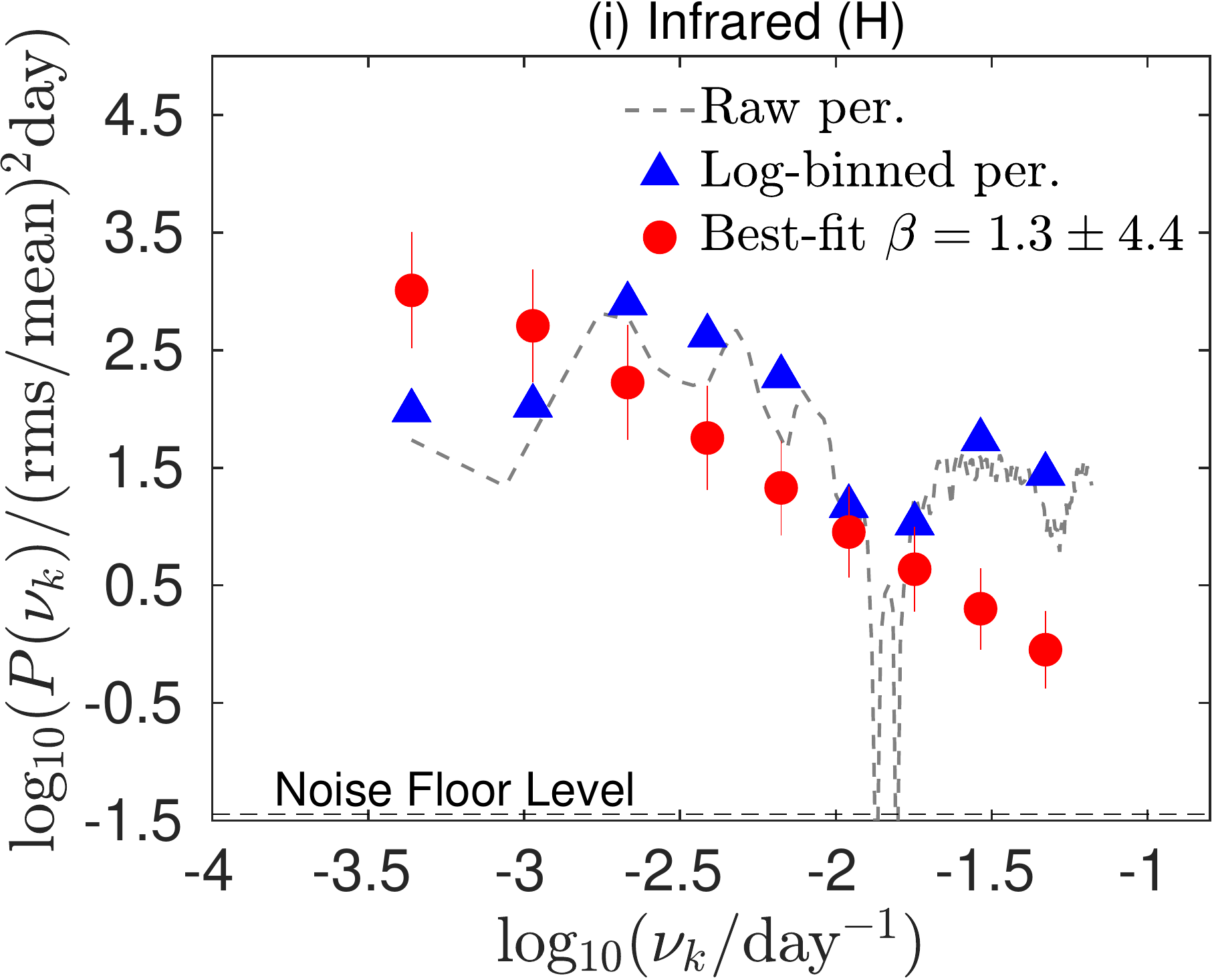}
\includegraphics[width=0.25\textwidth]{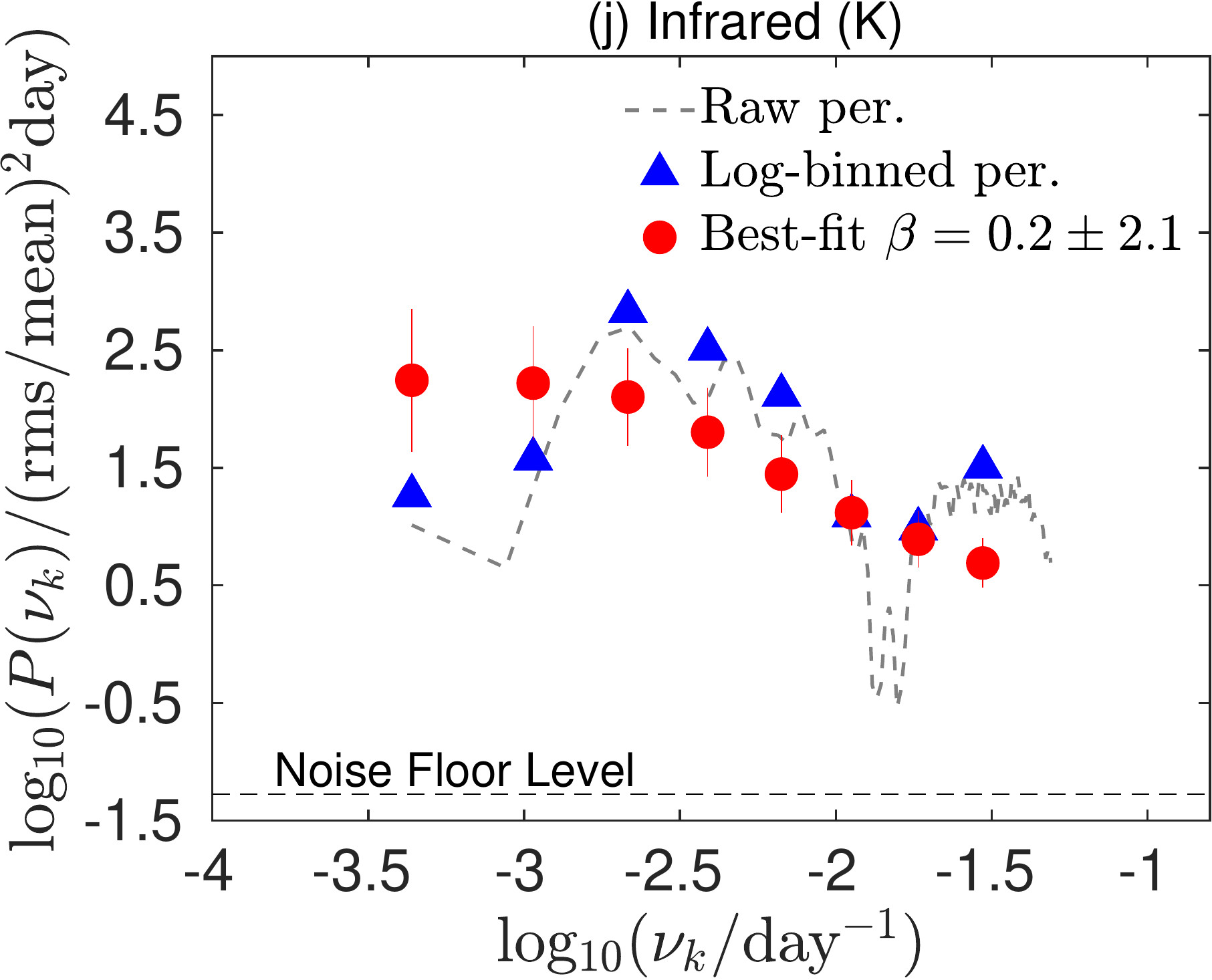}
\includegraphics[width=0.25\textwidth]{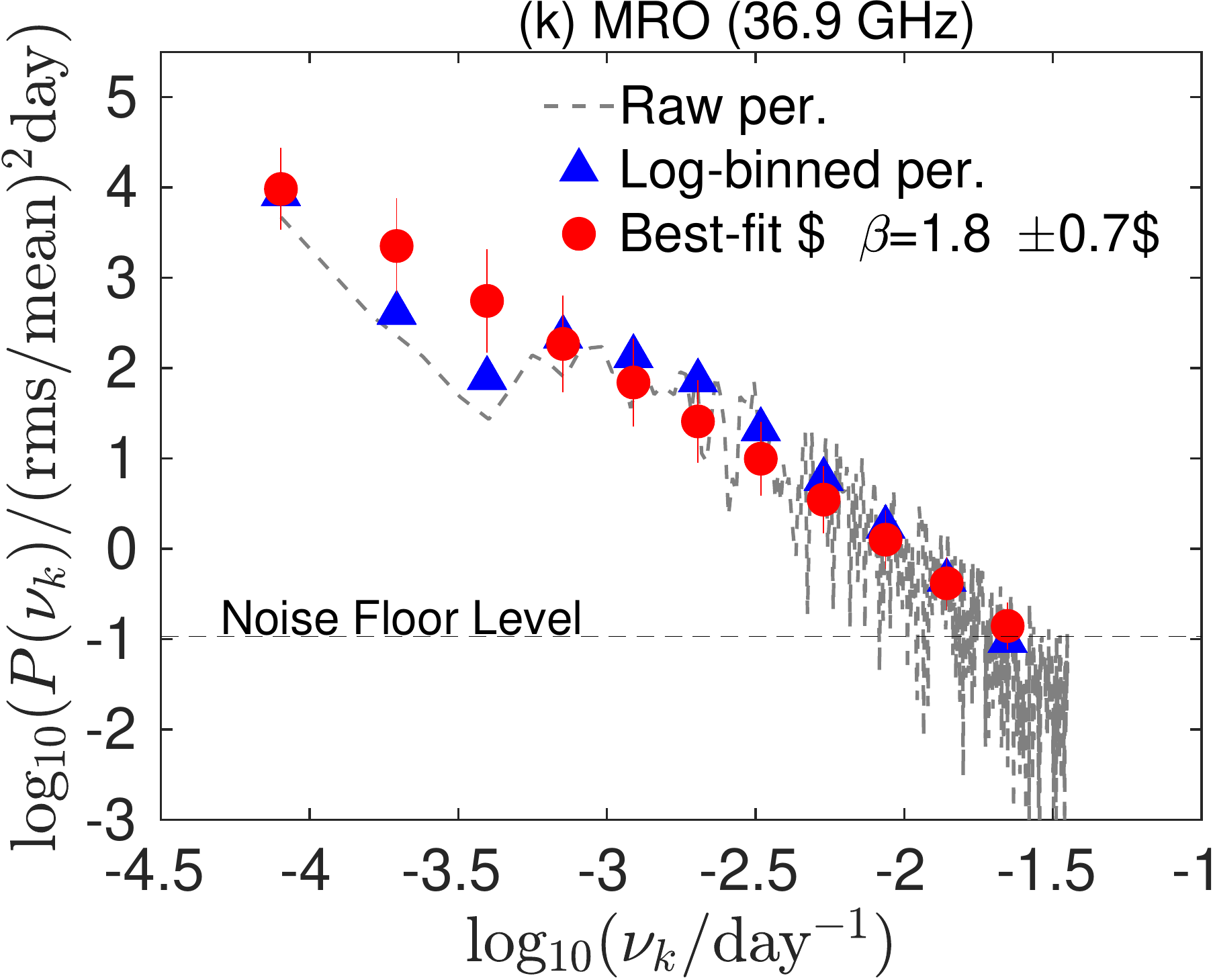}
\includegraphics[width=0.25\textwidth]{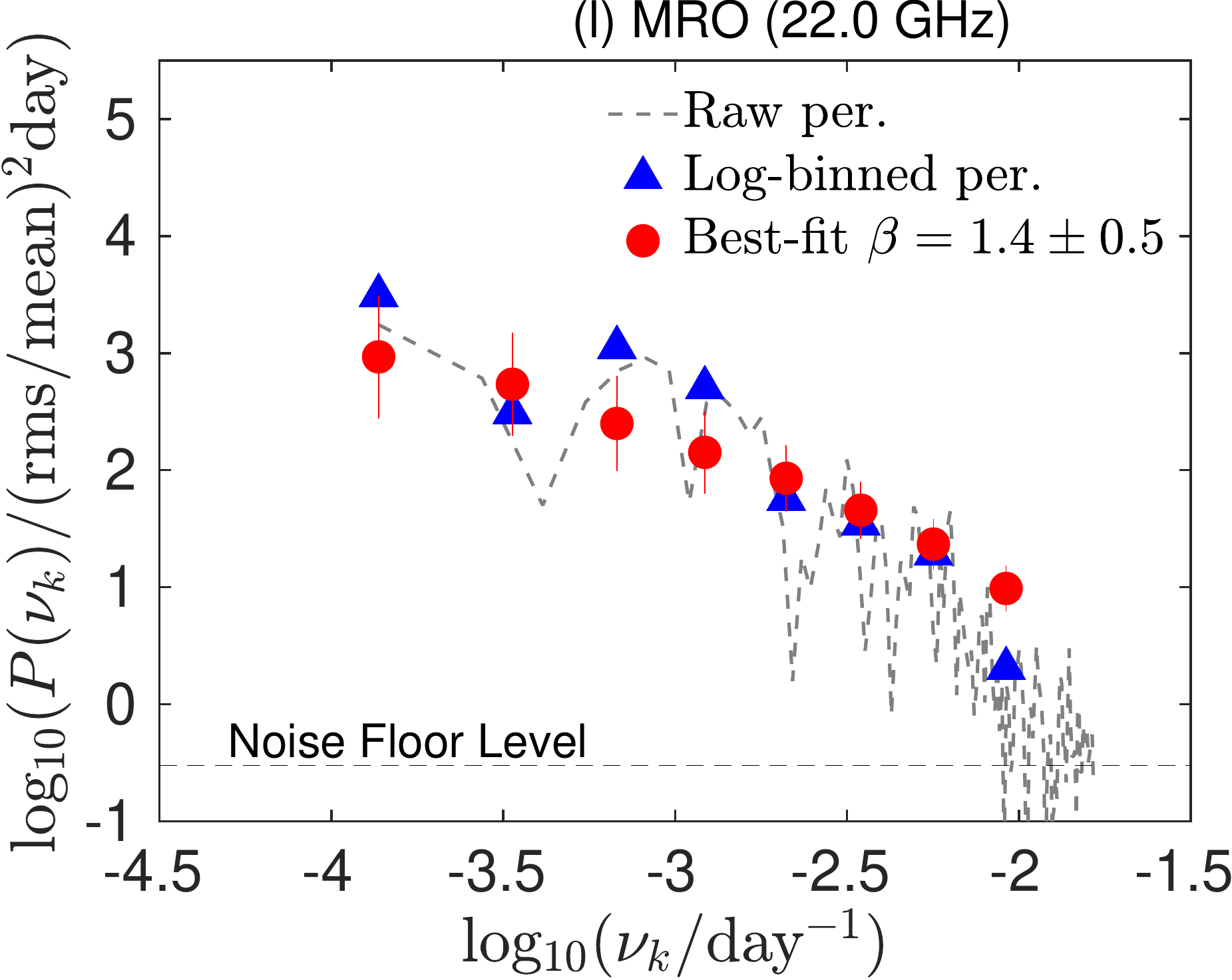}
}
\hbox{
\includegraphics[width=0.25\textwidth]{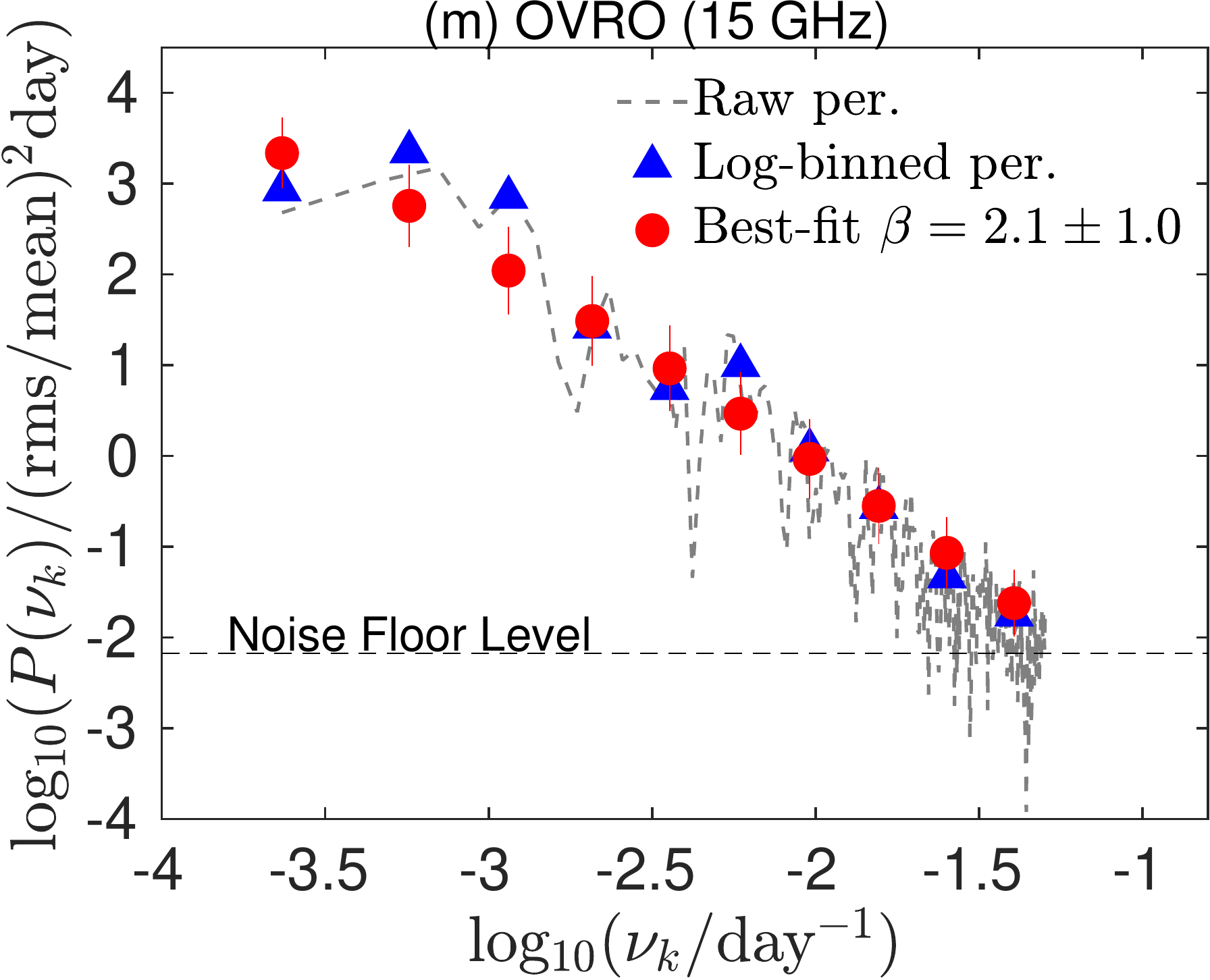}
\includegraphics[width=0.25\textwidth]{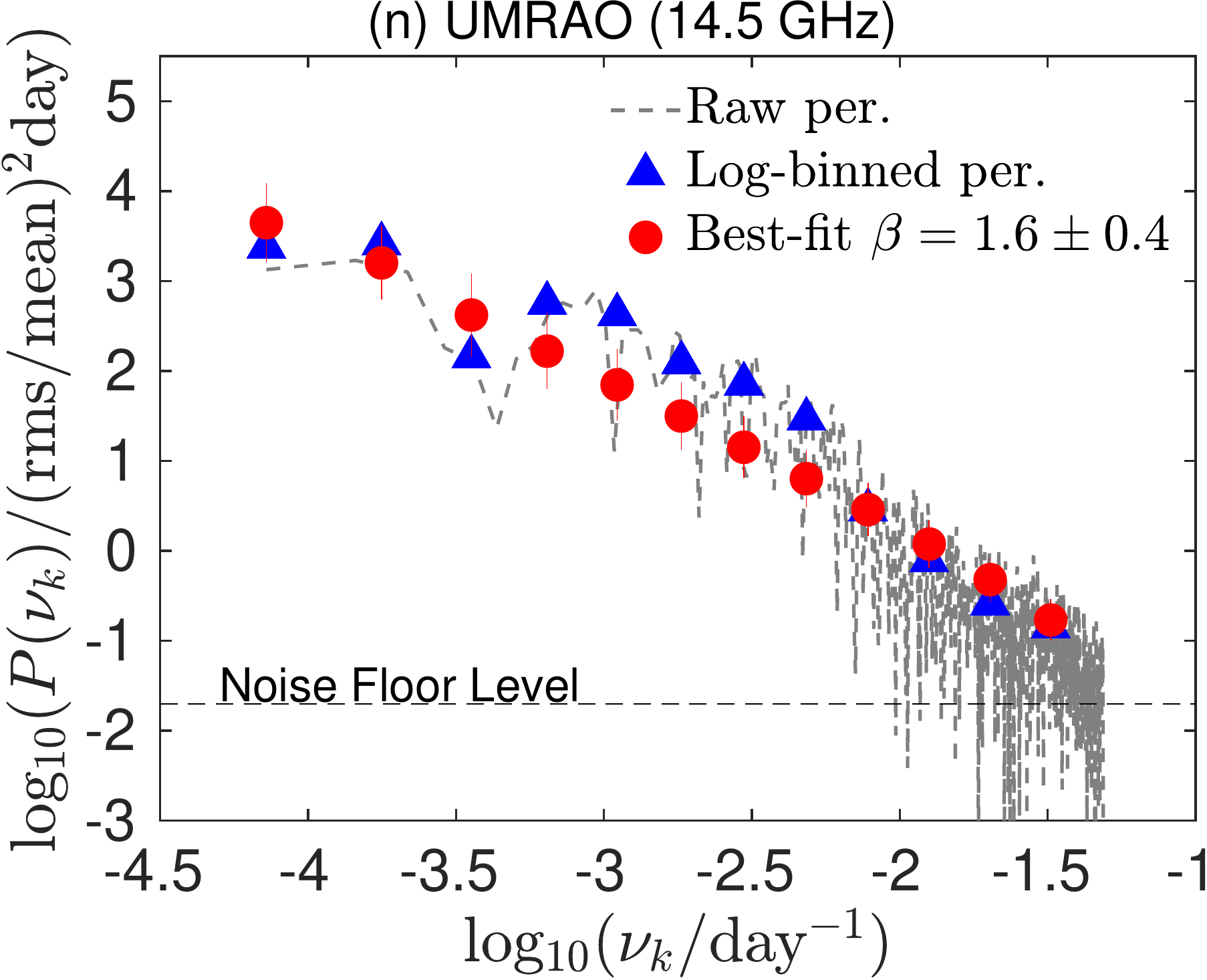}
\includegraphics[width=0.25\textwidth]{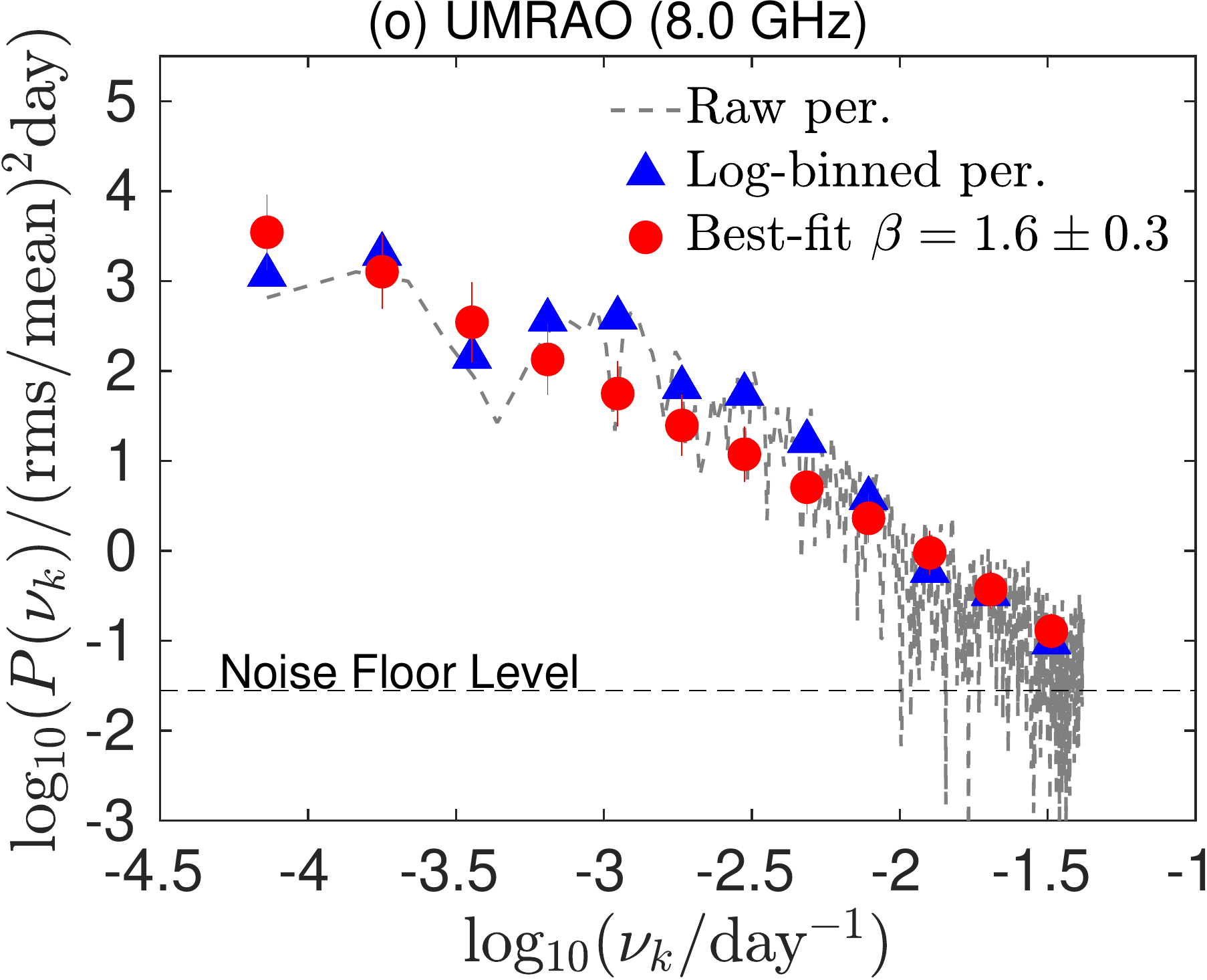}
\includegraphics[width=0.25\textwidth]{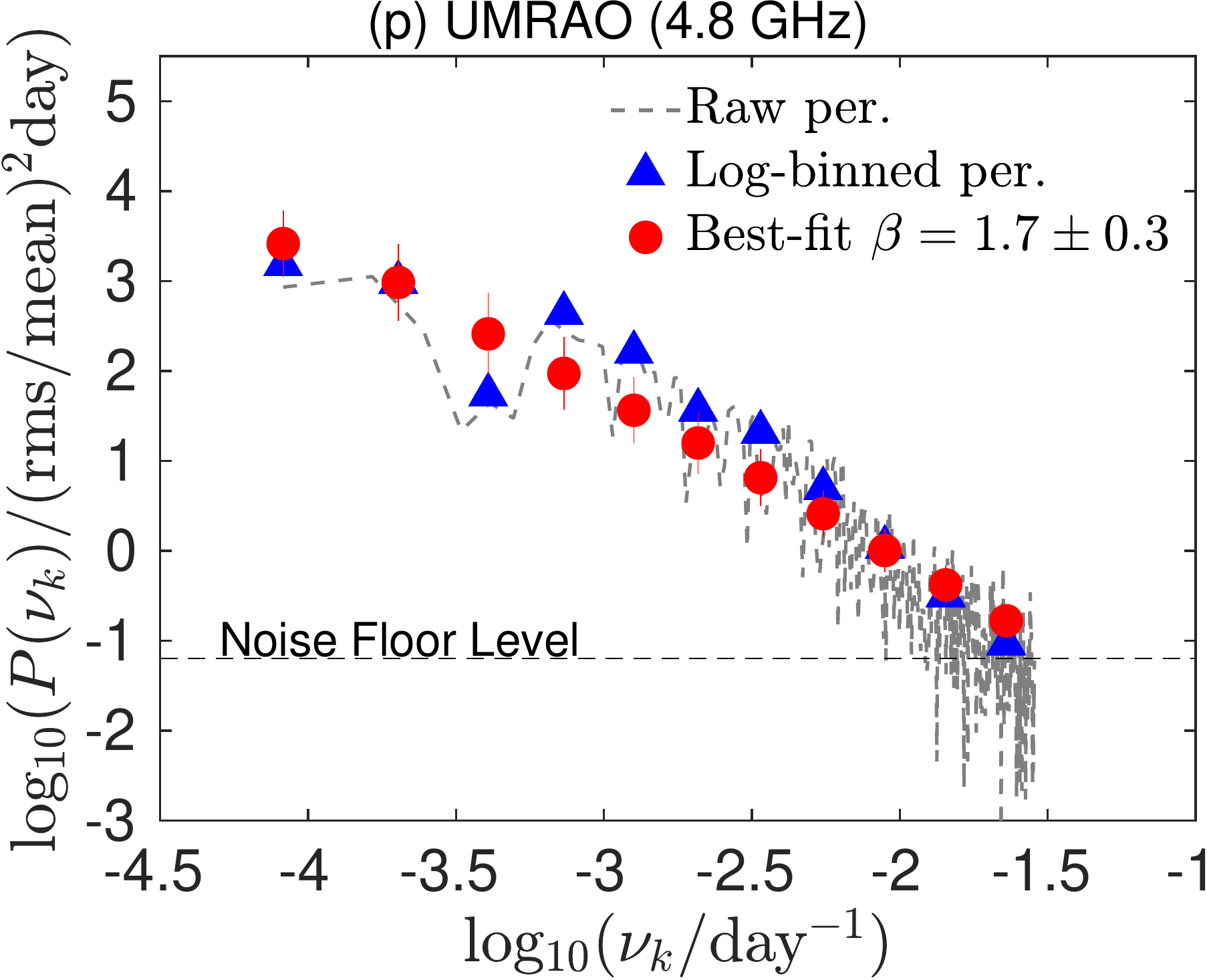}
}
\includegraphics[width=0.25\textwidth]{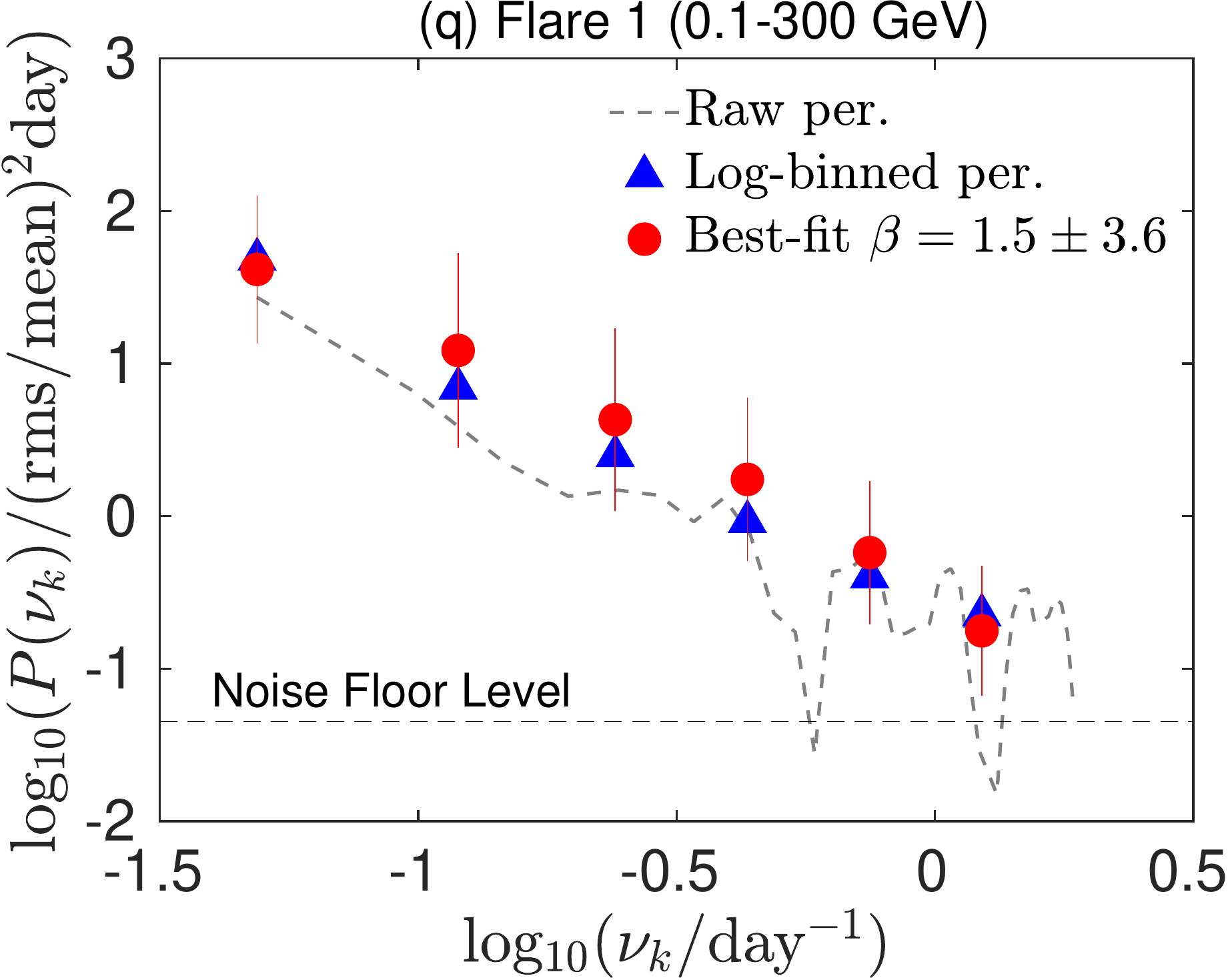}

\caption{Best-fit PSDs of the multiwavelength light curves shown in Figures~\ref{fig:lcpks1510}, ~\ref{fig:fermi3hr}(d), and ~\ref{fig:mwrad}(b) for the blazar PKS\,1510$-$089, derived using the PSRESP method, and displayed as in Figure~\ref{fig:psd3c279}. }
\label{fig:psdpks1510}
\end{figure*}

\begin{figure*}
\hbox{
\includegraphics[width=0.33\textwidth]{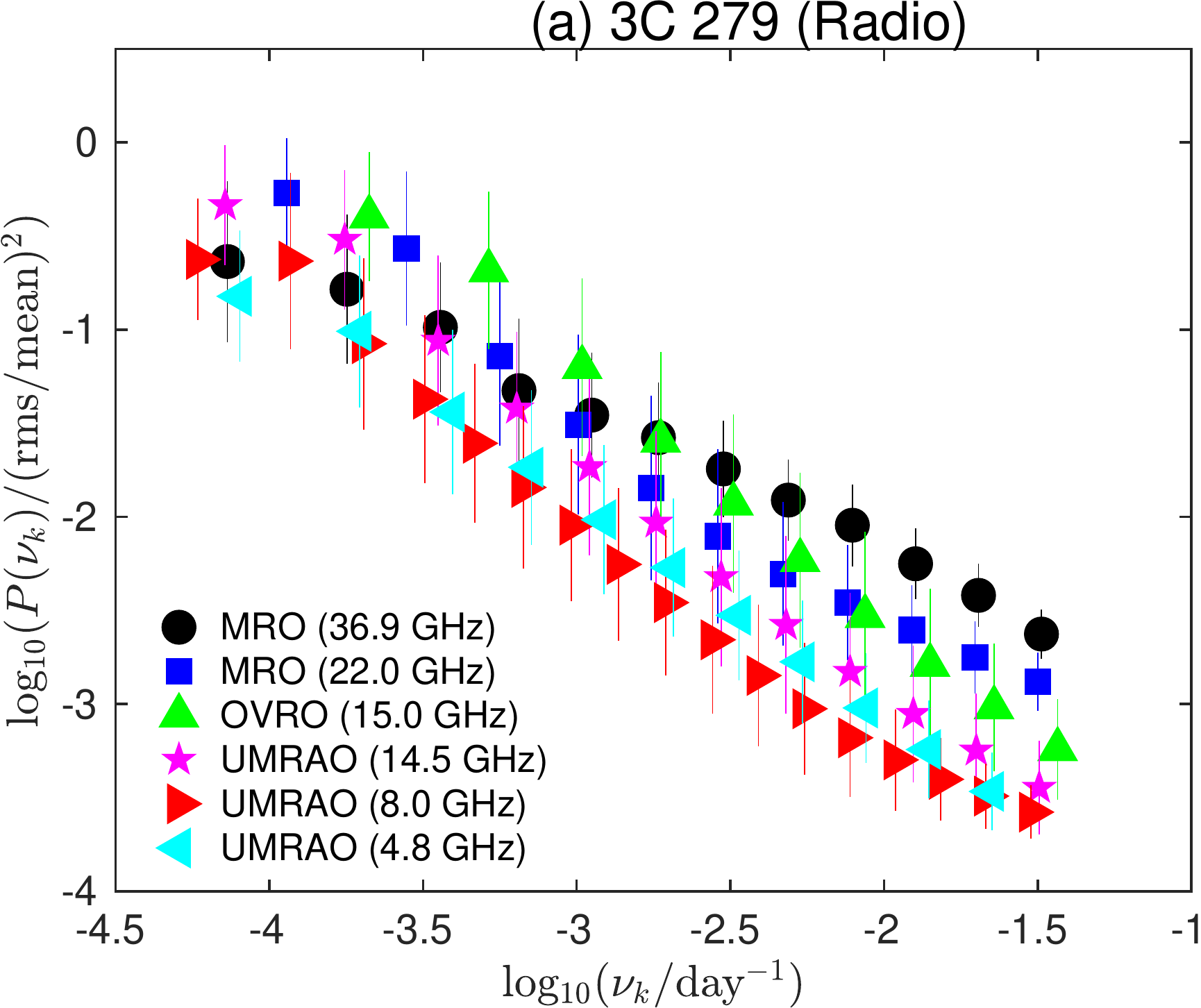}
\includegraphics[width=0.33\textwidth]{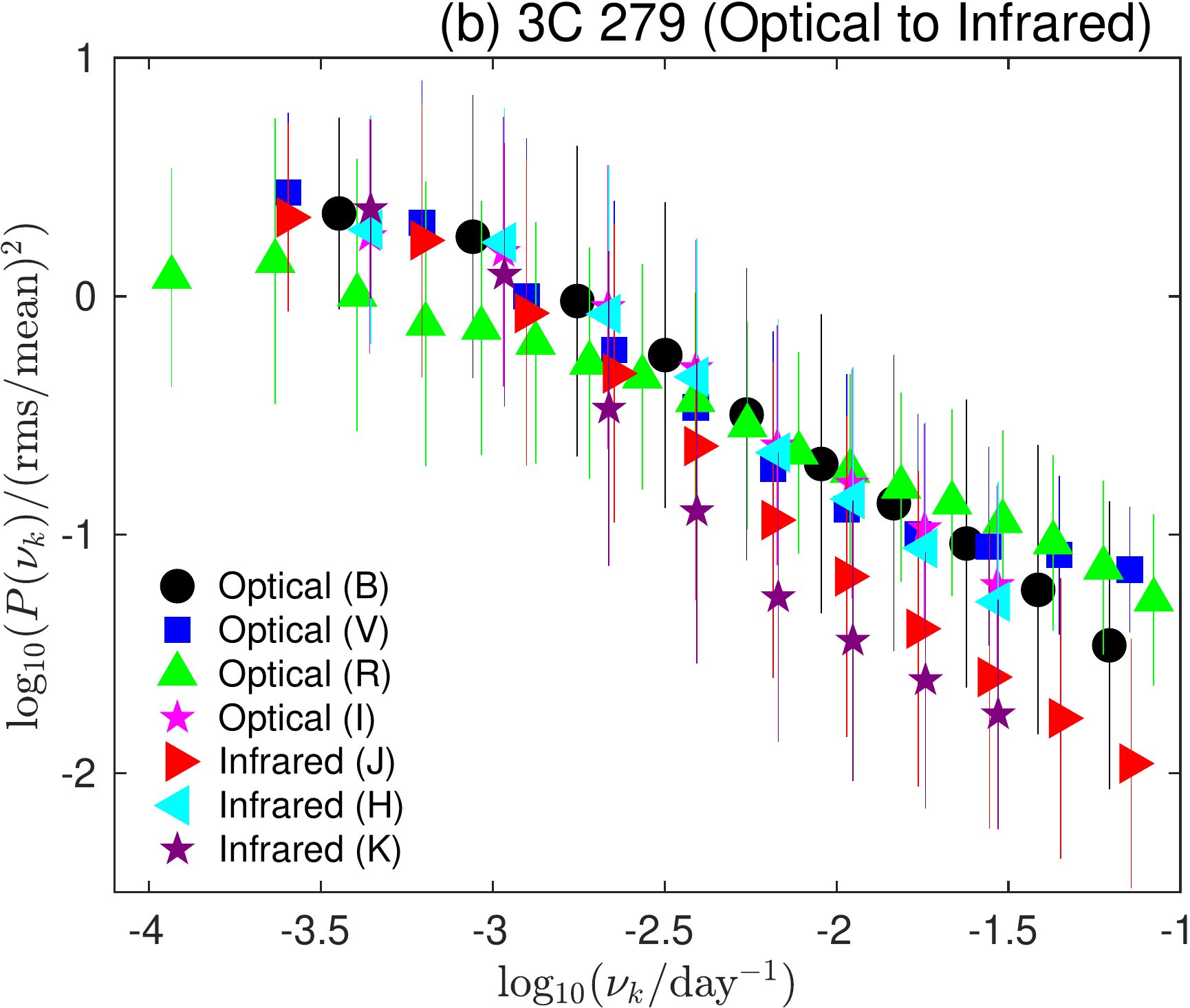}
\includegraphics[width=0.33\textwidth]{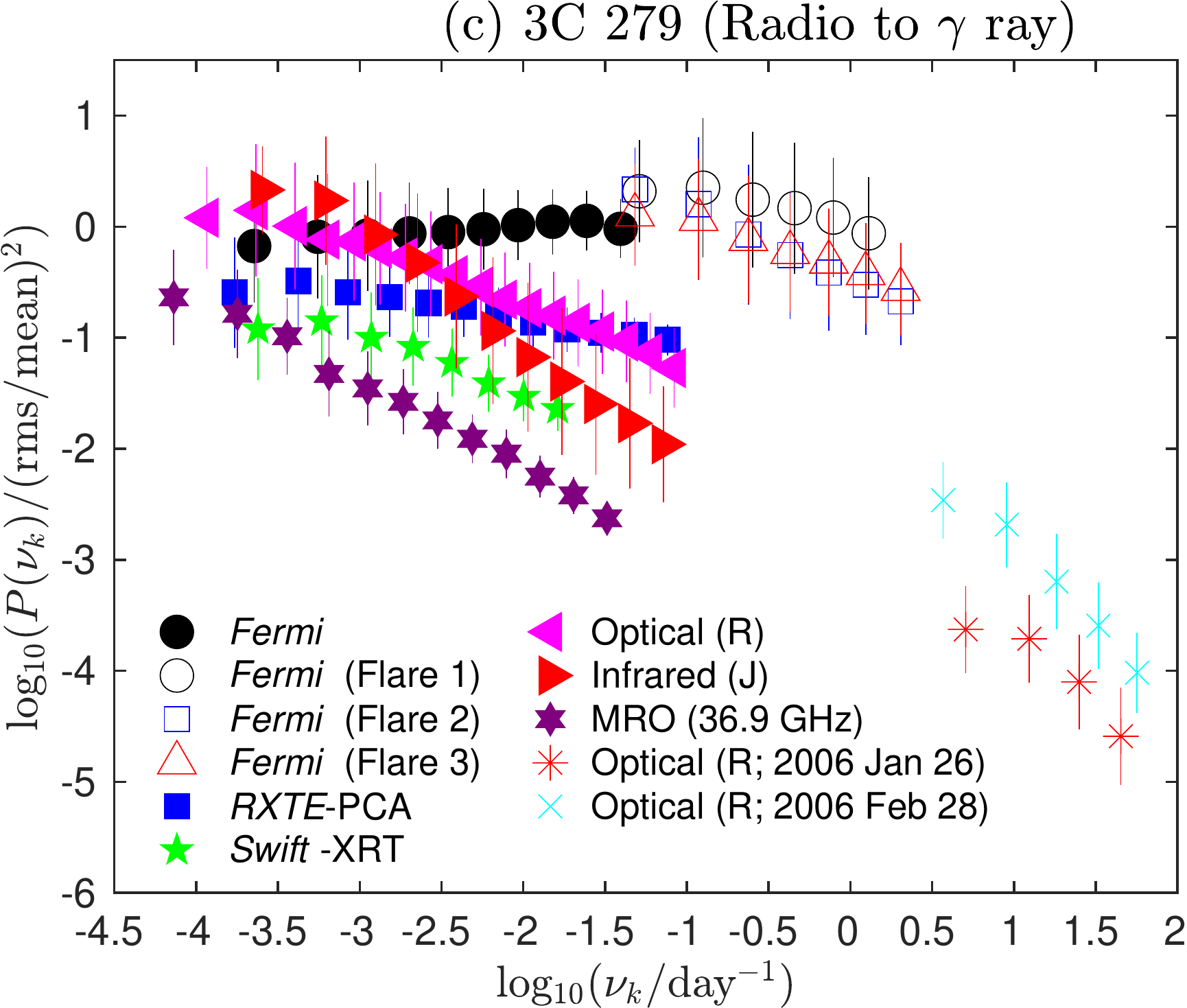}
}
\hbox{
\includegraphics[width=0.33\textwidth]{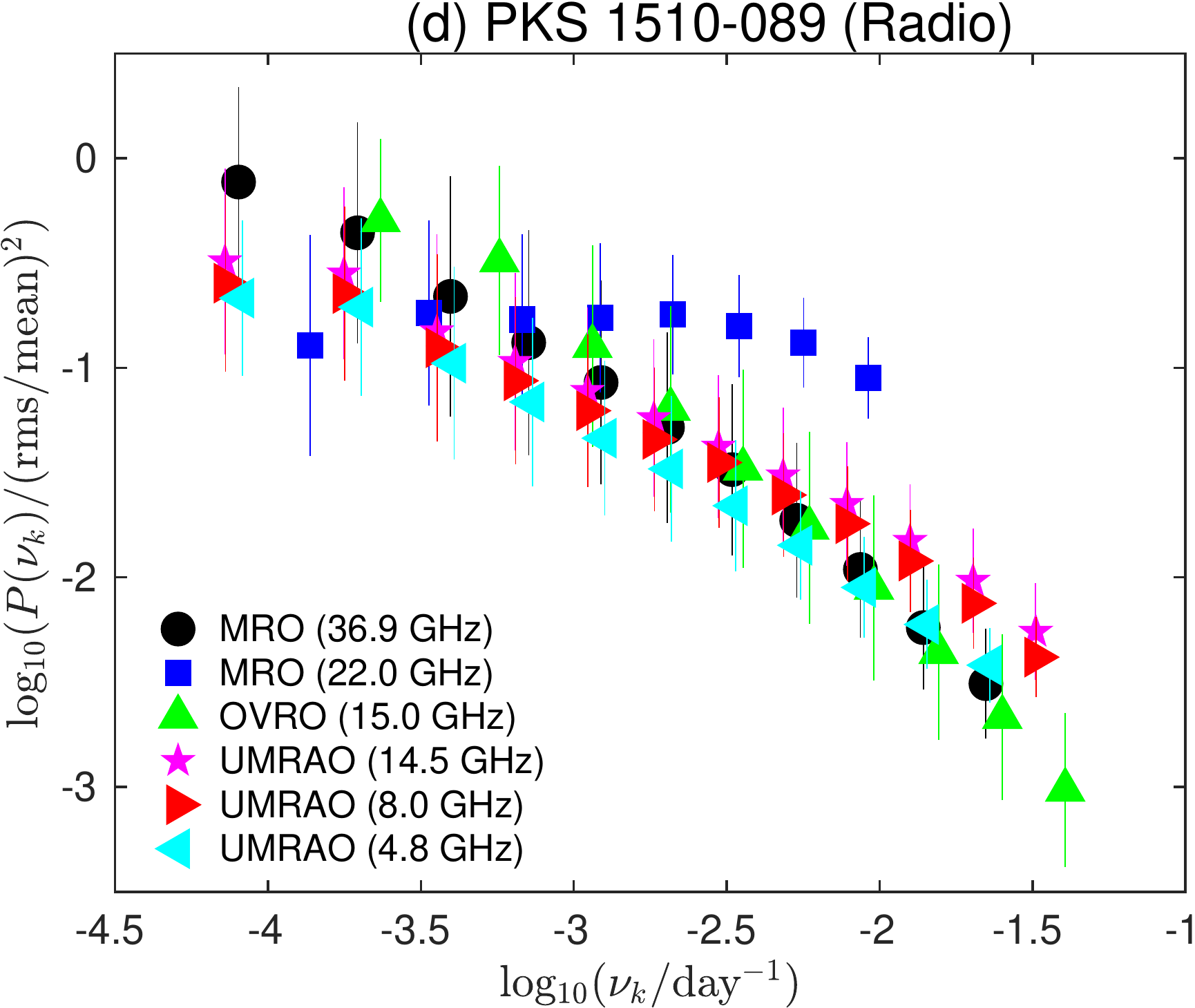}
\includegraphics[width=0.33\textwidth]{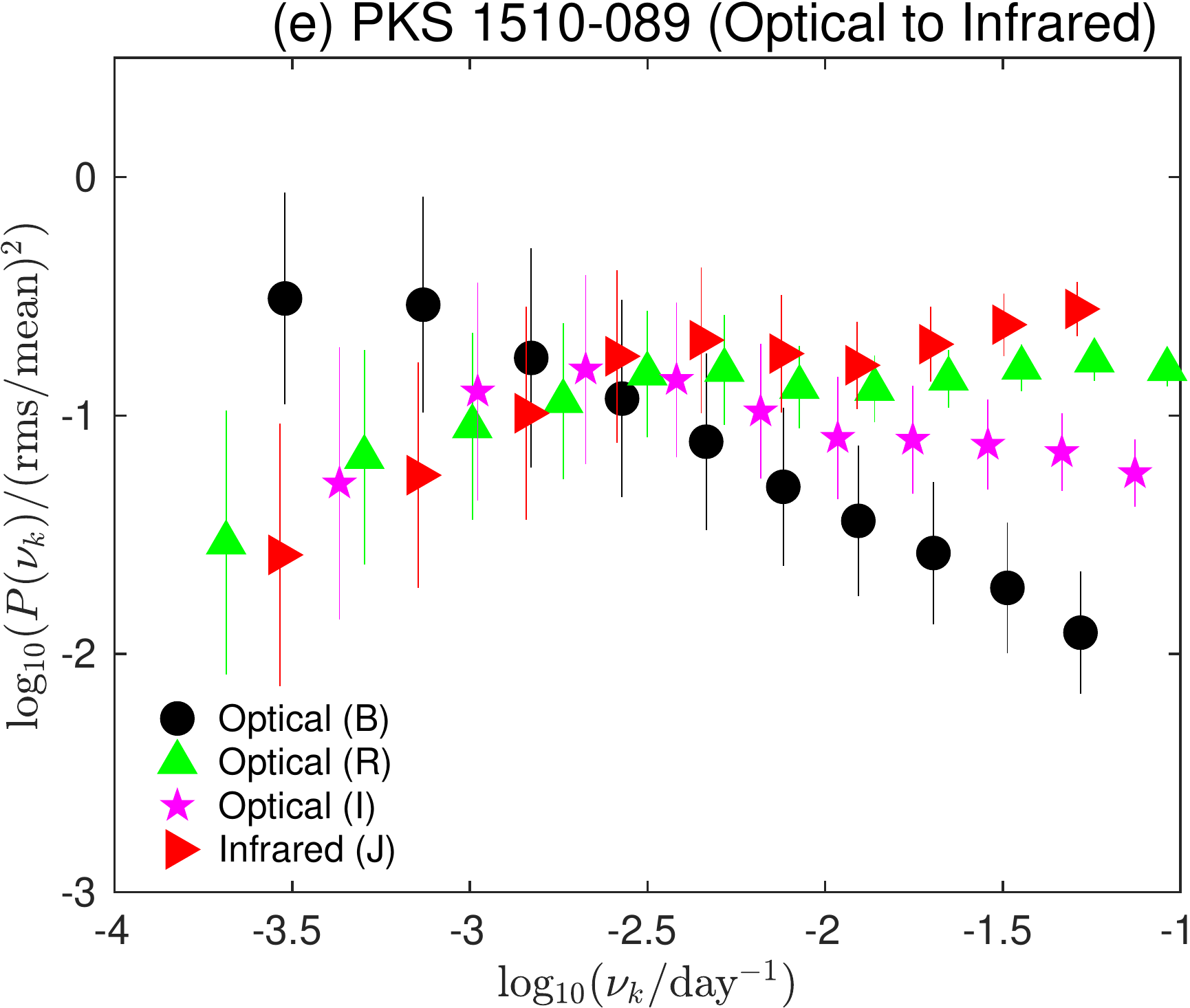}
\includegraphics[width=0.33\textwidth]{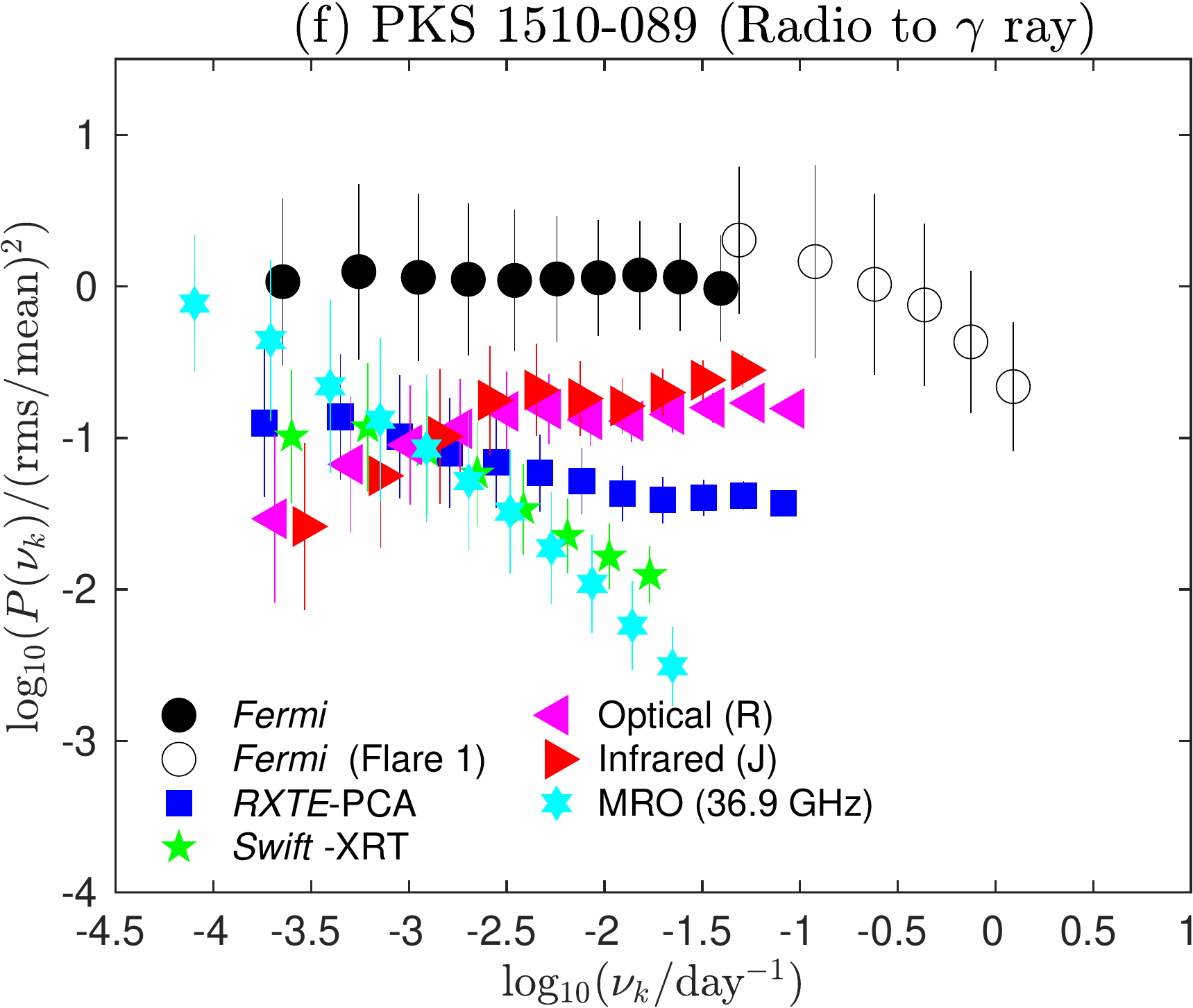}
}

\caption{Square fractional variability as a function of timescale for the blazars 3C\,279 (panels a, b, and c) and PKS\,1510$-$089 (d, e, and f). The filled symbols show mean and standard deviation of best-fit PSDs given by the PSRESP method each frequency, while open symbols show different gamma-ray flare epochs. In panel (c) we have included the $R$-band intranight PSDs presented in \citet{Goyal21} for the blazar 3C\,279.}
\label{fig:jointpsds}
\end{figure*}

Details on the light curves used for the analysis and the PSDs derived for the given light curve, along with the best-fit PSDs and the maximum $p_{\beta}$, are summarized in Table~\ref{tab:psd}. Figures~\ref{fig:psd3c279} and ~\ref{fig:psdpks1510} present the corresponding best-fit PSDs for the analyzed light curves. The corresponding $p_\beta$ distribution curves are given in Figures~\ref{appfig:beta3c} and ~\ref{appfig:betapks} for the blazars 3C\,279 and PKS\,1510$-$089, respectively. In our analysis, we have not subtracted the constant noise floor level. The PSRESP method also allows us to compute the rejection confidence for the input PSD shape; the maximum probability lower than $x$\% means that the rejection confidence ($1-p_{\beta}$) is higher than $(100-x)$\% for the (input) PSD model. In our analysis, we set the rejection confidence at 90\% for the input PSD model. Moreover, a direct comparison of the average fractional variability power for a given timescale is given by the $\nu_k P(\nu_k)$ vs. $\nu_k$ curves for the acceptable PSD fits in Figure~\ref{fig:jointpsds} for blazars 3C\,279 (panels a, b, and c) and PKS\,1510$-$089 (panels d, e, and f). The quantity $\nu_k P(\nu_k)$ is equivalent to the square fractional variability \citep[see][and references therein]{Goyal20}. In addition, we note that \citet{Goyal21} derived the $R$-band intranight variability PSDs for the blazar 3C\,279 (monitored on 2006 January 26, 2006 February 28, and 2009 April 20) and PKS\,1510$-$089 (monitored on 2009 May 1) using the PSRESP method. On two occasions (2006 January 26 and 2006 February 28), the intranight PSDs for the blazar 3C\,279 showed a good-fit to a power-law model ($p_\beta >$0.1) while the intranight PSD obtained on 2009 Apr 20 gave a bad fit ($p_\beta<$0.1). Similarly, the intranight PSD obtained for the blazar PKS\,1510$-$089 gave a bad fit. Therefore, we extend the frequency range of the optical power spectrum to minute timescales for the blazar 3C\,279 using the 2006 January 26 and 2006 February 28 intranight PSDs.  

It can be seen from Table~\ref{tab:psd} that the derived 98\% confidence limits on the best-fit PSD slopes using the PSRESP method are large in some cases. Therefore, it is fortunate that we can obtain representative PSD slopes (and the corresponding errors) on emission frequencies by averaging several PSD slopes at nearby frequencies. Such an approach is possible solely due to the availability of the large number of light curves analyzed by us in this study. For this, we average all the PSD slopes at different radio and optical frequencies; this gives the mean PSD slope at two different synchrotron frequencies while the combination of two X-ray PSD slopes and one GeV PSD slope gives an estimate of the mean PSD slope at IC frequencies of the emission spectrum (see Section~\ref{sec:intro}). As UMRAO 14.5\,GHz and OVRO 15.0\,GHz datasets are essentially the same \citep[][]{Aller85, Richards11}, for the calculation of mean PSD slope at radio-synchrotron frequencies we did not use the OVRO estimates since the UMRAO 14.5\,GHz PSD covers a wider range of variability frequencies. For averaging, we used the PSD slopes which were successfully represented by the single power-law model (i.e., $p_\beta> 0.1$). The mean PSD slope, $\beta_{\rm mean}$, was computed in a straightforward manner while the error was computed using the MC bootstrap method as follows. For a sample of PSD slopes, a slope is drawn from a Gaussian distribution of mean and standard deviation equal to $\beta$ and $\sigma$ of the best-fit PSD estimate, respectively (note that Table~\ref{tab:psd} reports errors equal to 2.354$\sigma$ of the fitted Gaussian function). The $\beta_{\rm mean}$ is computed. These two steps are repeated 1,000 times. The error is computed as the standard deviation of the distribution of $\beta_{\rm mean}$ values. Table~\ref{tab:psdmean} gives the summary of the mean PSD slopes and the corresponding standard deviation (1$\sigma$ error) for the synchrotron and IC portions of the emission continuum for these two blazars.

\section{Results}\label{sec:results}

We have obtained long-term multiwavelength variability PSDs of two luminous blazars, 3C\,279 and PKS\,1510$-$089, using well-sampled (days/weeks sampling interval), long-duration (years--decades), datasets with good photometric accuracy ($\sim$1--15\%). The variability PSDs cover $\sim$2-4 decades in temporal frequency, corresponding to decades/years/months/days timescales, depending on the waveband. The key results of our detailed analysis are as follows:

\begin{enumerate}

\item{For both of the targets, the radio variability PSDs are best described by the single power-law shapes with confidence ranging from $\sim$14\% to $\sim$97\% (Column 11; Table~\ref{tab:psd}). The best-fit $\beta$  range between $\sim$1.4 and $\sim$2.9 over the frequency range log$_{10}$$\nu_k$ (day$^{-1}$) between $-$4.2 and $-$1.2, indicative of flicker to red-noise type statistical characters of variability (Table~\ref{tab:psd}; Figures~\ref{fig:psd3c279} and ~\ref{fig:psdpks1510}). The normalizations of  these radio PSDs turn out to be consistent with one another within 1--2$\sigma$ uncertainties on overlapping variability frequencies (panels (a) and (d) of Figure~\ref{fig:jointpsds}).}

\item{The optical and infrared variability PSDs for the blazar 3C\,279 show acceptable fits to single power-law shapes with confidence in the range $\sim$56\%--$\sim$99\% (Column 11; Table~\ref{tab:psd}). The best-fit $\beta$ values show a range between $\sim$1.4 and $\sim$2.9 over the frequency range log$_{10}$$\nu_k$ (day$^{-1}$) between $-$3.93 and $-$1.09, indicative of flicker to red-noise variability (Table~\ref{tab:psd}; Figure~\ref{fig:psd3c279}). For the blazar PKS\,1510$-$089, the single power-law shape of the PSDs give acceptable fits with confidence $\geq$10\% to $B$, $R$, $I$, and $J$-band PSDs (Column 11; Table~\ref{tab:psd}). The $\beta$s range between $\sim$0.2 and $\sim$1.5 over the frequency range log$_{10}$$\nu_k$ (day$^{-1}$) between $-$3.68 and $-$1.03, closer to flicker noise. (Table~\ref{tab:psd}; Figure~\ref{fig:psdpks1510}). The variability PSDs at $V$, $H$, and $K$-bands do not provide acceptable fits to single power-law shapes for the blazar PKS\,1510$-$089 (Table~\ref{tab:psd}; Figure~\ref{fig:psdpks1510}). The normalizations of these PSDs again are consistent with one another  within 1-2$\sigma$ uncertainties where the  variability frequencies overlap (panels (b) and (e) of Figure~\ref{fig:jointpsds}).}

\item{For both blazars, the X-ray variability PSDs are  well described by single power-law shapes with confidence ranging from $\sim$48\% to $\sim$99\% (Column 11; Table~\ref{tab:psd}). The best-fit $\beta$s are between $\sim$1.2 and $\sim$1.5 over the frequency range log$_{10}$$\nu_k$ (day$^{-1}$) between $-$3.76 and $-$1.08, indicative of nearly flicker-noise variability (Table~\ref{tab:psd}; Figures~\ref{fig:psd3c279} and ~\ref{fig:psdpks1510}). The PSDs obtained using {\it RXTE-}PCA and {\it Swift-}XRT datasets for both of the targets complement each other well for overlapping variability frequencies (panels (c) and (f) of Figures~\ref{fig:jointpsds}).}

\item{For both FSRQs, the $\gamma$-ray variability PSDs are also very well described by single power-law shapes with confidence ranging from $\sim$30\% to $\sim$59\% (Column 11; Table~\ref{tab:psd}). The best-fit $\beta$ values are $\sim$1.0 and $\sim$0.9 for the blazars 3C\,279 and PKS\,1510$-$089, respectively, for the frequency range log$_{10}$ $\nu_k$ (day$^{-1}$) between $-$3.64 and $-$1.40 (1-week integration bin light curves), clearly indicative of flicker-noise (Table~\ref{tab:psd}; Figures~\ref{fig:psd3c279} and ~\ref{fig:psdpks1510}). The short-term variability PSD slopes derived using  3-hr integration bin light curves during bright phases (and of duration 3 weeks), identified as Flare 1, Flare 2, and  Flare 3 for the blazar 3C\,279 and Flare 1 for the blazar PKS\,1510$-$089, are $\sim$1.4, $\sim$2.0, $\sim$1.5, and $\sim$1.7, respectively, covering the frequency range log$_{10}$$\nu_k$ (day$^{-1}$) between $-1.31$ and $+0.51$ (Table~\ref{tab:psd}; Figures~\ref{fig:psd3c279} and ~\ref{fig:psdpks1510}). The inclusion of short-term $\gamma$-ray variability PSDs allows us to characterize the variability spectrum across four decades without any gaps for these blazars, albeit only during the flares, when the fluxes were high enough to allow measurements over those short periods. The normalizations of the short-term $\gamma$-ray variability PSDs appear to be smooth extrapolations from those at longer timescales (panels (c) and (f) of Figure~\ref{fig:jointpsds}).}

\item{The inclusion of $R$-band intranight PSDs from \citet{Goyal21} allows us to construct, for the first time, the power spectrum across $\sim$6 decades of variability frequencies for the blazar 3C\,279. The normalization of intranight PSDs appears to be a smooth extrapolation from longer timescales (panel (c) of Figure~\ref{fig:jointpsds}), indicating the dominance of a single variability process operating across 6 decades of time.}

\item{A direct comparison of variability power over different timescales probed by our analysis indicates more power ($\geq$2$\sigma$ level) on timescales shorter than 100 days in $\gamma$-rays, as compared to radio, infrared, optical and X-ray energies for the blazar 3C\,279 (panel (c) of Figure~\ref{fig:jointpsds}). The blazar PKS\,1510$-$089 also exhibits more variability power on timescales shorter than 100\,days in $\gamma$-ray, as compared to radio and X-ray energies (panel f of Figure~\ref{fig:jointpsds}). However, for this blazar, the variability power at optical and infrared energies is indistinguishable from those at $\gamma$-ray energies for overlapping variability frequencies (panel (f) of Figure~\ref{fig:jointpsds}).}

\item{The derivation of variability PSDs at multiple frequencies of the emission spectrum presented here allows us to obtain mean PSD slopes and the associated 1$\sigma$ uncertainties at two widely separated portions of the synchrotron part of the SED as well as at IC emission frequencies for the two targets (see Table~\ref{tab:psd}). For 3C\,279, the mean PSD slope at radio-synchrotron frequencies is 2.34$\pm$0.16 while at optical/infrared-synchrotron frequencies it is 1.95$\pm$0.22; however, at IC frequencies it is 1.20$\pm$0.15. For the blazar PKS\,1510$-$089, the mean PSD slope at radio-synchrotron frequencies is 1.60$\pm$0.08, at optical/infrared-synchrotron frequencies it is 0.77$\pm$0.23, and at IC frequencies is 1.30$\pm$0.16 (Table~\ref{tab:psdmean}). The mean $\gamma$-ray PSD slope for short-term variability turns out to be 1.60$\pm$0.70 after averaging the four PSD slope estimates for the targets (Flare 1,  Flare 2,  Flare 3 for 3C\,279 and Flare 1 for PKS\,1510$-$089). The mean PSD slopes at synchrotron frequencies and IC frequency are different from each other at the $\geq$2$\sigma$ level for the blazar 3C\,279. The mean PSD slopes at synchrotron and IC frequencies turns out to be consistent with each other within 2$\sigma$ uncertainty for the blazar  PKS\,1510$-$089. On the whole, the synchrotron PSD slopes of 3C 279 are steeper than those of PKS\,1510$-$089 while their IC PSD slopes are similar.}

\end{enumerate}

\section{Discussion and Summary}\label{sec:discussion}

\subsection{Comparison of PSD slopes with the literature}
The $\gamma$-ray PSD slopes computed with the PSRESP technique used here for the blazars 3C\,279 and PKS\,1510$-$089 employing the decade-long light curves are consistent with those given in previous studies \citep[$\beta$$\sim$0.6 and $\sim$1.0 for 3C\,279 and PKS\,1510$-$089, respectively;][]{Meyer19}. \citet{Nilsson18} reported the optical $R$-band PSD slopes $\sim$1.5 and $\sim$0.9 for the blazars 3C\,279 and PKS\,1510$-$089, respectively which are also consistent within errors with our estimates. \citet{Chatterjee08} reported the PSD slopes for the blazar 3C\,279 at  X-ray, $R$-band optical, and radio frequencies as $\sim$2.3, $\sim$1.7, and $\sim$2.3, respectively, using the decade-long {\it RXTE-}PCA data, $R$-band, and 14.5\,GHz UMRAO datasets for the period 1996-2007. Note that our longer duration X-ray, optical $R$-band and 14.5\,GHz light curves include datasets reported in \citet{Chatterjee08}. The X-ray PSD slope obtained in their analysis of {\it RXTE} data is significantly steeper, $\beta = 2.3\pm0.3$, as compared to our result of $1.2\pm0.3$. Our X-ray PSD slope is corroborated by the {\it Swift-}XRT dataset over a similar timescale ($\beta =1.4\pm0.7$); therefore, this discrepancy cannot be due to the different lengths of light curves used in \citet{Chatterjee08} and in our analyses for the PCA data. Good matches (within uncertainty), however, are found between the optical and radio PSD slopes between our analysis and that of \citet{Chatterjee08}. Using the full-duration UMRAO light curves for both blazars, \citet{Park17} reported the PSD slopes, $\beta$$\sim$2.5, $\sim$2.2, and $\sim$1.6 at 14.5\,GHz, 8.0\,GHz, and 4.8\,GHz, respectively, for 3C\,279 and $\beta$$\sim$1.5, $\sim$1.3, and $\sim$1.4 at 14.5\,GHz, 8.0\,GHz, and 4.8\,GHz, respectively, for PKS\,1510$-$089; these are consistent within errors with our results. 

We note that \citet[][]{Ryan19} performed direct modeling of 9.5-year long, weekly-binned $\gamma$-ray light curves for a sample of blazars, including the targets studied here, using the CARMA process \citep[][]{Kelly14}. Their analysis revealed a break on a year-like timescale in the PSD of the blazar PKS\,1510$-$089 while the PSD of 3C\,279 showed no hint of a break (meaning that the relaxation timescale could be longer than the length of the time series for this source). Our PSD analysis for the blazar PKS\,1510$-$089 contradicts the results of \citet[][]{Ryan19} in the sense that our $\gamma$-ray PSD is best-fitted with a single power-law model with a reasonably good probability ($p_\beta$=0.3; Table~\ref{tab:psd}). For the blazar 3C\,279, \citet[][]{Ryan19} reports no hint of a break, indicating a good match between our PSD modeling and that reported from the CARMA modeling.

\subsection{Comment on reported quasi-periodicities}
We now mention the reported QPOs on year-like timescales in $\gamma$-ray and optical/infrared light curves for the period 2005-2014 claimed by \citet{Sandrinelli16} for the blazars 3C\,279 ($V$, $R$, and $K$-bands) and PKS\,1510$-$089 ($V$, $R$, $J$, and $K$-bands). We also use the same dataset presented in \citet{Sandrinelli16} and supplement it with other datasets to increase the length of the analyzed light curves. Our $\gamma$-ray PSDs are well-represented by the single power-law spectral shapes with high confidence ($p_\beta$$\geq$0.76). Moreover, the optical/infrared PSDs of the blazar 3C\,279 are represented by the single power-law spectral form adopted in our analysis with high confidence ($p_\beta >$0.6). For the blazar PKS\,1510$-$089, the PSDs obtained at $B$, $R$, $I$, and $J$-bands are also well represented by a single power-law spectral form with decent confidence ($p_\beta > 0.18$, meaning that the rejection confidence for the single power-law spectral shape is $<$82\%). The PSDs obtained at $V$, $H$, and $K$-bands, where the data are more limited, give poor fits, with rejection confidence $\geq$95\%. The single power-law PSDs obtained in our analysis argue against the presence of reported QPOs at $V$, $R$, and $K$-bands for the blazar 3C\,279 and at $R$ and $J$-bands for the blazar PKS\,1510$-$089. This discrepancy between our results and those of \citet{Sandrinelli16}  could arise from the use of different methods of PSD estimation (Lomb-Scargle Periodogram in their analysis vs.\ DFT in our analysis) as well as the longer lengths of the light curves used in our study. If a QPO were always present, a longer duration light curve should reveal a higher significance for it, but if a QPO were transitory then additional data would weaken its significance. The lack of reproducibility is not uncommon in blazar studies as many previously claimed QPOs in the literature could not be confirmed when different methods or longer duration light curves were used for periodicity analyses \citep[for recent discussions, see][]{Goyal18, Covino19, Tarnopolski20, Covino20, Yang21}. Further investigation of this point is beyond the scope of this paper.  

\subsection{Possible interpretation for different statistical characters of low and high energy variability}\label{sec:ourmodel}
As the single-zone models do not explain the statistical properties of long-term multiwavelength variability adequately (Section~\ref{sec:intro}), based on the different statistical characters of multiwavelength variability exhibited by the BL Lac object PKS\,0735$+$178 ($\beta\sim$2 for radio/optical emissions but $\sim$1 at HE\,$\gamma$-ray energies), we hypothesized in \citet{Goyal17} that the entire broadband emission is produced in an extended and turbulent jet. We noted there that these results would arise if a single stochastic process is responsible for the variability at synchrotron frequencies while a linear superposition of two or more stochastic processes with widely different relaxation timescales produce variability at IC frequencies. We speculated that the drivers behind the synchrotron variability can be stochastic fluctuations in the jet plasma conditions (such as variations in the magnetic field or bulk Lorentz factor) leading to dissipation of energy over wide spatial scales. The radiative response of the accelerated particles will be delayed with respect to the input perturbations, thereby producing the red-noise variability at synchrotron frequencies on timescales smaller than the relaxation timescales of the process. Because we did not recover the flattening of PSDs on longer timescales ($\geq$1,000\,days), we noted that the driver should be related to some global magnetohydrodynamic timescale. At IC frequencies, however, due to inhomogeneities in the population of photons available for upscattering, an additional stochastic process with relaxation timescale equal to a light crossing time $\sim$1\,day (a valid approximation for a  boosted jet with bulk Lorentz factor 30) was envisaged. IC variability is expected to show red-noise character on timescales shorter than $\sim$day in our model. Our hypothesis was supported in \citet{Goyal20} where the VHE\,$\gamma$-rays also exhibited $\beta$$\sim$1 for the blazars Mrk\,421 and PKS\,2155$-$304 whose TeV emission is often concluded to be due to SSC processes \citep[][]{Abdo11a, Hess12, Aleksic12, Madejski16b, Abdalla20, Dmytriiev21}. 

The best-fit PSD slopes for all of the individual emission frequencies that we report here range between $\beta$$\sim$0.2-2.9; however, the statistically significant slopes span a narrower range, 0.7 to 2.9, which correspond to flicker and red-noise statistical characters of variability. We note that in many cases, the reported 98\% confidence limit on the PSD estimate is large (Table~\ref{tab:psd}) which renders a strict statistical classification of the variability unfeasible. The mean PSD slopes, however, obtained at radio-synchrotron and optical-synchrotron frequencies correspond  to statistical characters consistent with strict red-noise $\beta$$\sim$2 (except for PKS\,1510$-$089 for which the mean $\beta$ at optical/infrared frequency is $\sim$0.8) and to a strict flicker-noise ($\beta$$\sim$1) statistical character at IC emission frequency on timescales ranging from decades/years to weeks/days. As such their broadband variability behaviors basically can be understood within our model, as summarized in the  previous paragraph and given in more detail in \citet{Goyal17}. 

\subsection{Flicker-noise type variability at optical/IR energies of PKS\,1510$-$089: jet + accretion disk}
For PKS\,1510$-$089 the optical/infrared variability seems to be better represented by flicker noise whenever the single power-law form was found to be an adequate fit. This is not expected within our model (Section~\ref{sec:ourmodel}) as the emission at optical/infrared frequencies is presumed to be dominated by synchrotron processes for such FSRQ type blazars, although the radio variability is well-represented by red-noise  processes for the same source and, of course, that is synchrotron radiation. Therefore, our model cannot be considered inadequate based on the flicker-noise character at synchrotron frequencies because in such a case we expect the radio frequencies to mimic the infrared/optical frequencies. 

However, the FSRQ PKS\,1510$-$089 is well-known to have a prominent thermal accretion disk component in its SED at UV/infrared frequencies, not only  during quiescent states but even when flaring \citep[e.g.,][]{DAmmando11, Nalewajko12, Aleksic14, Ahnen17}; therefore, one expects different variability properties if the jet emission \citep[as the source shows significant optical polarization variability at times;][]{Itoh16, Beaklini18} is diluted by the accretion disk component. Assuming that the majority of the disk emission originates within 100\, Schwarzschild radii of the accretion disk \citep[][]{Shakura76} with viscosity parameter $\alpha=0.01$, the characteristic light-crossing, gas orbital, and thermal/viscous timescales for this blazar turn out to be $\leq$1\,day, $\sim$31--73\,days, and $\sim$1.4--3.2\,yrs, respectively \citep{Kelly09}, for a black hole of mass range (3--7)$\times$10${^7}\,M_\odot$ (Table~\ref{tab:sample}). Since the optical/infrared PSDs cover 6--12\,yrs\, down to weeks timescale, such instabilities operating on these timescales seem to be valid candidates for the production of fluctuations in the disk emission in this source \citep[e.g.,][]{Wiita06}. Therefore, the flatter optical/infrared PSD slopes for this source could be reconciled if the emission at optical/infrared frequencies is driven by a linear combination of stochastic process(es) driving the disk emission at shorter timescales (days to years) and the jet emission which could include longer relaxation timescales ($\approx$1,000\,days) \citep[see, in this regard,][]{Kelly11, Sobolewska14}.

\subsection{Relaxation/characteristic timescales}
The combination of  $\gamma$-ray measurements made over the long-term (obtained with 1-week integrations) and short-term (obtained with 3-hr integrations for short durations only when source fluxes were high) allowed us to characterize the $\gamma$-ray variability PSDs across 4 decades. The mean PSD slope at short timescales is consistent with those at longer timescales, within the large uncertainties of the former and as such does not require the presence of a break as apparently seen in some blazars around $\sim$days timescales \citep[][]{Sobolewska14, Ryan19}. The optical PSD for the blazar 3C\,279 could be characterized across 6 decades due to the inclusion of intranight PSDs on two occasions \citep[2006 January 26 and 2006 February 28;][]{Goyal21}. The continuity of normalized power spectra across 6 decades indicates that a continuous (single) stochastic process appears to drive the variability at optical frequencies. 

In none of the analyzed light curves, however, do we recover the relaxation timescale of the process driving the variability, which should entail seeing $\beta$ changing from $\geq$1 to $\sim$0 at frequencies below the inverse of that timescale. The $\gamma$-ray and optical/infrared PSDs are in particular well represented by single power-law spectral shapes with high confidence up to timescales of the order of a decade or more. As the majority of HE jet emission is believed to originate at $\sim$1\,pc\, distance from the supermassive black hole \citep[see,][and references therein]{Sikora09, Harvey20}, the corresponding relaxation timescales for the $\gamma$-ray variability should turn out to be a few months in the observer's frame \citep[assuming a typical bulk Lorentz factor of 10;][]{Lister16}. The absence of PSD breaks in these blazars implies that the $\gamma$-ray emission originates, like the optical/infrared emission, from a rather extended region, in accordance with the expectation of our hypothesis discussed in Section~\ref{sec:ourmodel}. 

\subsection{Confronting the PSD slopes with multi-zone models}
The noise-like appearance of blazar light curves initiated efforts to model the jet emission due to many cells behind a shock \citep{Marscher14, Calafut15, Pollack16}. The detailed model developed by \citet{Marscher14} provides plausible flux light curves at both synchrotron and SSC emission frequencies but did not provide the PSDs, making a direct comparison difficult with these multiband variability observations. In contrast, the modeling done by \citet[][]{Calafut15} and \citet[][]{Pollack16} computes light curves as well as PSDs, but only at synchrotron emission frequencies.  For relativistic hydrodynamic simulations of 2D jets where the emission and its variability are shaped by the combination of bulk Lorentz factor fluctuations and the turbulence within the jets, PSD slopes in the range 2.1 to 2.9 due to bulk Lorentz factor changes and in the range 1.7 to 2.3 due to turbulence are expected \citep[][]{Pollack16}.  In this scenario, the majority of radio-optical PSD slopes (synchrotron emission frequencies) obtained in our study can be reconciled with those attributed primarily to changes in bulk Lorentz factors for the blazar 3C\,279 and  to turbulence for the blazar PKS\,1510$-$089. However, such models would need to be extended to show that they naturally can produce significantly flatter PSD slopes at IC energies, before considering them actually supported by our analysis.

\subsection{Key conclusions}
In  summary, we have explored the statistical properties of blazar light curves by means of variability PSDs using good photometric quality ($\sim$1-15\% flux accuracy), extensive multiwavelength datasets ($\geq$15 different frequencies covering 13 decades of the emission spectrum) available for the blazars 3C\,279 and PKS\,1510$-$089 over timescales exceeding a decade. The majority of PSDs show a good fit to single power-law spectral shapes over temporal frequencies ranging between $\sim$10$^{-4}$ and $\sim$10$^{-1}$ (day$^{-1}$). However, we note that PSDs on temporal frequencies longer than 10$^{-2.7}$ (day$^{-1}$) are poorly sampled, providing only a few up to 10 cycles in the PSDs. 

The mean PSD slopes at radio and optical-synchrotron emission frequencies are steeper from those at IC emission frequencies at $\geq$2$\sigma$ level on timescales ranging from decades to weeks/days for the blazar 3C\,279. On similar broad timescales, the mean PSD slopes at radio and optical-synchrotron emission frequencies and at IC emission frequencies turn out to be statistically indistinguishable ($\leq$2$\sigma$ level) from one another for the blazar PKS\,1510$-$089. The $\gamma$-ray variability PSDs covers $\sim$4 decades of the variability spectrum for these two blazars where the short-term variability PSDs connect smoothly to the long-term variability PSDs. The optical variability power spectrum covers $\sim$6 decades for the blazar 3C\,279 and the normalization of intranight PSDs appears to be a simple extrapolation from longer timescales. 

\begin{acknowledgments}

We thank the journal referee for insightful and constructive comments, which improved the clarity and content of the manuscript. We thank our internal {\it Fermi}-LAT reviewer, C.\,C. Cheung, for careful reading and several constructive comments on the manuscript. We also thank the {\it Fermi}-LAT publication board members, Philippe Bruel and Matthew Kerr, for useful comments. A.\ G. acknowledges the financial support from the Polish National Science Centre (NCN) through the grant 2018/29/B/ST9/02298. {\L}.\,S. was supported by the Polish NCN grant 2016/22/E/ST9/00061.
 S.\ Z. acknowledges NCN grant 2018/29/B/ST9/01793. S.\,K. acknowledges support from the European Research Council (ERC) under the European Unions Horizon 2020 research and innovation programme under grant agreement No.~771282. A.\,G. acknowledges many discussions with FSSC help desk contact Nestor Mirabel regarding the LAT analysis. The Monte-Carlo  simulations of the light curves have been performed at the Prometheus cluster of the Cyfronet PL grid under the computing grants `lcsims2' and `plglcsims'. A.\,G. thanks Micha{\l}~Ostrowski, Marek Sikora, and Stefan Wagner for useful discussions on the manuscript.

This research has used data from the University of Michigan Radio Astronomy Observatory which was supported by the University of Michigan; research at this facility was supported in part by NSF grants AST-8021250,  AST-8301234, AST-8501093, AST-8815678, AST-9120224, AST-9421979, AST-9617032, AST-9900723, AST-0307629, AST-0607523 and earlier awards, and by NASA under awards NNX09AU16G, NNX10AP16G, NNX11AO13G, and NNX13AP18G. This publication makes use of data obtained at Mets\"ahovi Radio Observatory, operated by Aalto University in Finland. This research has made use of data from the OVRO 40-m monitoring program which was supported in part by NASA grants NNX08AW31G, NNX11A043G and NNX14AQ89G, and NSF grants AST-0808050 and AST-1109911 and private funding from Caltech and the MPIfR. This paper has made use of up-to-date SMARTS optical/near-infrared light curves that are available at www.astro.yale.edu/smarts/glast/home.php. The research at Boston University was supported by a number of NASA {\it Fermi} Guest Investigator grants, most recently 80NSSC20K1567. The material is based upon work supported by NASA under award number 80GSFC21M0002.

The \textit{Fermi}-LAT Collaboration acknowledges generous ongoing support from a number of agencies and institutes that have supported both the development and the operation of the LAT as well as scientific data analysis. These include the National Aeronautics and Space Administration and the Department of Energy in the United States, the Commissariat \'a l'Energie Atomique  and the Centre National de la Recherche Scientifique/Institut National de Physique Nucl\'eaire et de Physique des Particules in France, the Agenzia Spaziale Italiana and the Istituto Nazionale di Fisica Nucleare in Italy, the Ministry of Education,Culture, Sports, Science and Technology (MEXT), High Energy Accelerator Research Organization (KEK) and Japan Aerospace Exploration Agency (JAXA) in Japan, and the K.~A.~Wallenberg Foundation, the Swedish Research Council and the Swedish National Space Board in Sweden.
Additional support for science analysis during the operations phase is gratefully acknowledged from the Istituto Nazionale di Astrofisica in Italy and the Centre National d\'Etudes Spatiales in France. This work performed in part under DOE Contract DE-AC02-76SF00515.

\end{acknowledgments}

\vspace{5mm}
\facilities{{\it Fermi}, {\it RXTE}, {\it Swift}, SMARTS, REM, Tuorla, VLBA-BU-BLAZAR, SKYNET, MRO, OVRO, UMRAO}


\appendix
\counterwithin{figure}{section}
\section{Spectral window function of the Fourier transformation on the observed datasets}\label{app:A}
An efficient diagnostic to discern the deleterious effects of Fourier transformation on the time series is to derive its spectral window function. The spectral window function measures the response of a Fourier transformation when applied on {\it unevenly} sampled time series. Specifically, it is 1 at the zeroth Fourier frequency and 0 at other frequencies for an {\it evenly} sampled time series \citep[e.g.,][]{Deeming75, Goyal20}. Figures~\ref{appfig:swf3c} and ~\ref{appfig:swfpks} present the sampling distribution and the corresponding spectral window functions for the long-term light curves of 3C\,279 and PKS\,1510$-$089, respectively. The spectral window functions corresponding to the {\it Fermi-}LAT curves for the targets show a uniform response across the frequency range as the LAT data are precisely sampled in 1-week intervals, and the other sampling values (14 and 21 days) are due to the missing TS$<$10 data points. The majority of spectral window function curves corresponding to optical, infrared, and radio light curves exhibit broad/narrow peaks around $\sim$300\,days timescale due to seasonal gaps in observation because of the source's proximity to the Sun. The sharpness of the peak in the spectral window function depends on the number of cycles covered. Therefore, those are more prominently seen in the few decades-long MRO and UMRAO radio light curves as compared to the nearly decade-long optical and infrared light curves. The spectral window function corresponding to the {\it RXTE-}PCA and {\it Swift-}XRT datasets also shows prominent peaks around the $\sim$300\,days timescale, due to the scheduling constraints of these observations with the satellites due to the position of the Sun. Moreover, as expected, the spectral window functions related to the 3-hr integration bin show a rather uniform response across the frequency range as the light curves are without any periodic gaps,
 like those of weekly-binned LAT light curves.             

\begin{figure*}[ht!]
\hbox{
\includegraphics[width=0.25\textwidth]{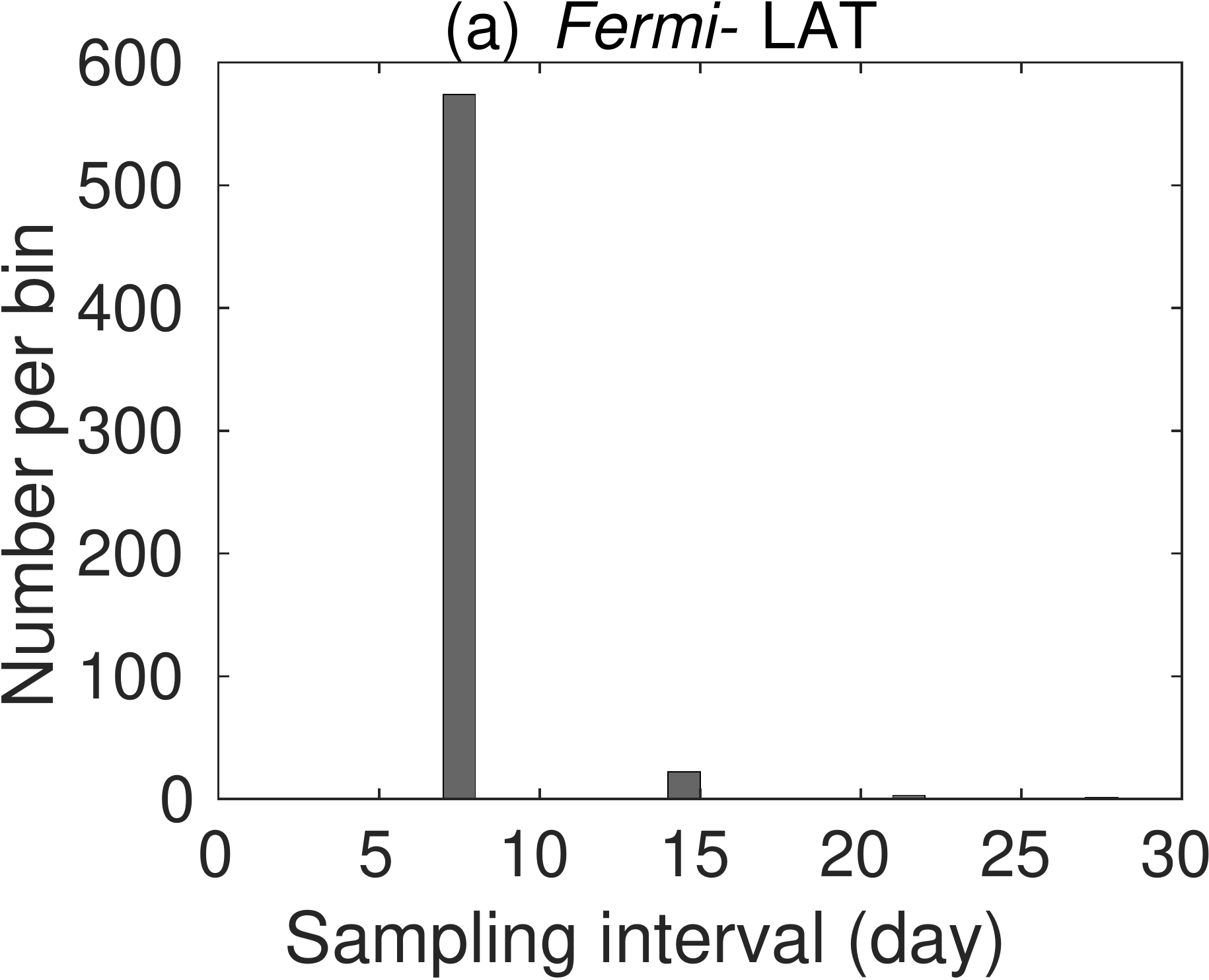}
\includegraphics[width=0.25\textwidth]{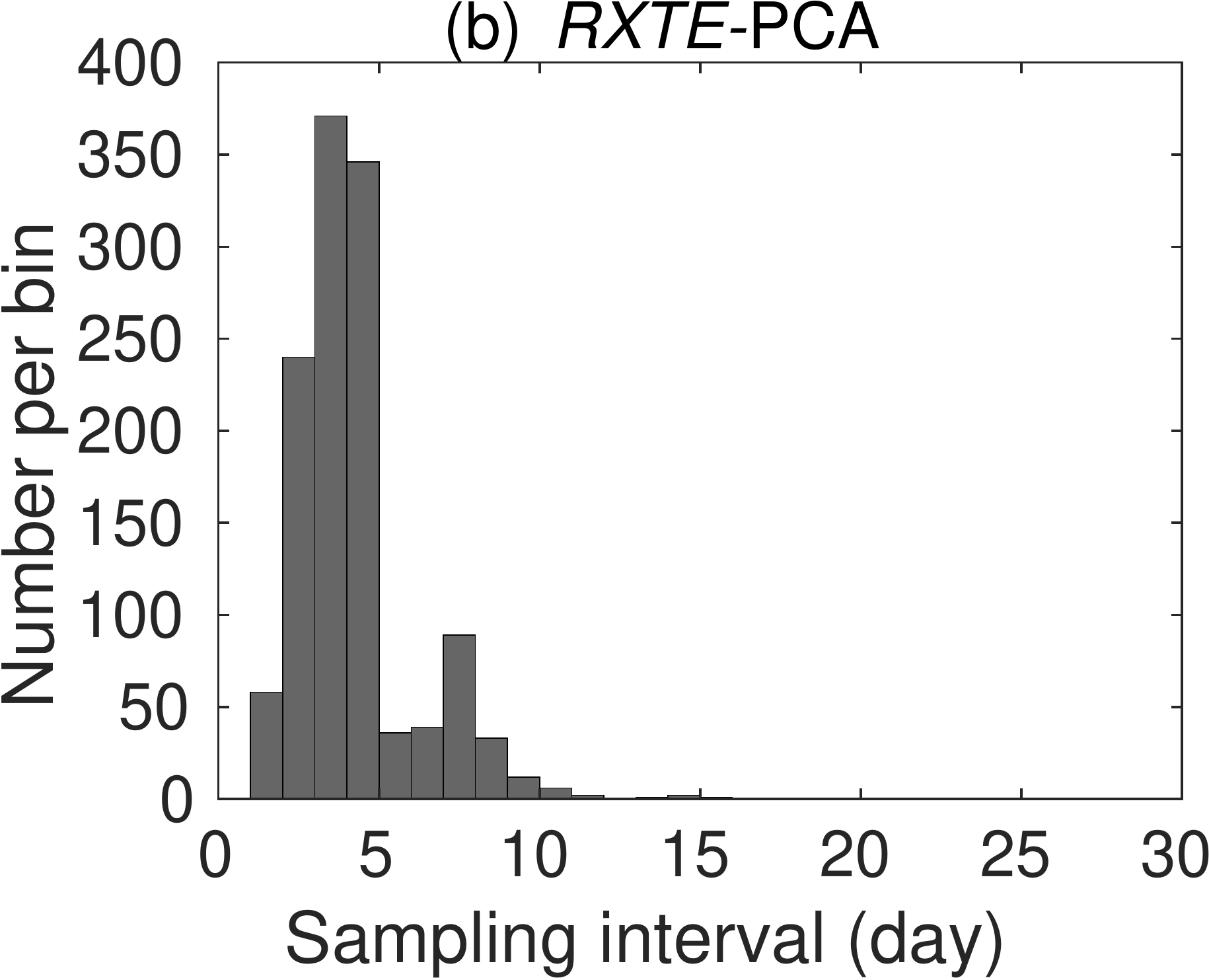}
\includegraphics[width=0.25\textwidth]{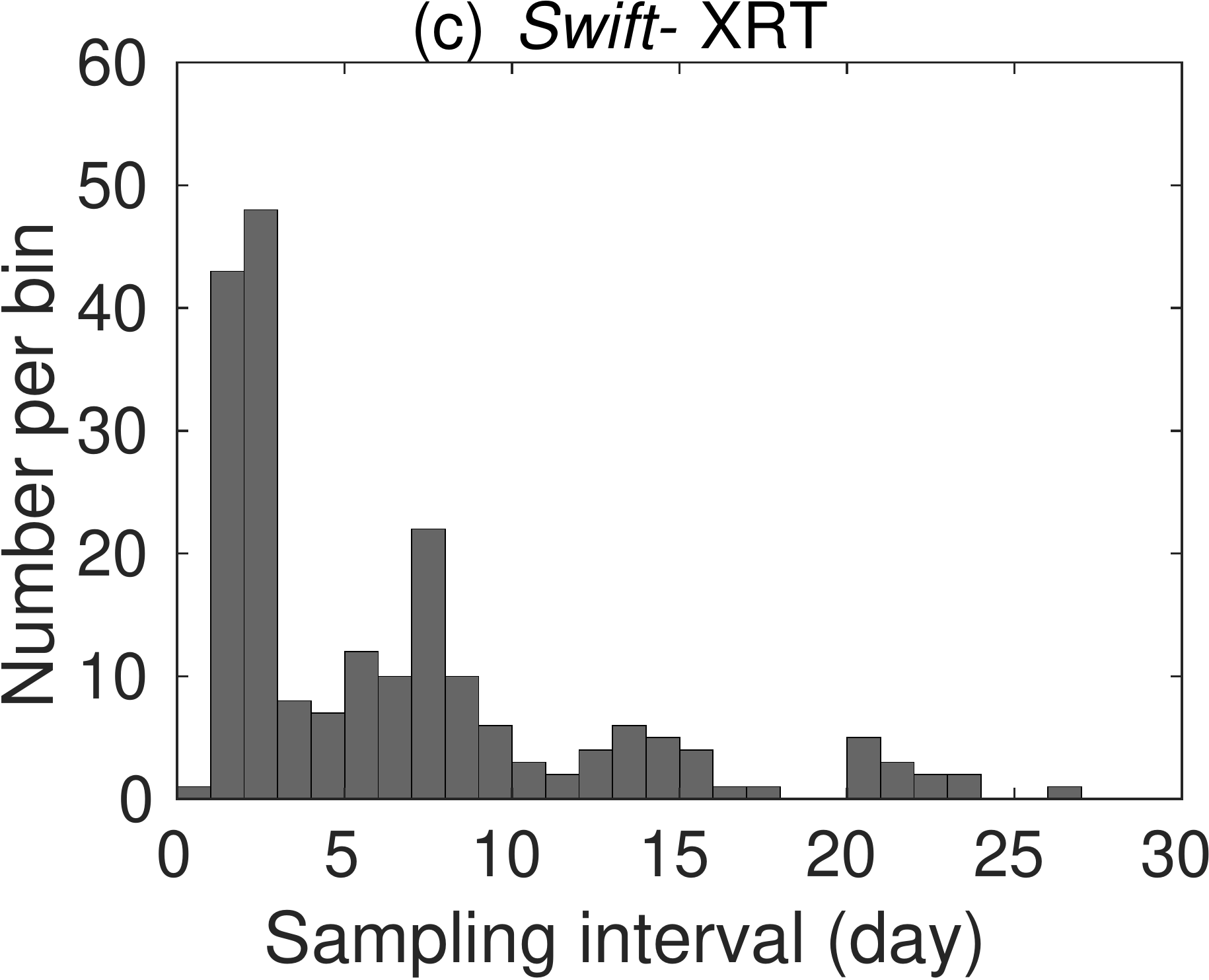}
\includegraphics[width=0.25\textwidth]{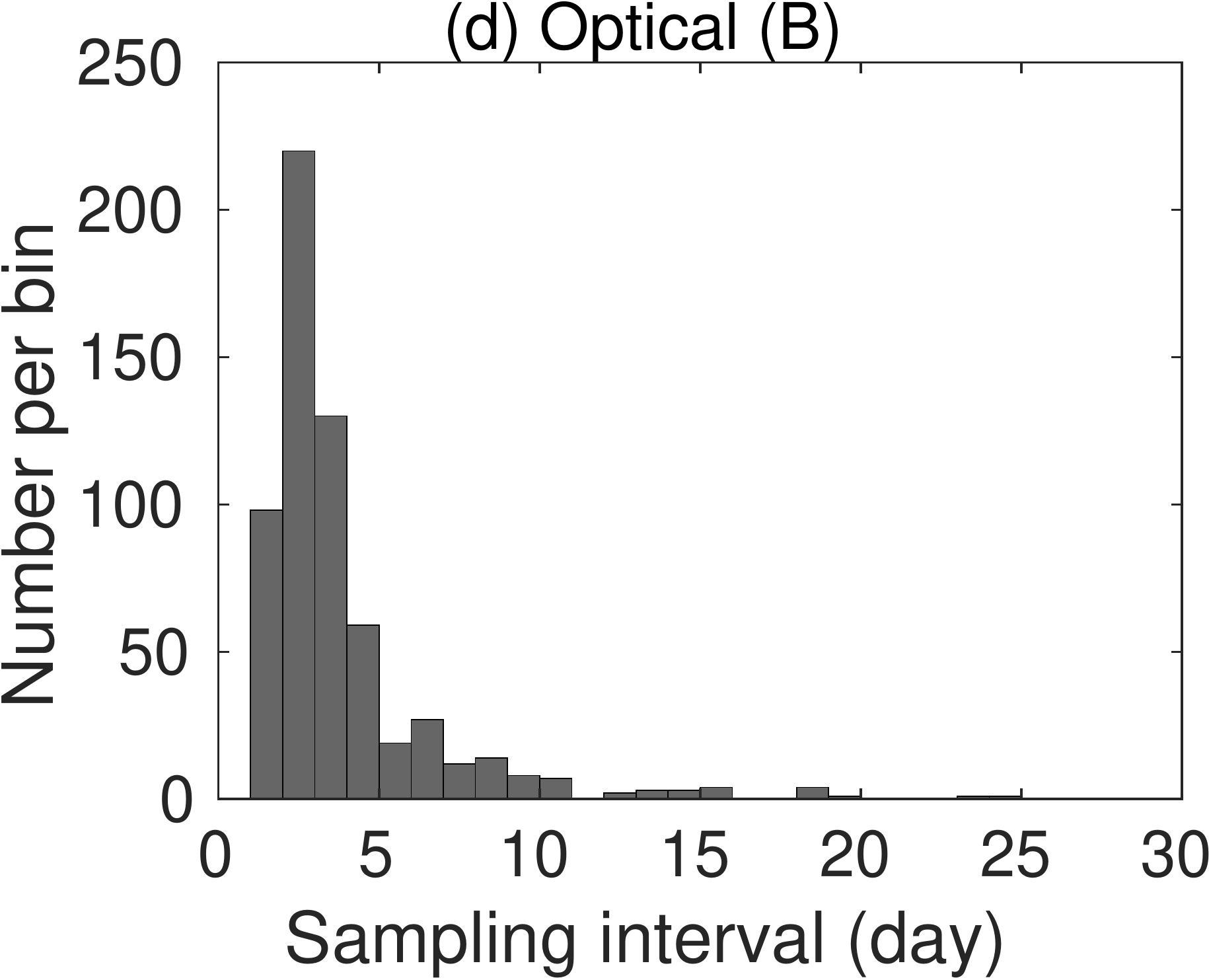}
}
\hbox{
\includegraphics[width=0.25\textwidth]{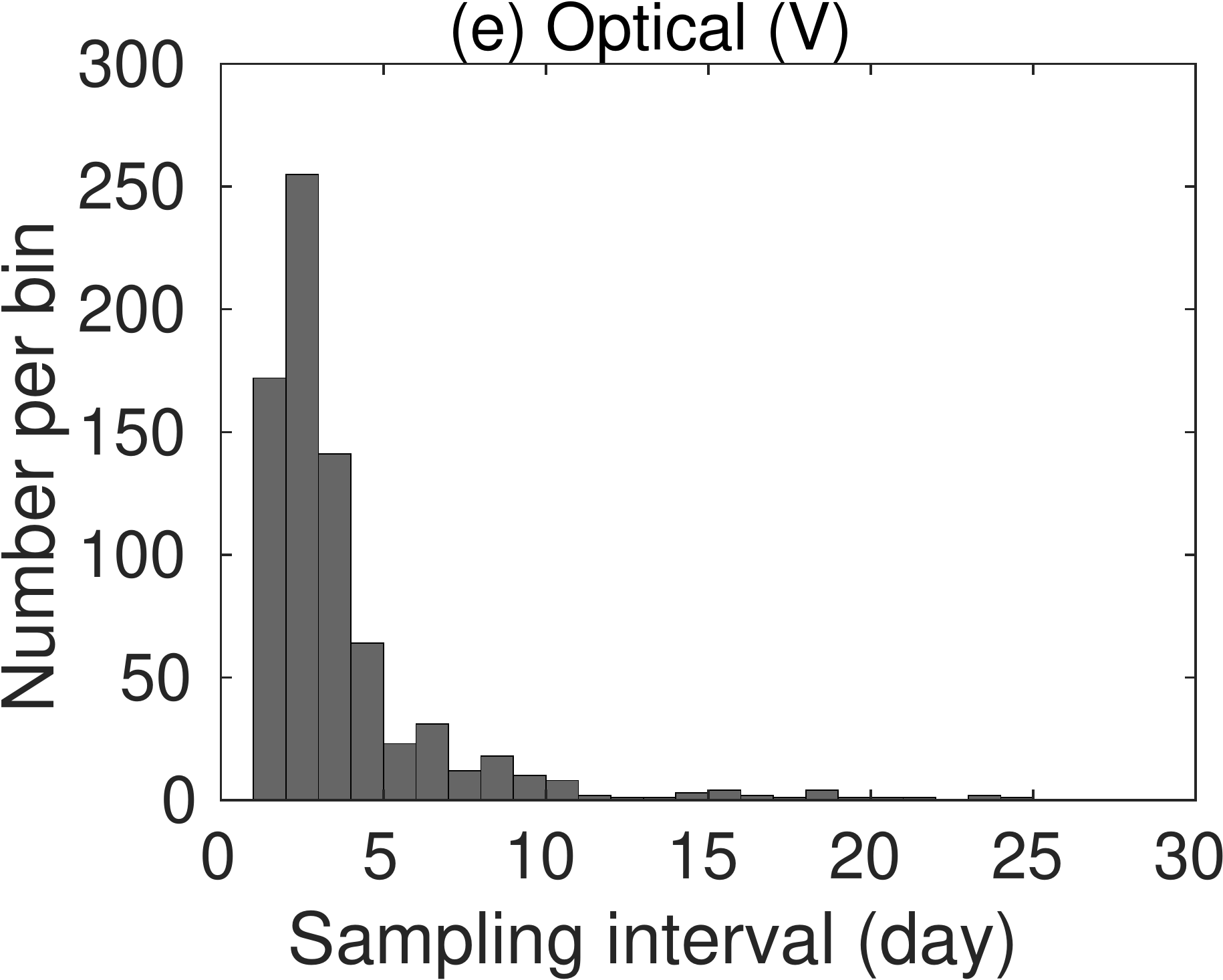}
\includegraphics[width=0.25\textwidth]{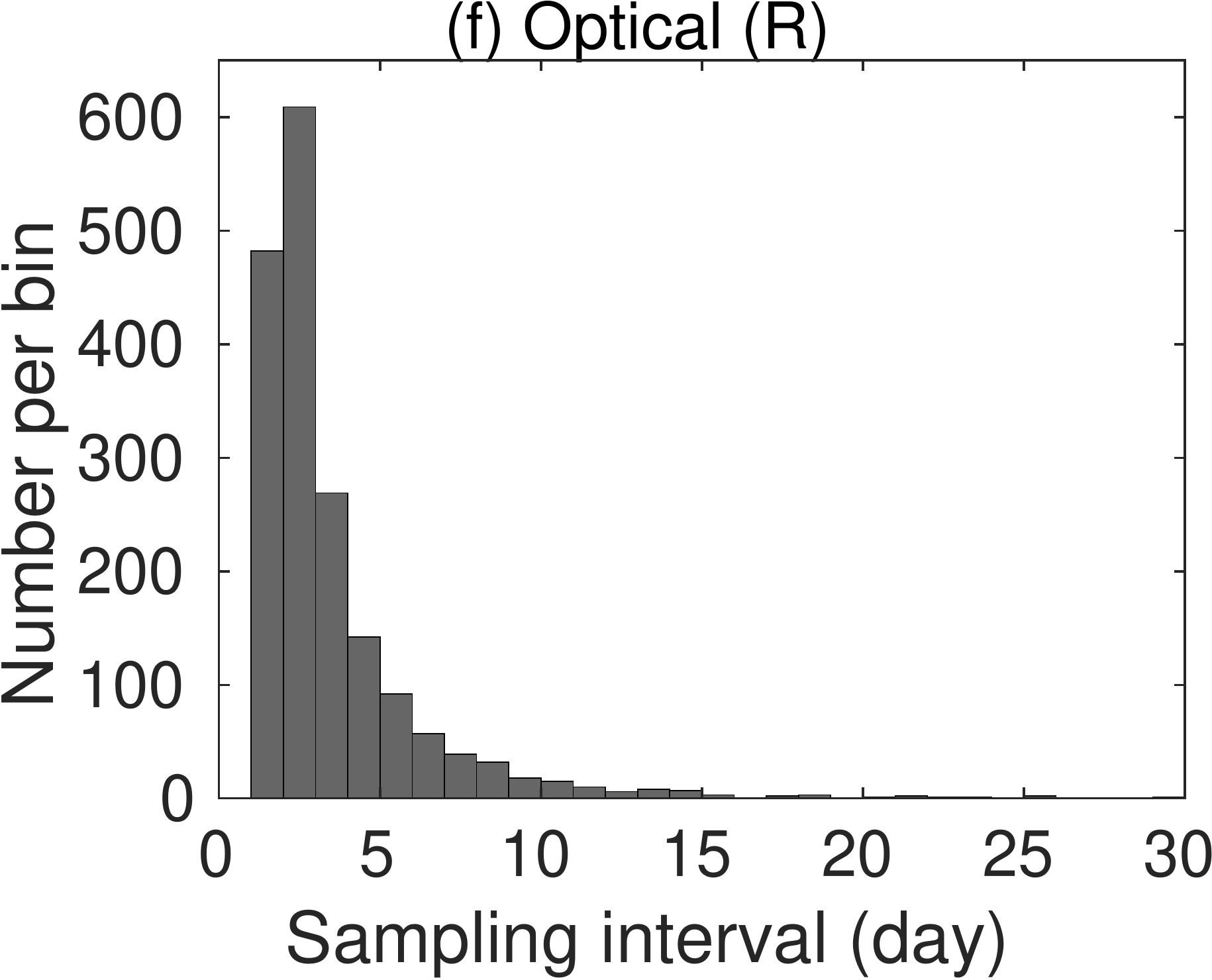}
\includegraphics[width=0.25\textwidth]{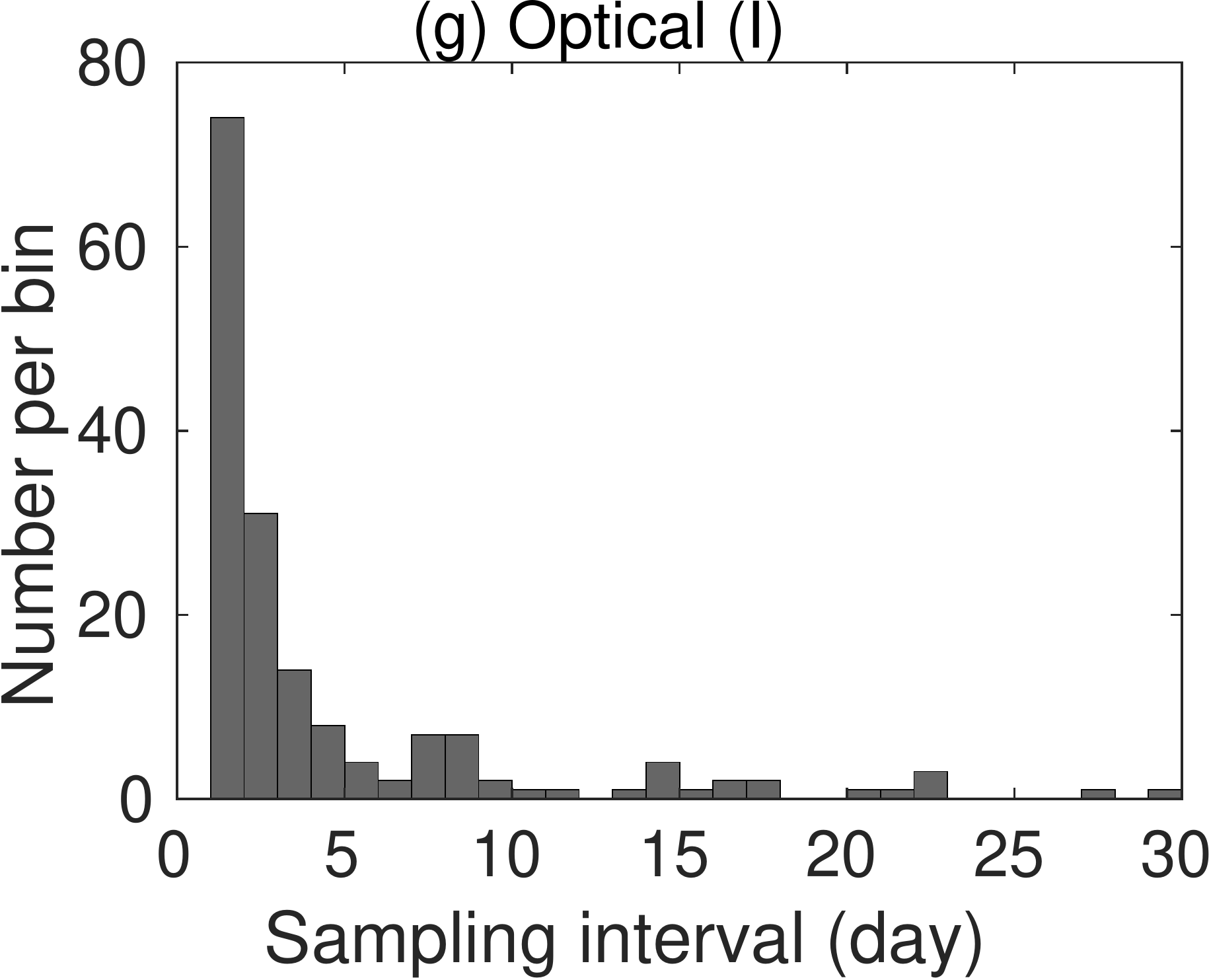}
\includegraphics[width=0.25\textwidth]{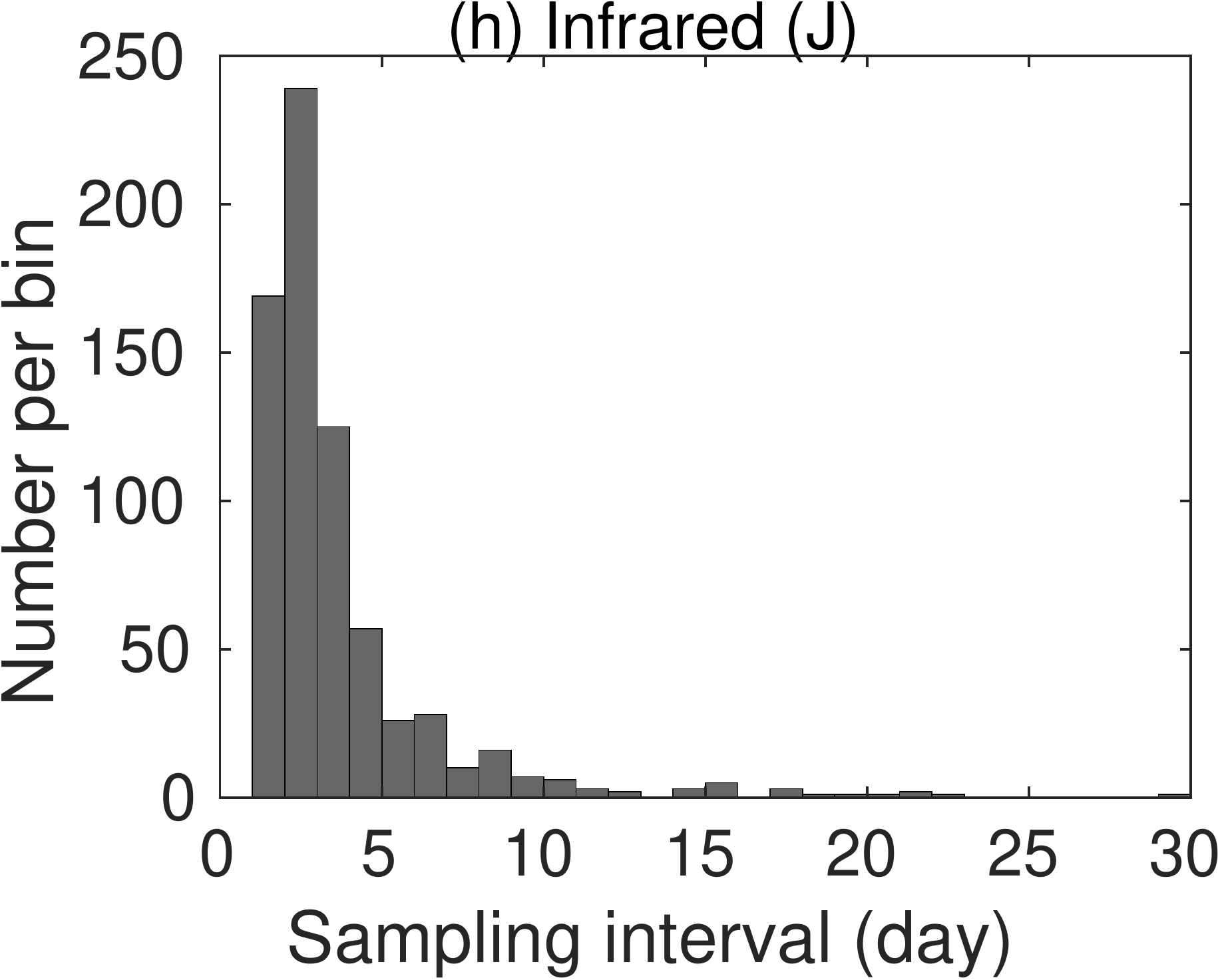}
}
\hbox{
\includegraphics[width=0.25\textwidth]{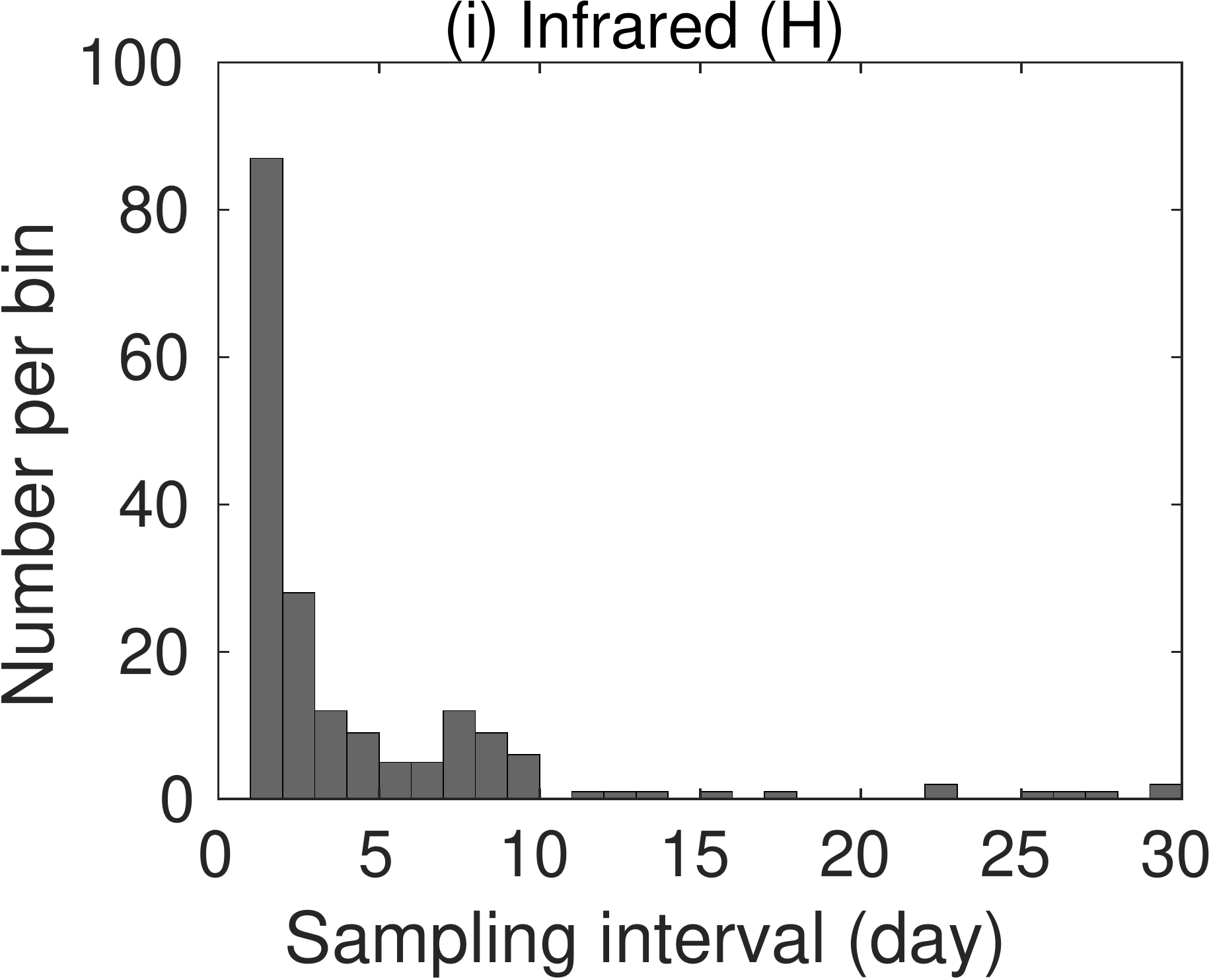}
\includegraphics[width=0.25\textwidth]{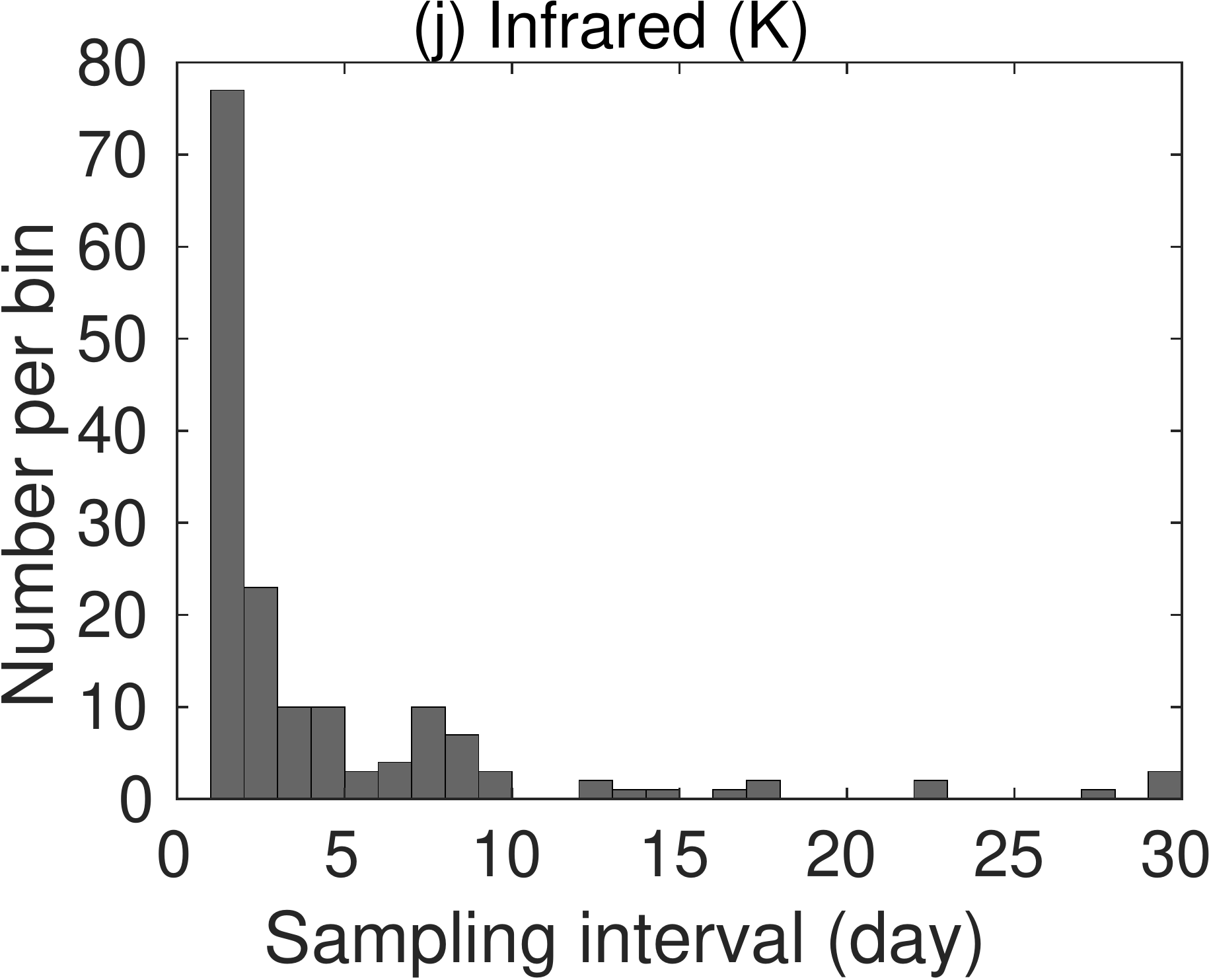}
\includegraphics[width=0.25\textwidth]{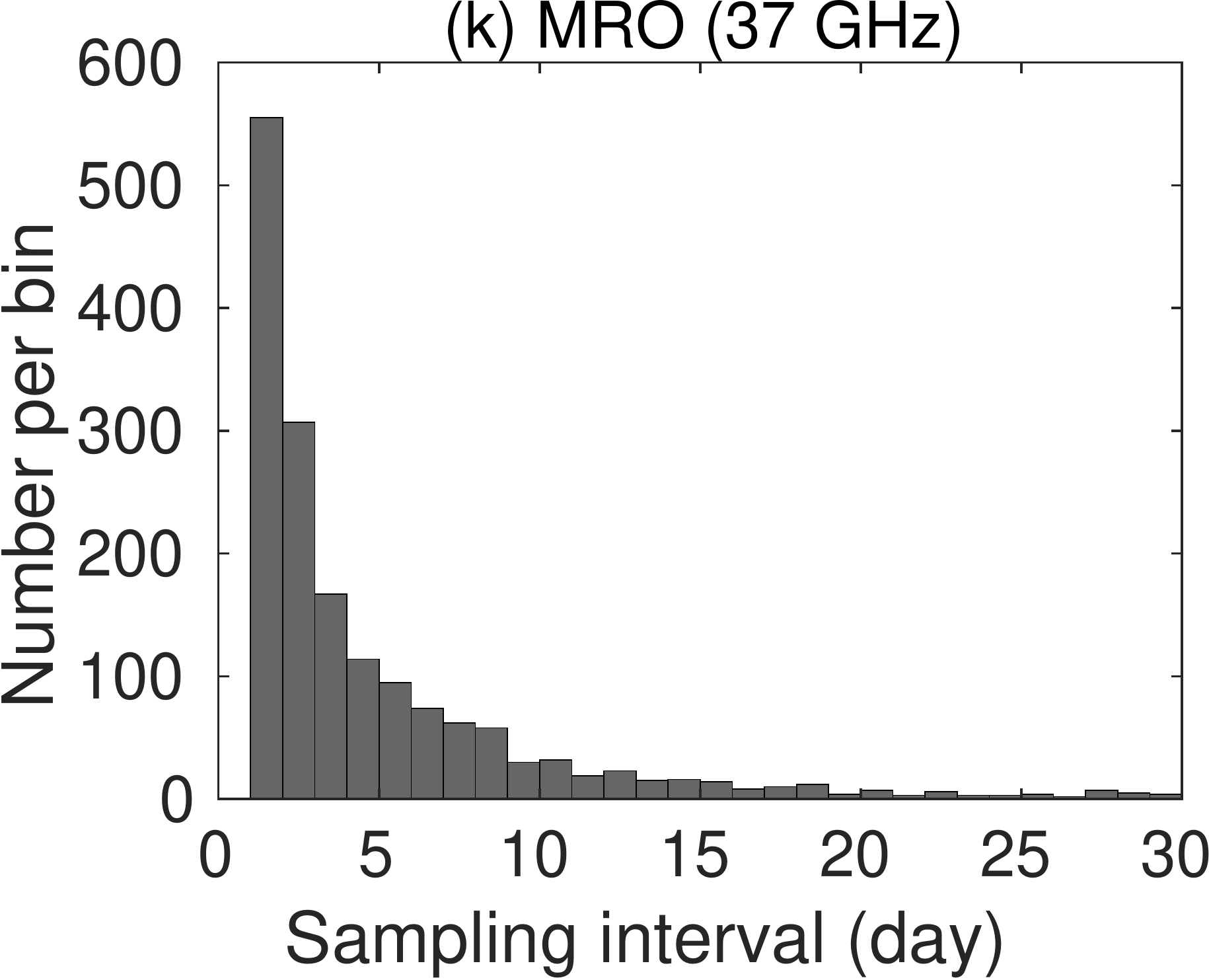}
\includegraphics[width=0.25\textwidth]{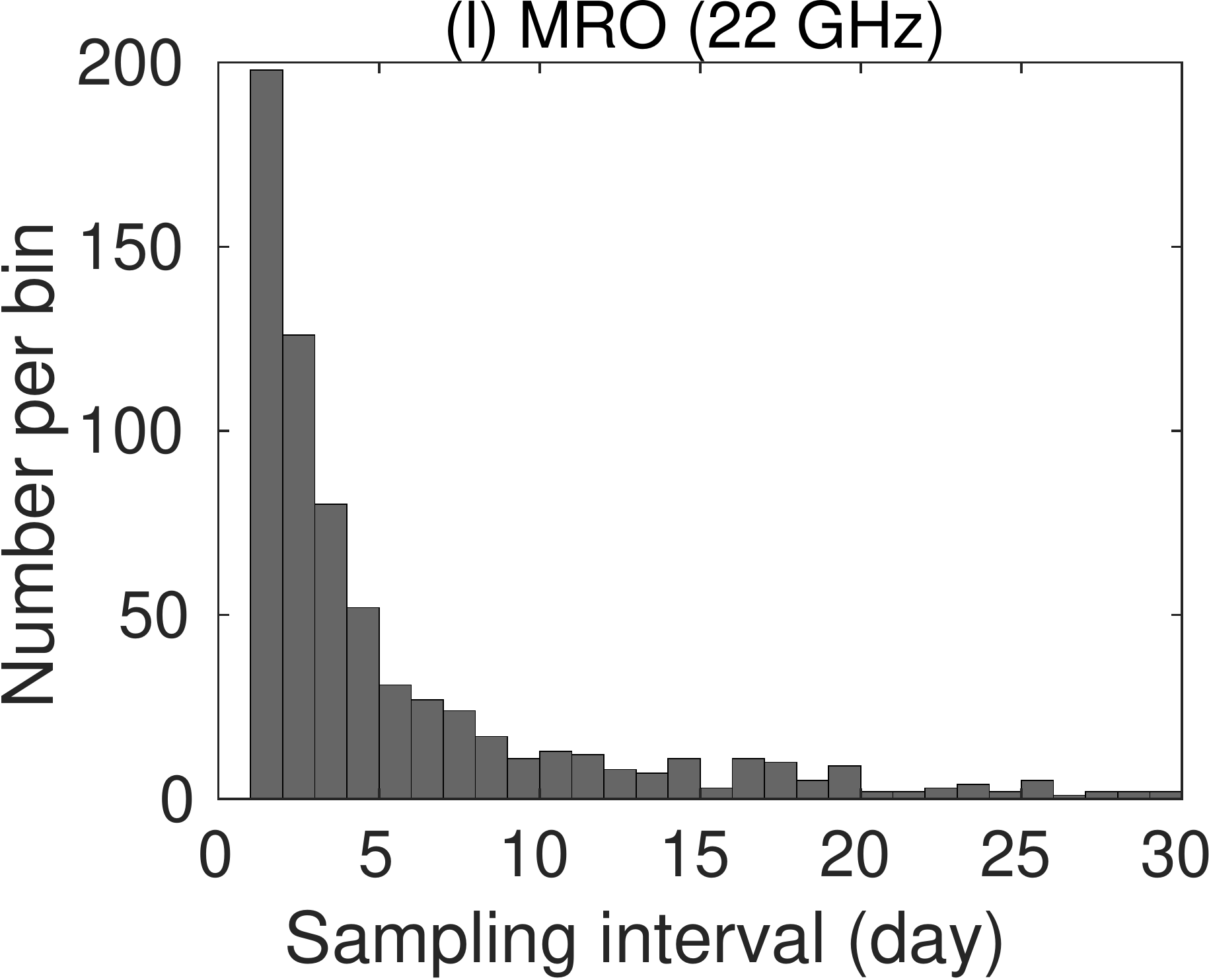}
}
\hbox{
\includegraphics[width=0.25\textwidth]{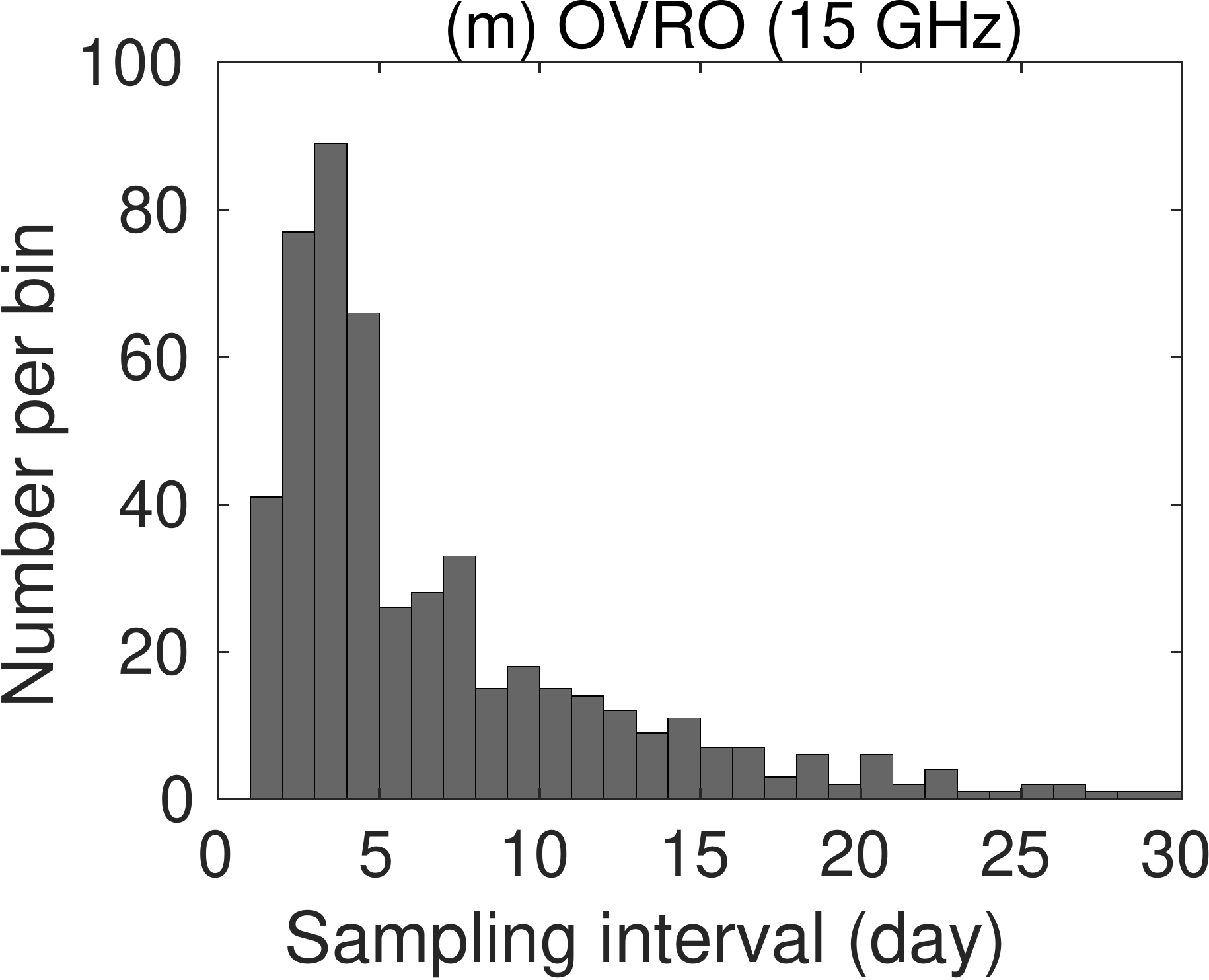}
\includegraphics[width=0.25\textwidth]{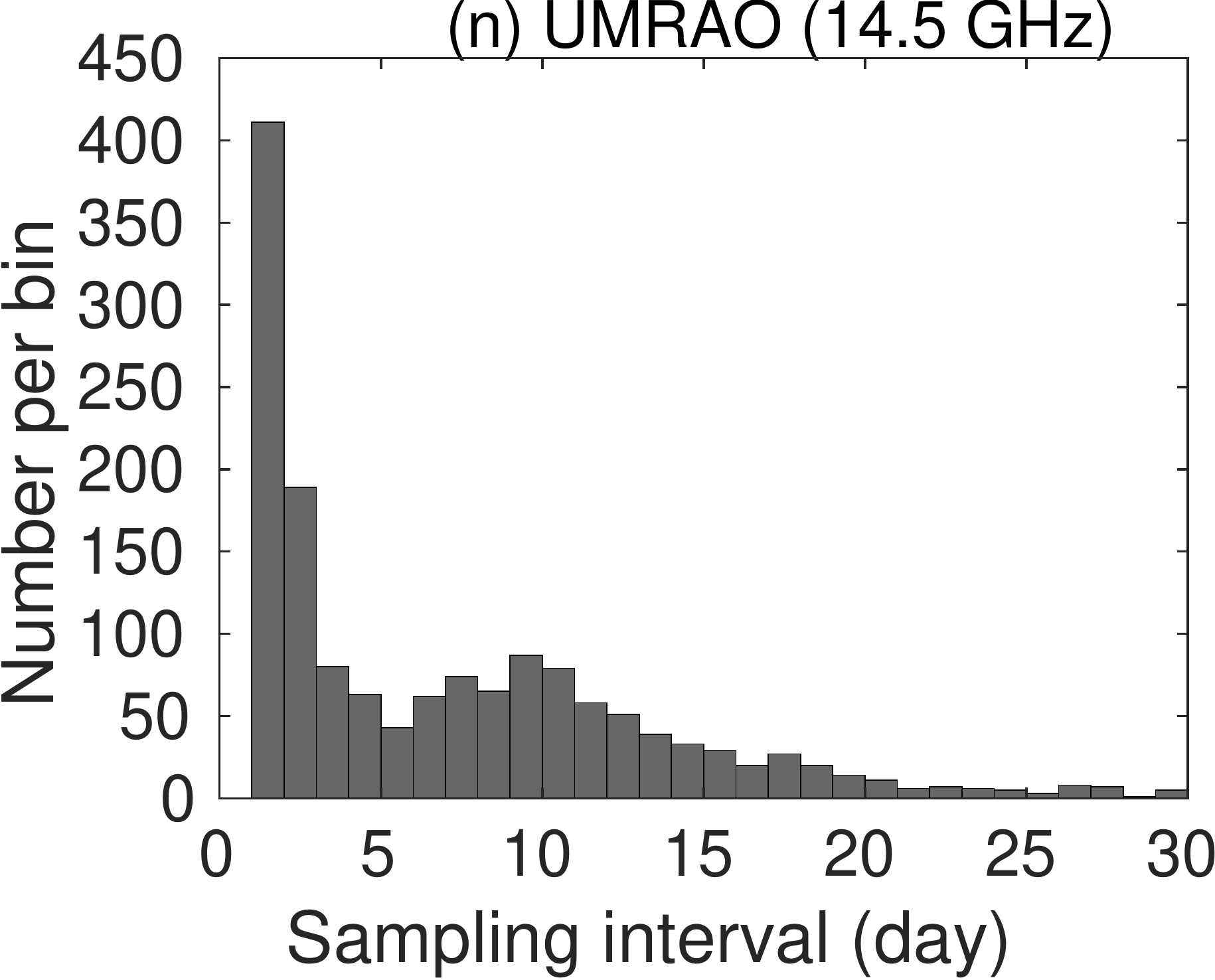}
\includegraphics[width=0.25\textwidth]{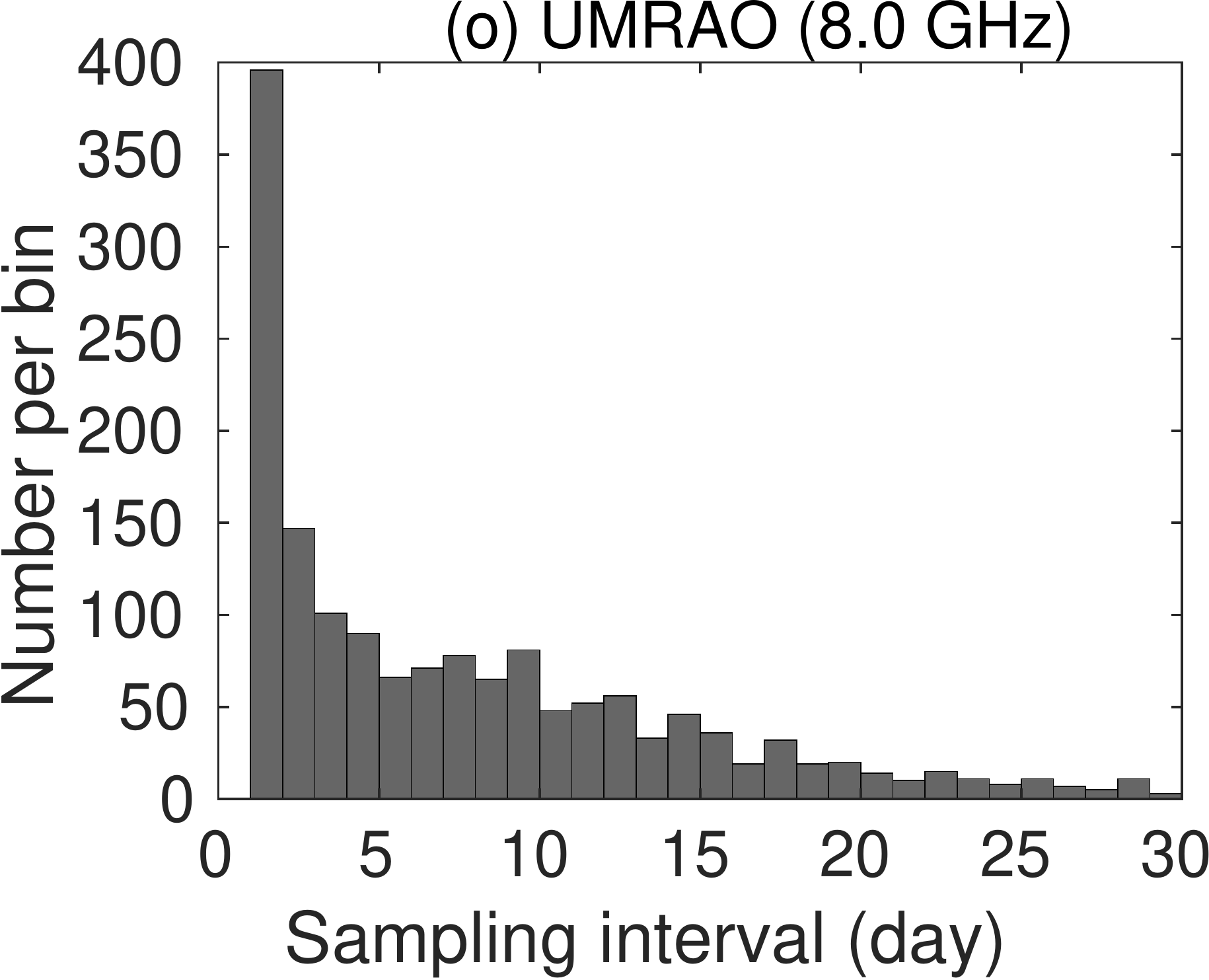}
\includegraphics[width=0.25\textwidth]{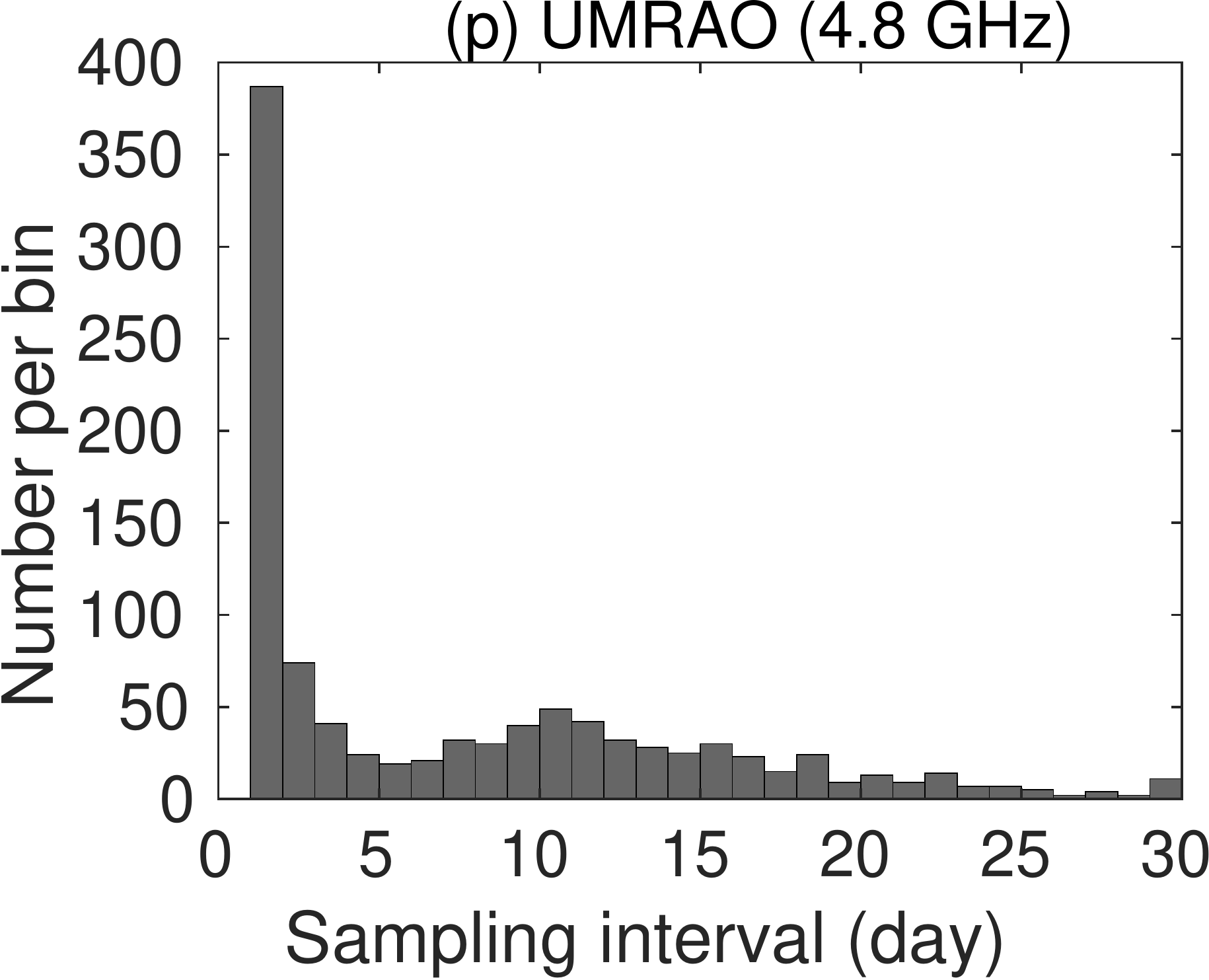}
}
\hbox{
\includegraphics[width=0.25\textwidth]{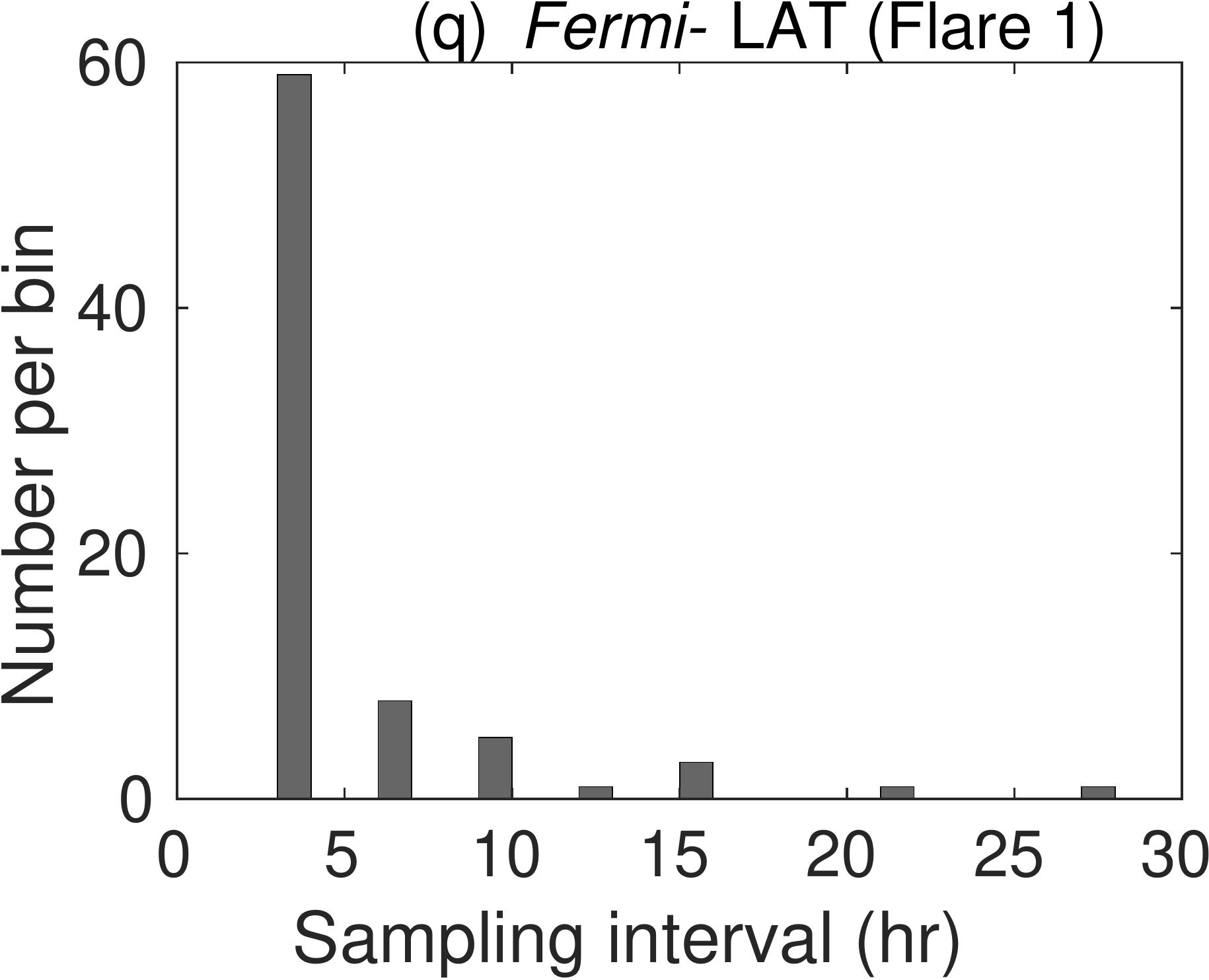}
\includegraphics[width=0.25\textwidth]{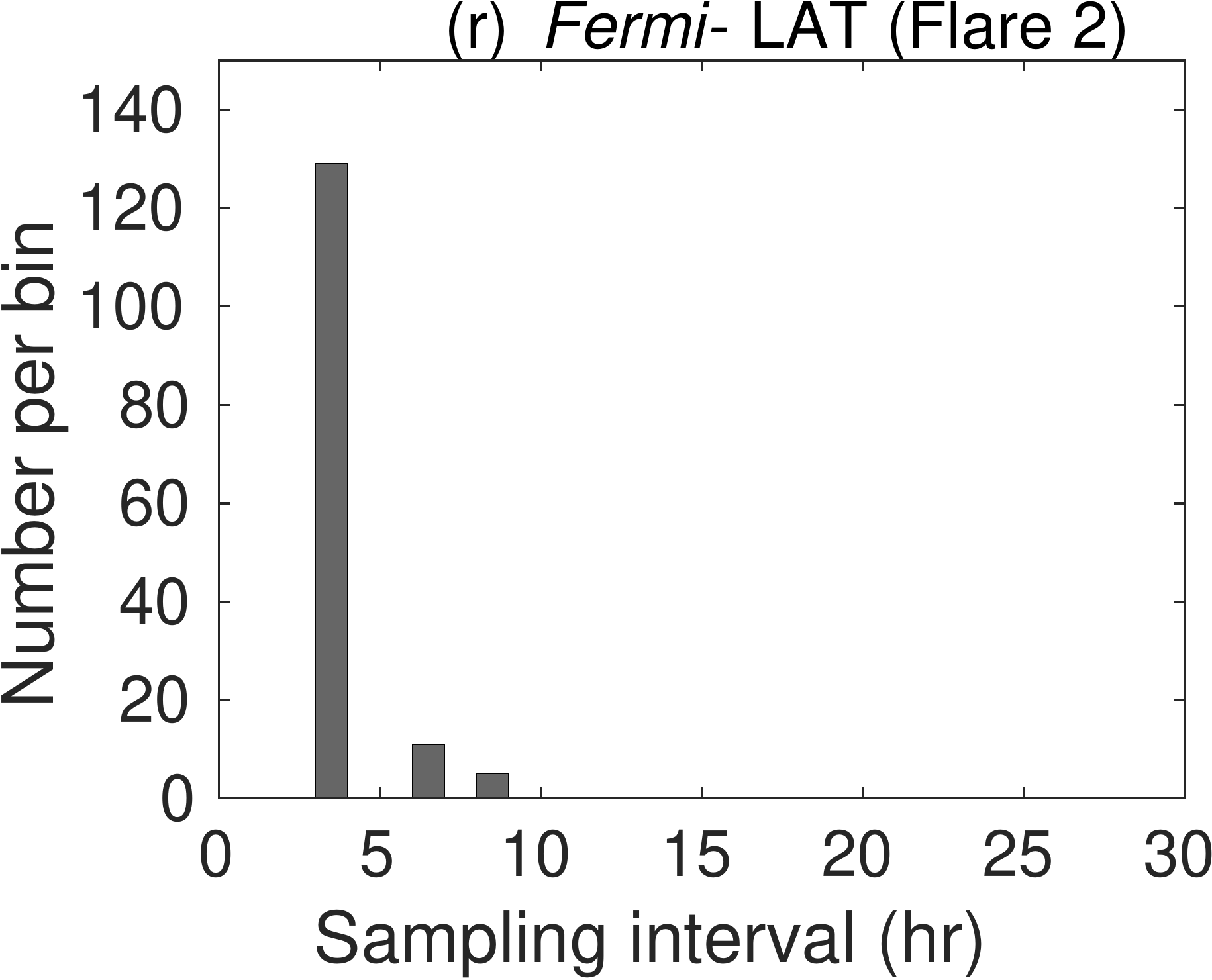}
\includegraphics[width=0.25\textwidth]{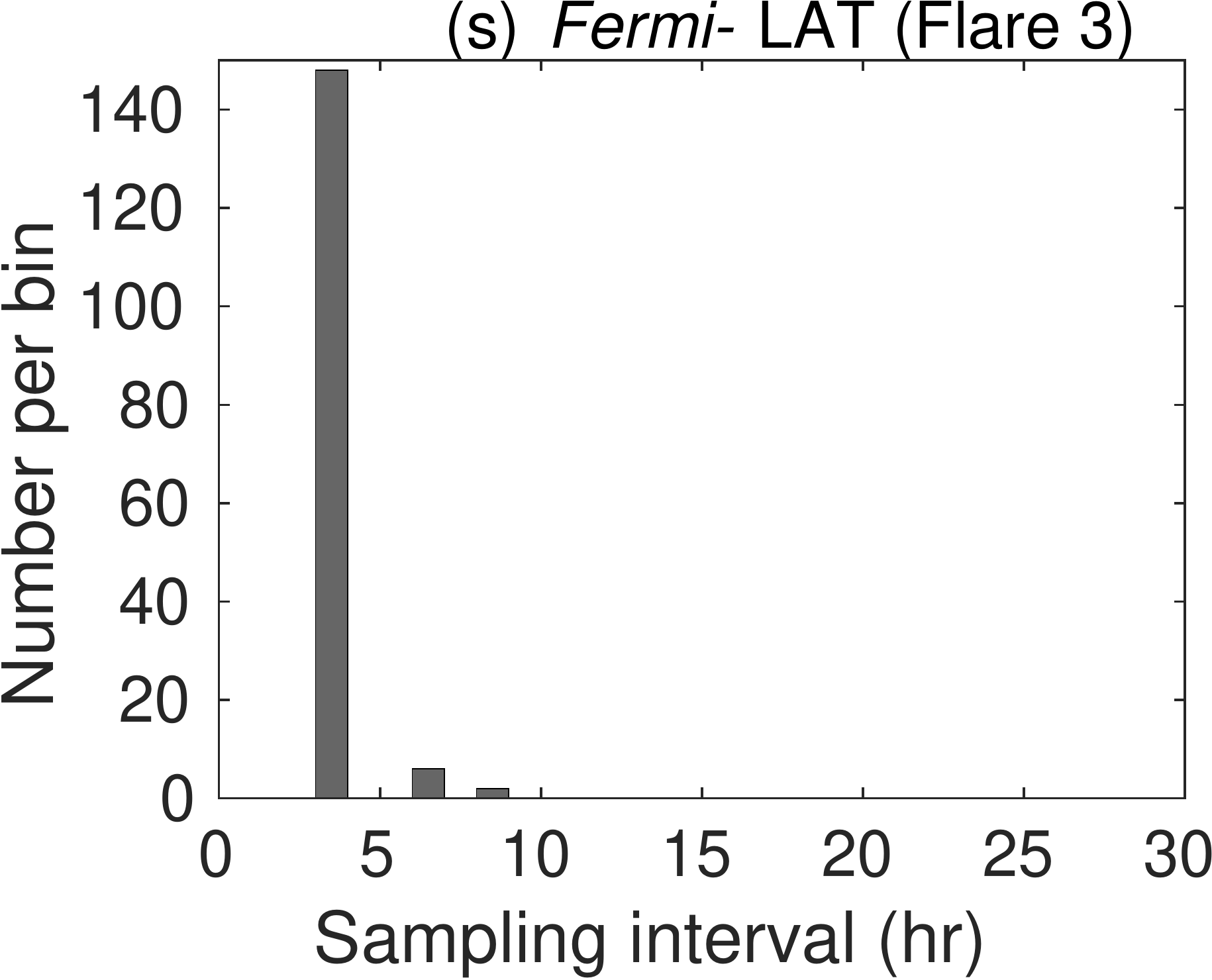}
}
\caption{Sampling distribution of the analyzed light curves (panels a-s) and the corresponding spectral window of DFT (panels a1-s1) are shown for the blazar 3C\,279. Sampling intervals larger than 30 days are not shown for clarity as they usually include less than 20 data points for the $\gamma$-ray, X-ray, optical, and infrared light curves and less than 60 data points for radio light curves (except for the OVRO light curve for which it is less than 20 data points).}
\label{appfig:swf3c}
\end{figure*}

\addtocounter{figure}{-1}

\begin{figure*}[ht!]

\hbox{
\includegraphics[width=0.25\textwidth]{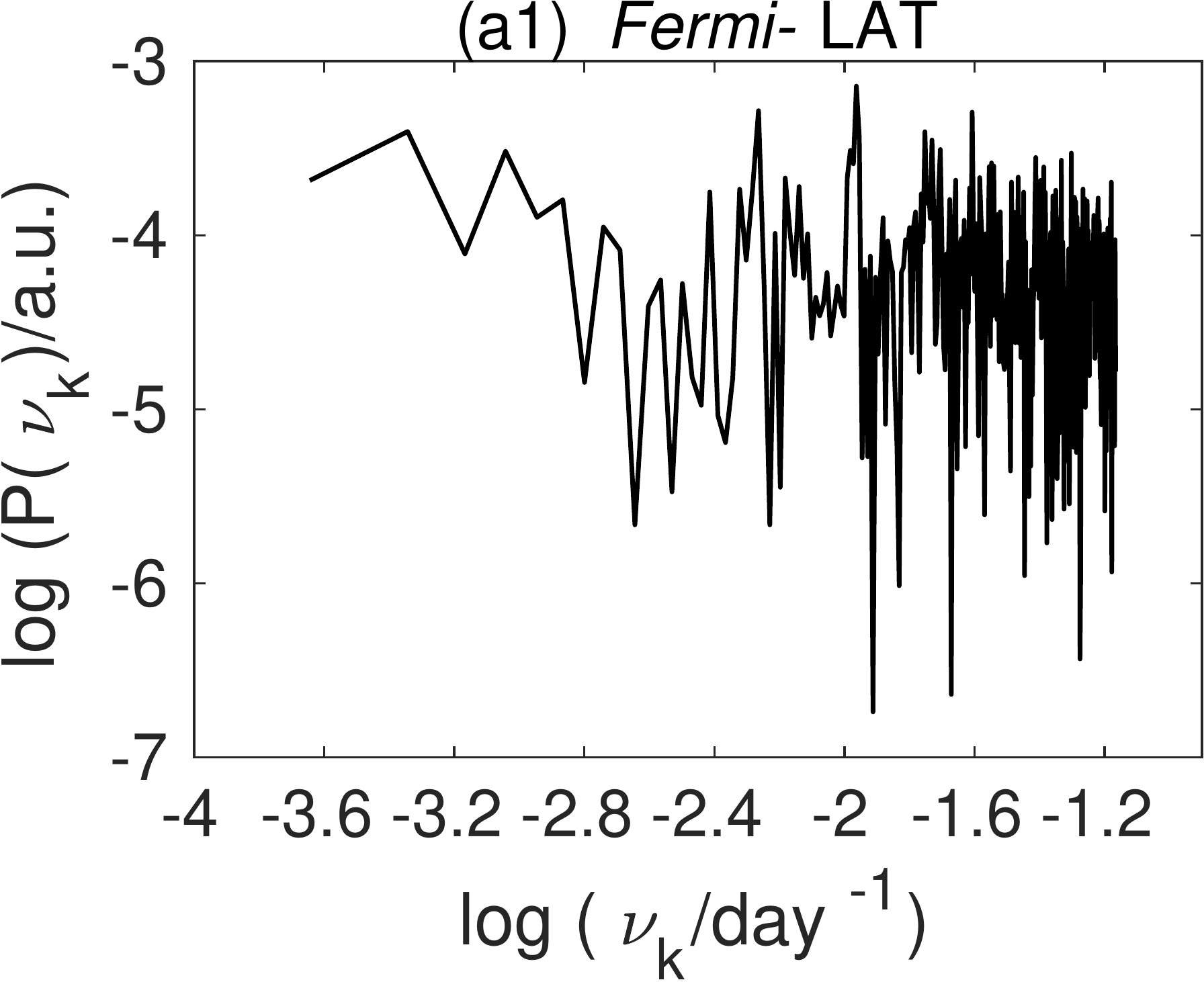}
\includegraphics[width=0.25\textwidth]{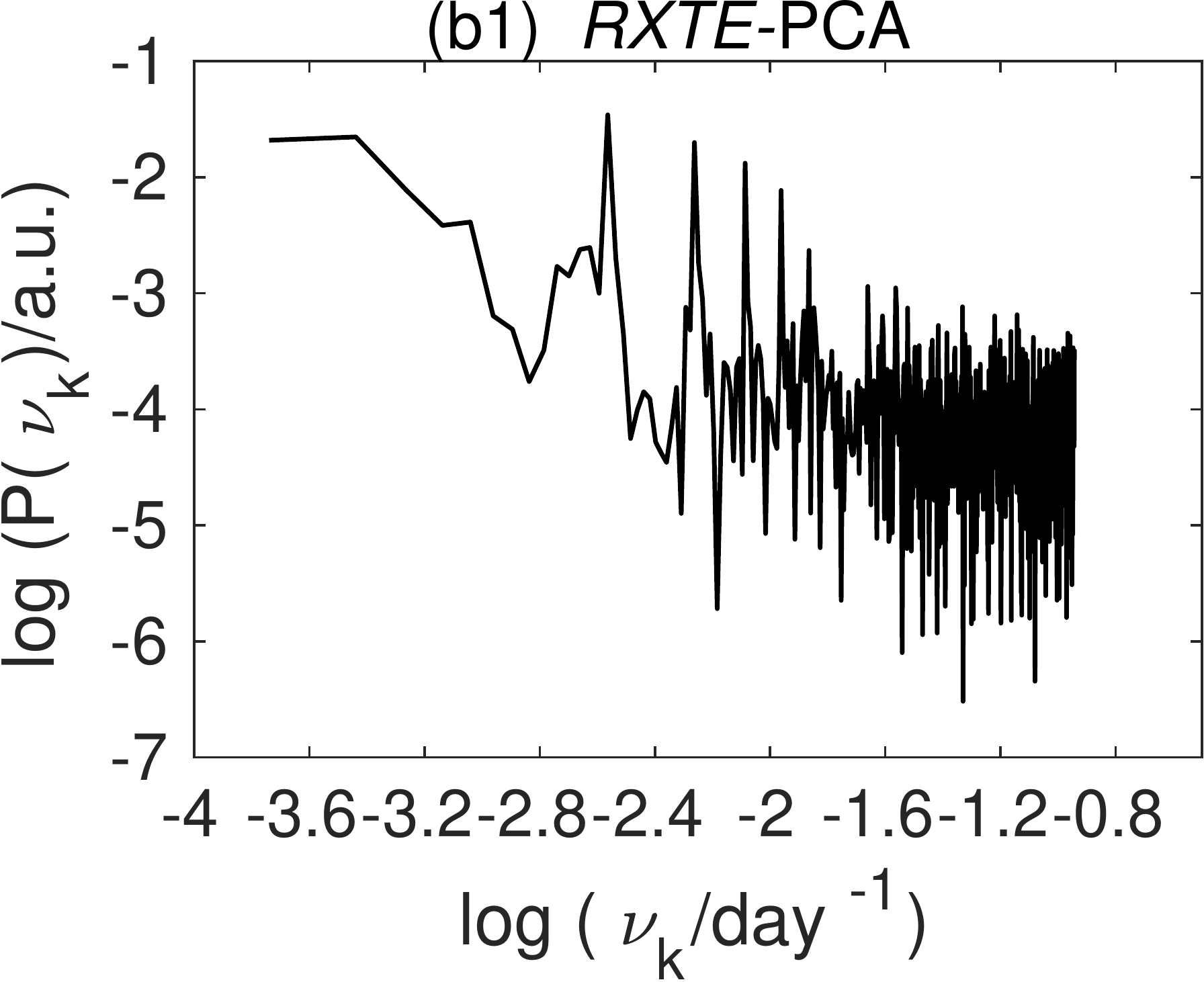}
\includegraphics[width=0.25\textwidth]{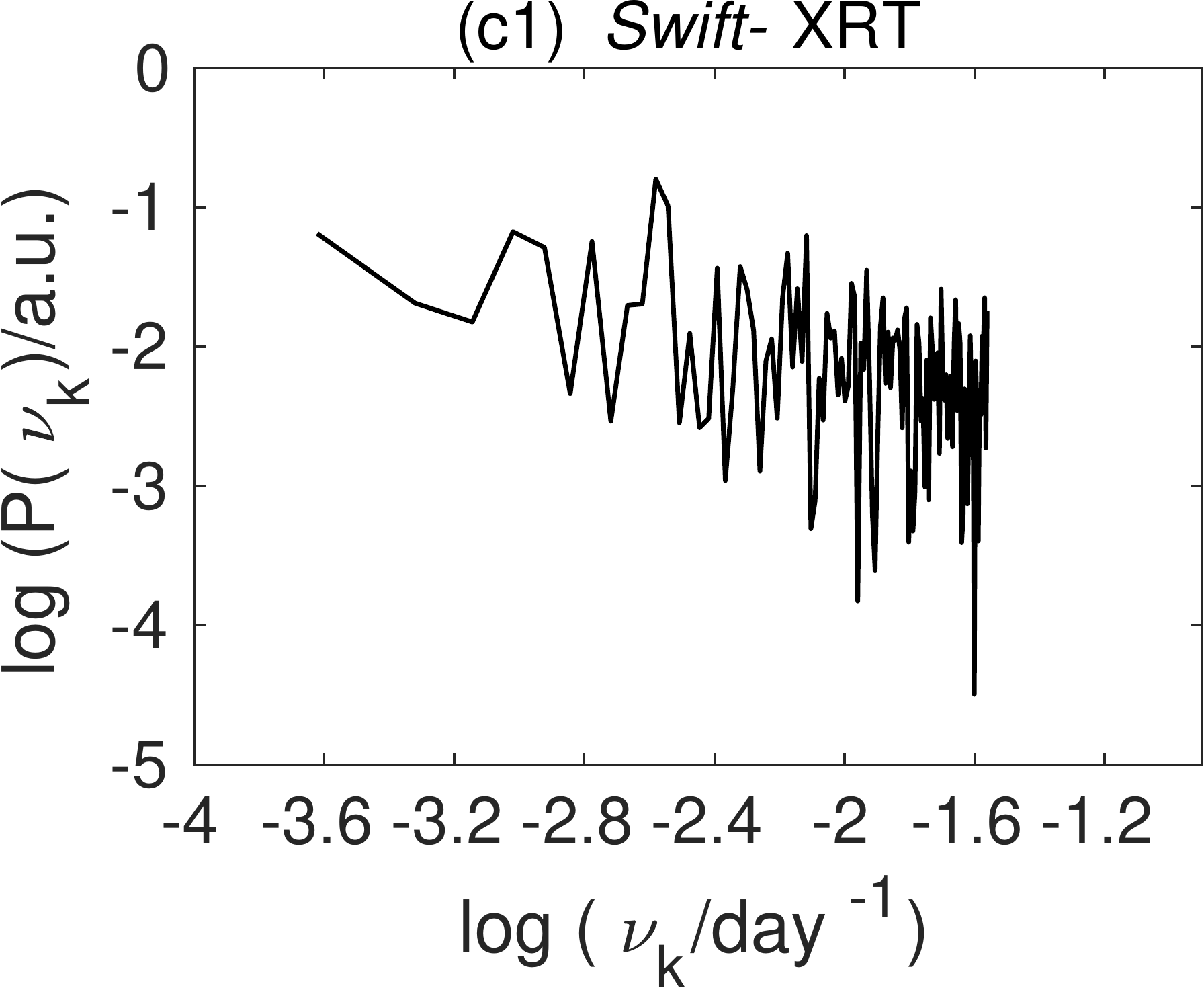}
\includegraphics[width=0.25\textwidth]{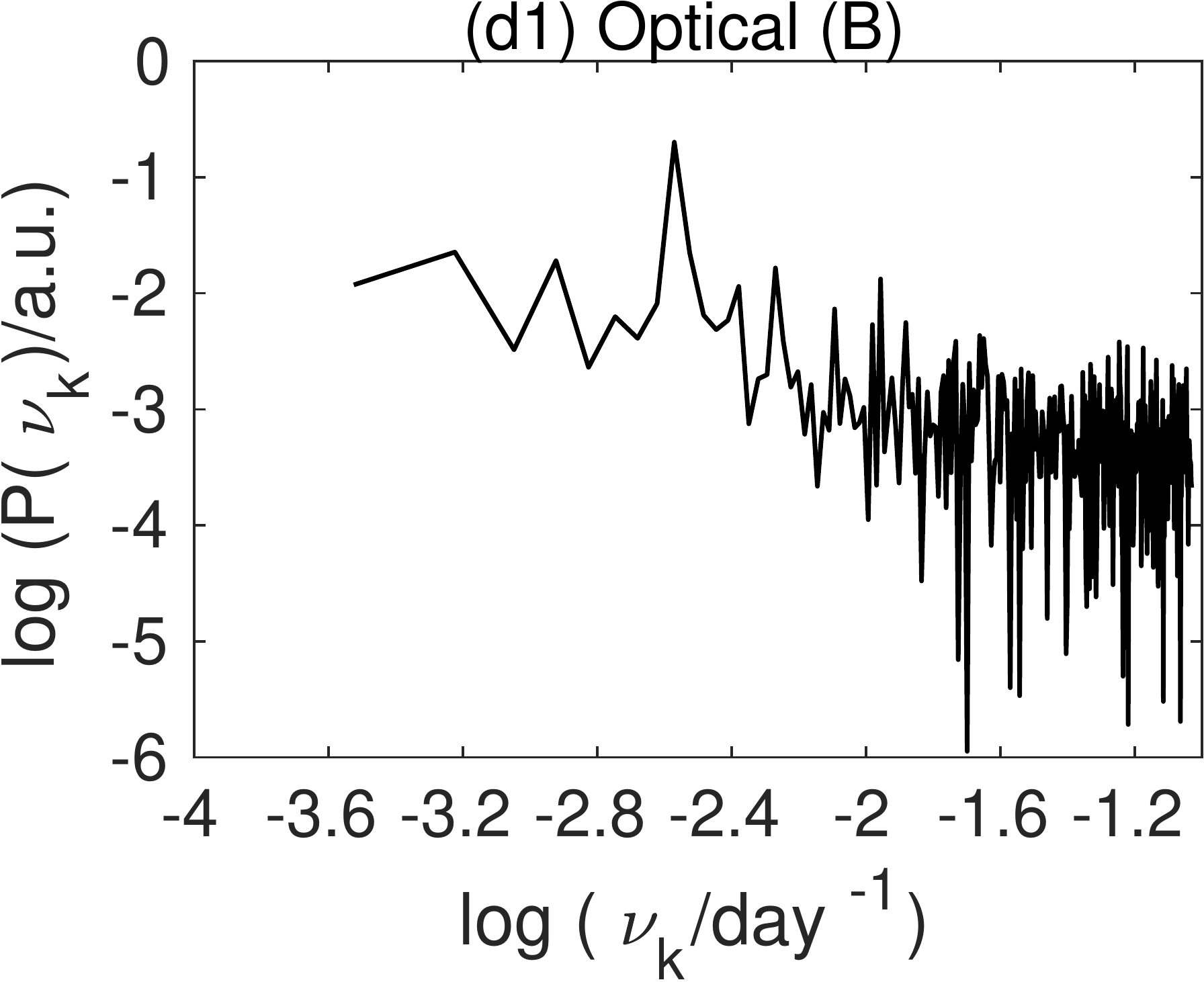}
}
\hbox{
\includegraphics[width=0.25\textwidth]{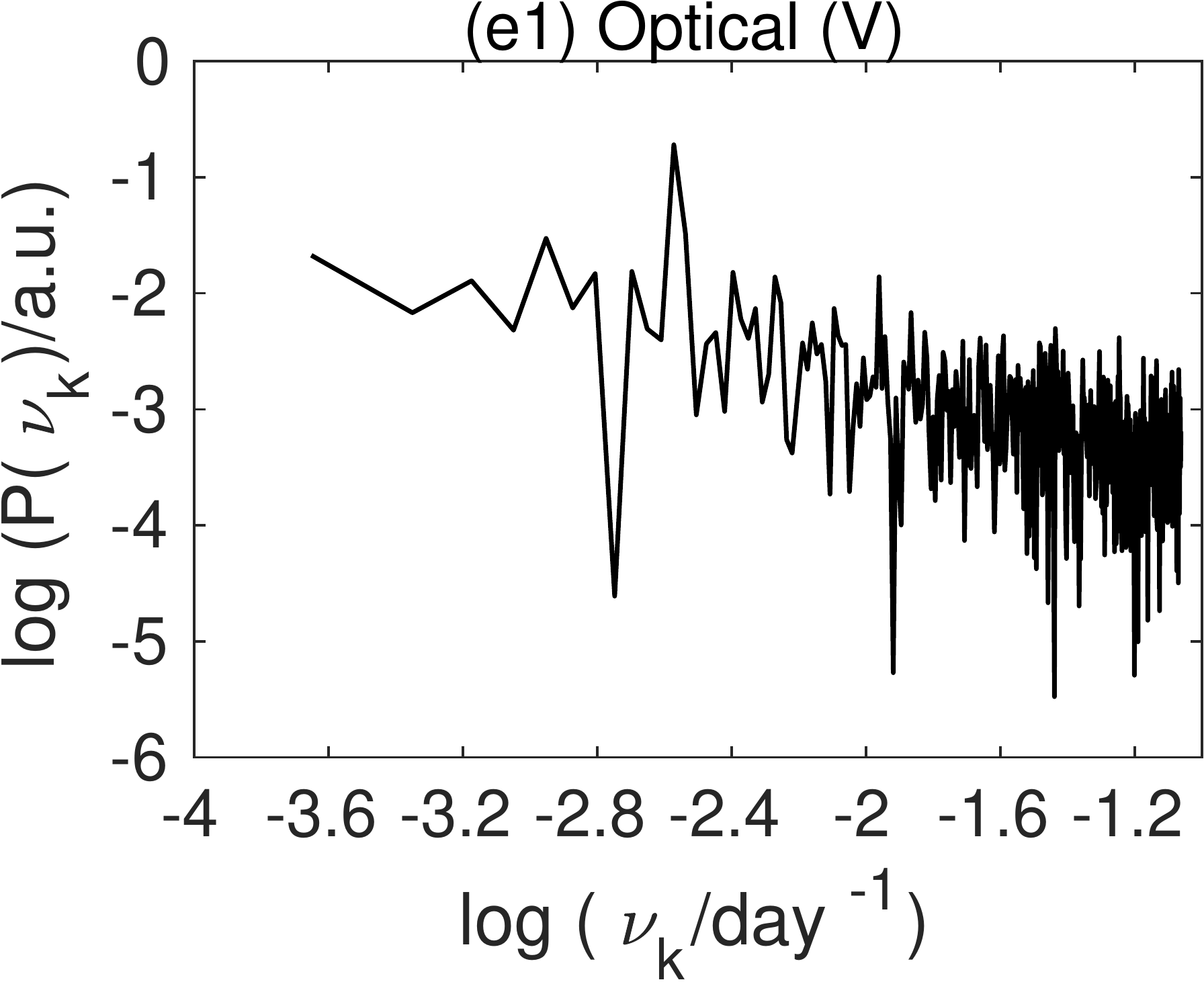}
\includegraphics[width=0.25\textwidth]{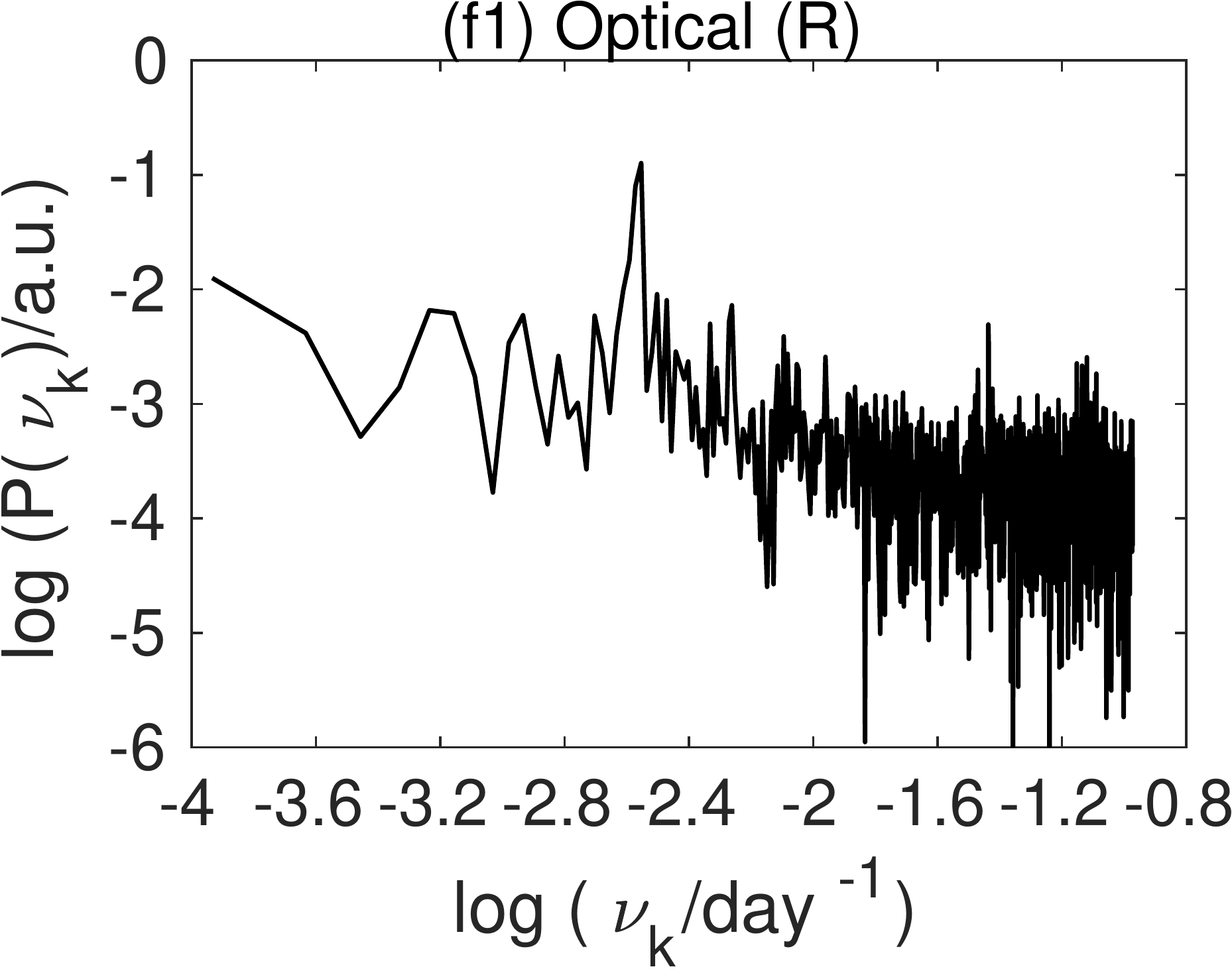}
\includegraphics[width=0.25\textwidth]{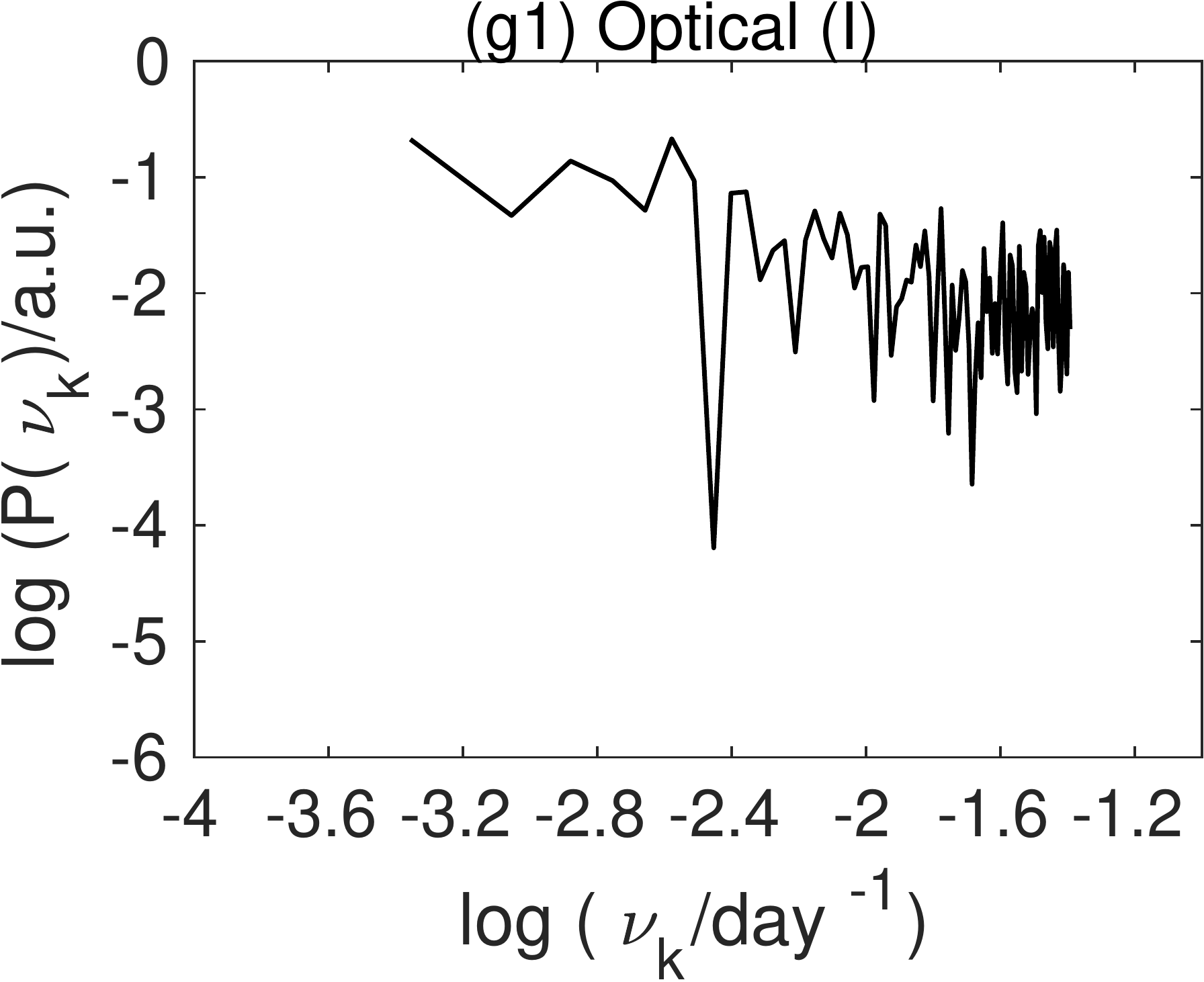}
\includegraphics[width=0.25\textwidth]{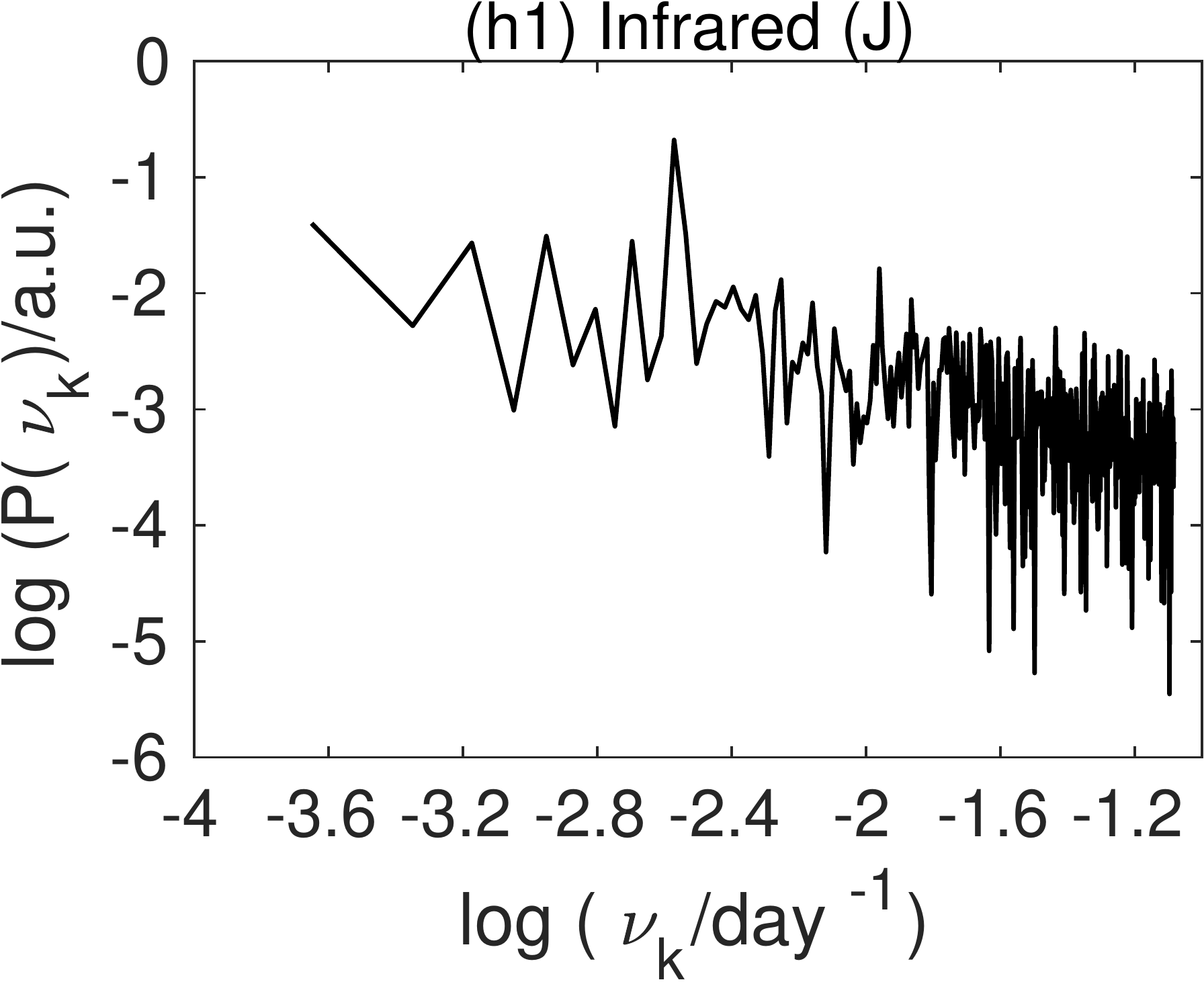}
}
\hbox{
\includegraphics[width=0.25\textwidth]{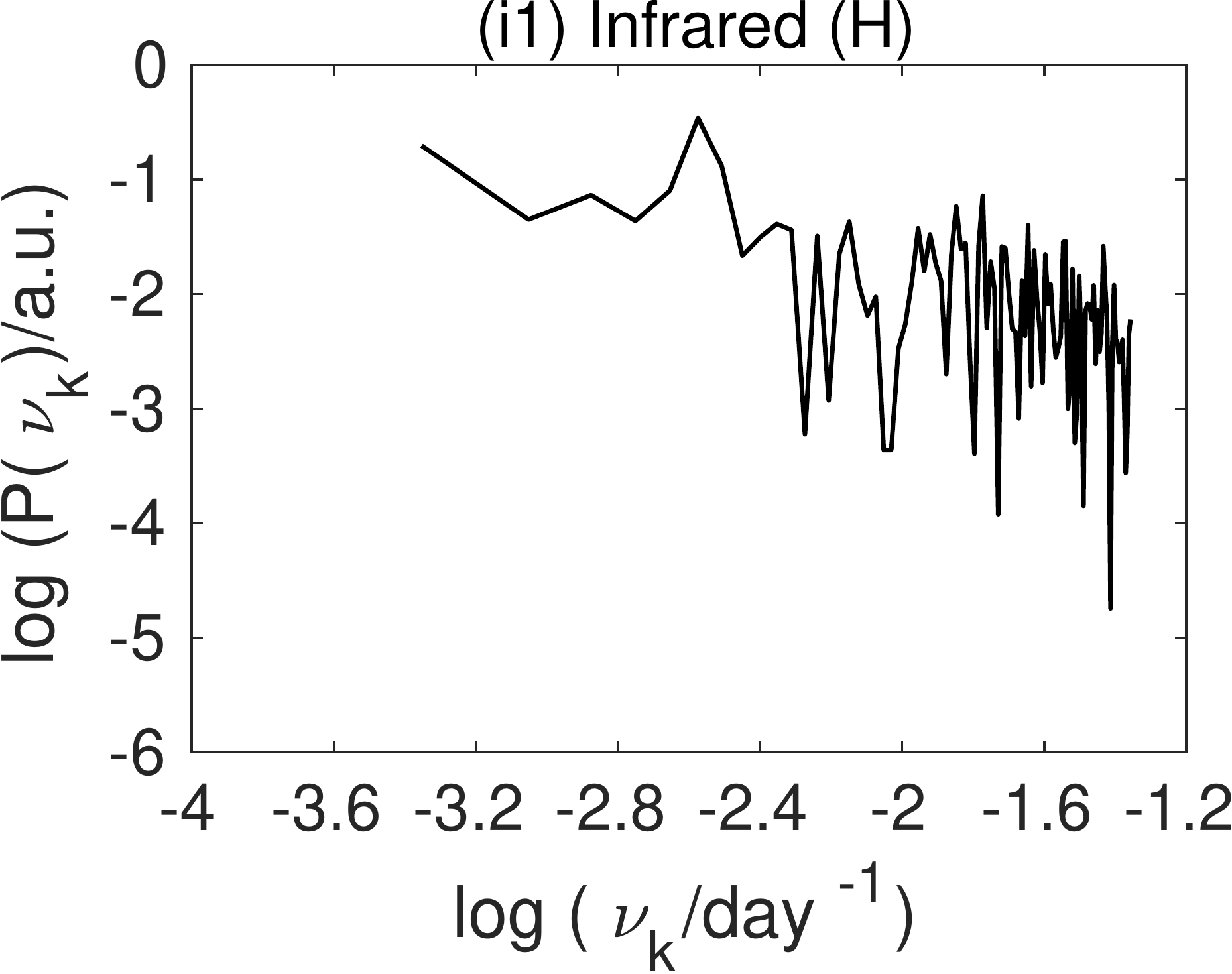}
\includegraphics[width=0.25\textwidth]{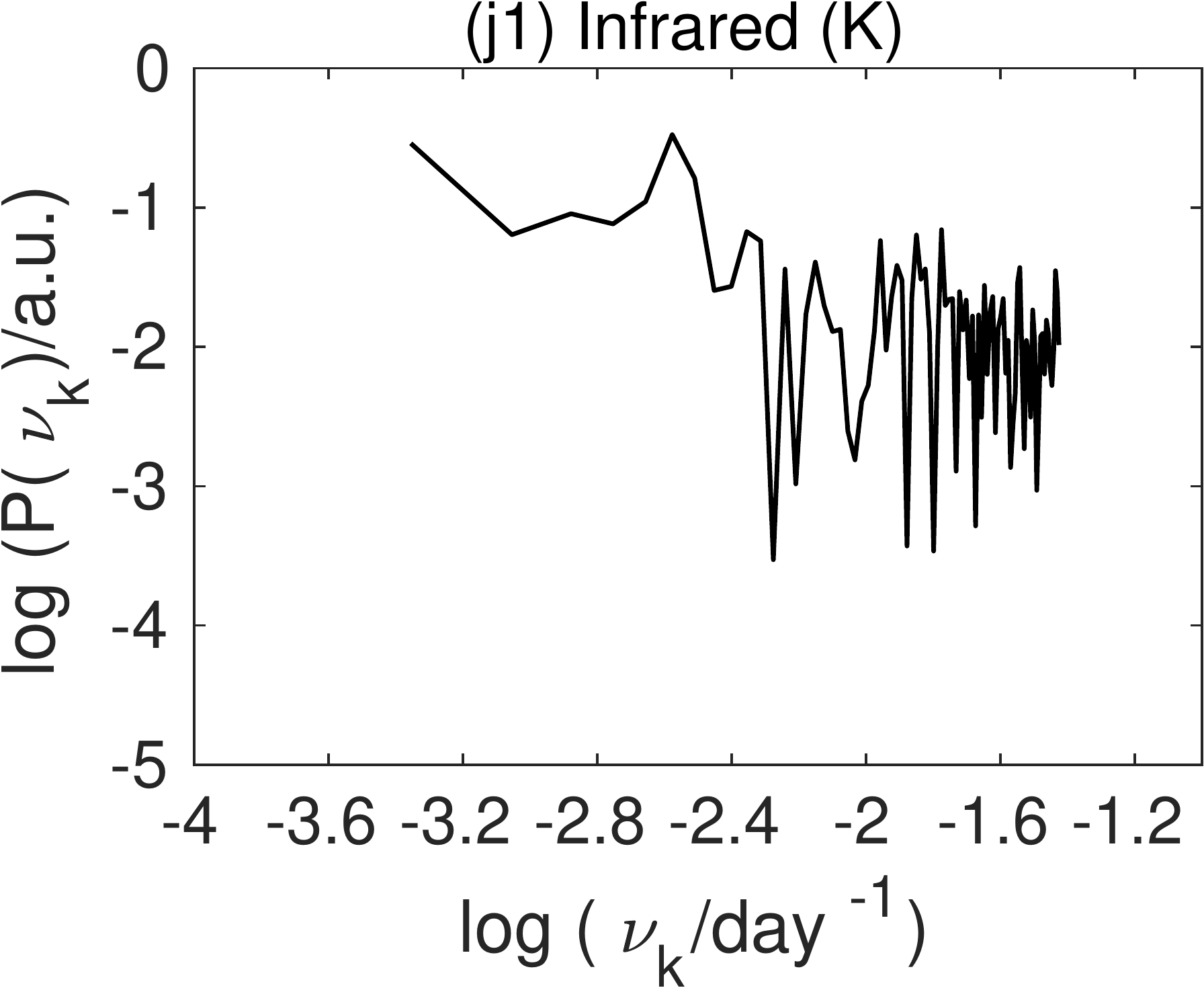}
\includegraphics[width=0.25\textwidth]{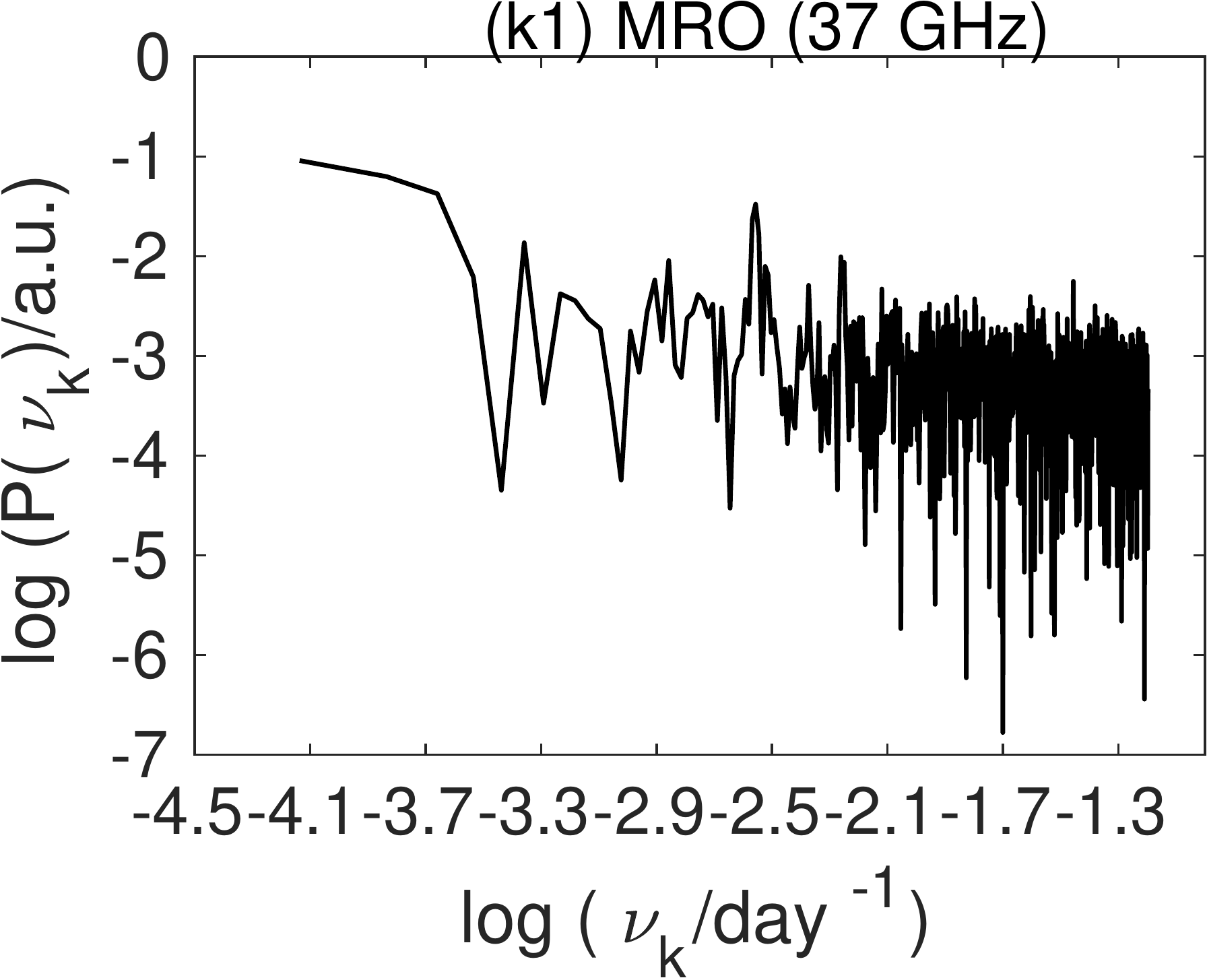}
\includegraphics[width=0.25\textwidth]{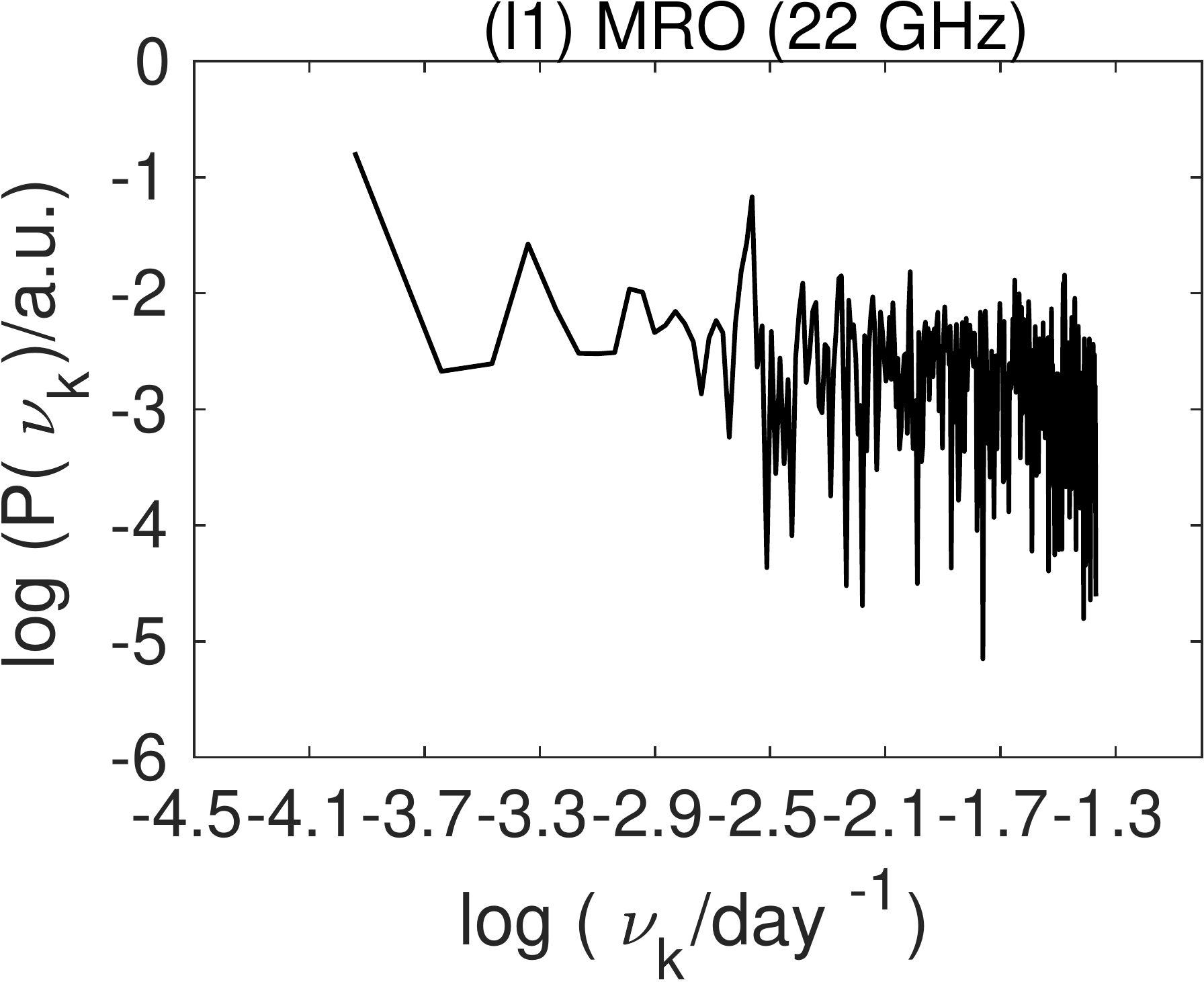}
}
\hbox{
\includegraphics[width=0.25\textwidth]{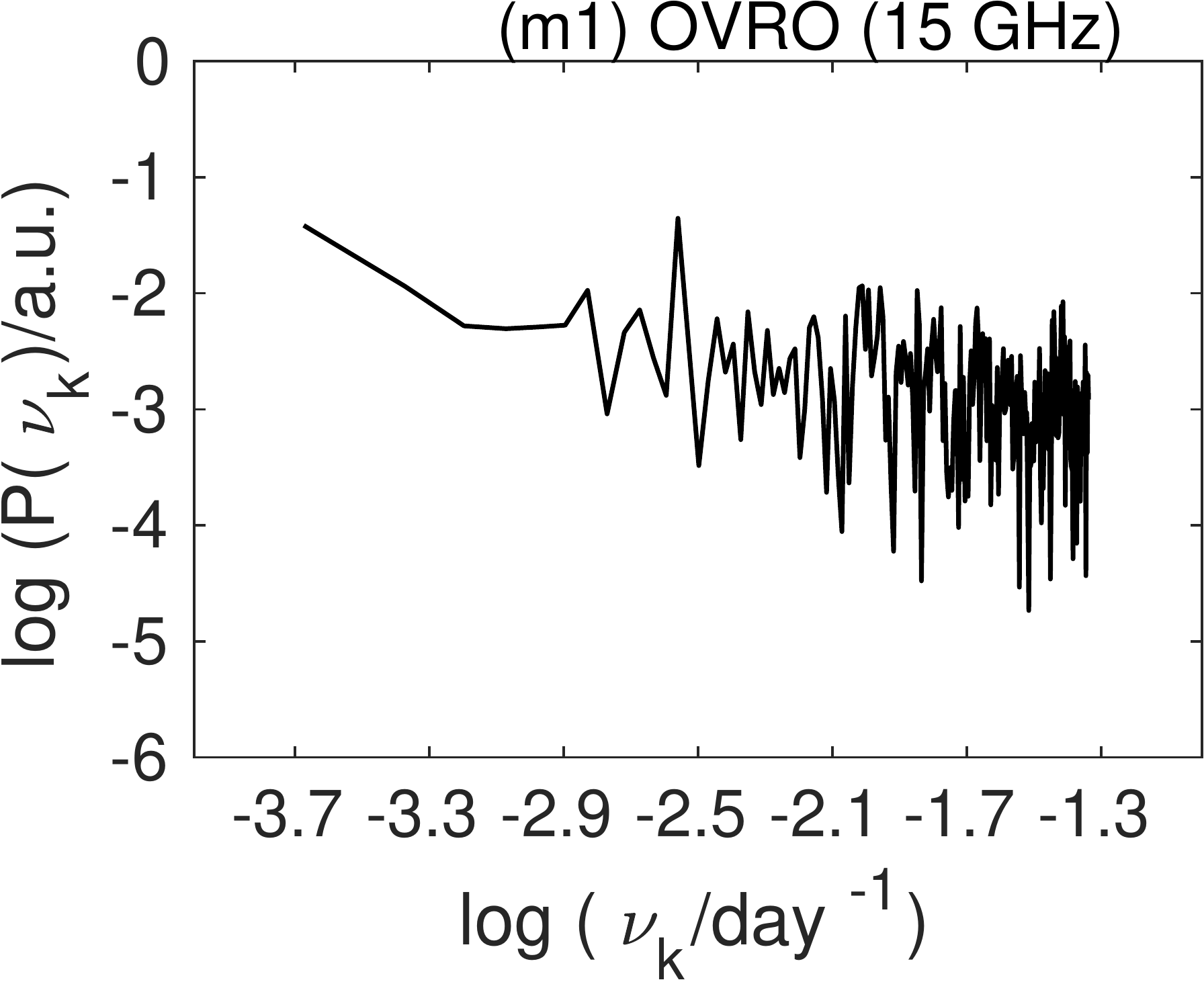}
\includegraphics[width=0.25\textwidth]{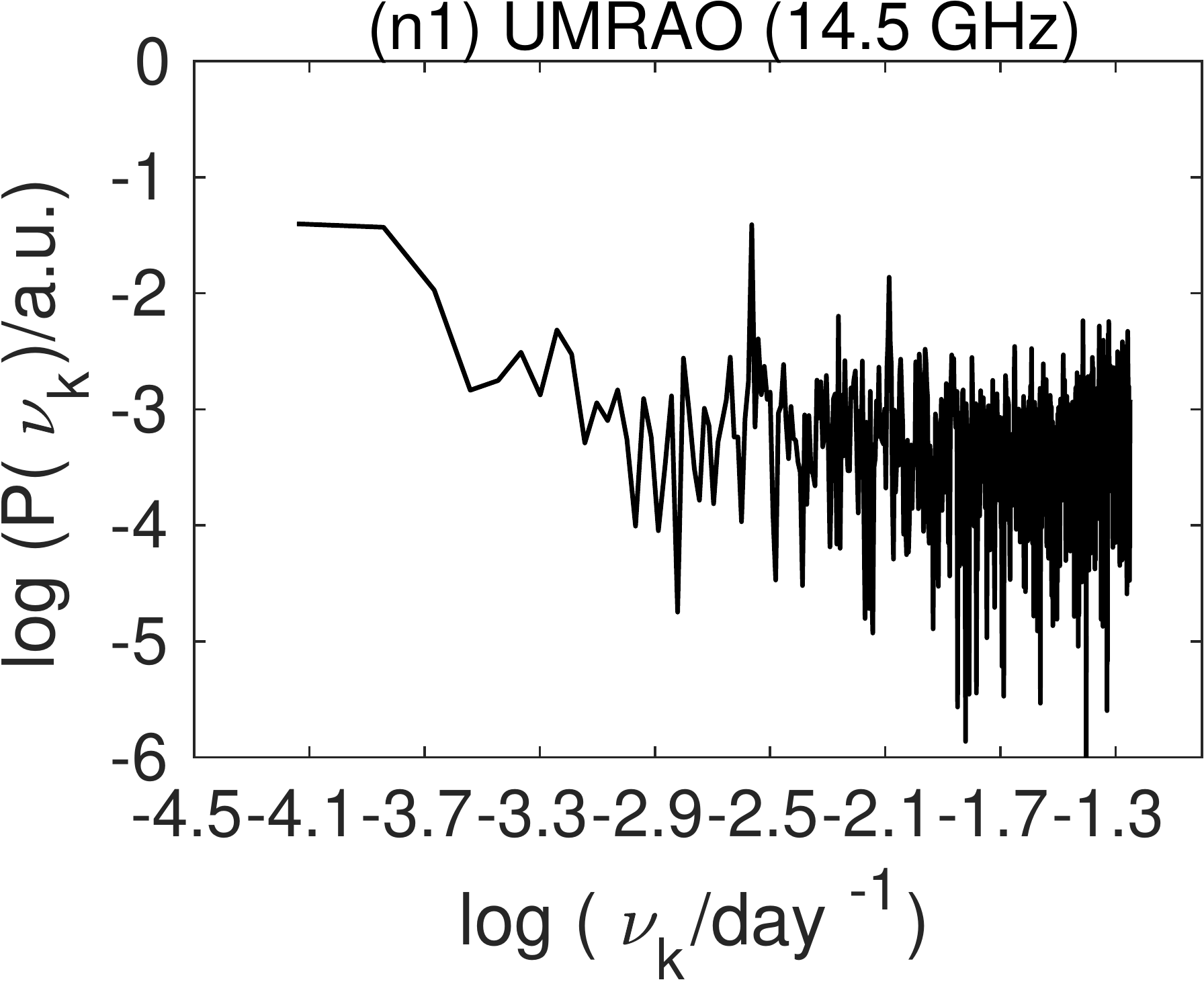}
\includegraphics[width=0.25\textwidth]{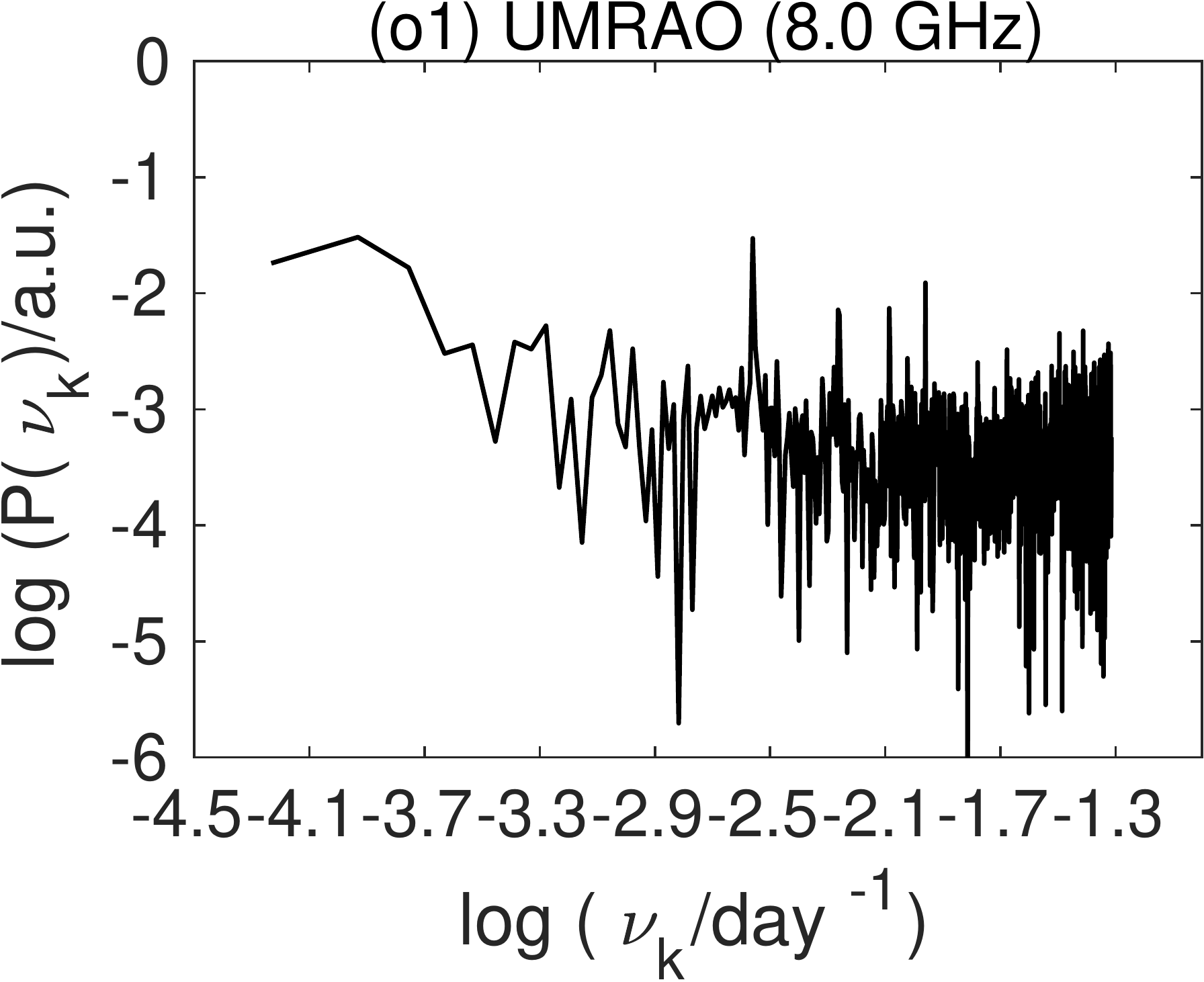}
\includegraphics[width=0.25\textwidth]{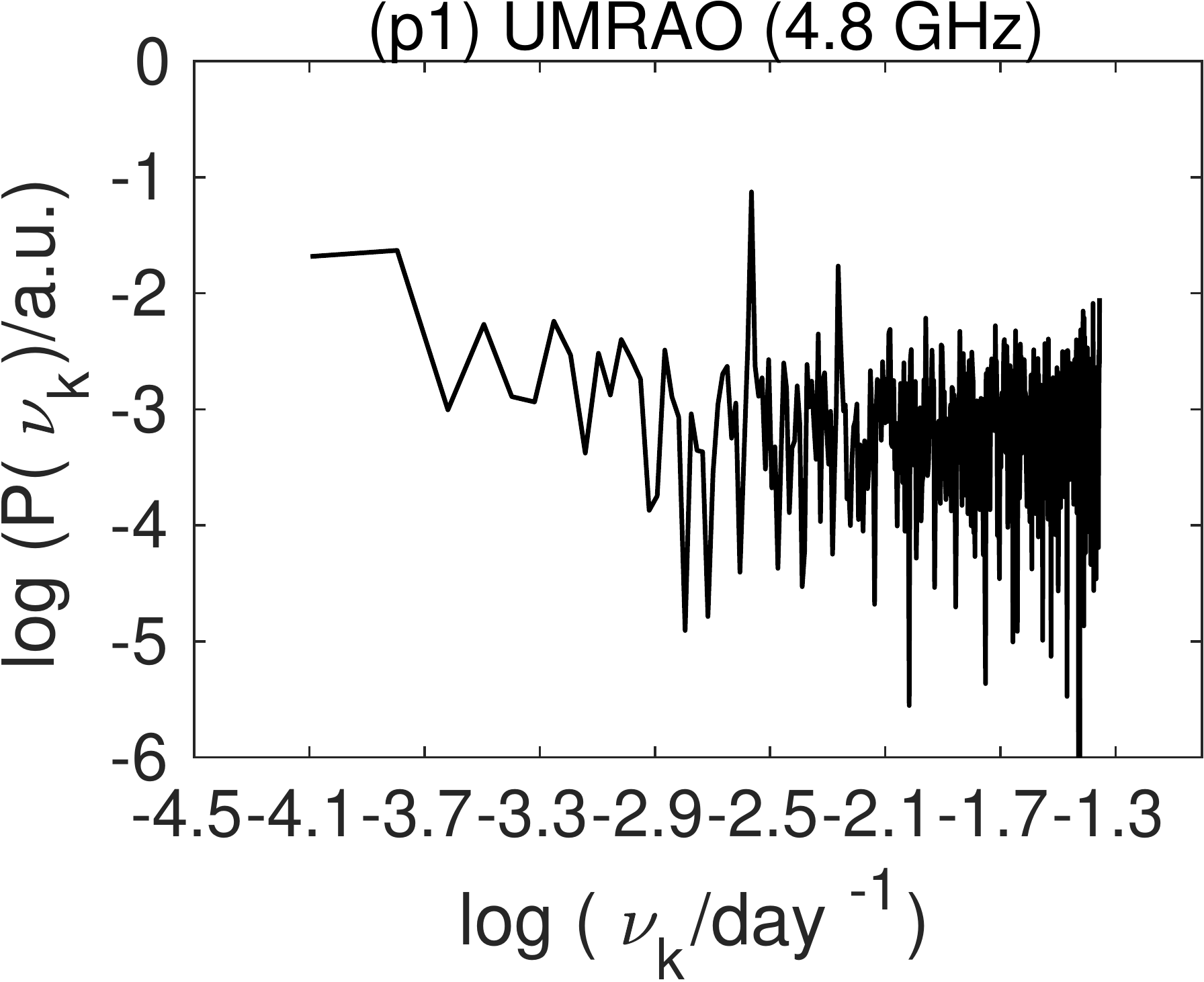}
}
\hbox{
\includegraphics[width=0.25\textwidth]{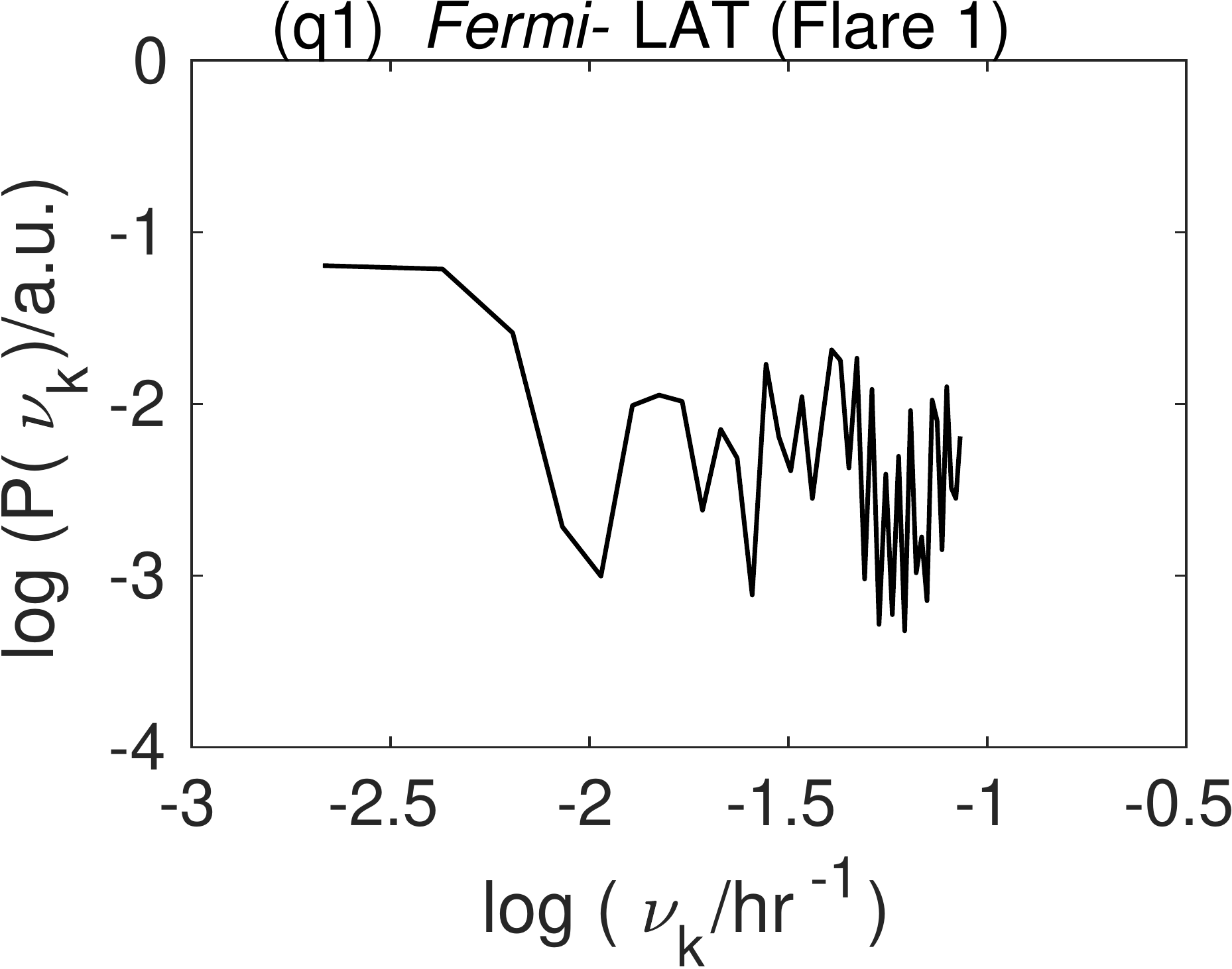}
\includegraphics[width=0.25\textwidth]{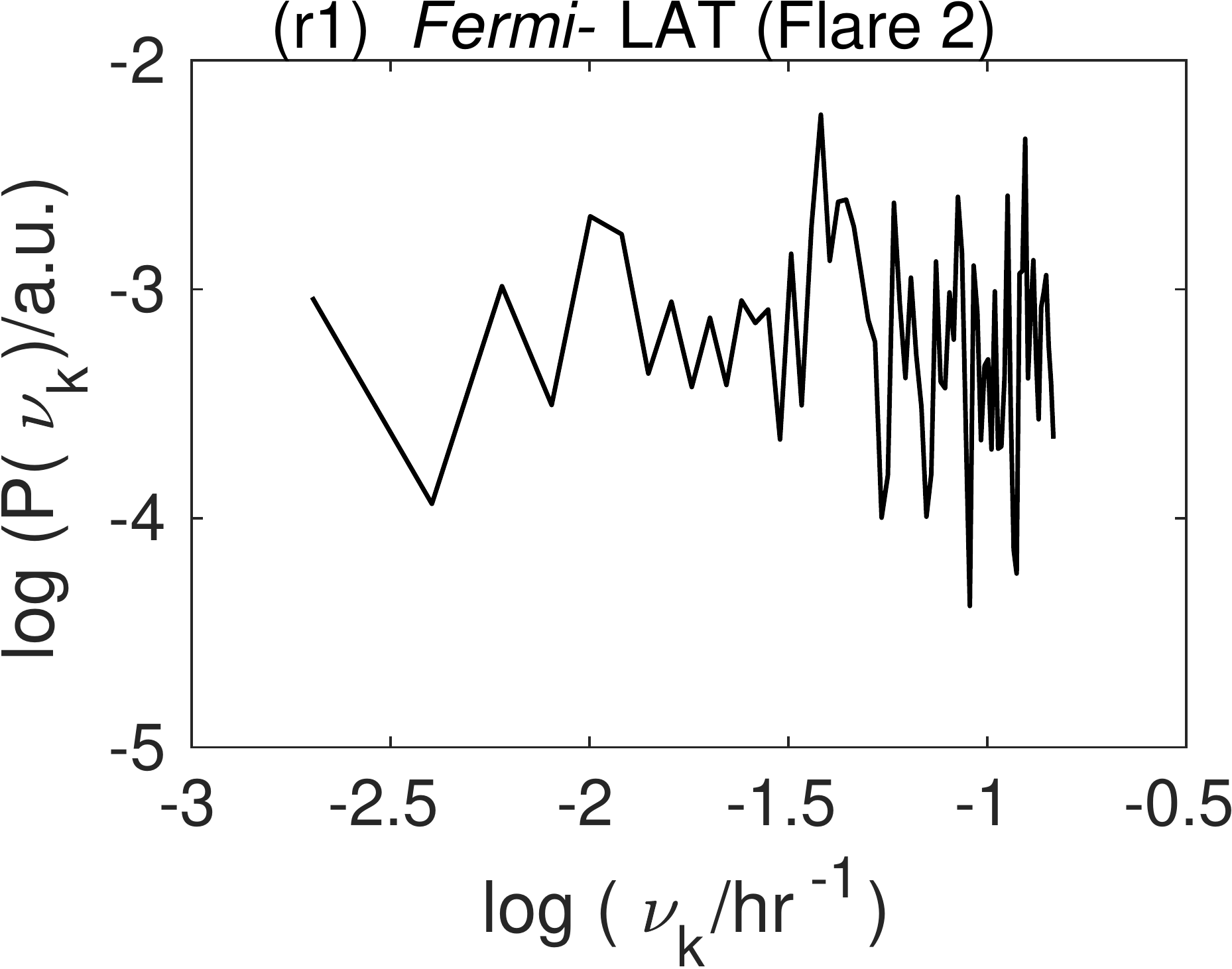}
\includegraphics[width=0.25\textwidth]{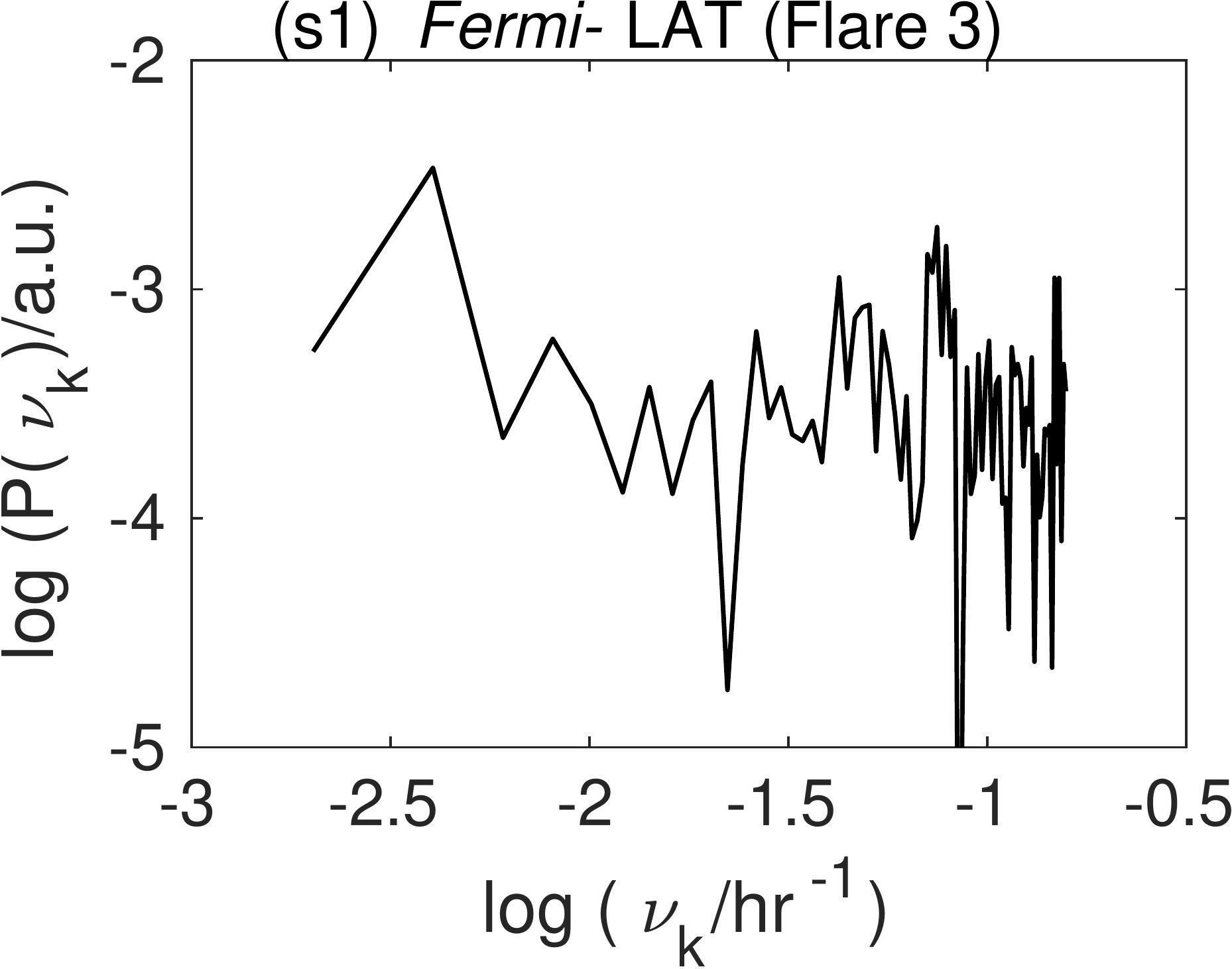}
}
\caption{(continued) }
\end{figure*}

\begin{figure*}[ht!]
\hbox{
\includegraphics[width=0.25\textwidth]{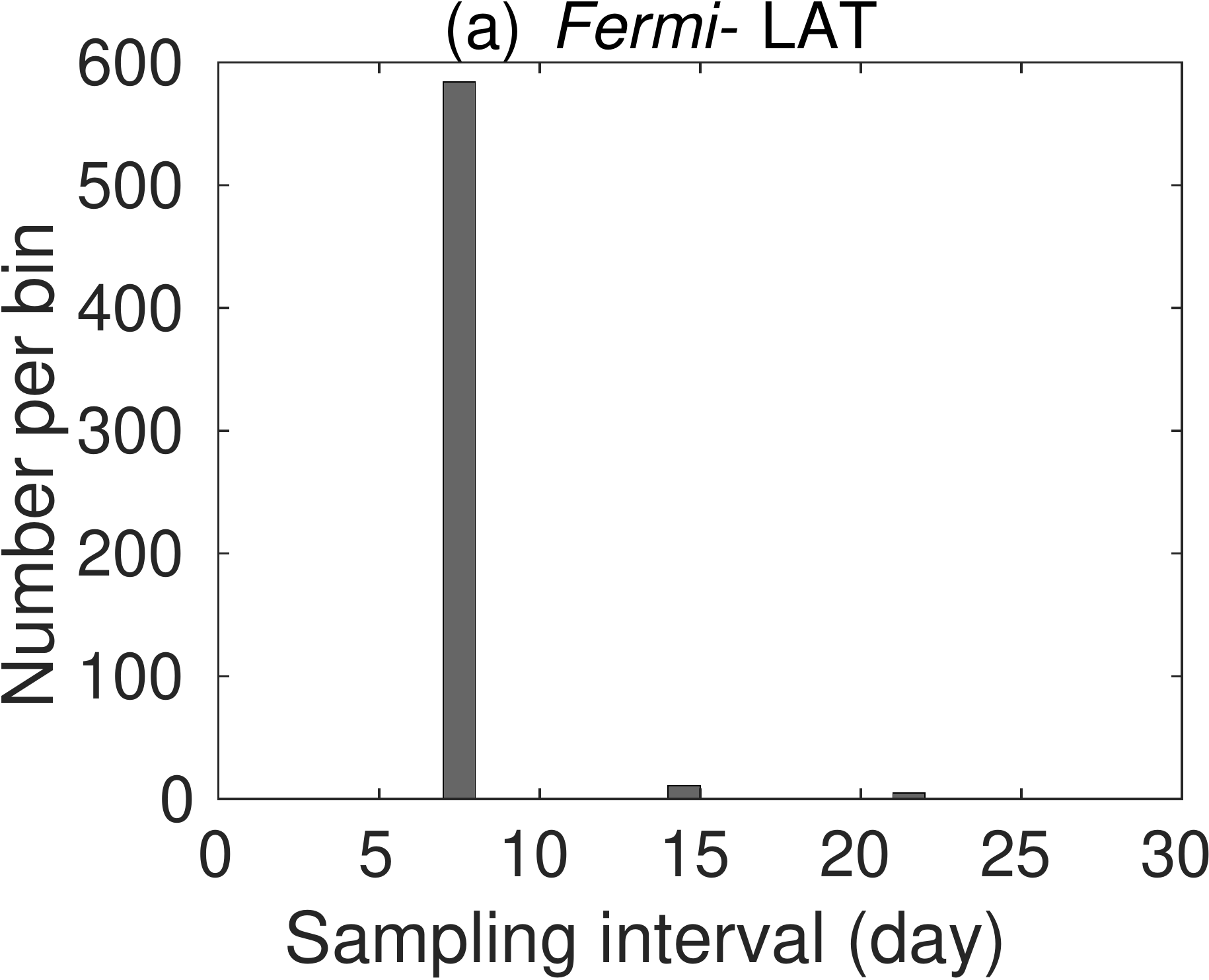}
\includegraphics[width=0.25\textwidth]{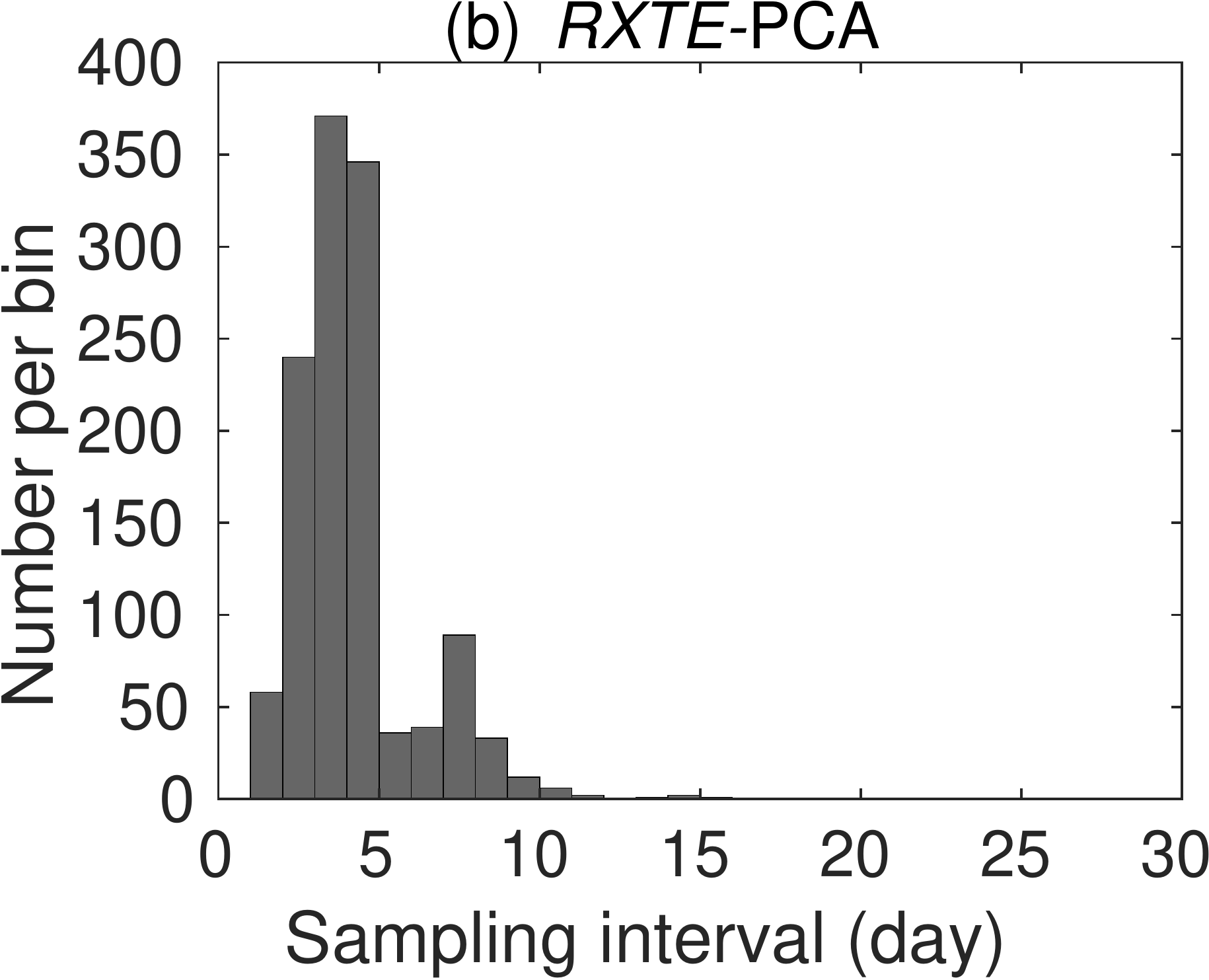}
\includegraphics[width=0.25\textwidth]{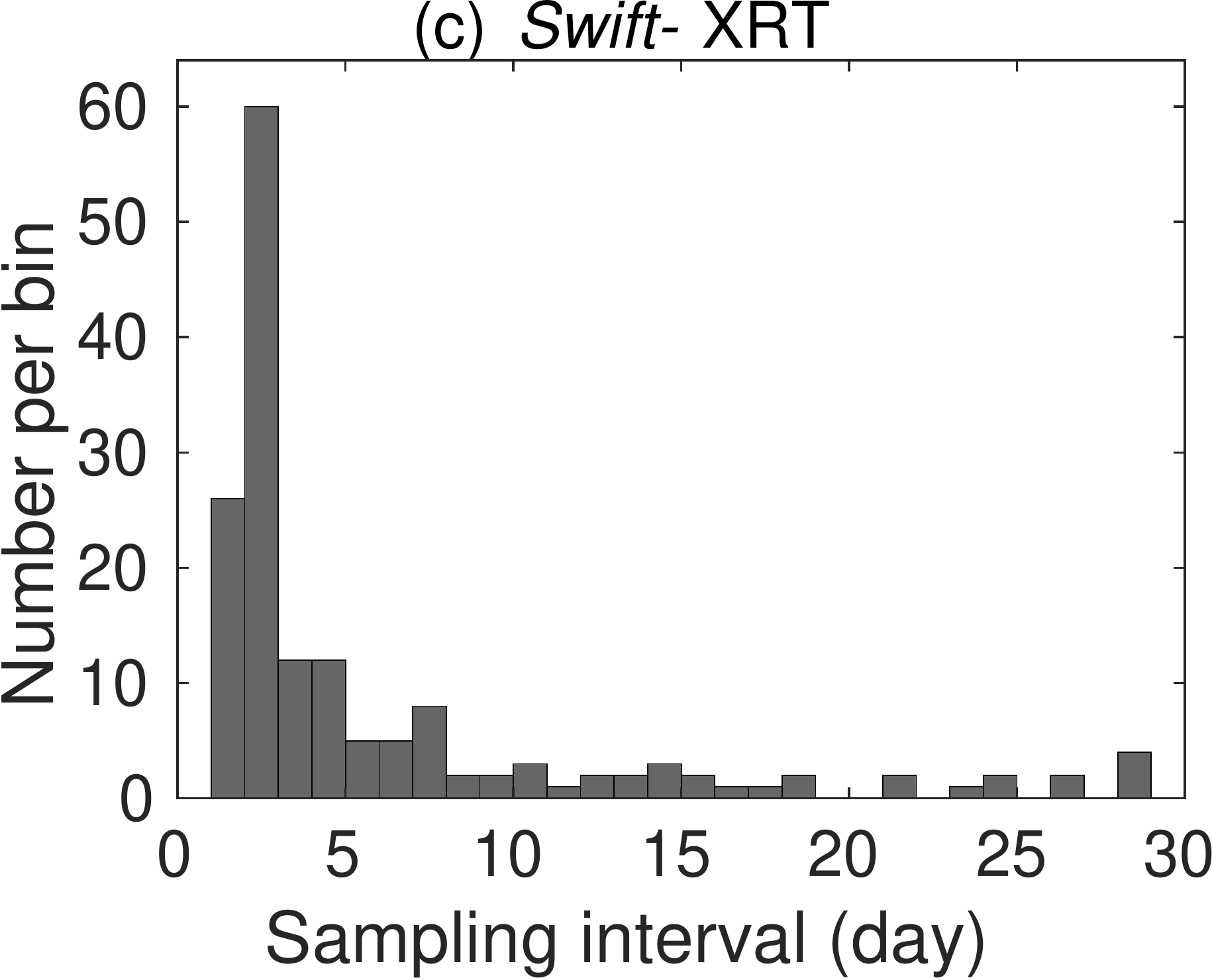}
\includegraphics[width=0.25\textwidth]{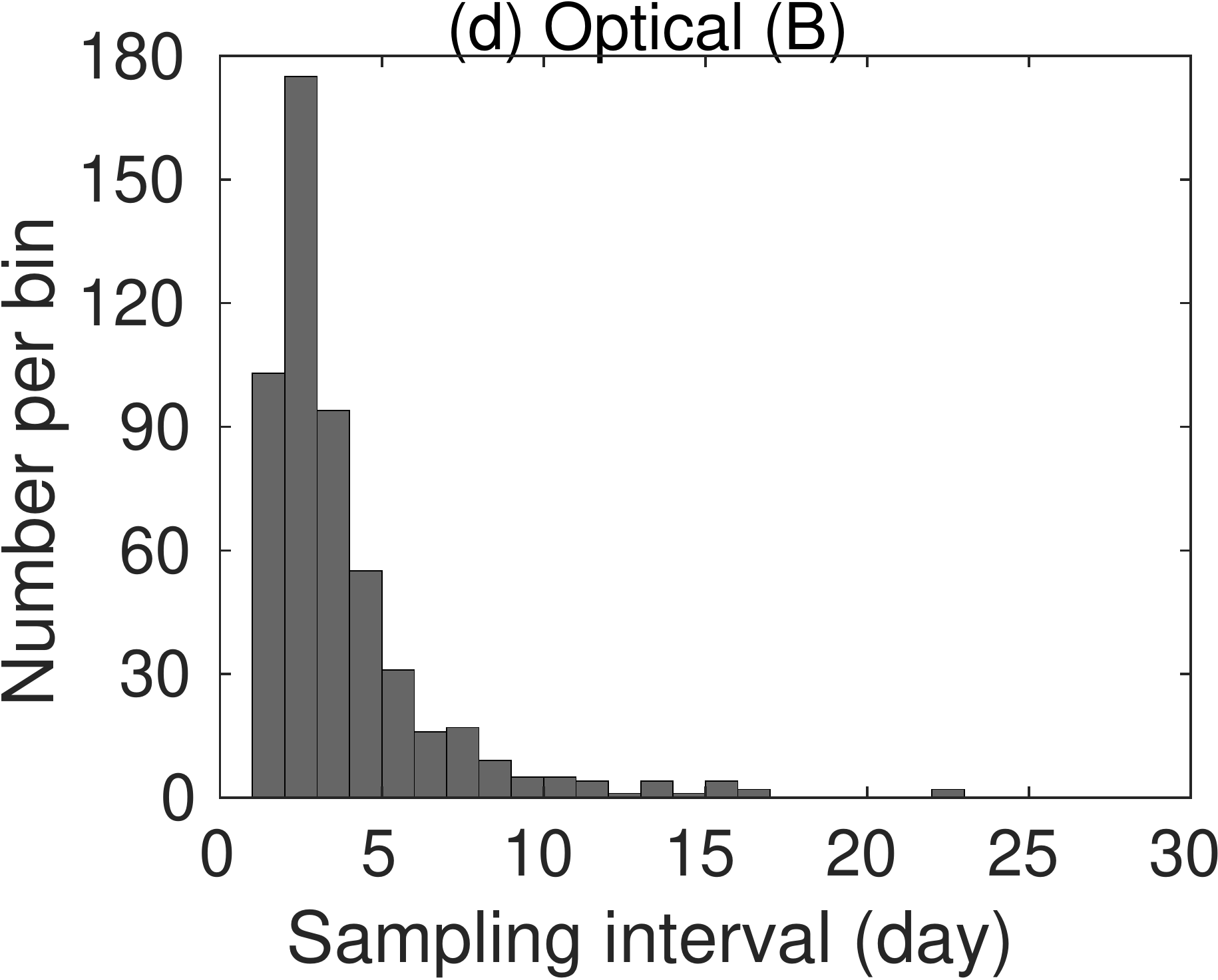}
}
\hbox{
\includegraphics[width=0.25\textwidth]{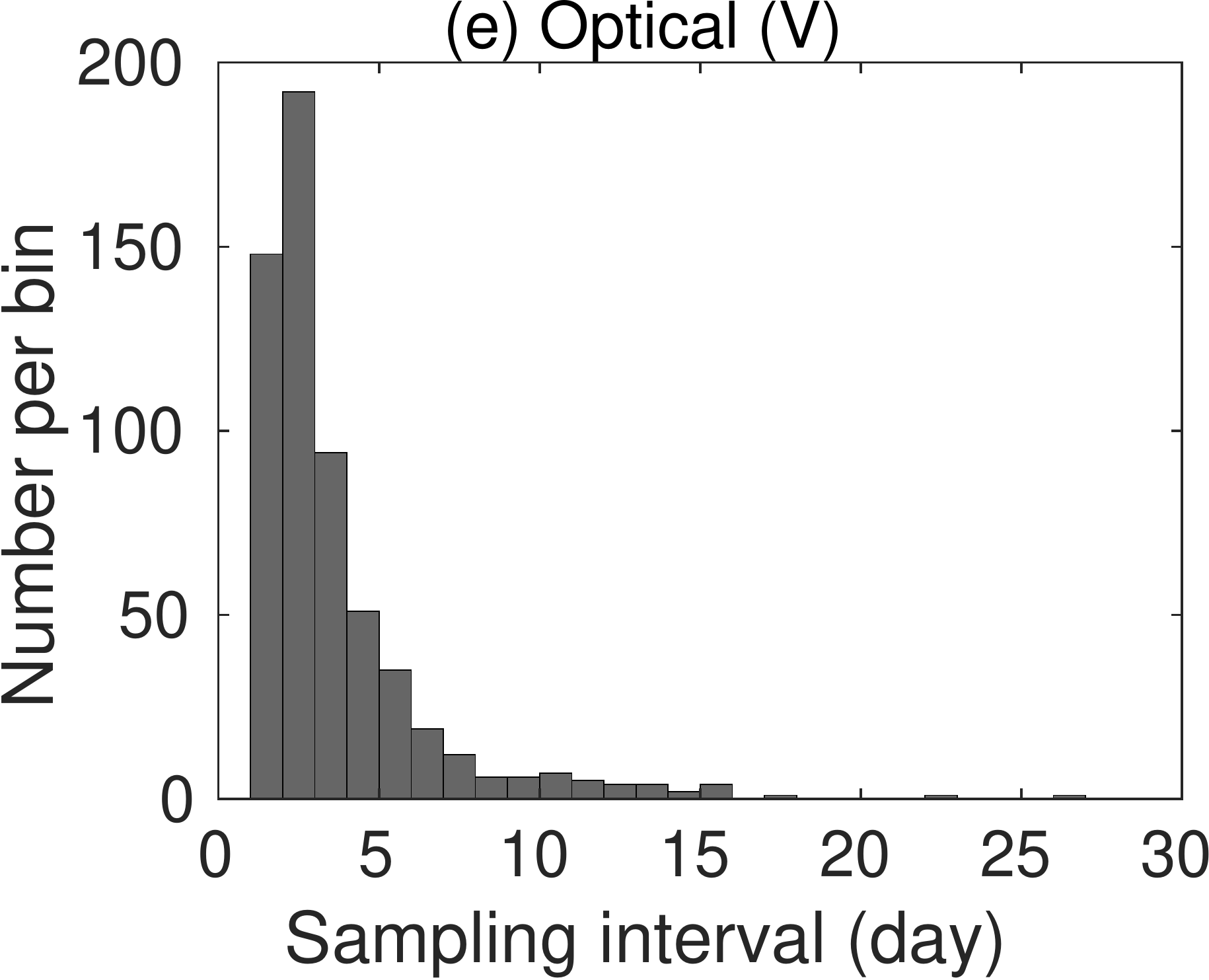}
\includegraphics[width=0.25\textwidth]{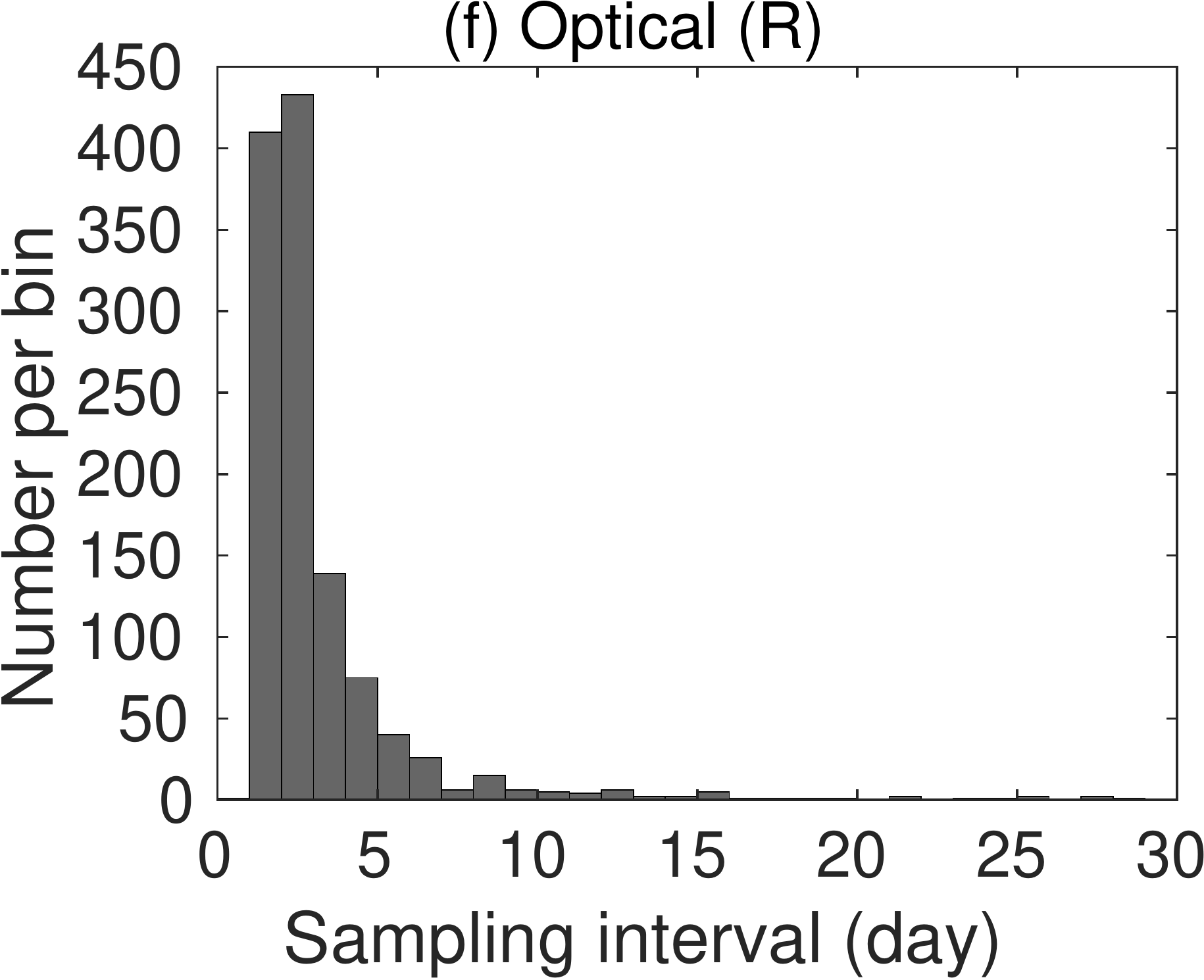}
\includegraphics[width=0.25\textwidth]{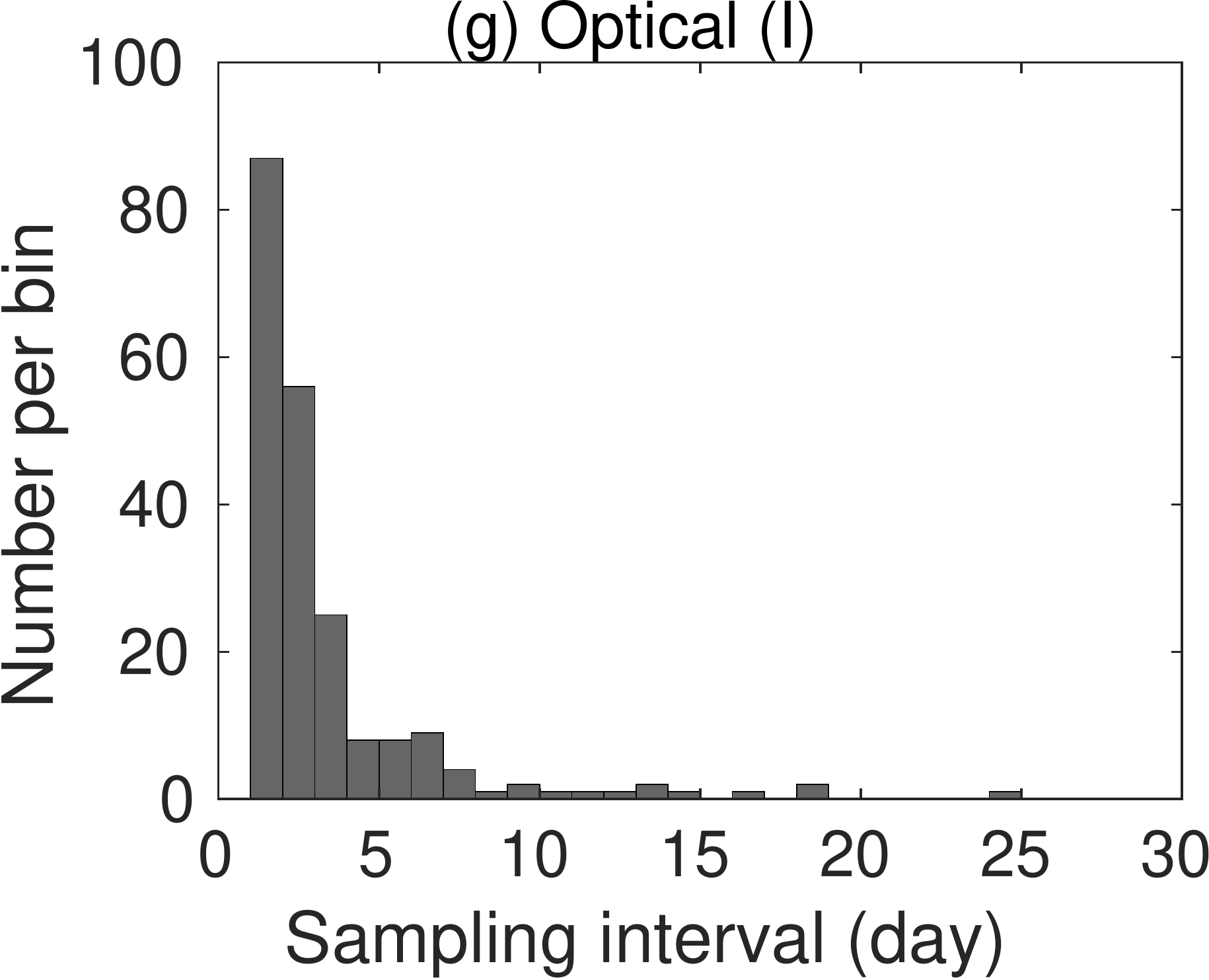}
\includegraphics[width=0.25\textwidth]{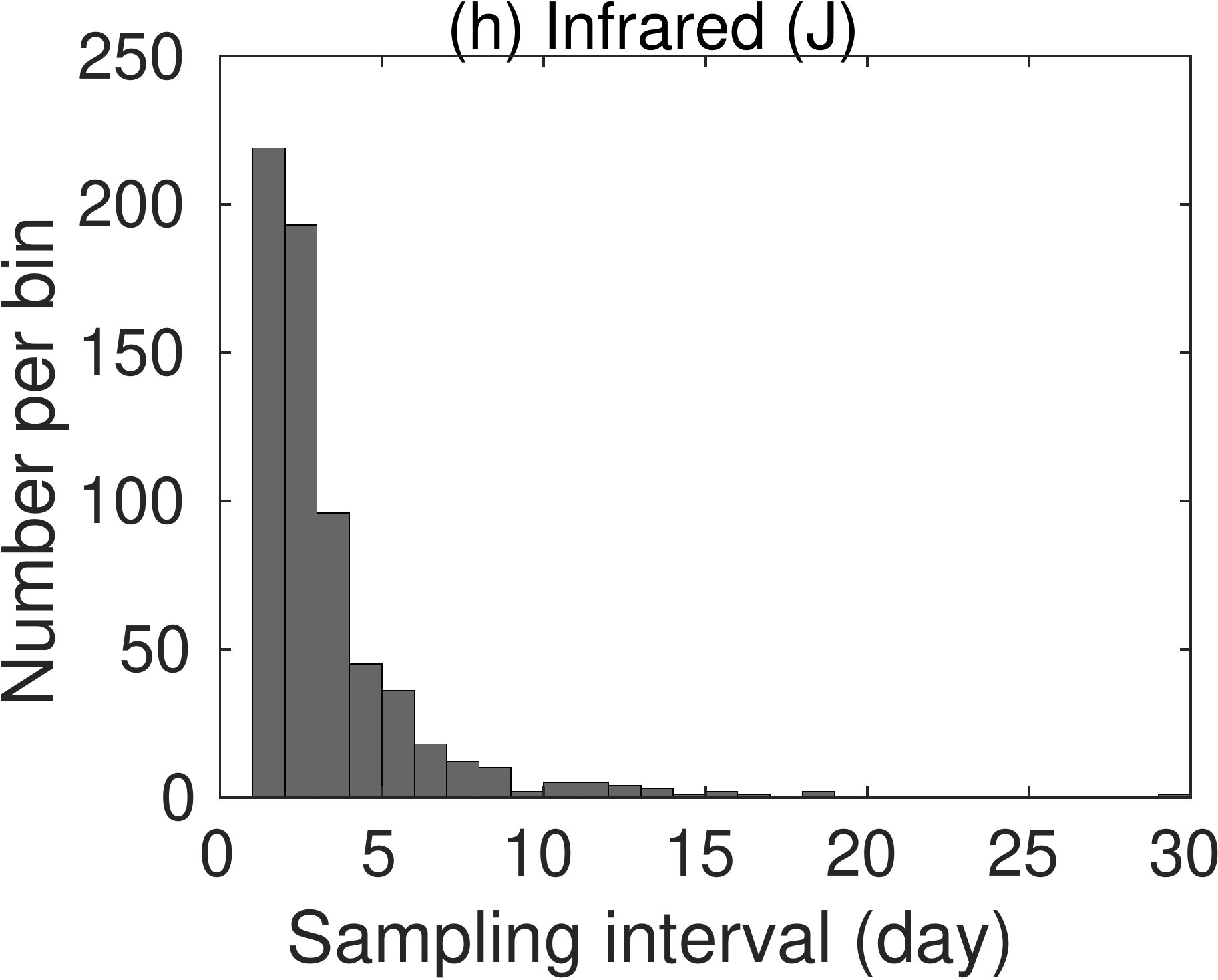}
}
\hbox{
\includegraphics[width=0.25\textwidth]{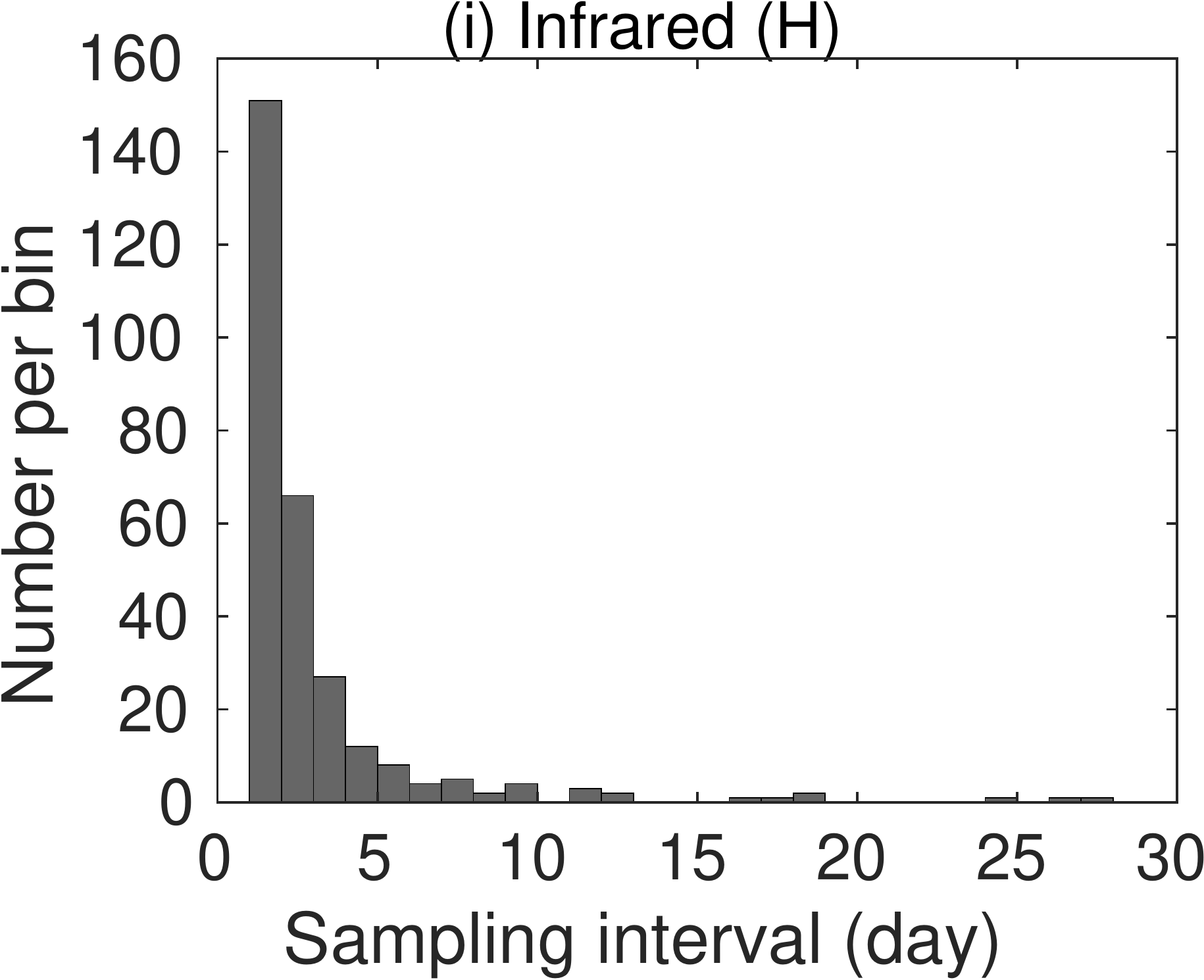}
\includegraphics[width=0.25\textwidth]{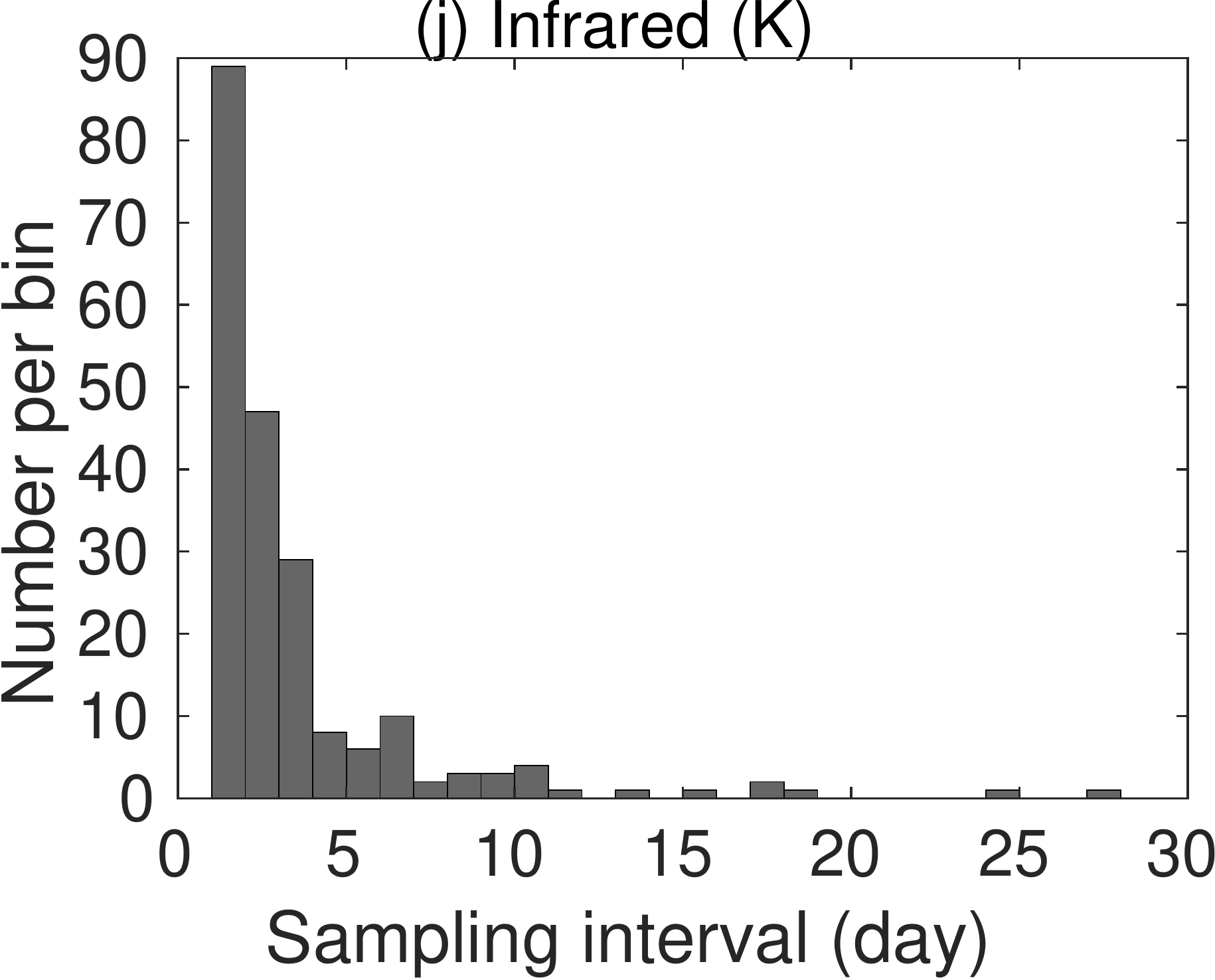}
\includegraphics[width=0.25\textwidth]{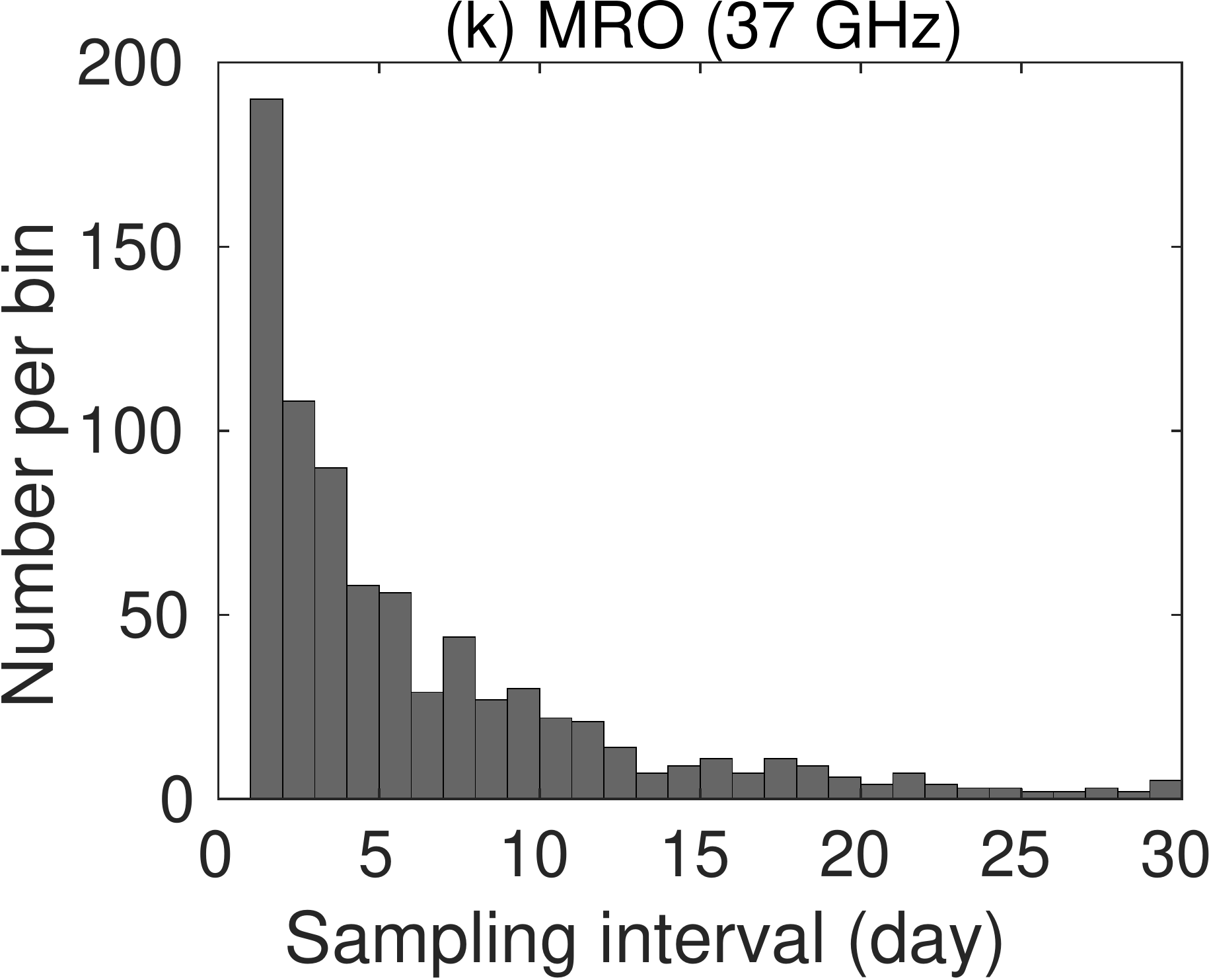}
\includegraphics[width=0.25\textwidth]{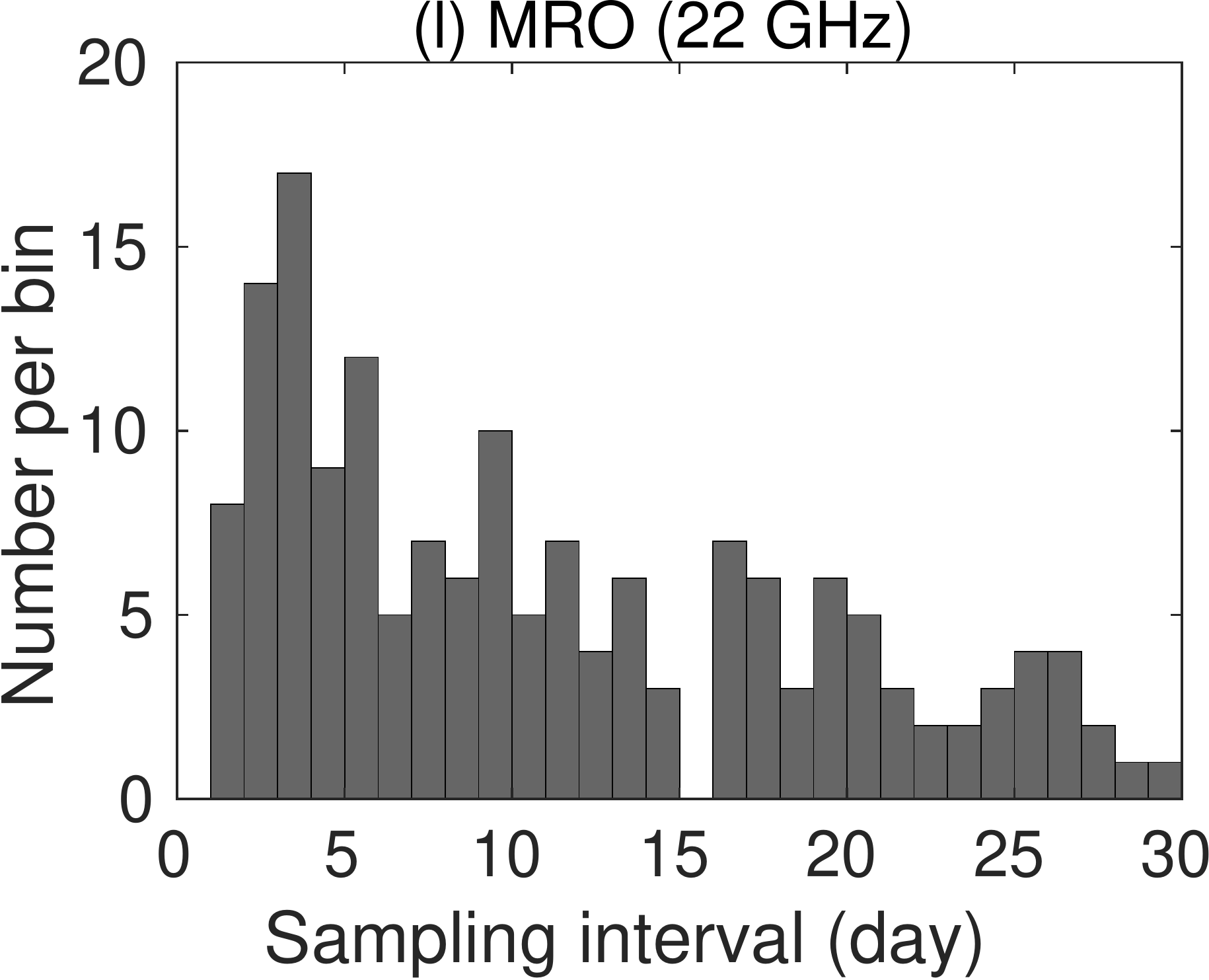}
}
\hbox{
\includegraphics[width=0.25\textwidth]{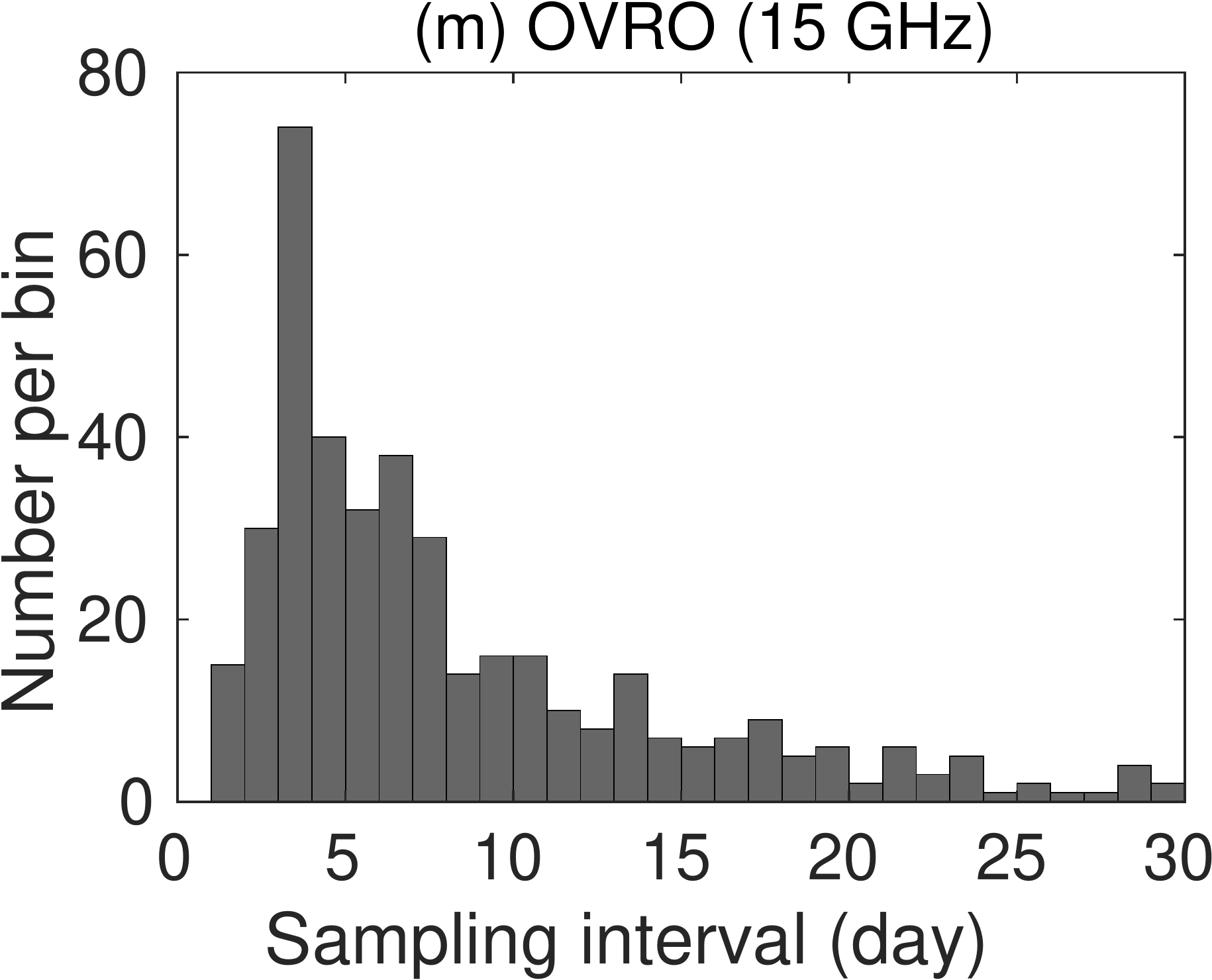}
\includegraphics[width=0.25\textwidth]{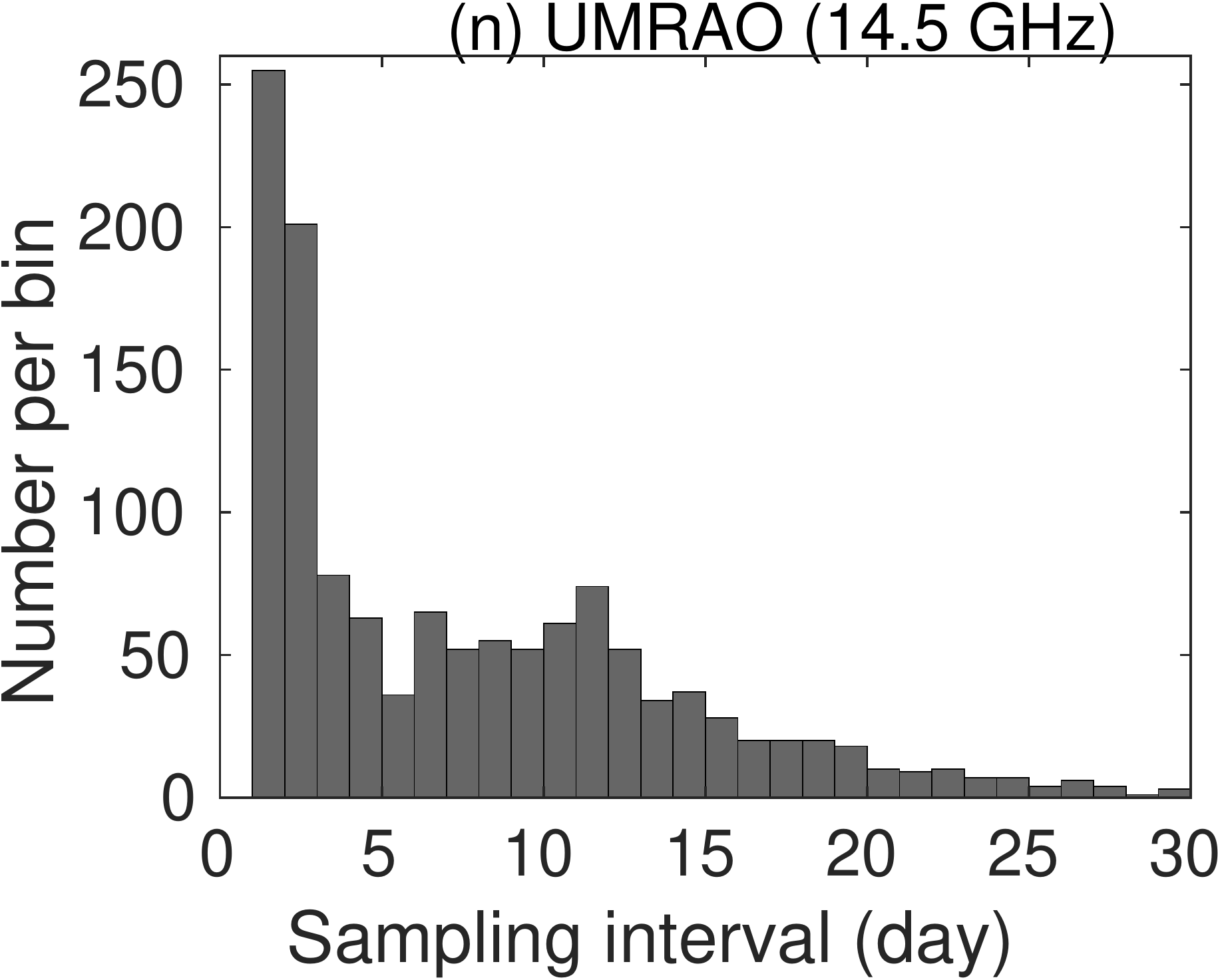}
\includegraphics[width=0.25\textwidth]{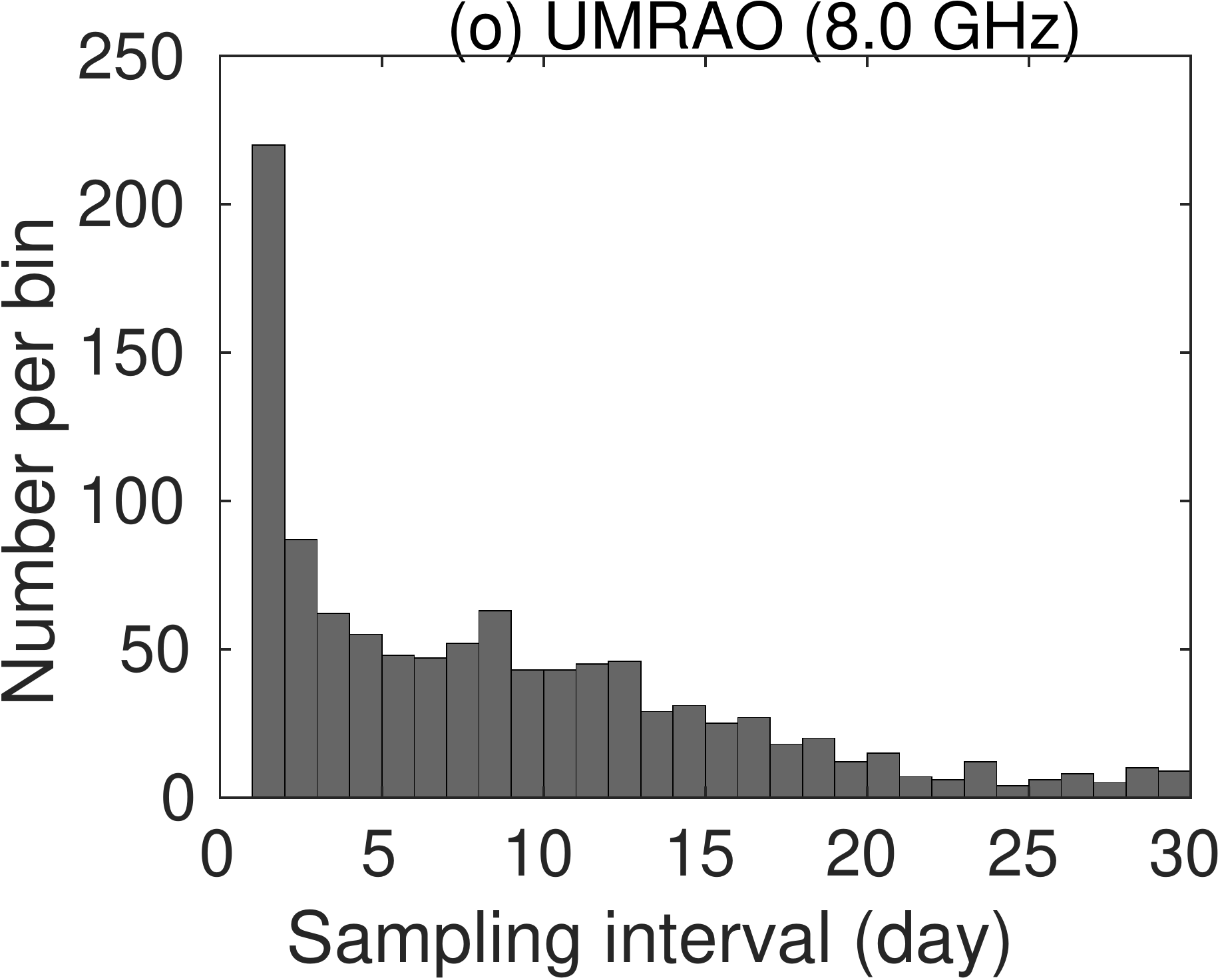}
\includegraphics[width=0.25\textwidth]{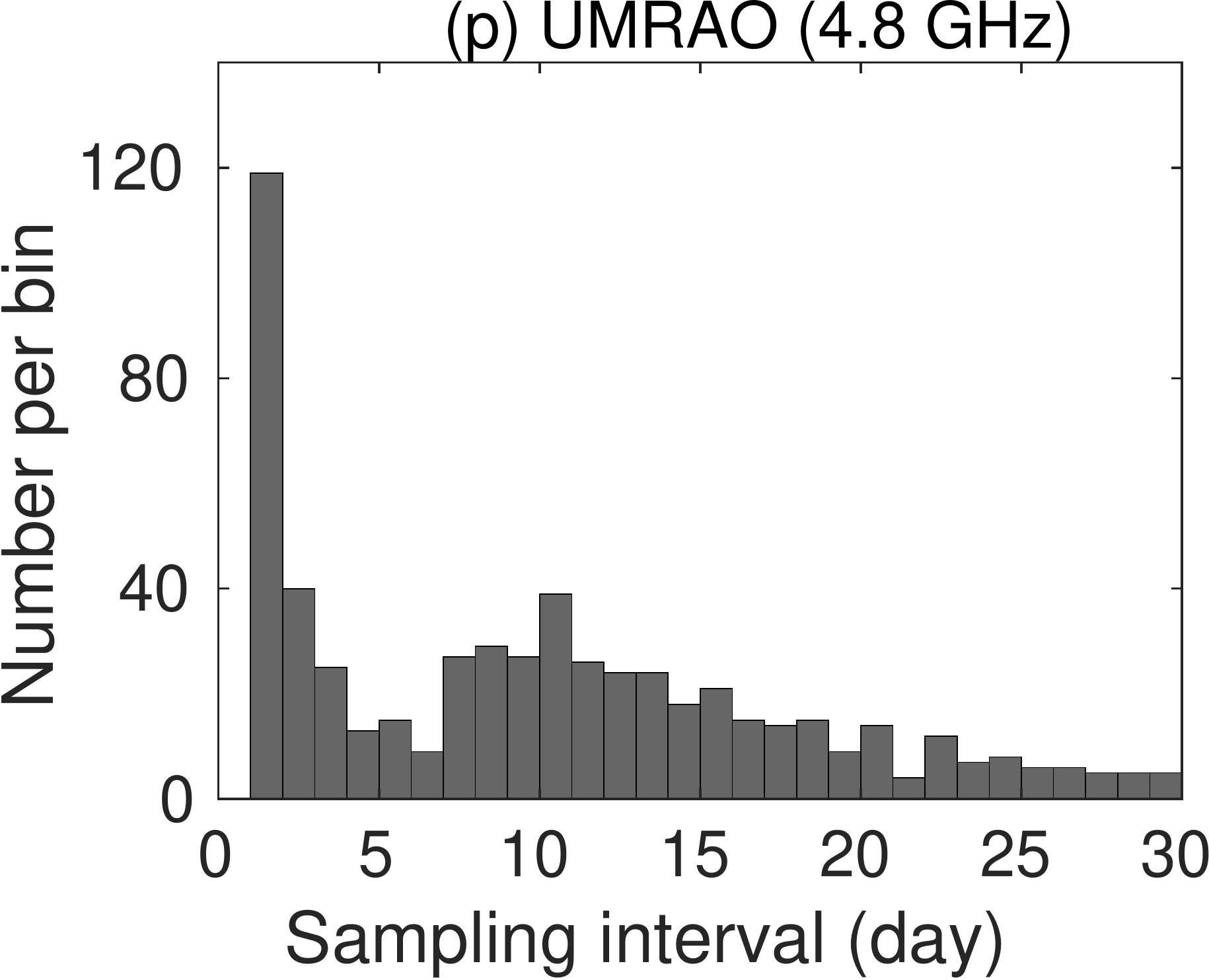}
}
\hbox{
\includegraphics[width=0.25\textwidth]{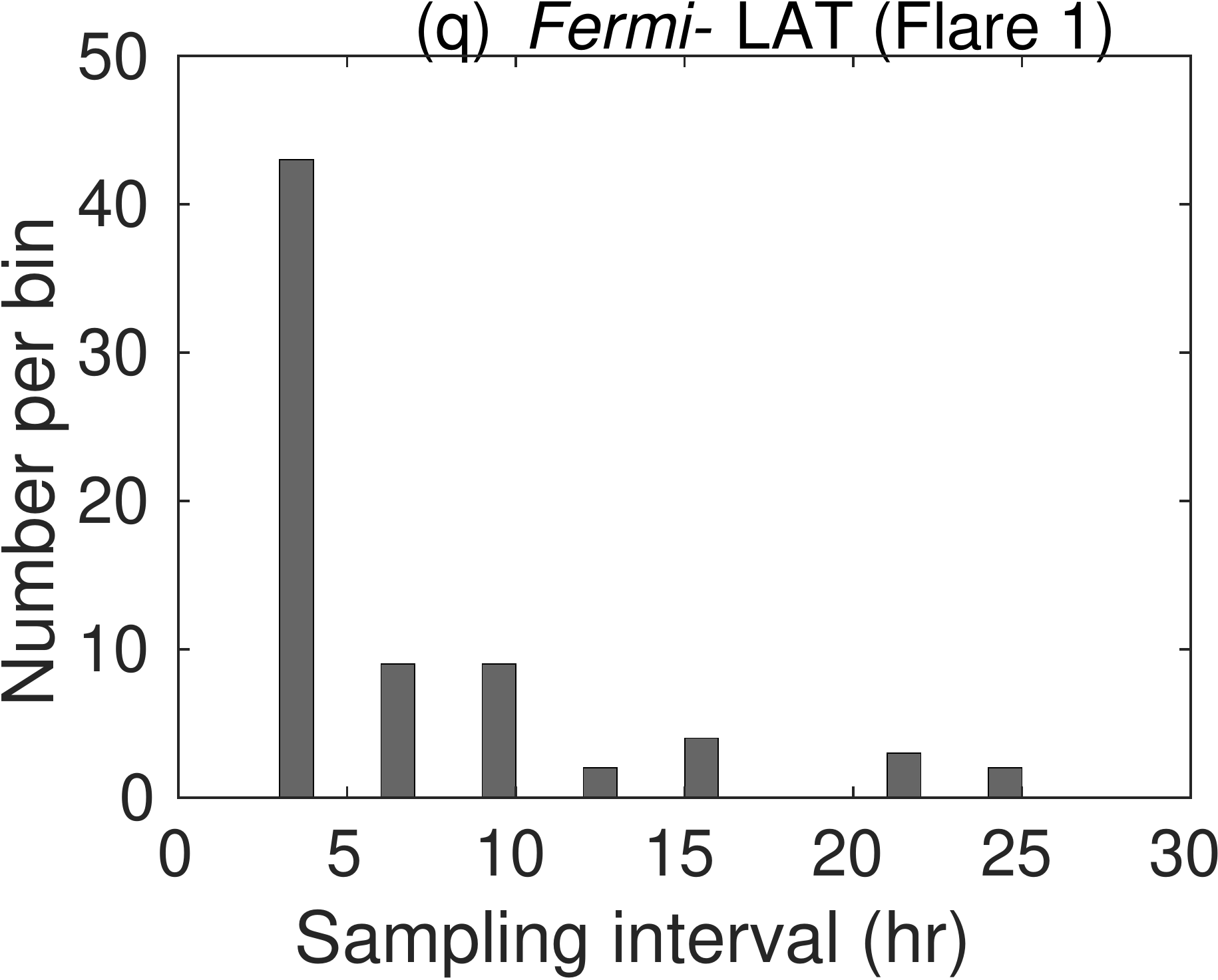}
\includegraphics[width=0.25\textwidth]{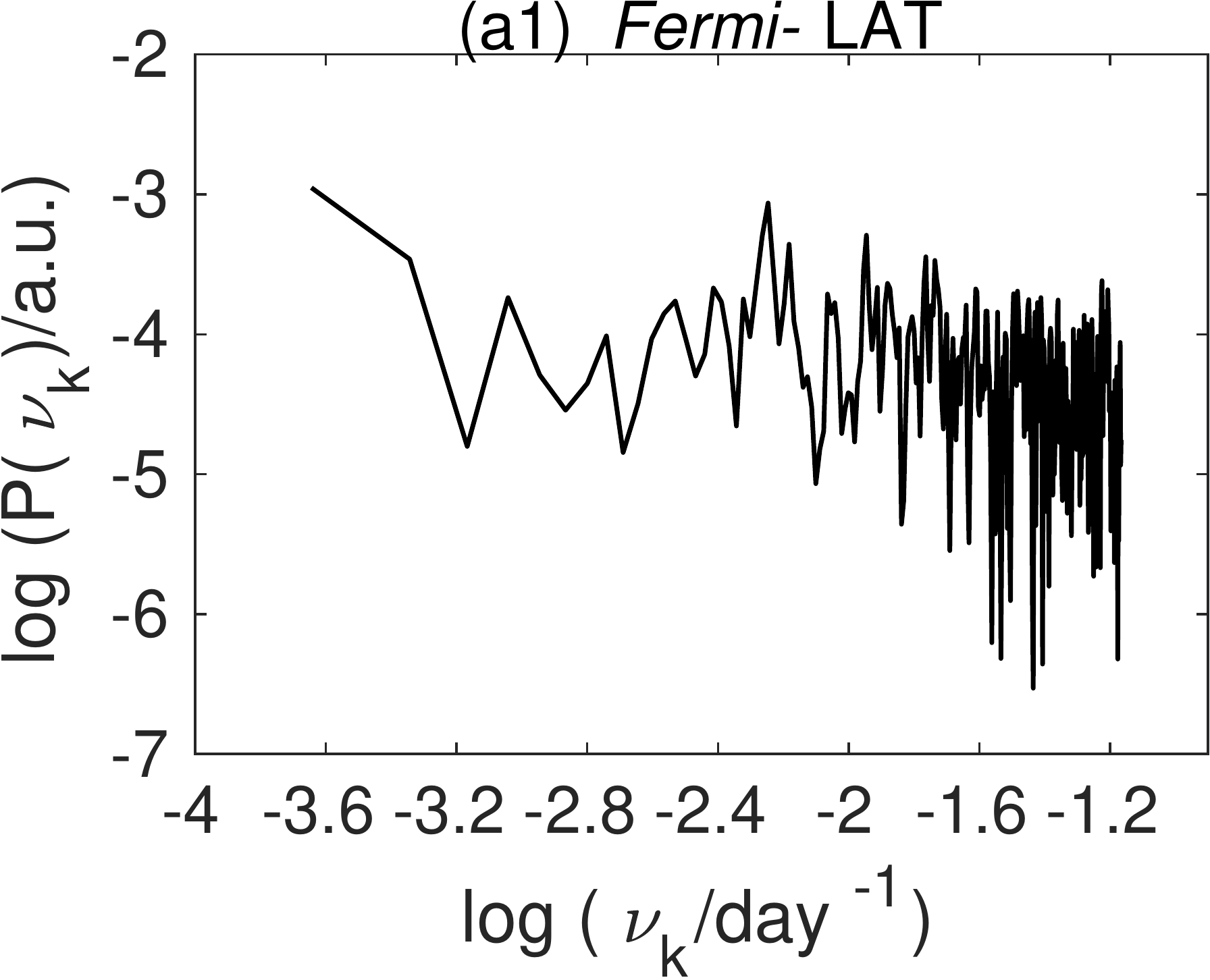}
\includegraphics[width=0.25\textwidth]{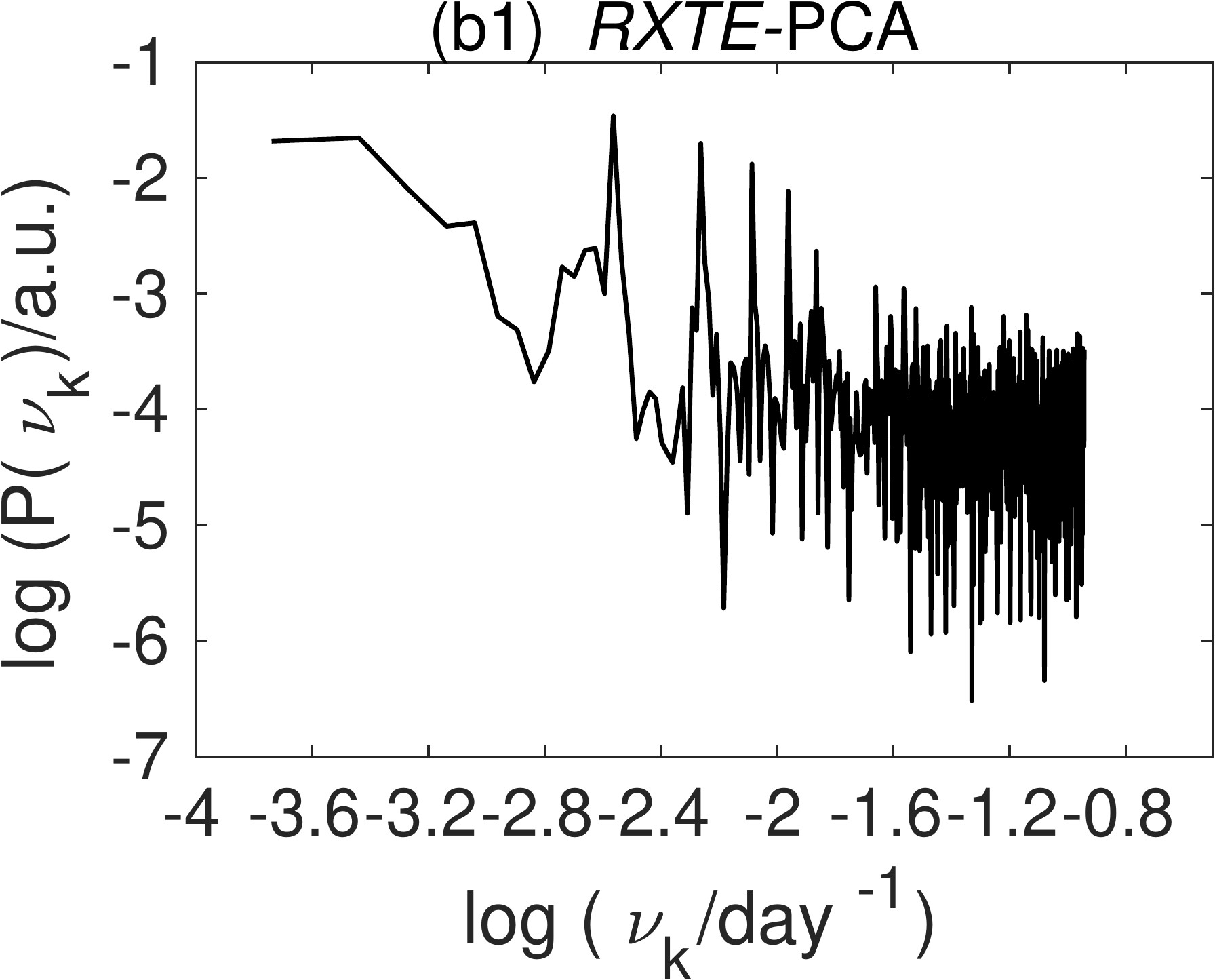}
\includegraphics[width=0.25\textwidth]{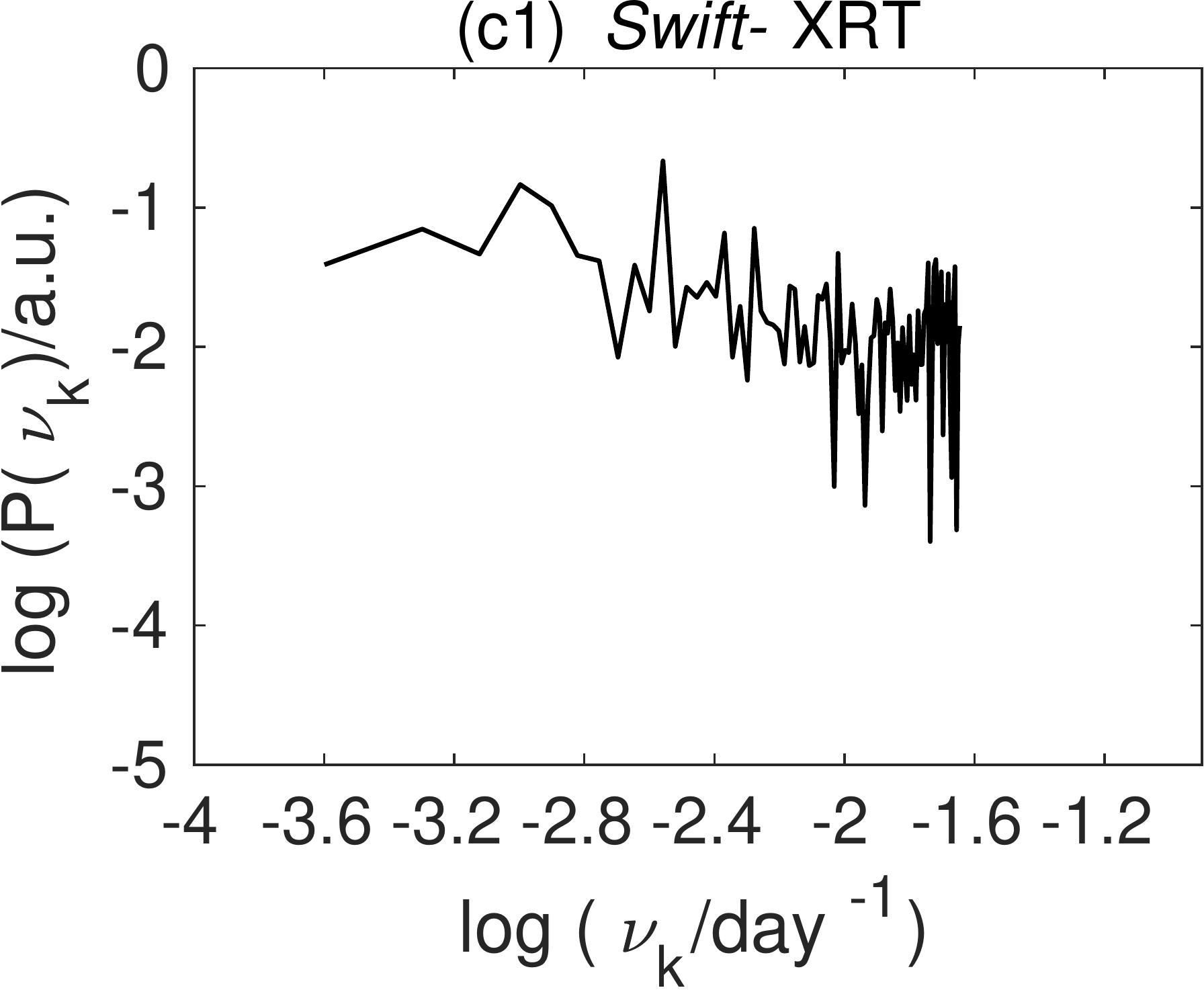}
}
\hbox{
\includegraphics[width=0.25\textwidth]{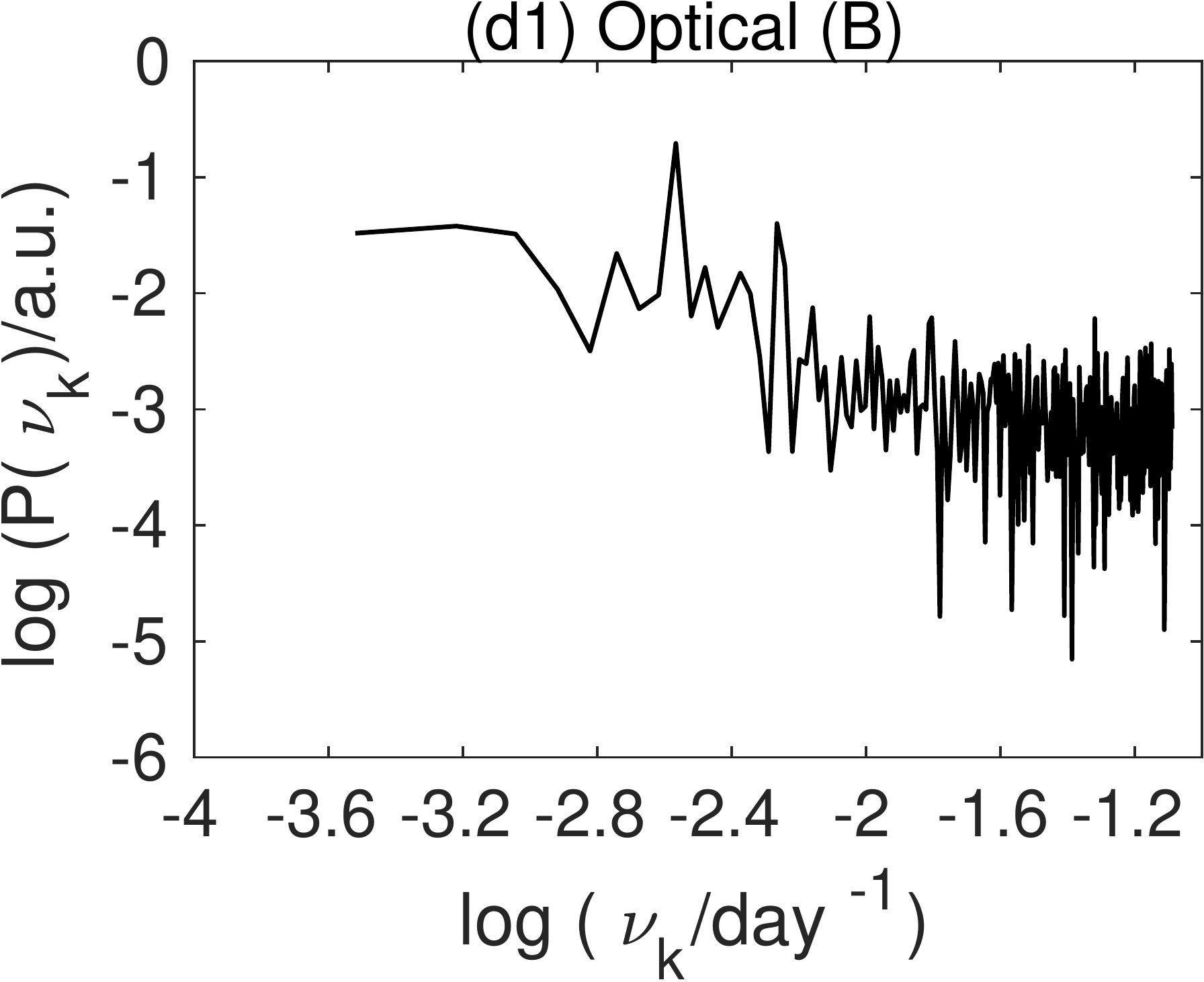}
\includegraphics[width=0.25\textwidth]{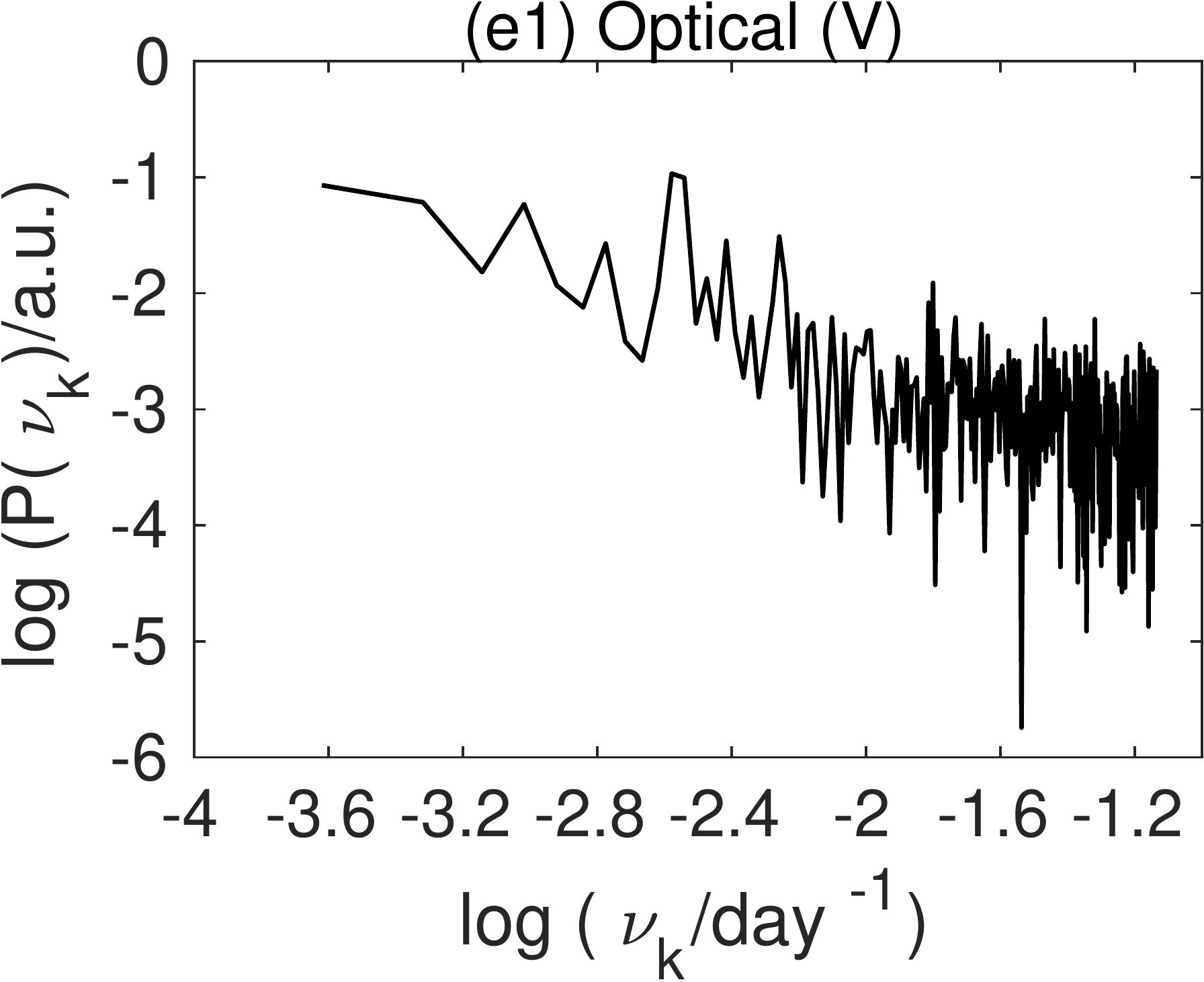}
\includegraphics[width=0.25\textwidth]{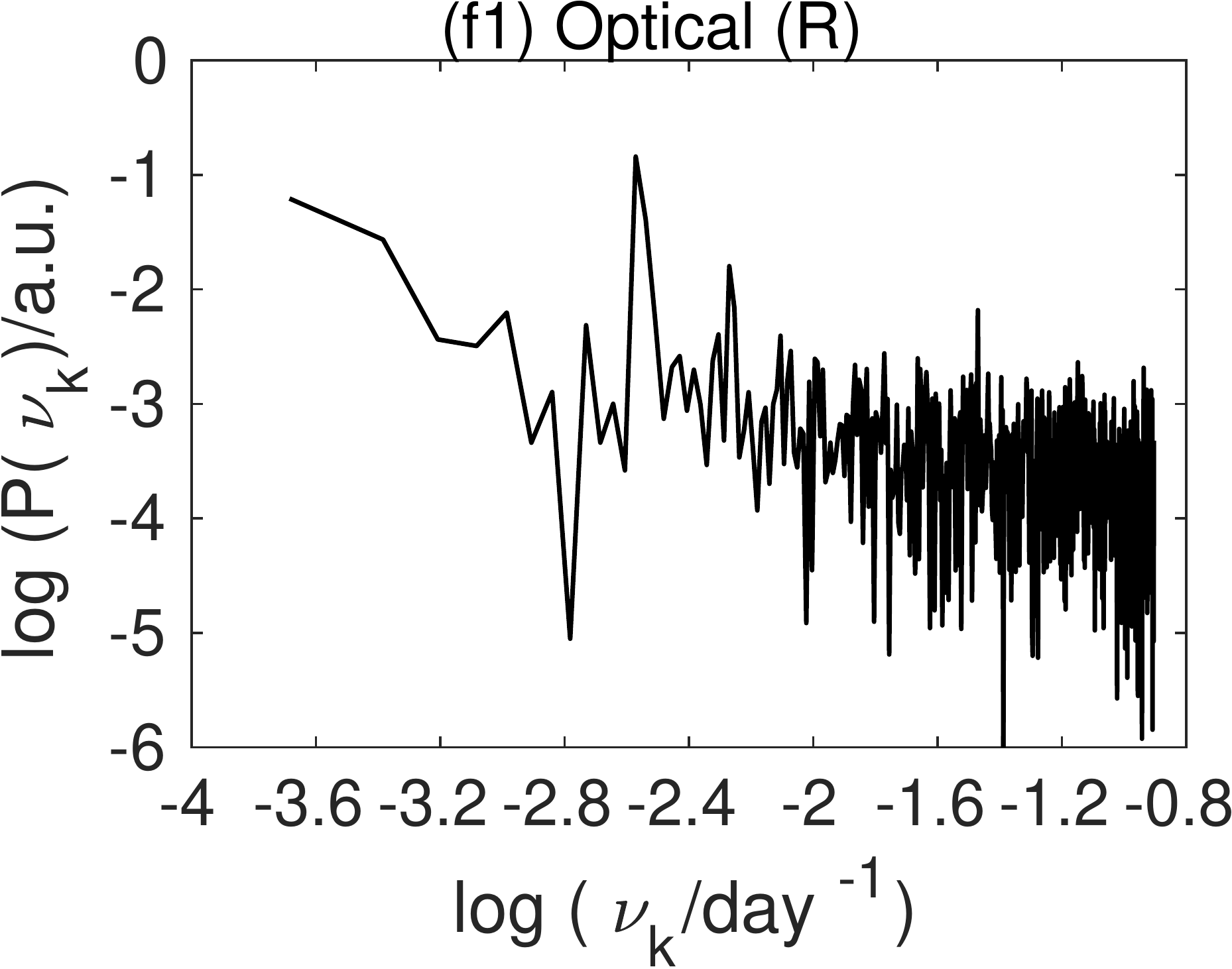}
\includegraphics[width=0.25\textwidth]{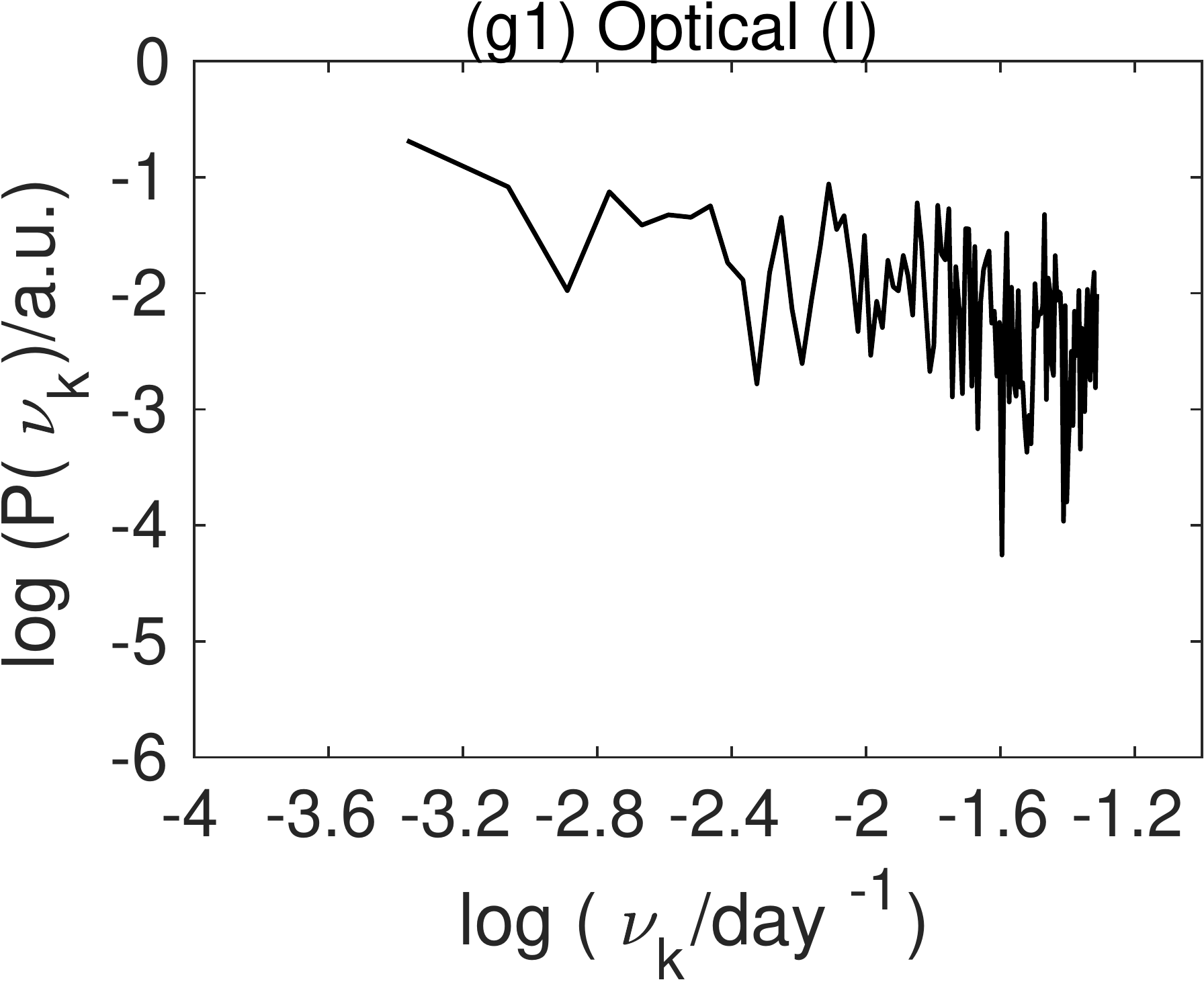}
}

\caption{As in Figure~\ref{appfig:swf3c}, but for the blazar PKS\,1510$-$089.}
\label{appfig:swfpks}
\end{figure*}

\addtocounter{figure}{-1}

\begin{figure*}[ht!]

\hbox{
\includegraphics[width=0.25\textwidth]{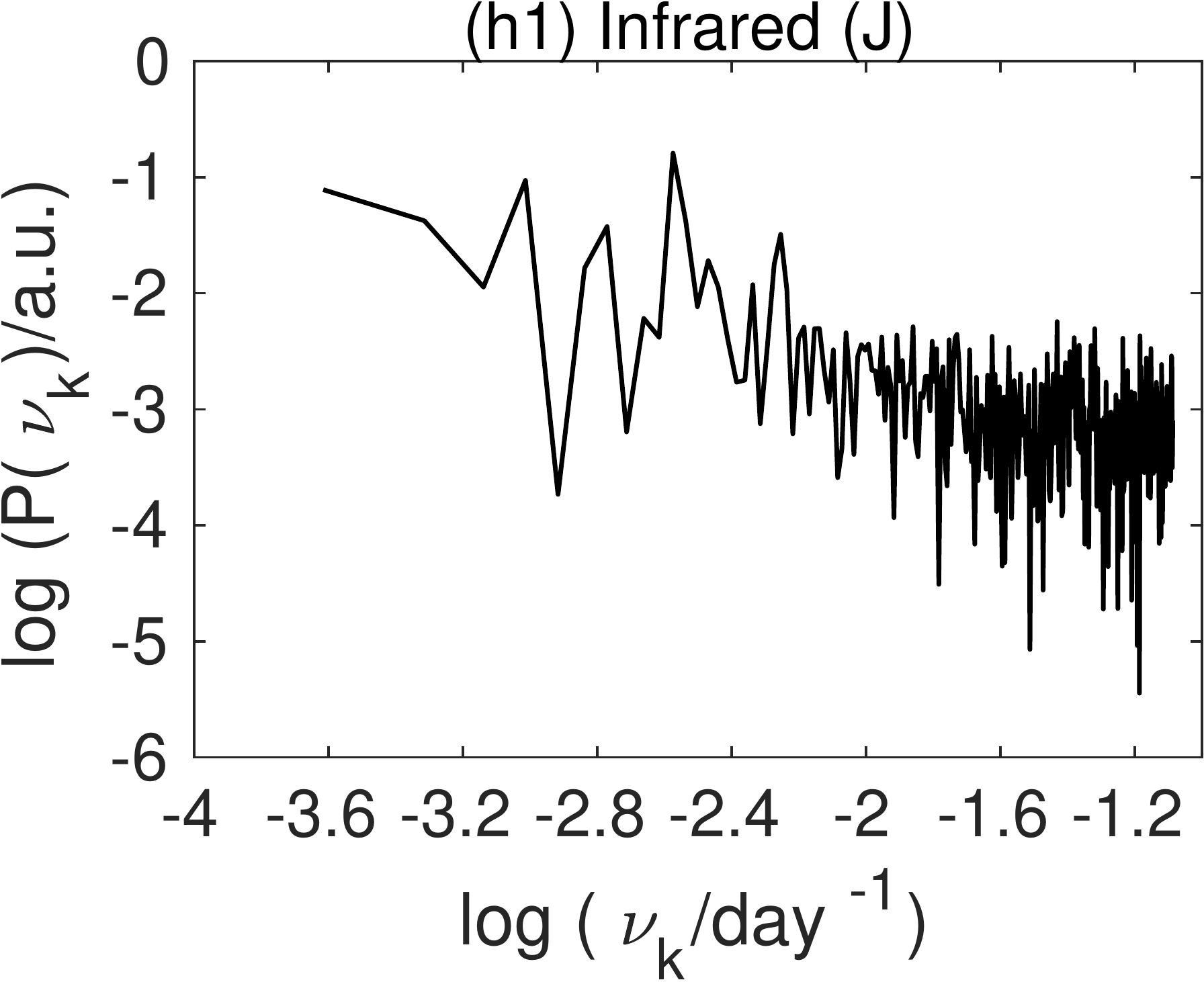}
\includegraphics[width=0.25\textwidth]{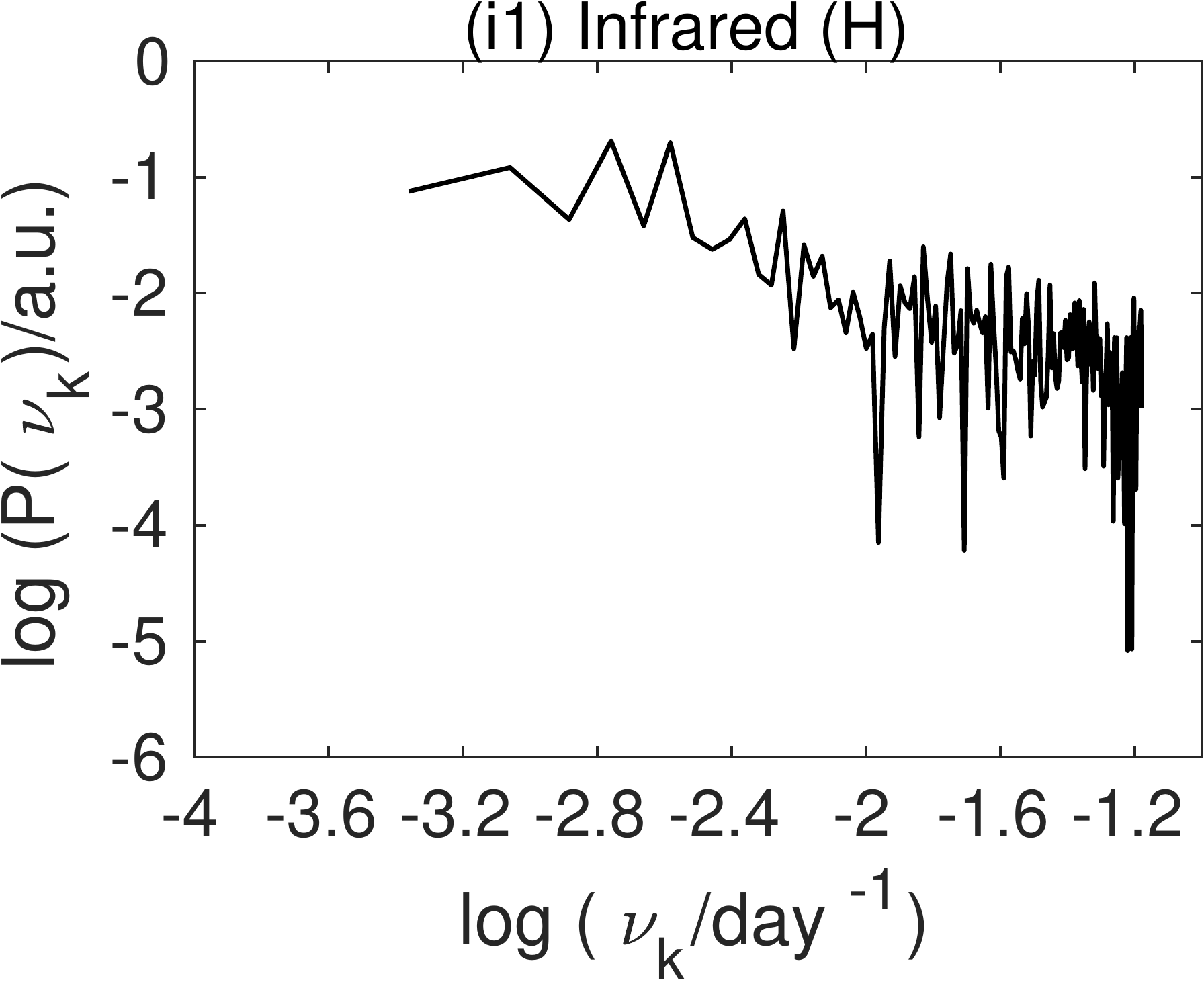}
\includegraphics[width=0.25\textwidth]{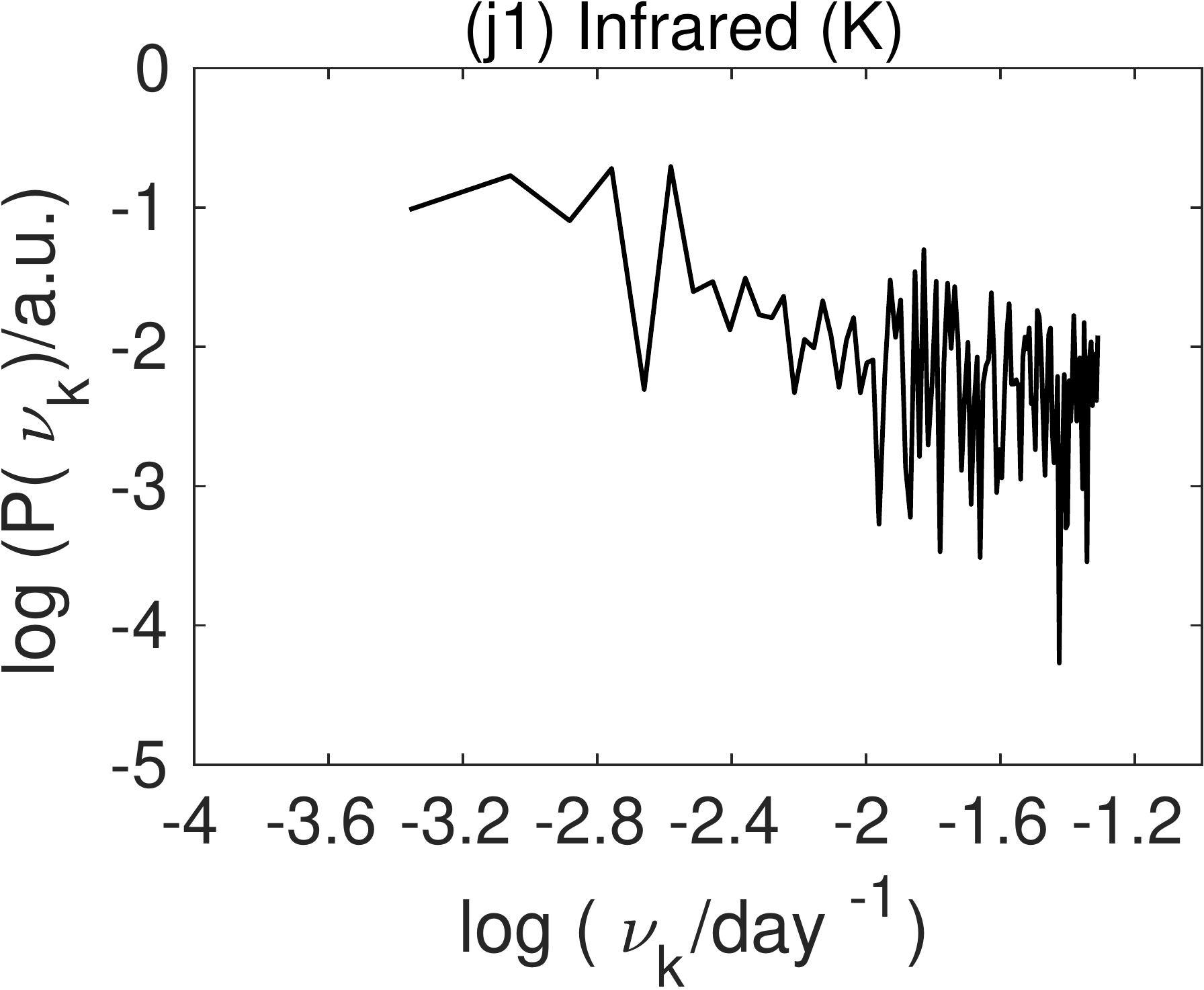}
\includegraphics[width=0.25\textwidth]{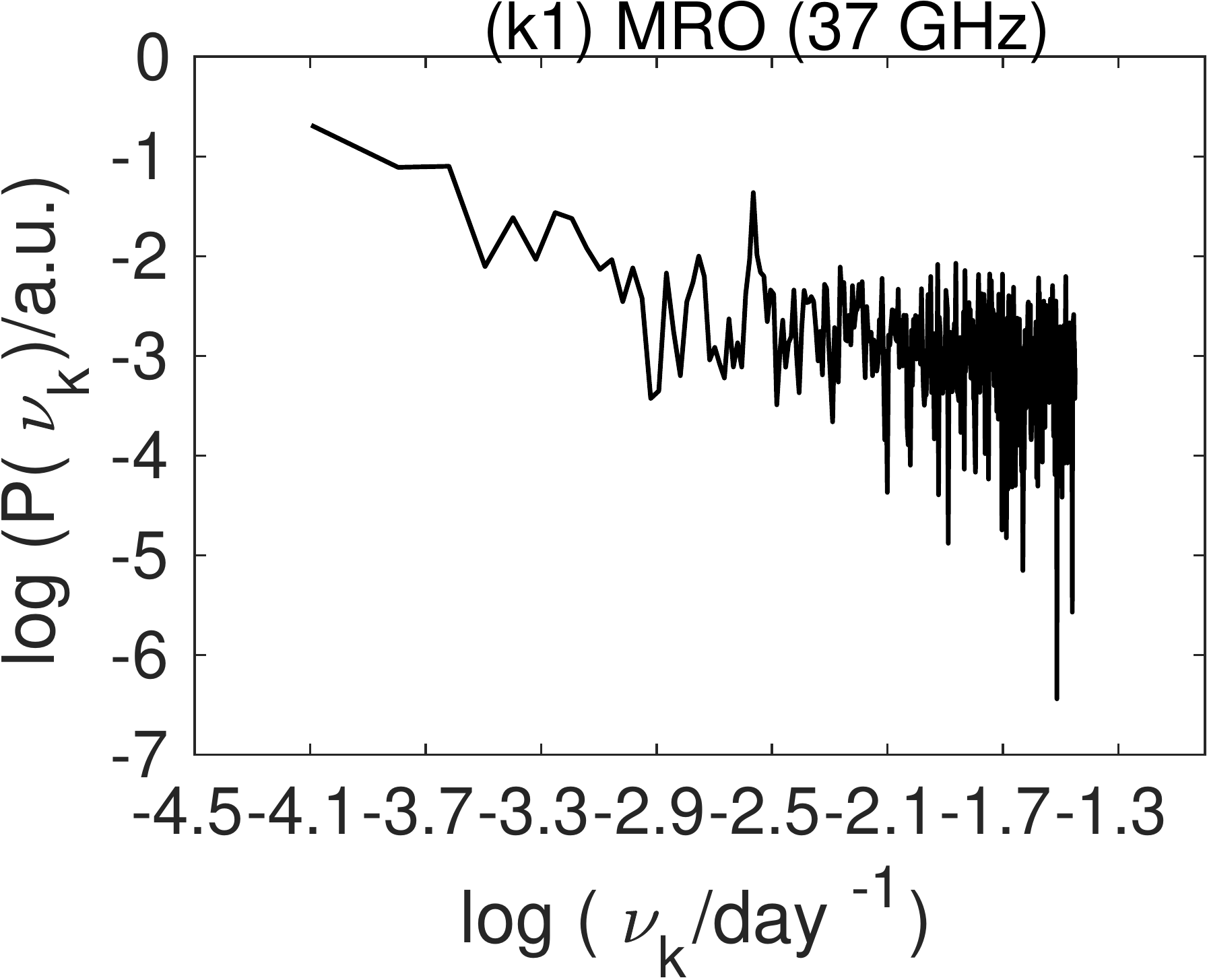}
}
\hbox{
\includegraphics[width=0.25\textwidth]{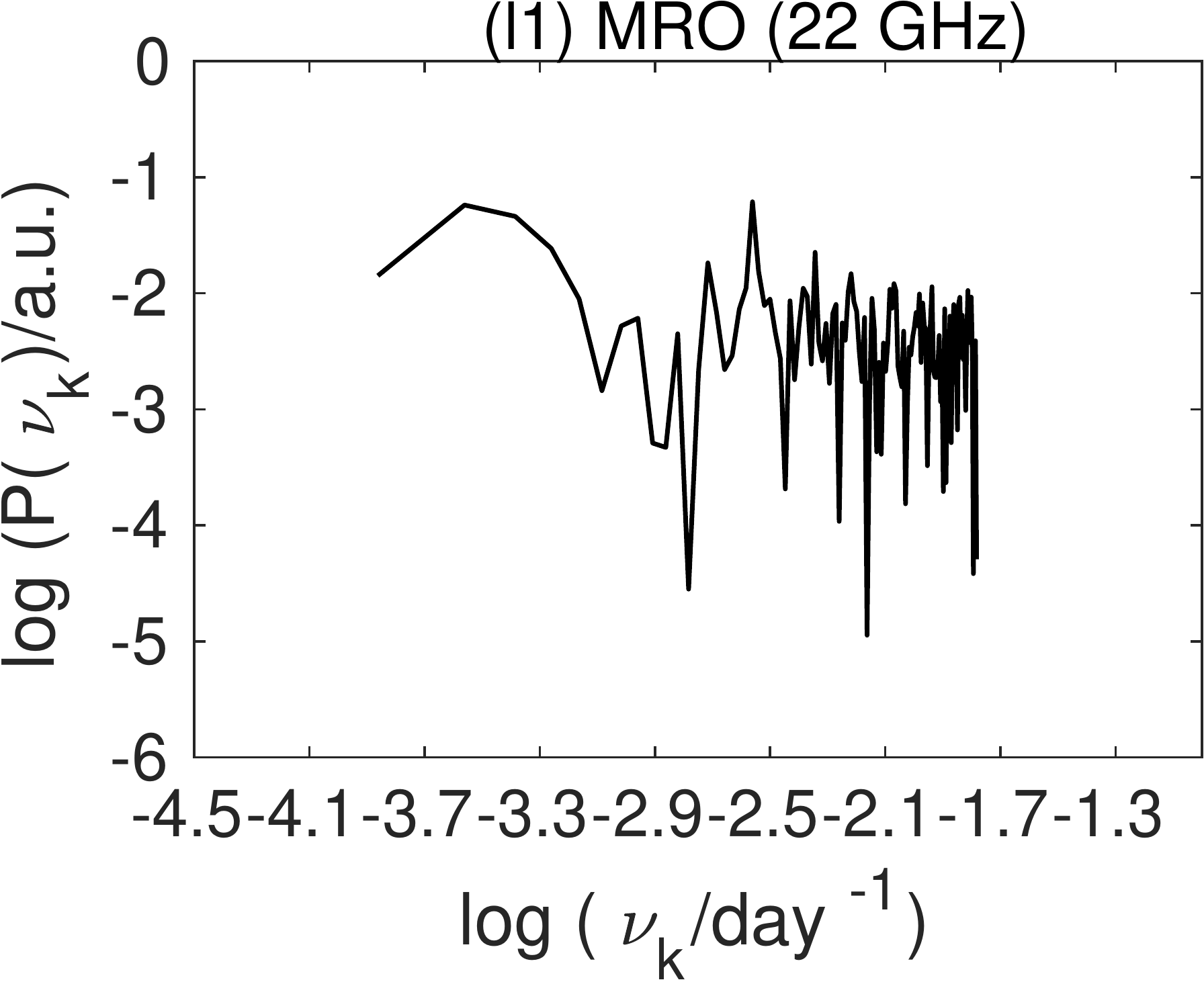}
\includegraphics[width=0.25\textwidth]{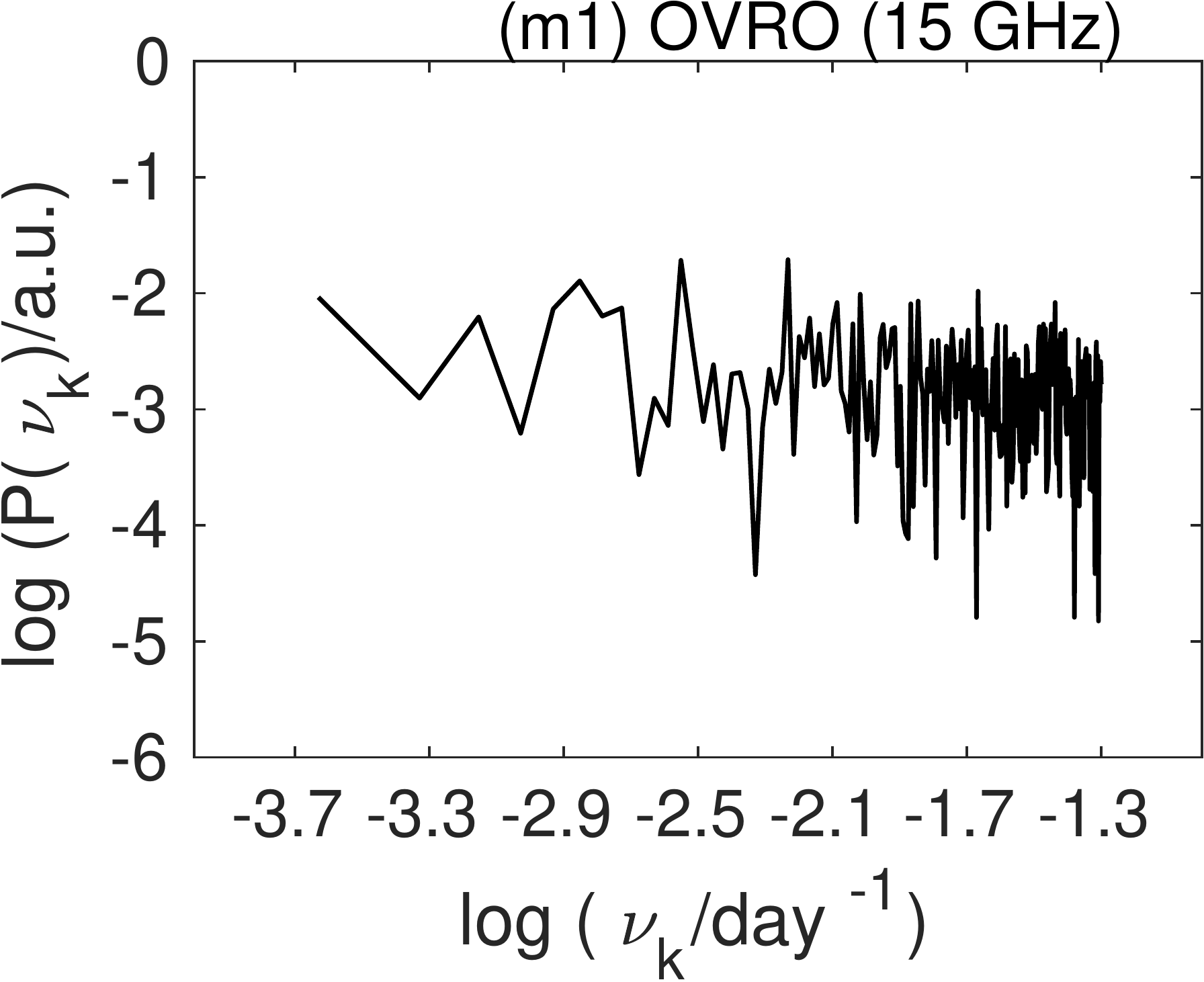}
\includegraphics[width=0.25\textwidth]{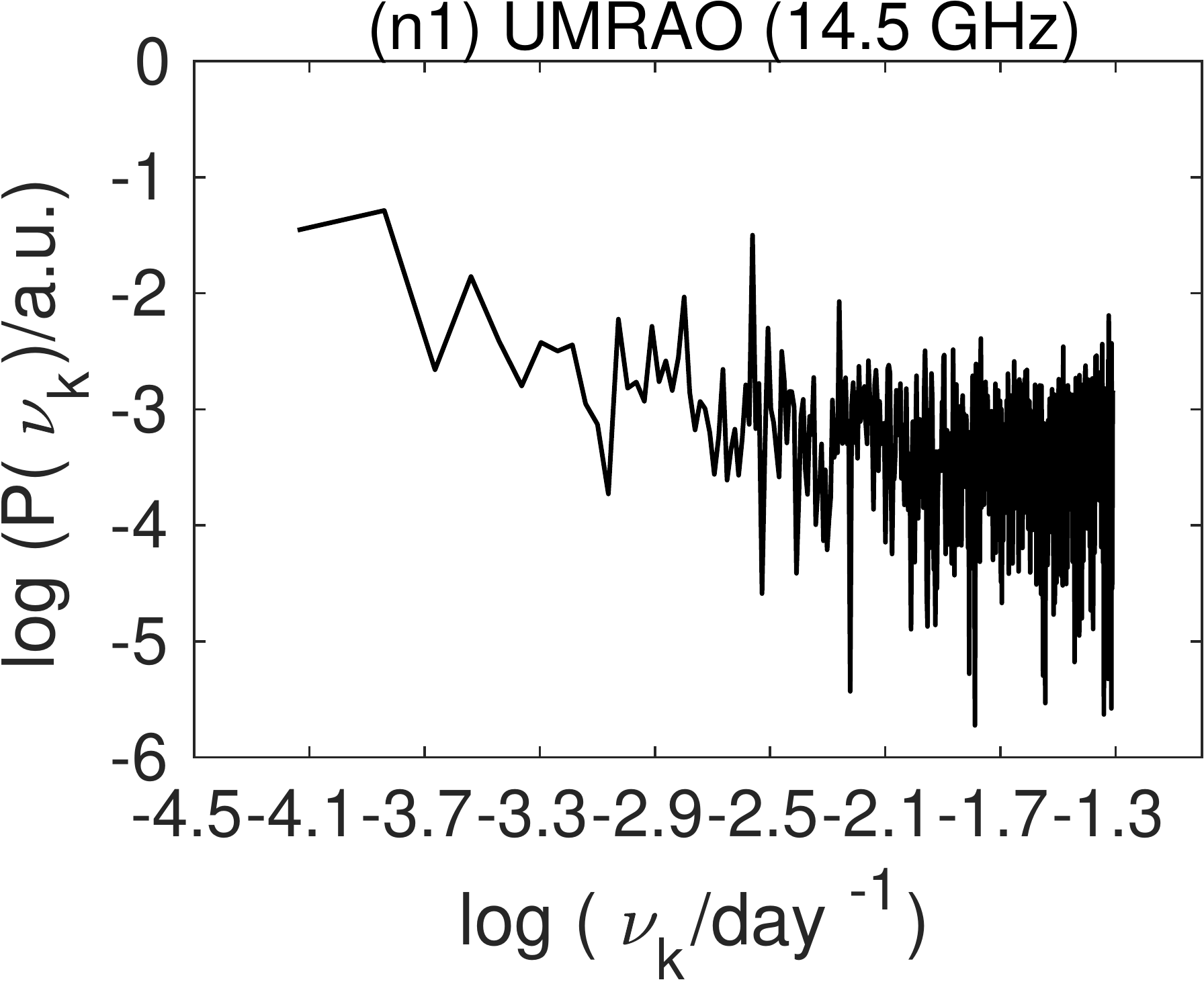}
\includegraphics[width=0.25\textwidth]{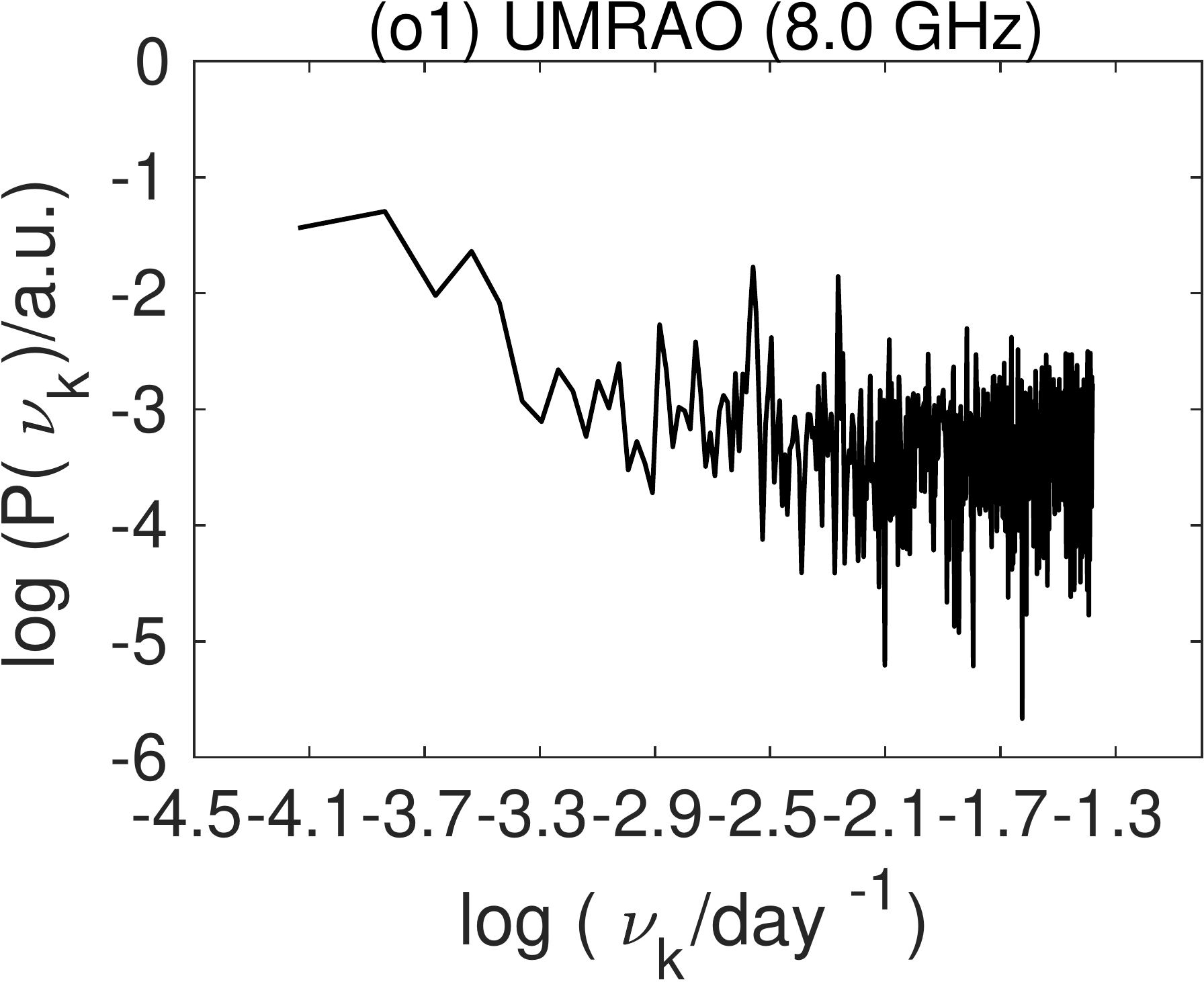}
}
\hbox{
\includegraphics[width=0.25\textwidth]{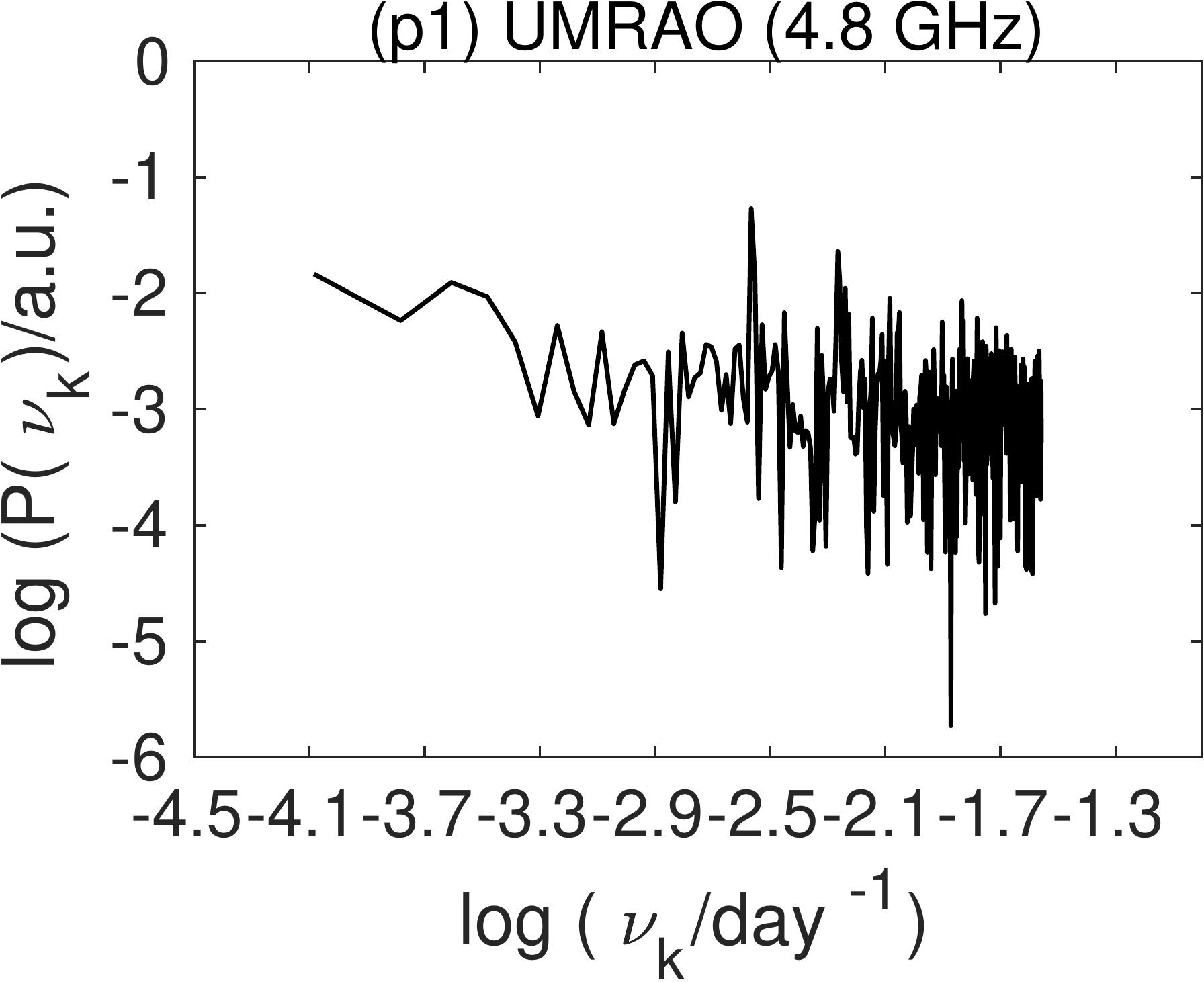}
\includegraphics[width=0.25\textwidth]{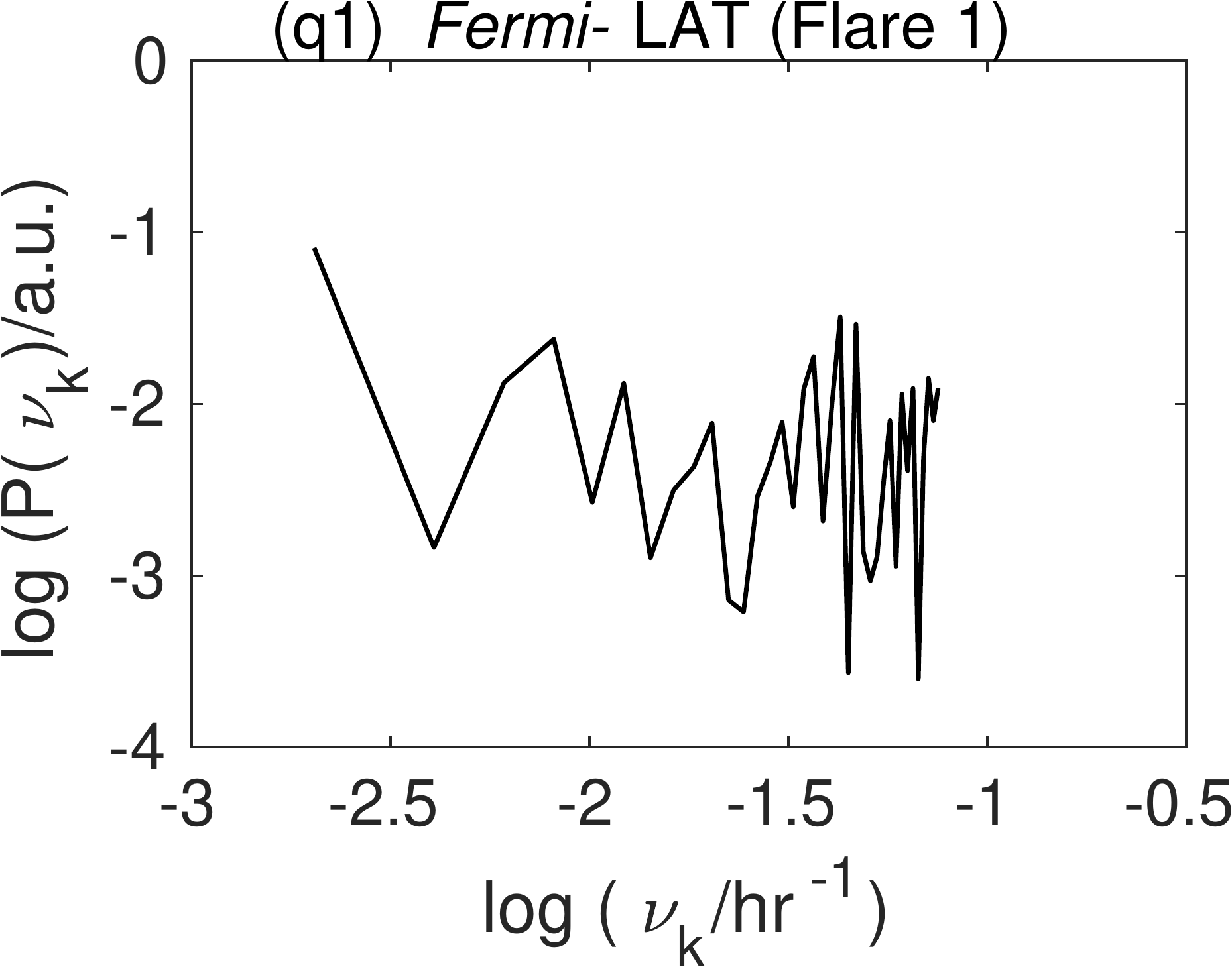}
}
\caption{(continued)}
\end{figure*}

\clearpage
\newpage
\section{Distribution curves of the analyzed PSDs}

\begin{figure*}[ht!]
\hbox{
\includegraphics[width=0.25\textwidth]{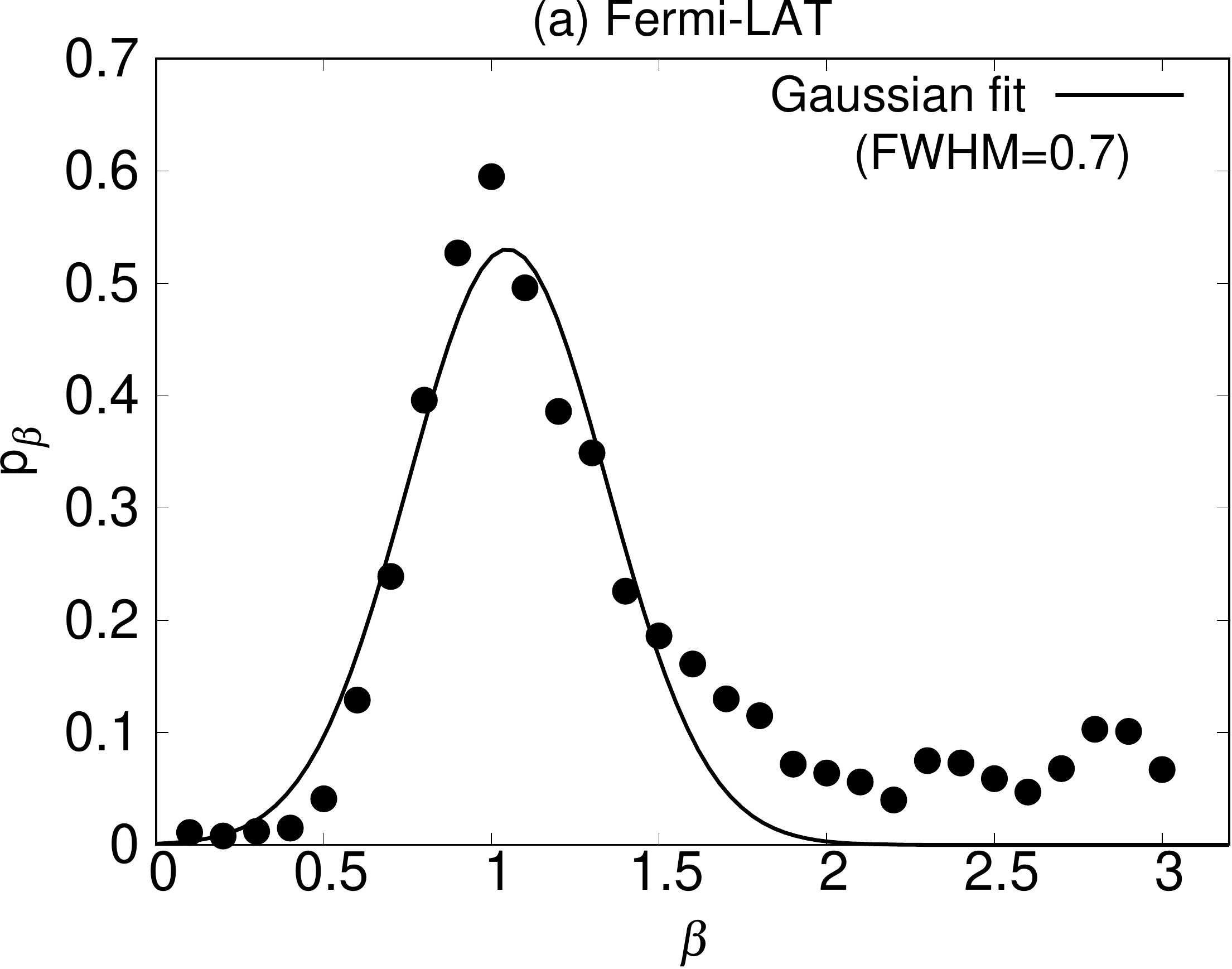}
\includegraphics[width=0.25\textwidth]{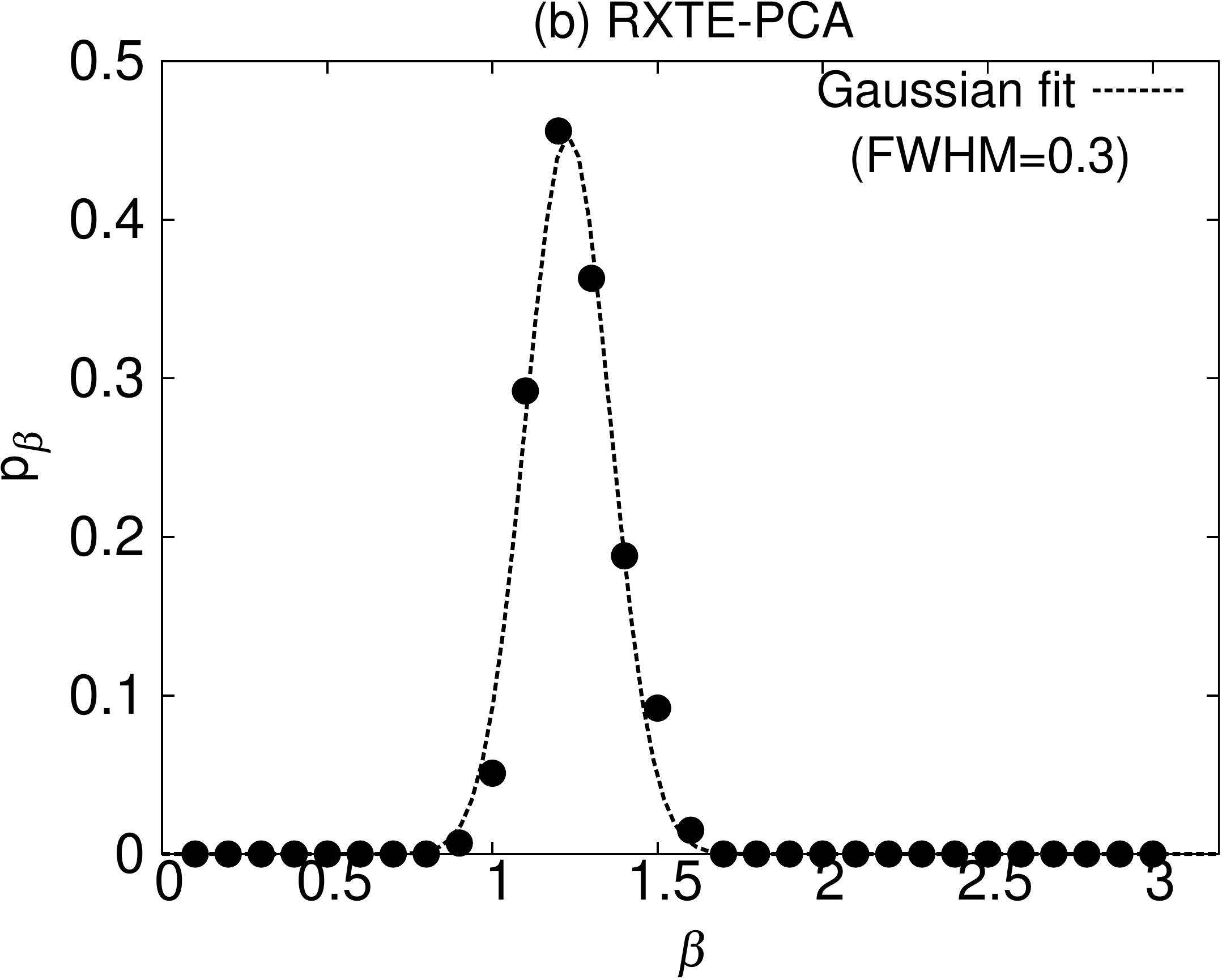}
\includegraphics[width=0.25\textwidth]{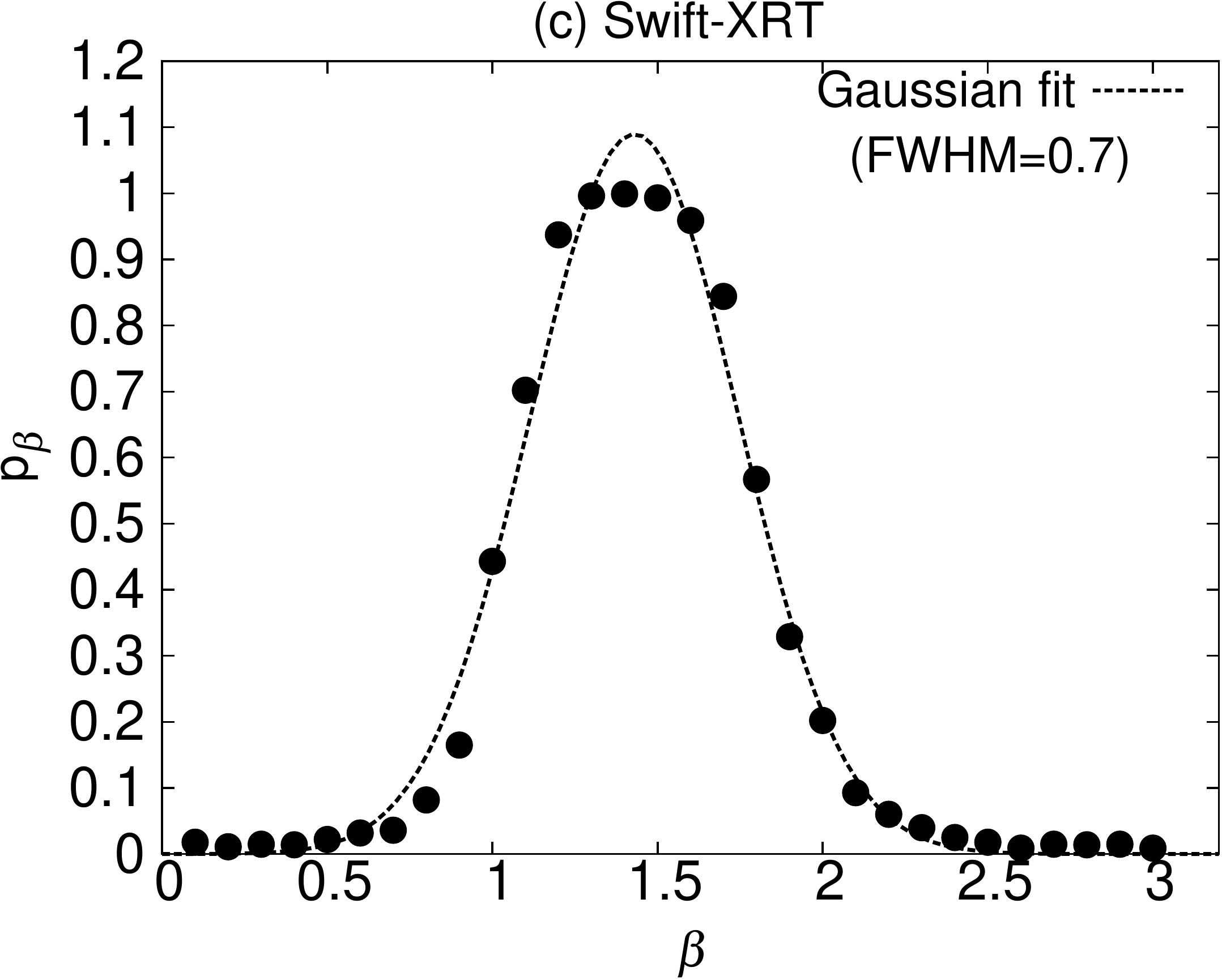}
\includegraphics[width=0.25\textwidth]{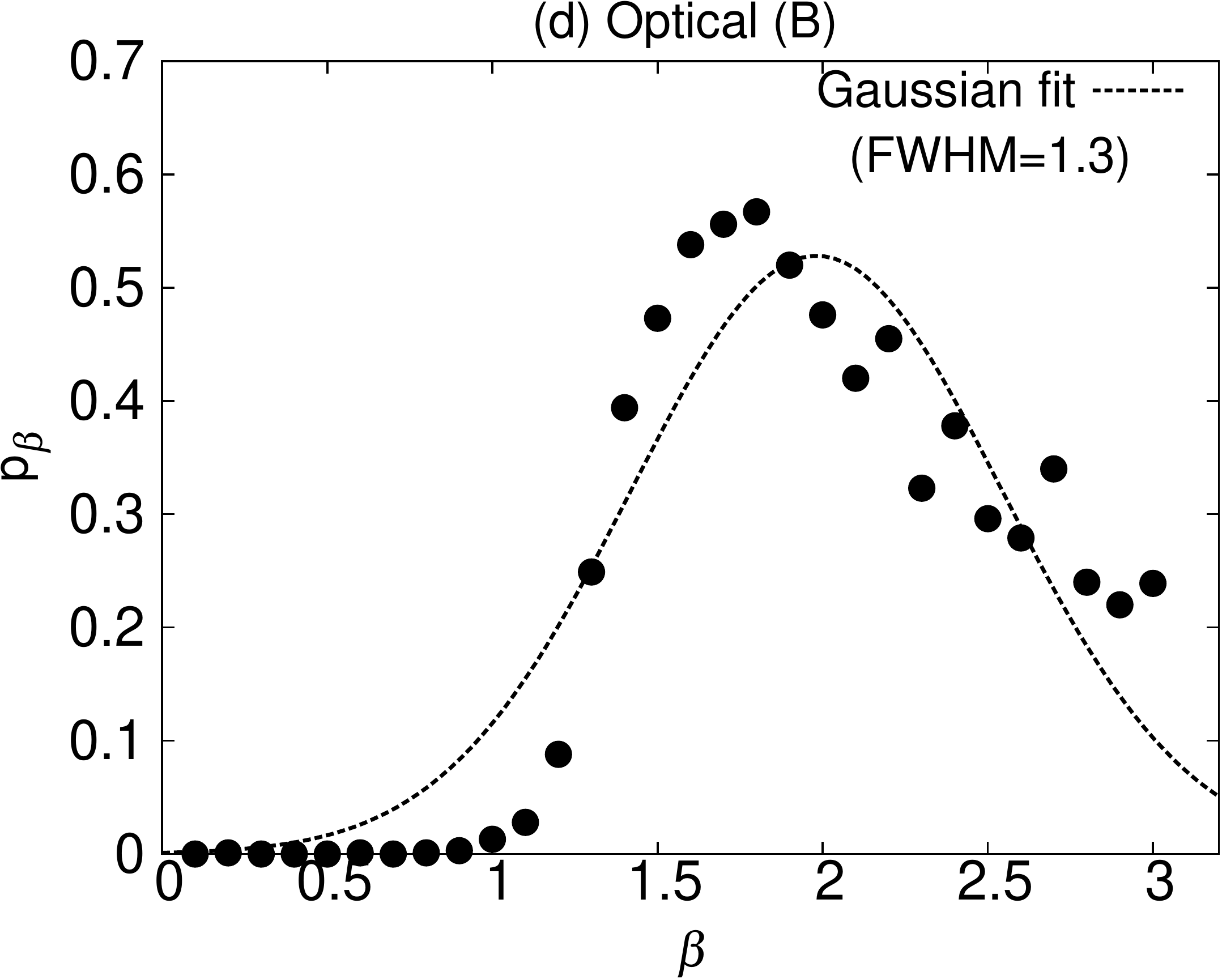}
}
\hbox{
\includegraphics[width=0.25\textwidth]{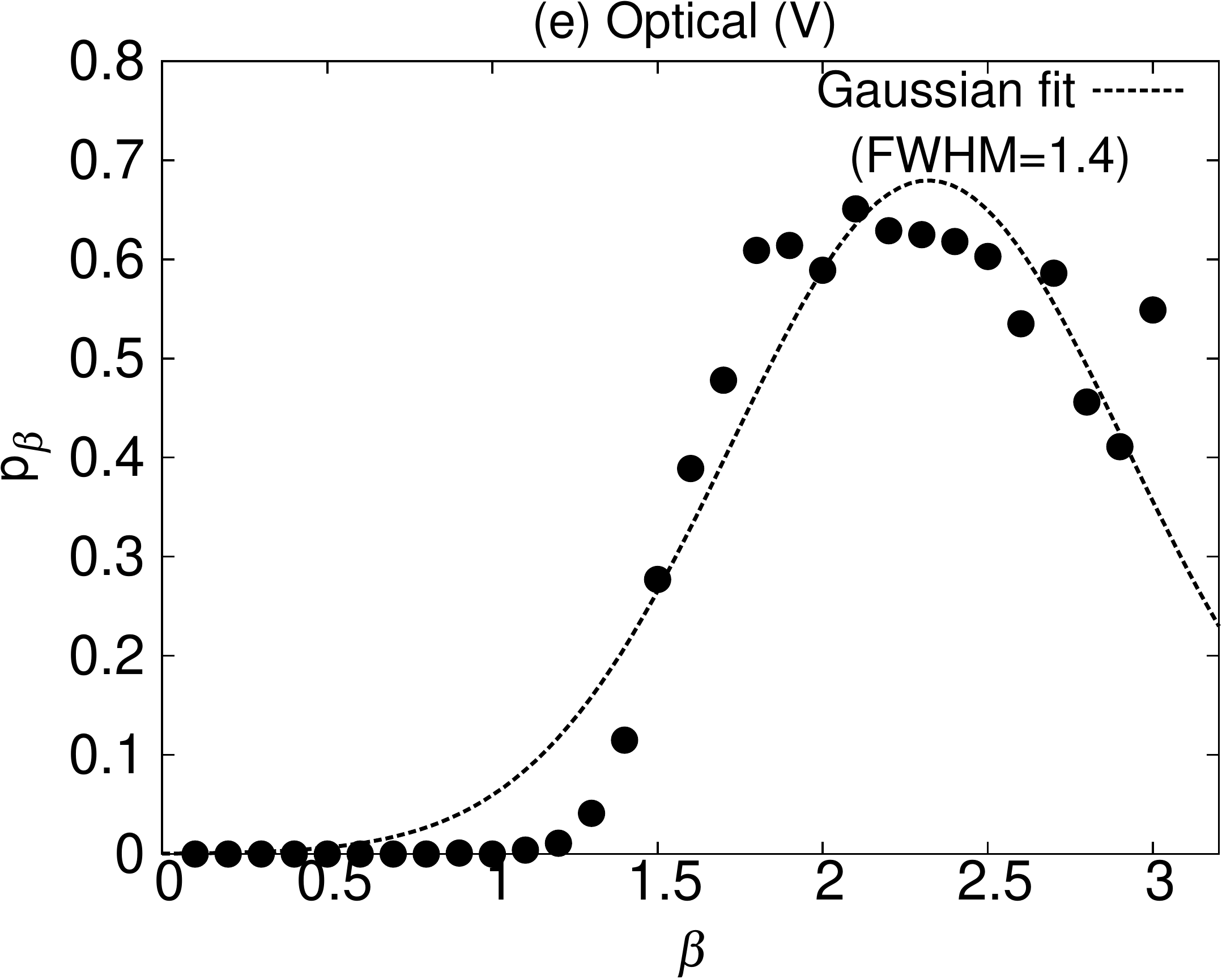}
\includegraphics[width=0.25\textwidth]{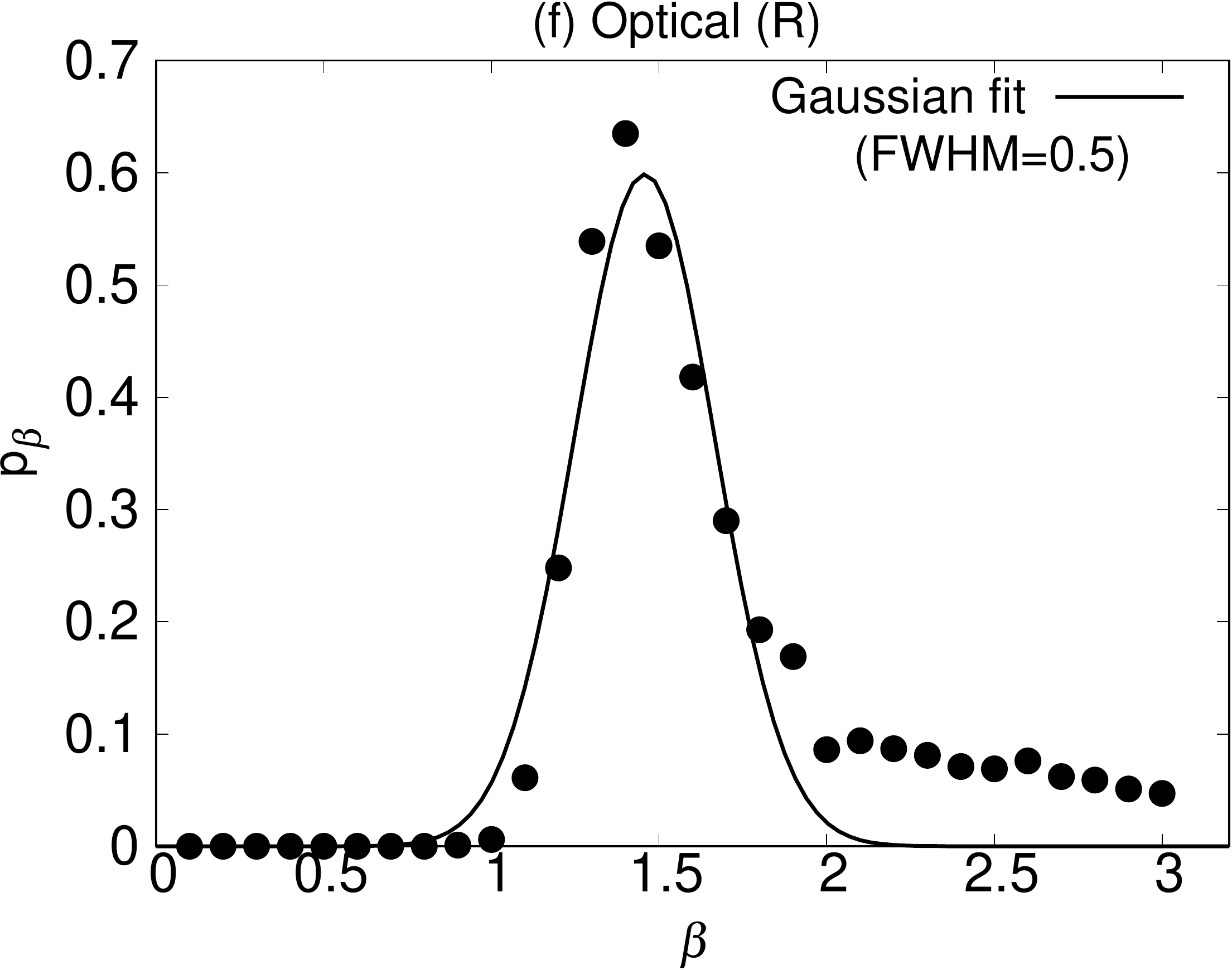}
\includegraphics[width=0.25\textwidth]{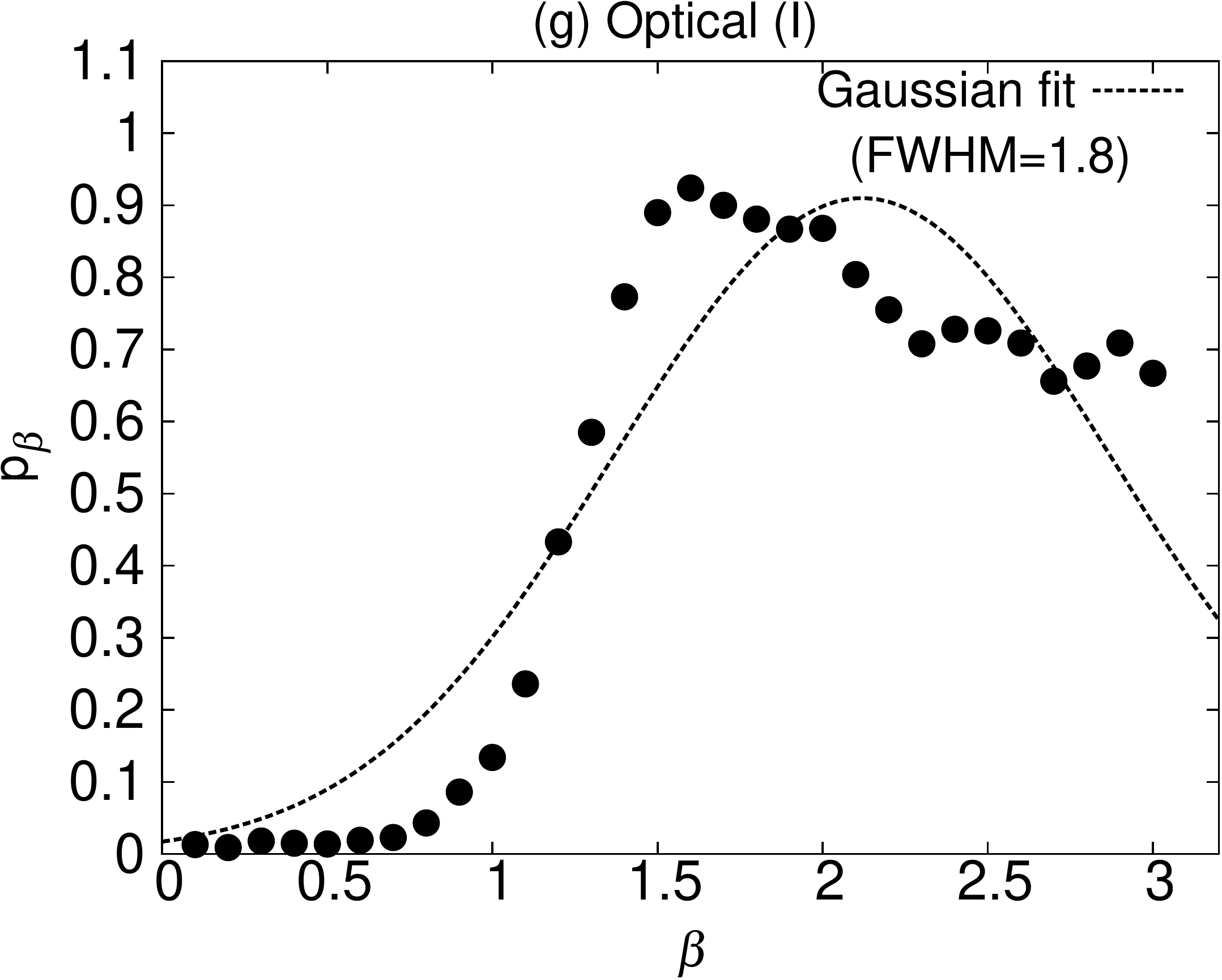}
\includegraphics[width=0.25\textwidth]{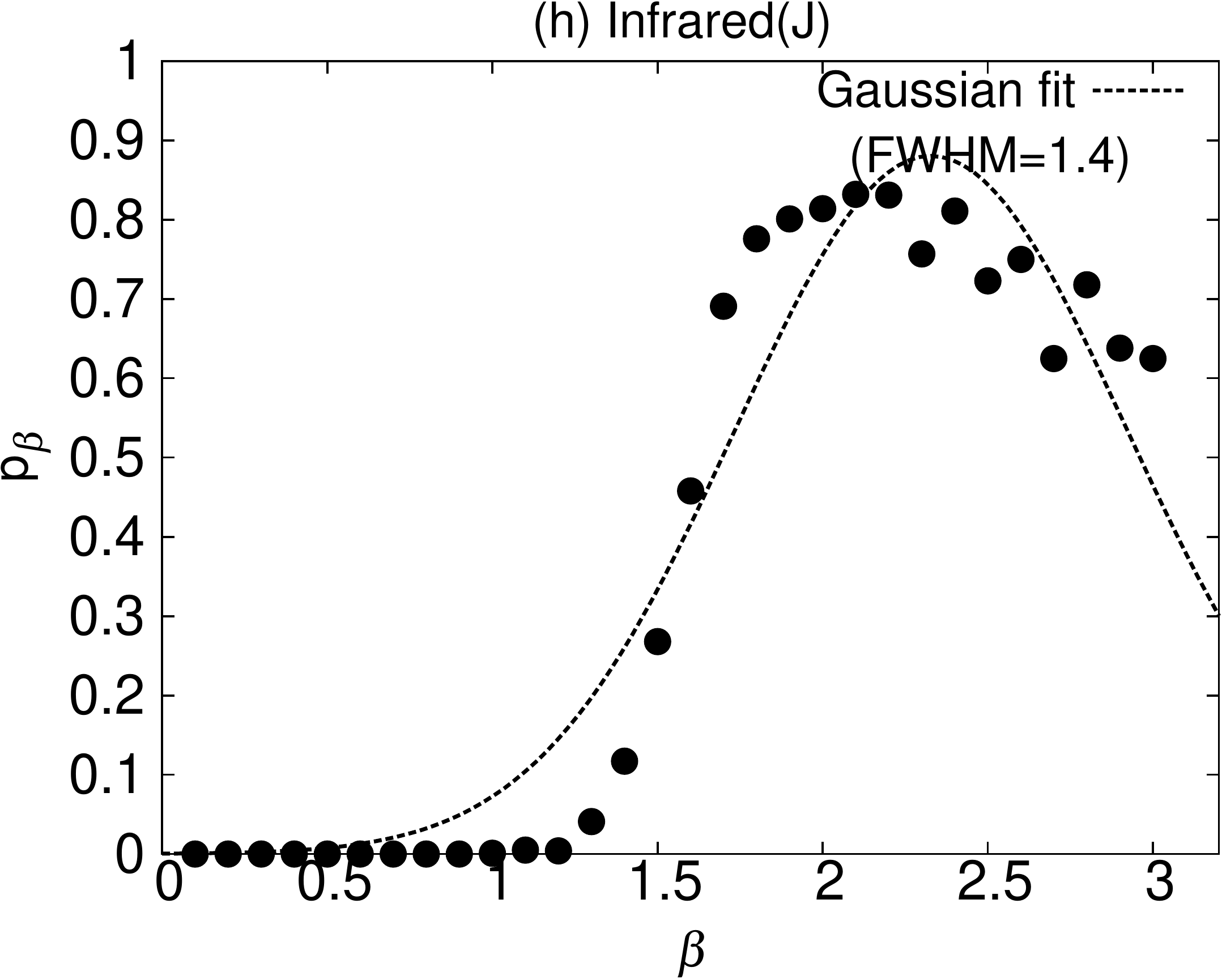}
}
\hbox{
\includegraphics[width=0.25\textwidth]{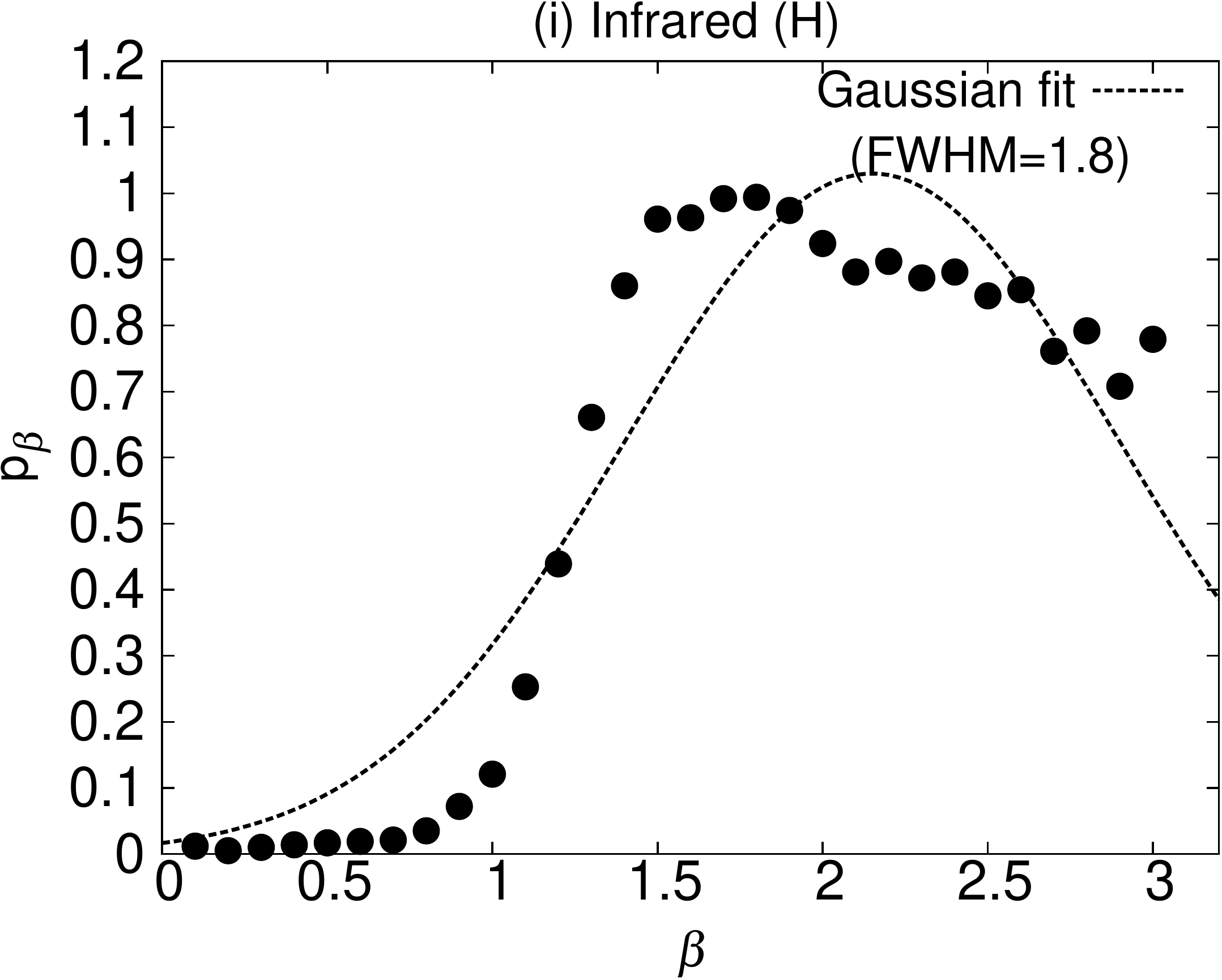}
\includegraphics[width=0.25\textwidth]{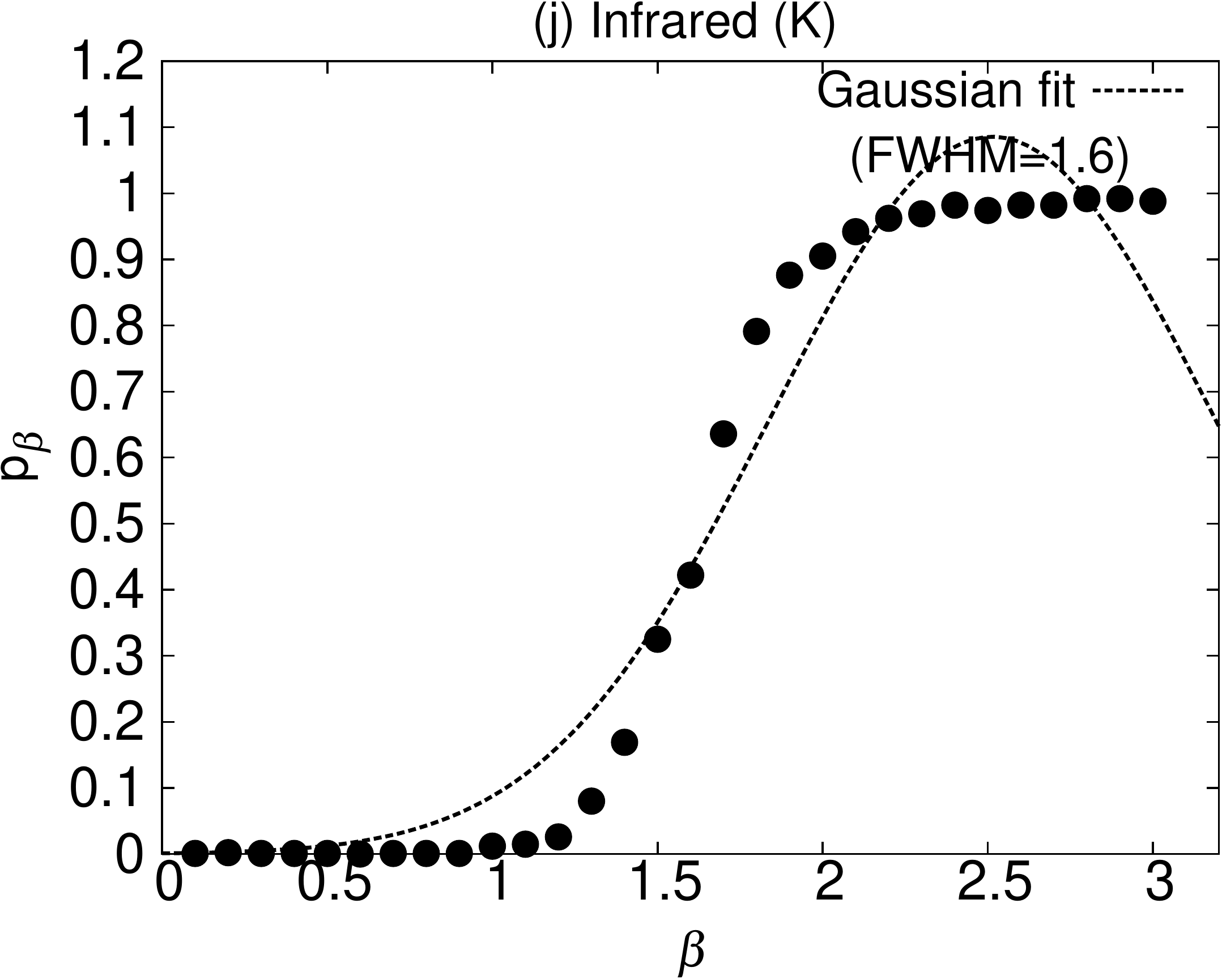}
\includegraphics[width=0.25\textwidth]{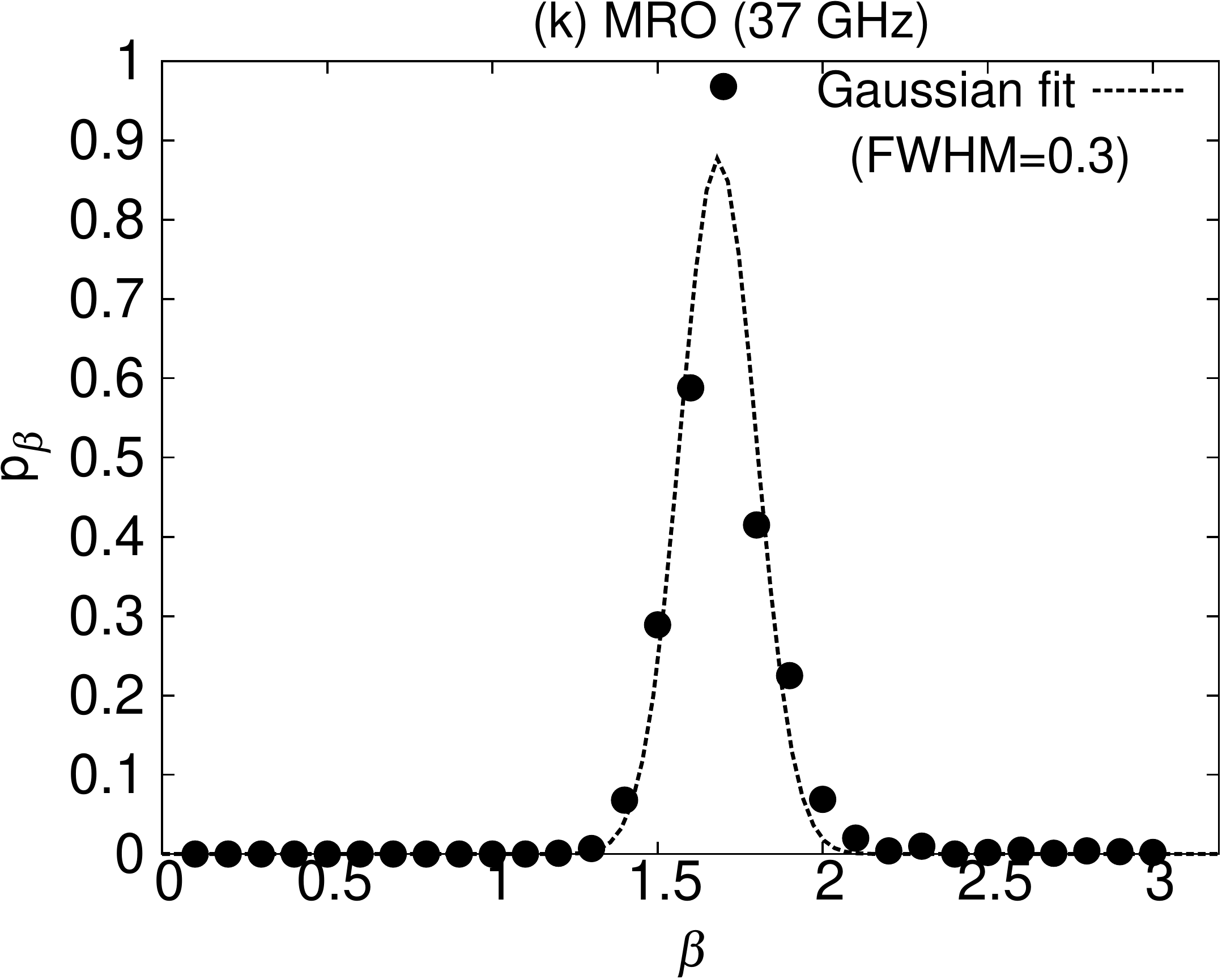}
\includegraphics[width=0.25\textwidth]{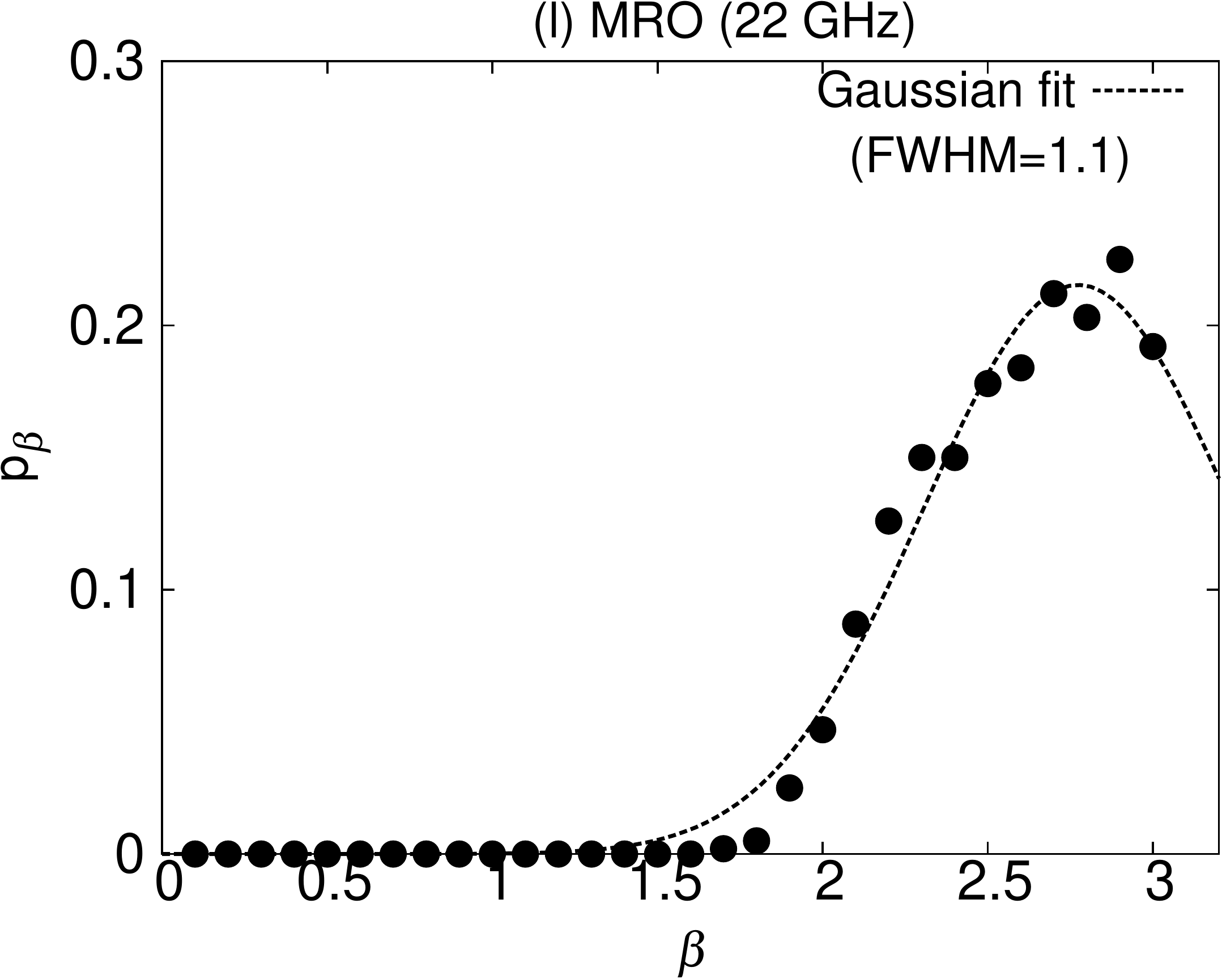}
}
\hbox{
\includegraphics[width=0.25\textwidth]{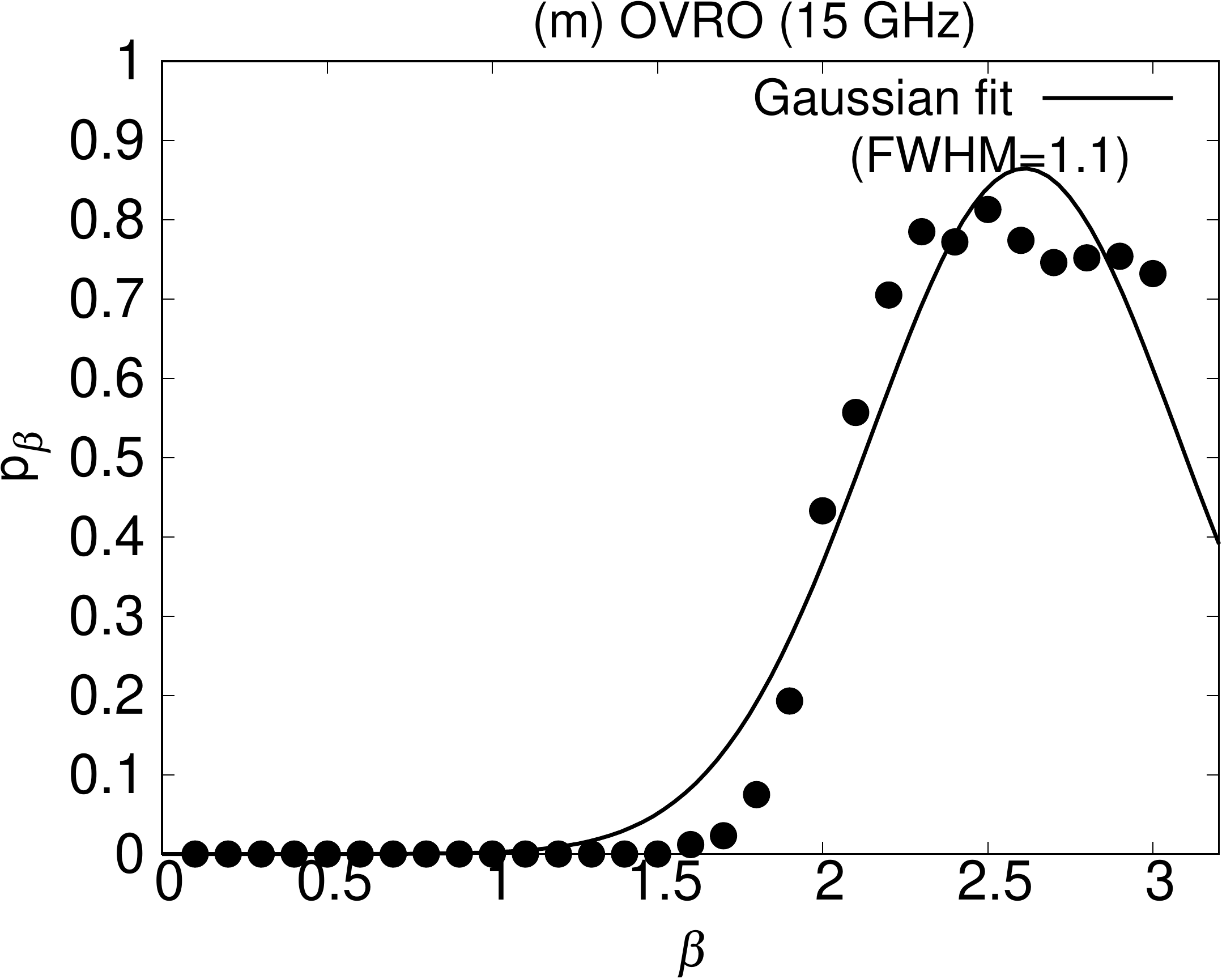}
\includegraphics[width=0.25\textwidth]{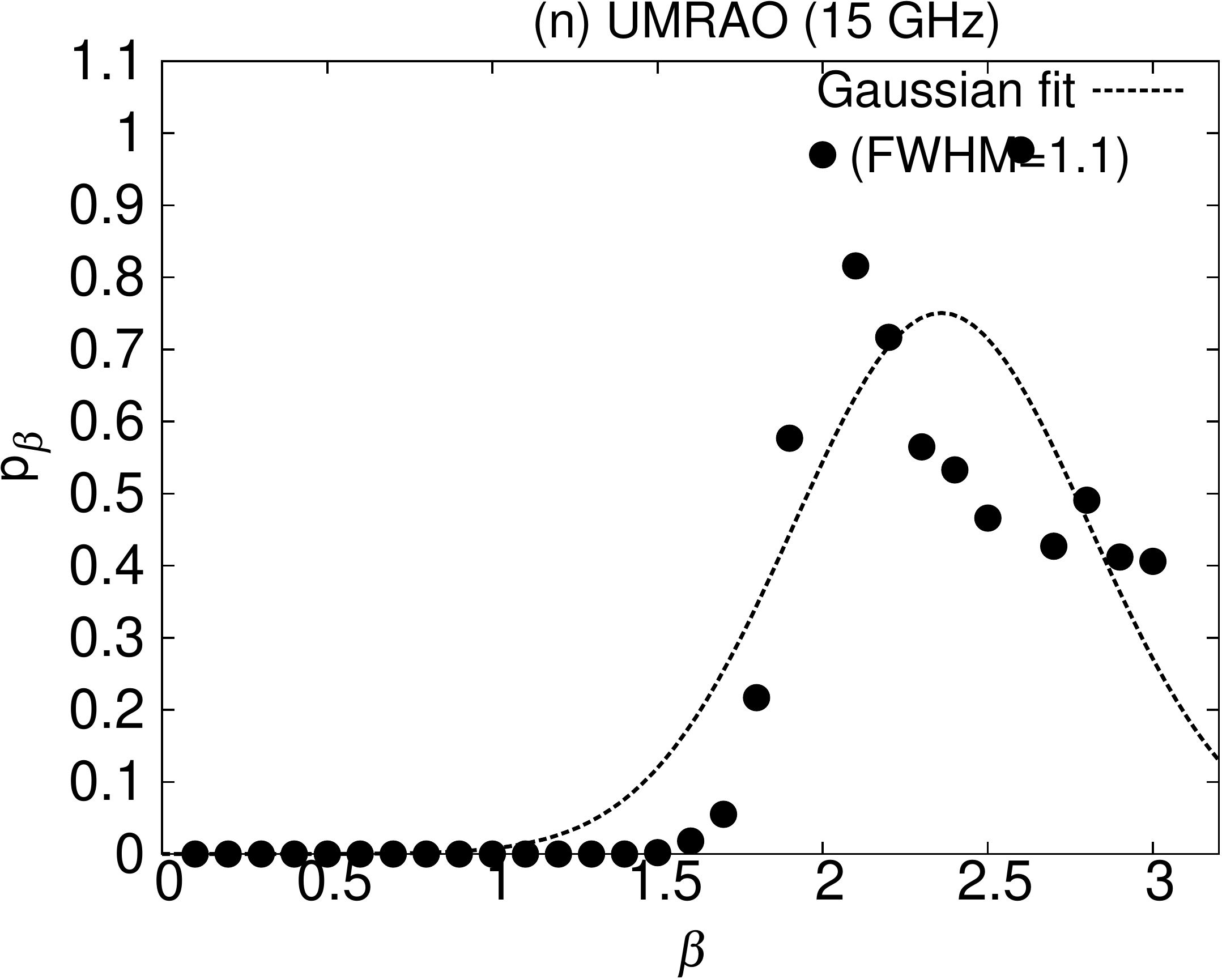}
\includegraphics[width=0.25\textwidth]{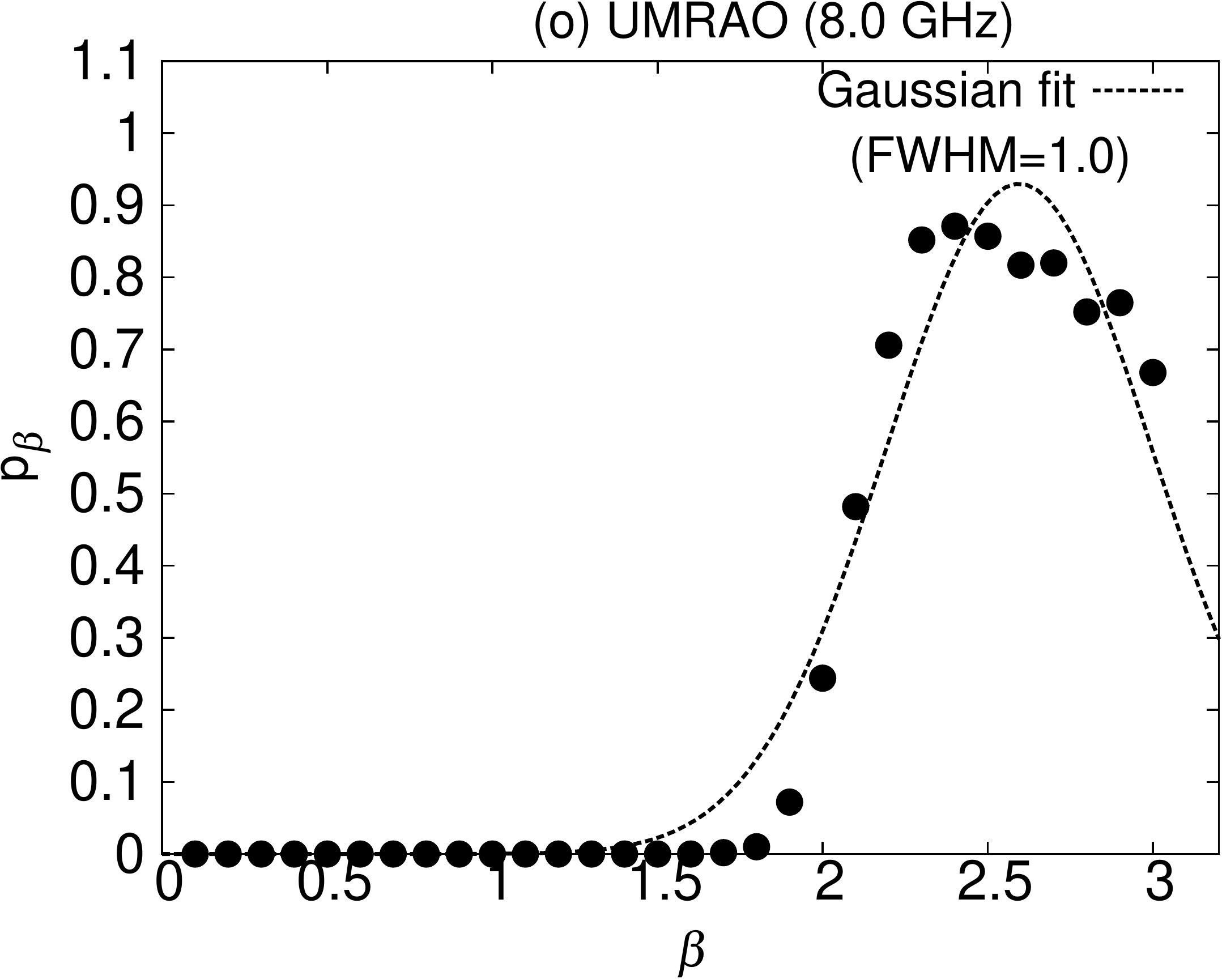}
\includegraphics[width=0.25\textwidth]{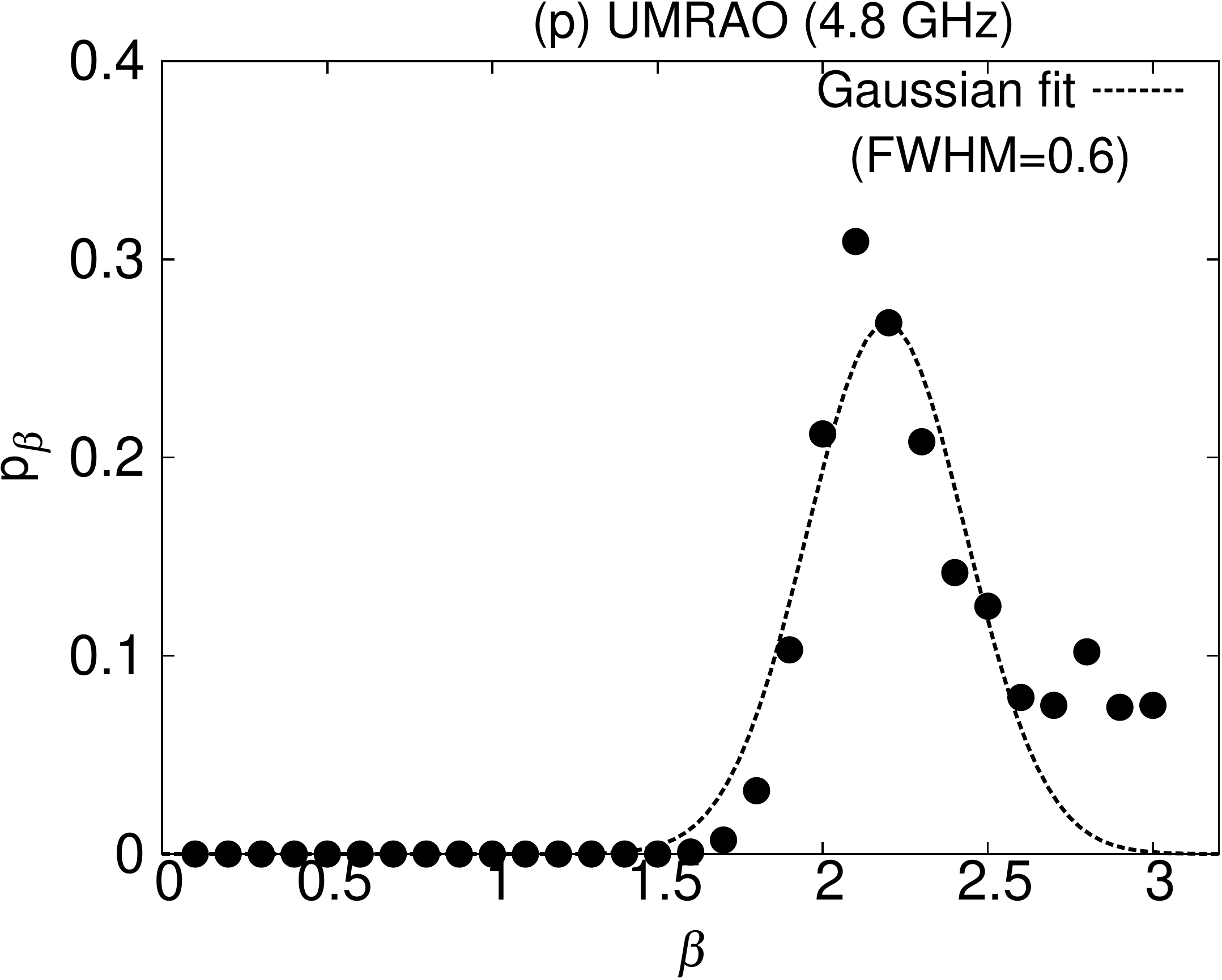}
}
\hbox{
\includegraphics[width=0.25\textwidth]{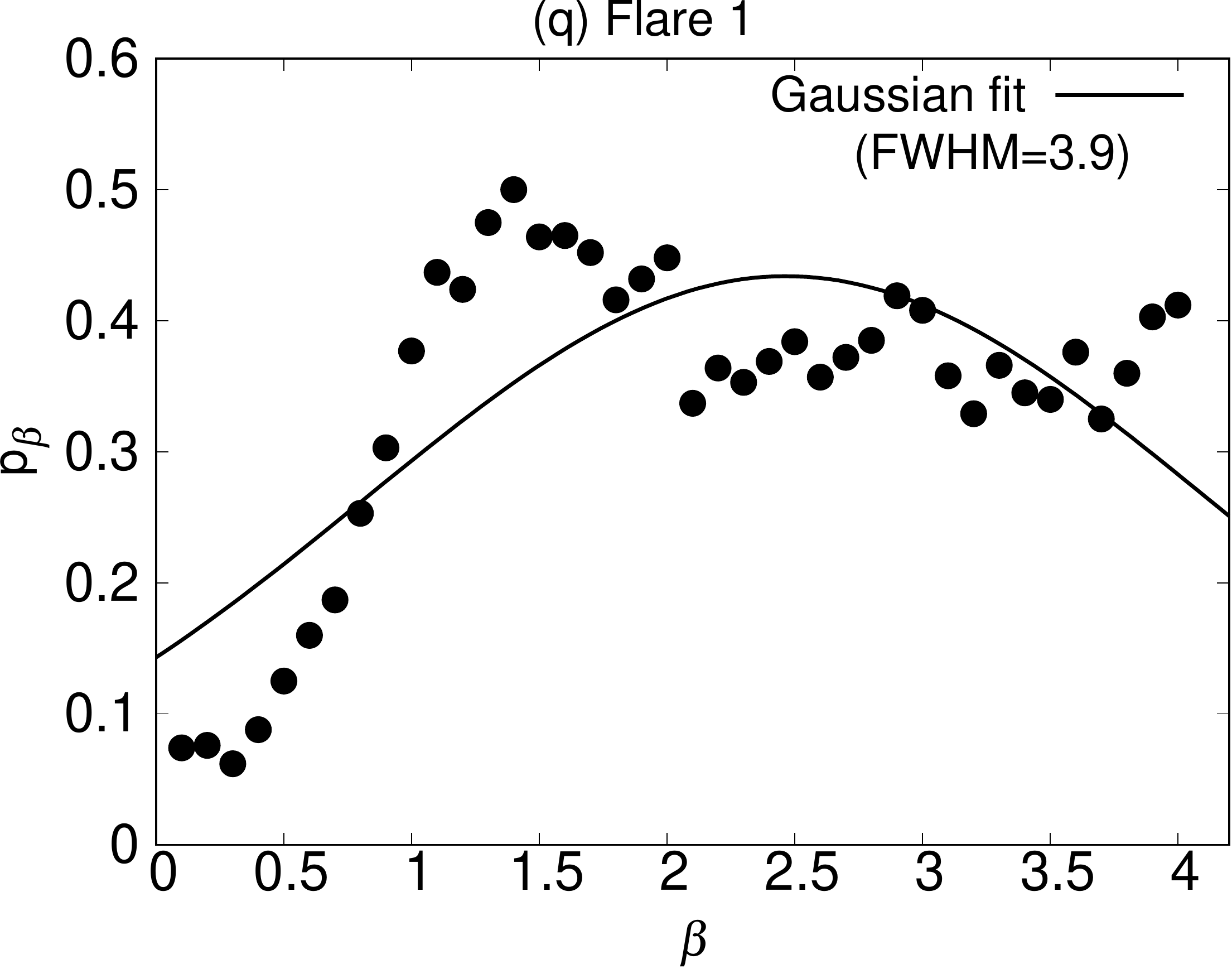}
\includegraphics[width=0.25\textwidth]{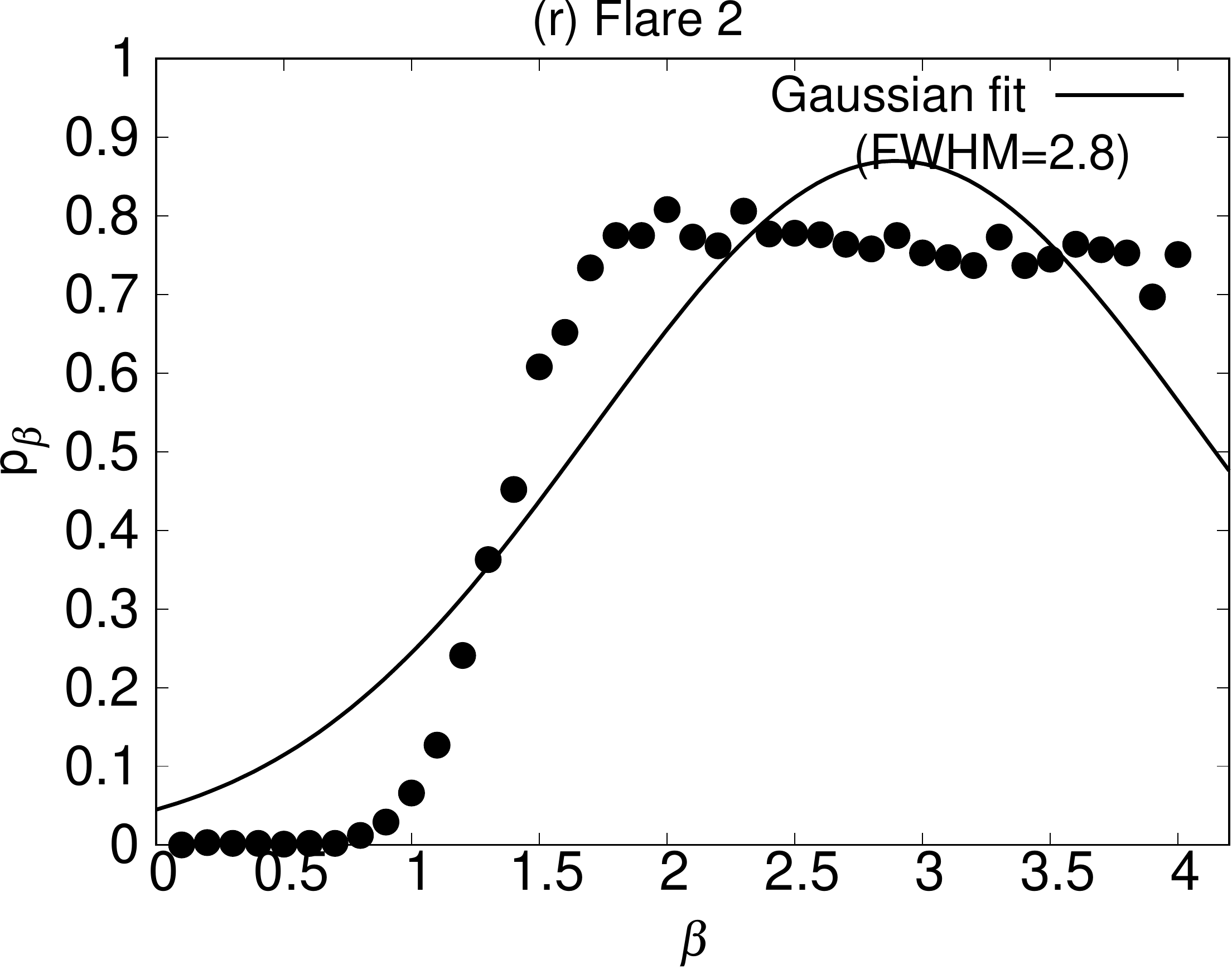}
\includegraphics[width=0.25\textwidth]{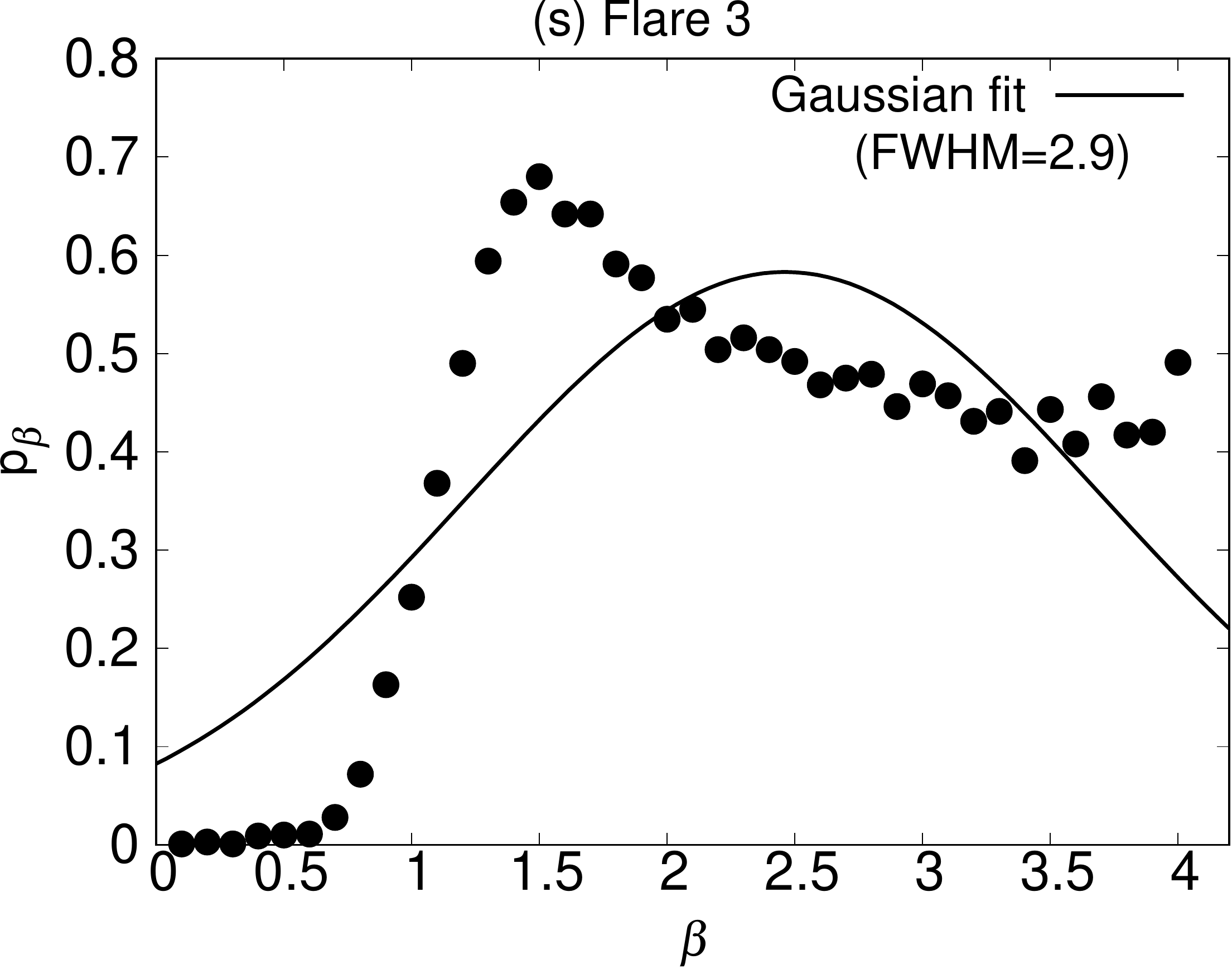}
}

\caption{Distribution curves for the spectral parameters investigated using the PSRESP method for the blazar 3C\,279. The best-fitting value of $\beta$ is taken as the one with the highest $p_\beta$. The dashed curve is a Gaussian function fitted to the data with full-width at half maxima (FWHM), related to the standard deviation as 2.354$\sigma$.}
\label{appfig:beta3c}
\end{figure*}

\begin{figure*}
\hbox{
\includegraphics[width=0.25\textwidth]{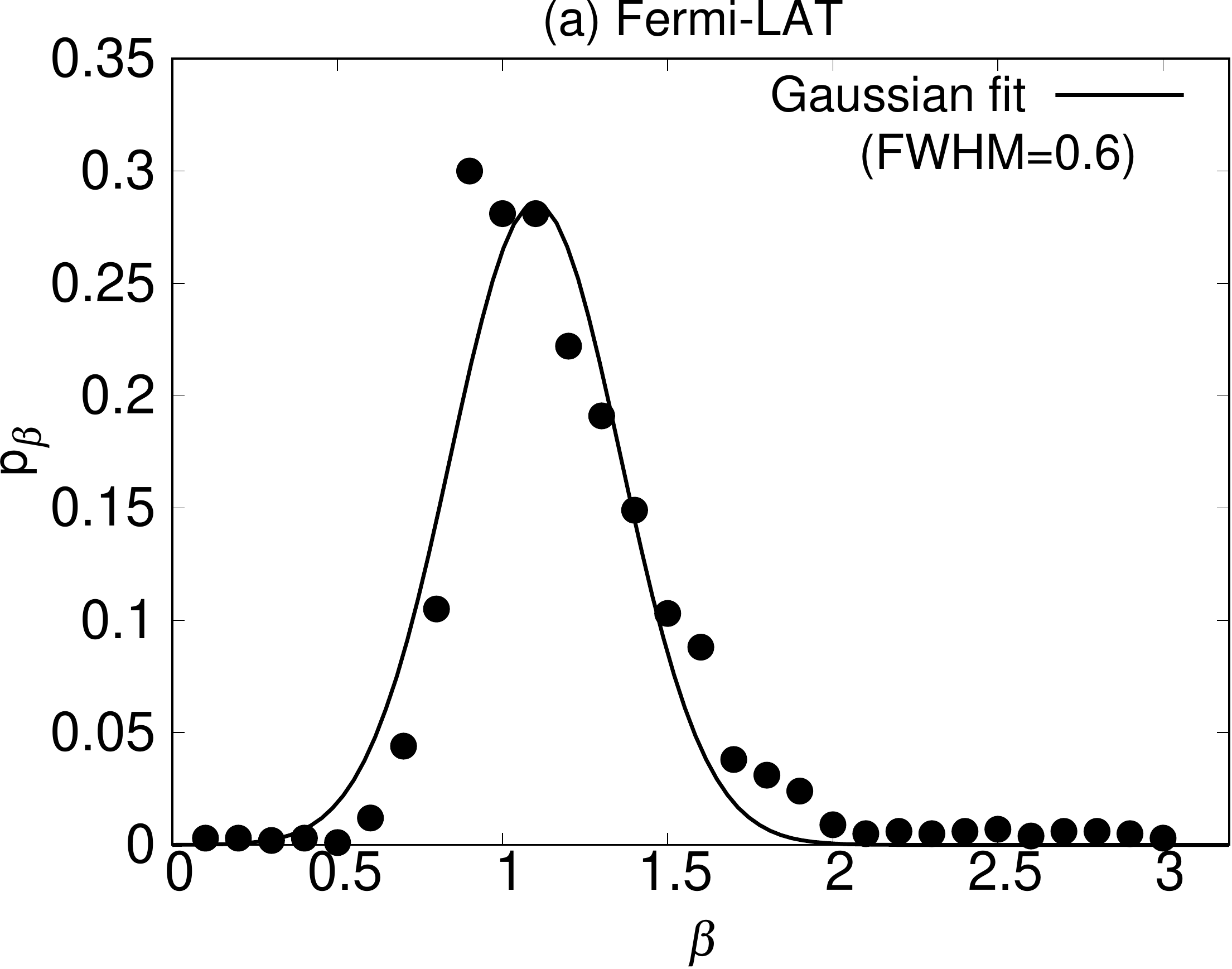}
\includegraphics[width=0.25\textwidth]{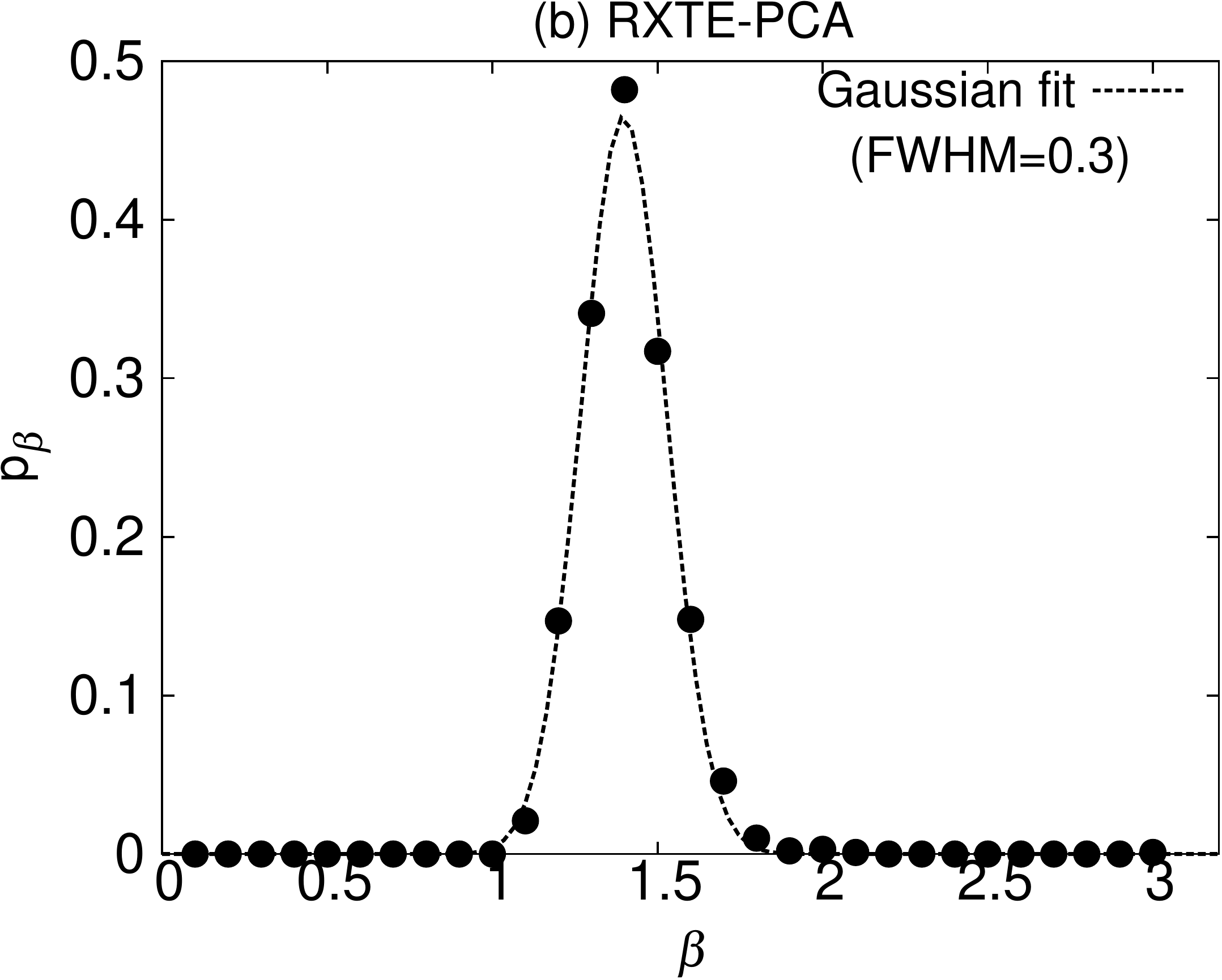}
\includegraphics[width=0.25\textwidth]{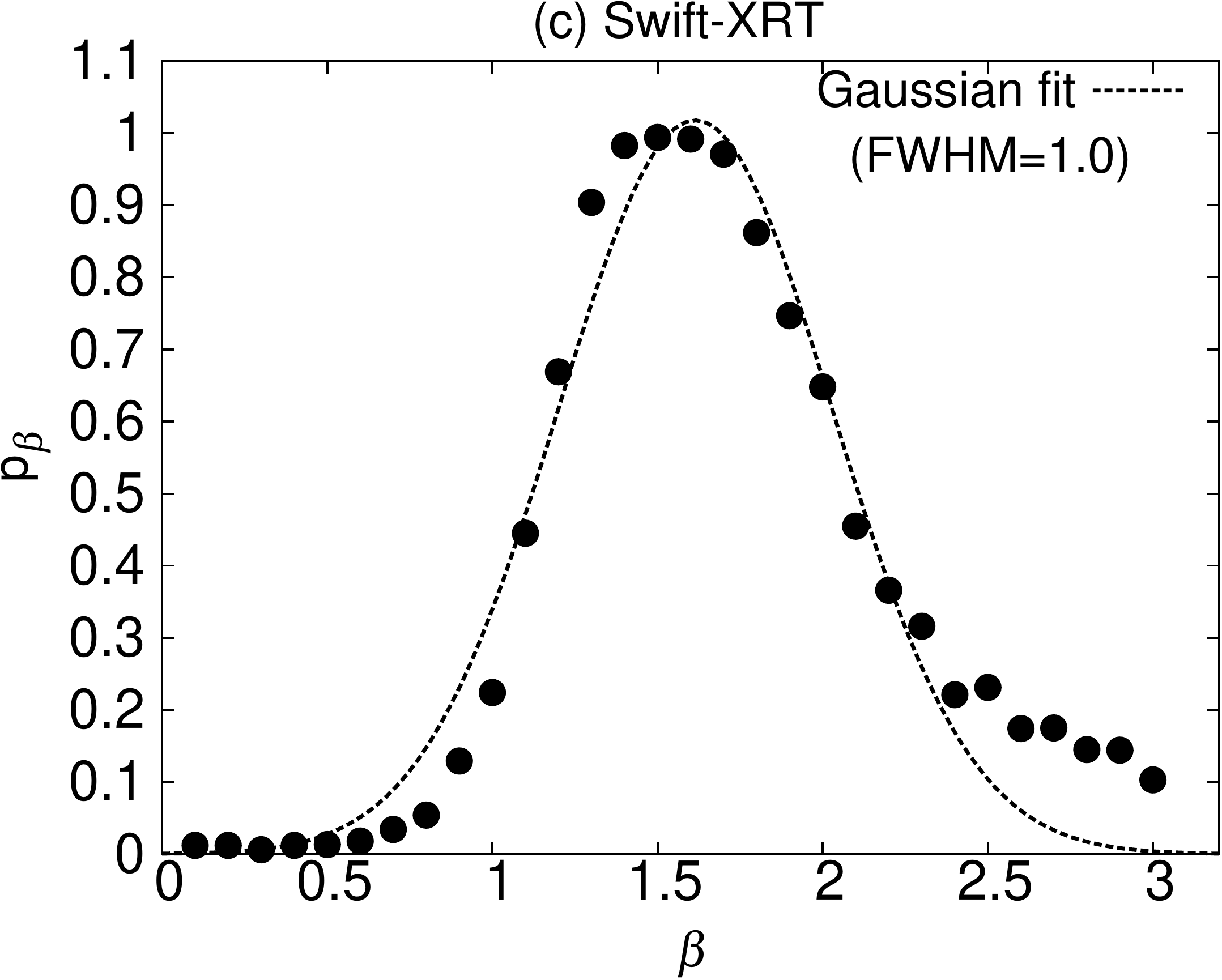}
\includegraphics[width=0.25\textwidth]{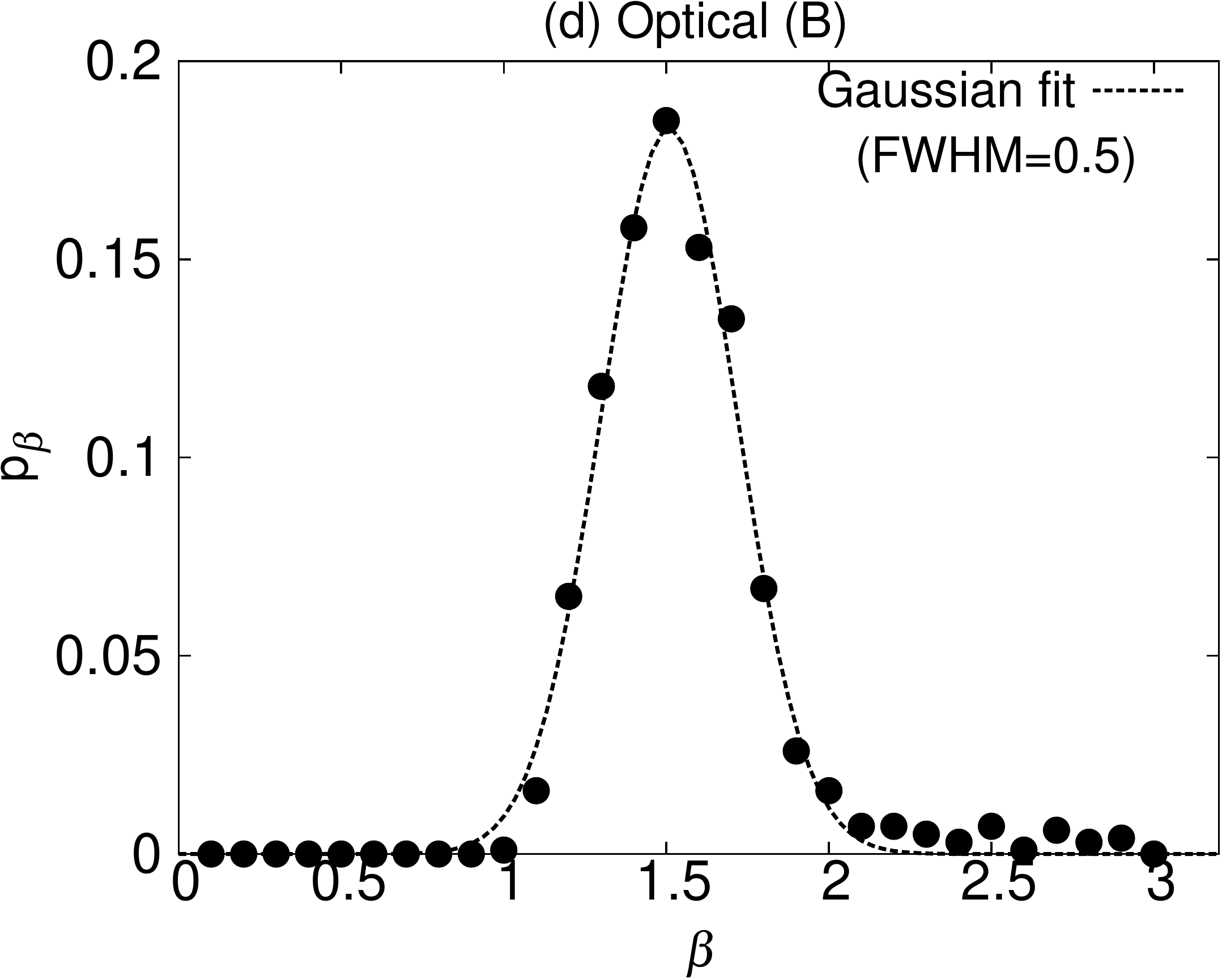}
}
\hbox{
\includegraphics[width=0.25\textwidth]{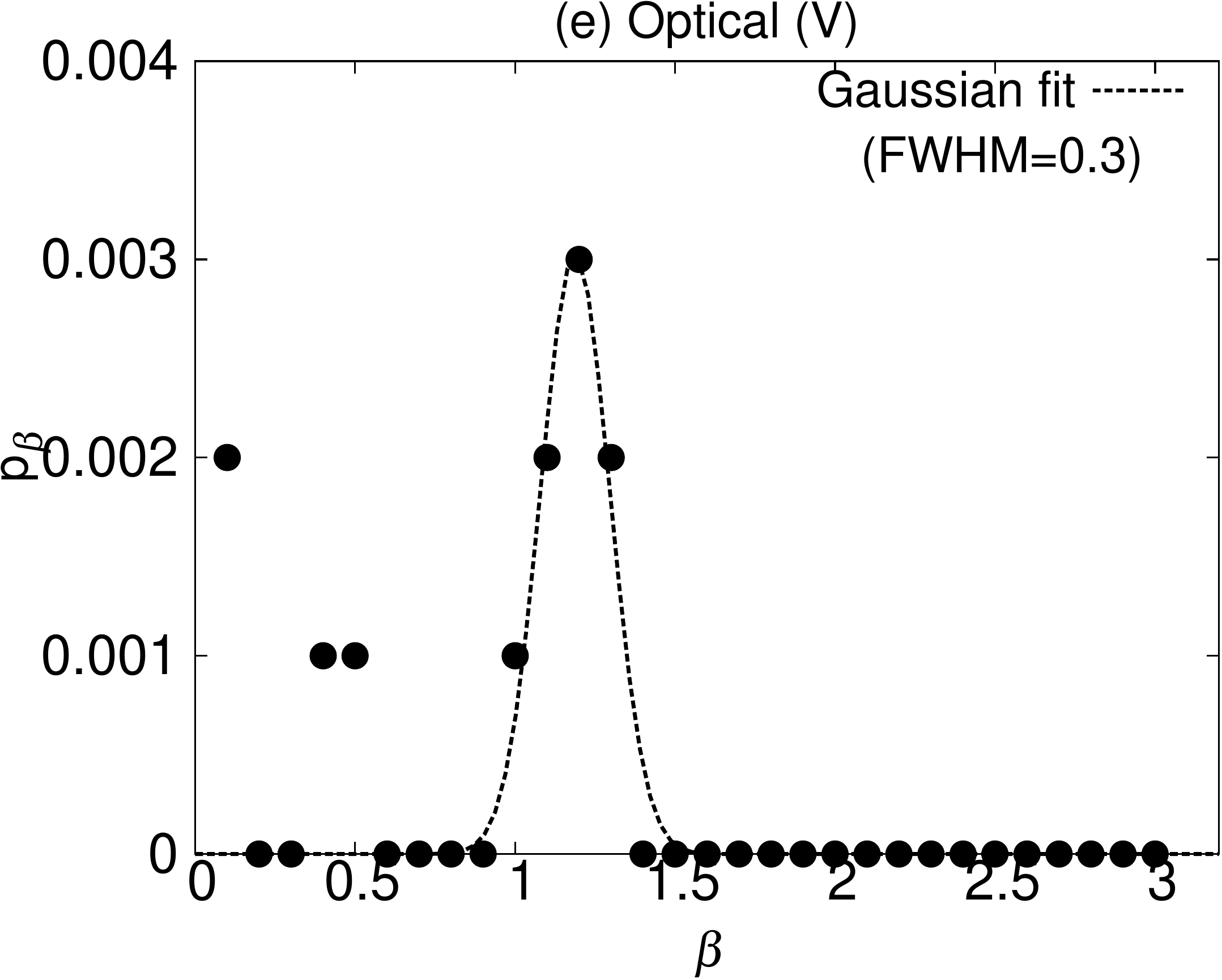}
\includegraphics[width=0.25\textwidth]{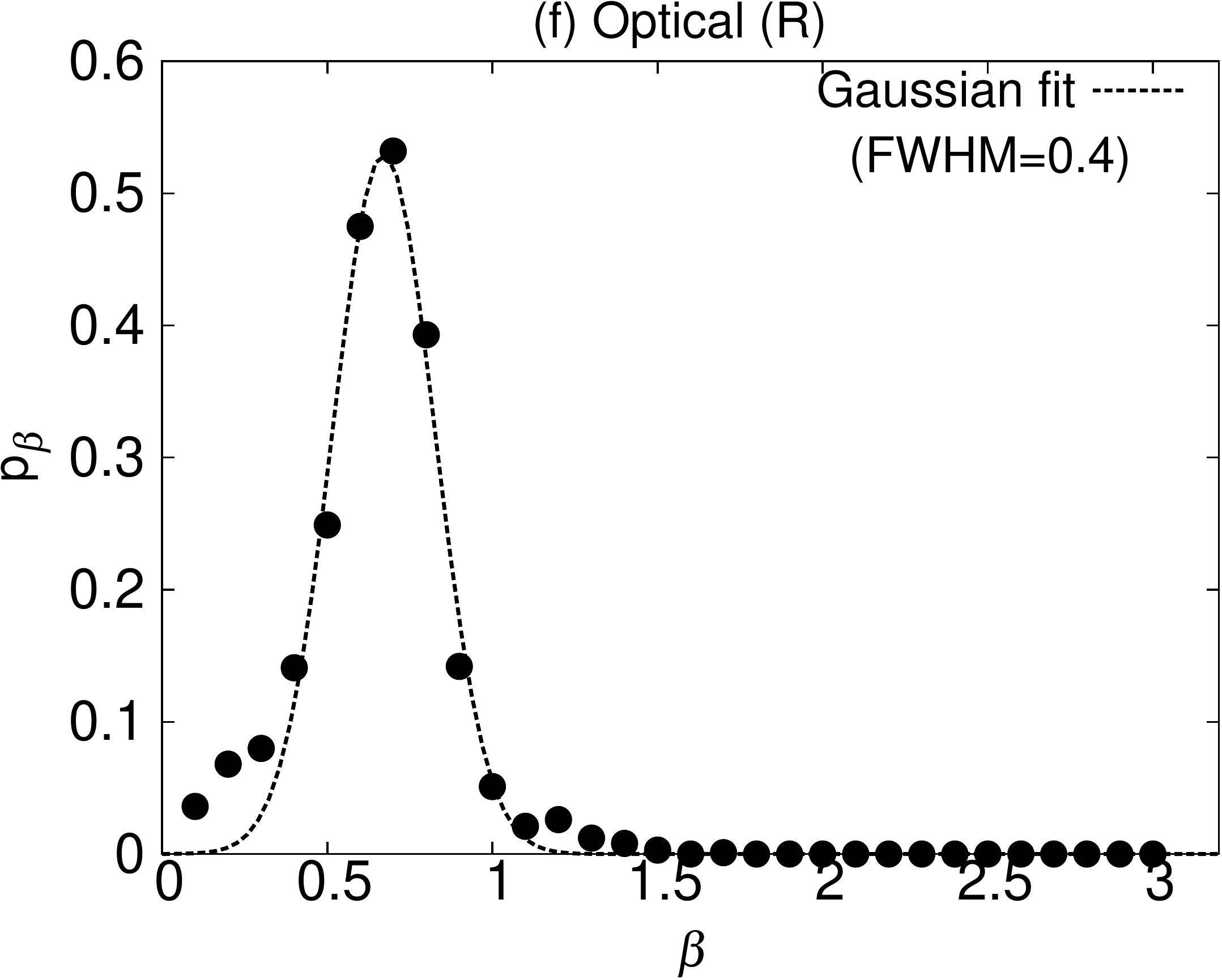}
\includegraphics[width=0.25\textwidth]{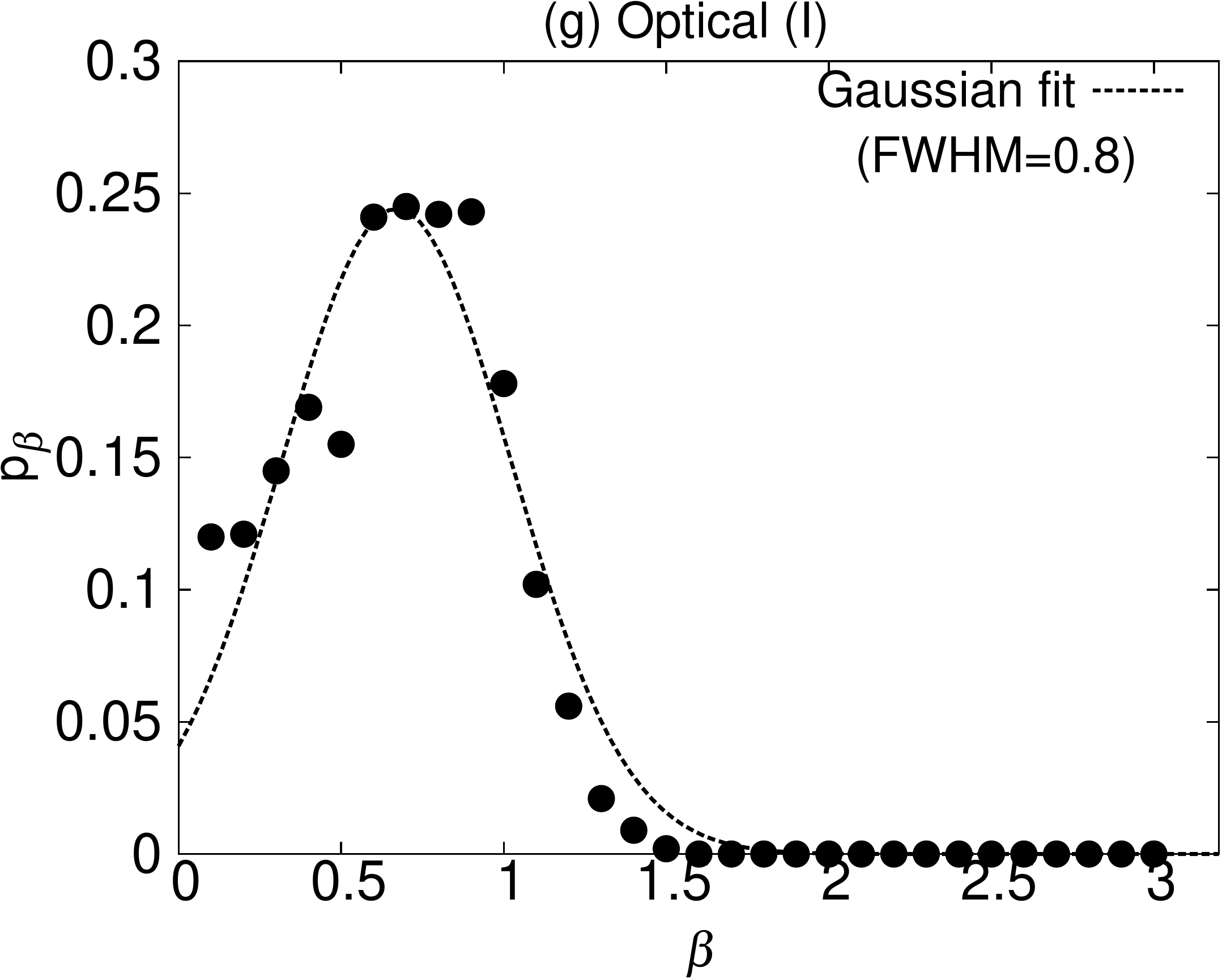}
\includegraphics[width=0.25\textwidth]{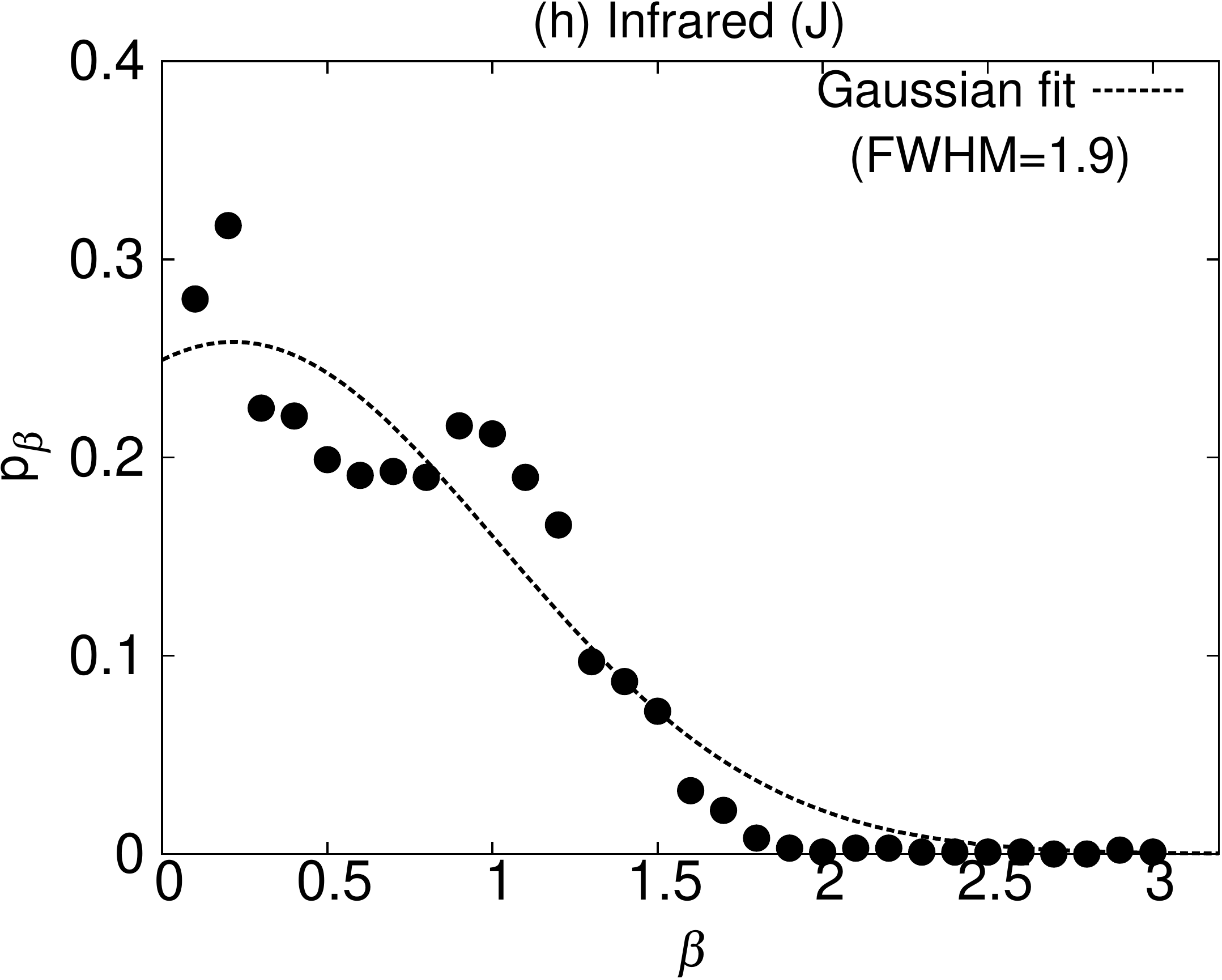}
}
\hbox{
\includegraphics[width=0.25\textwidth]{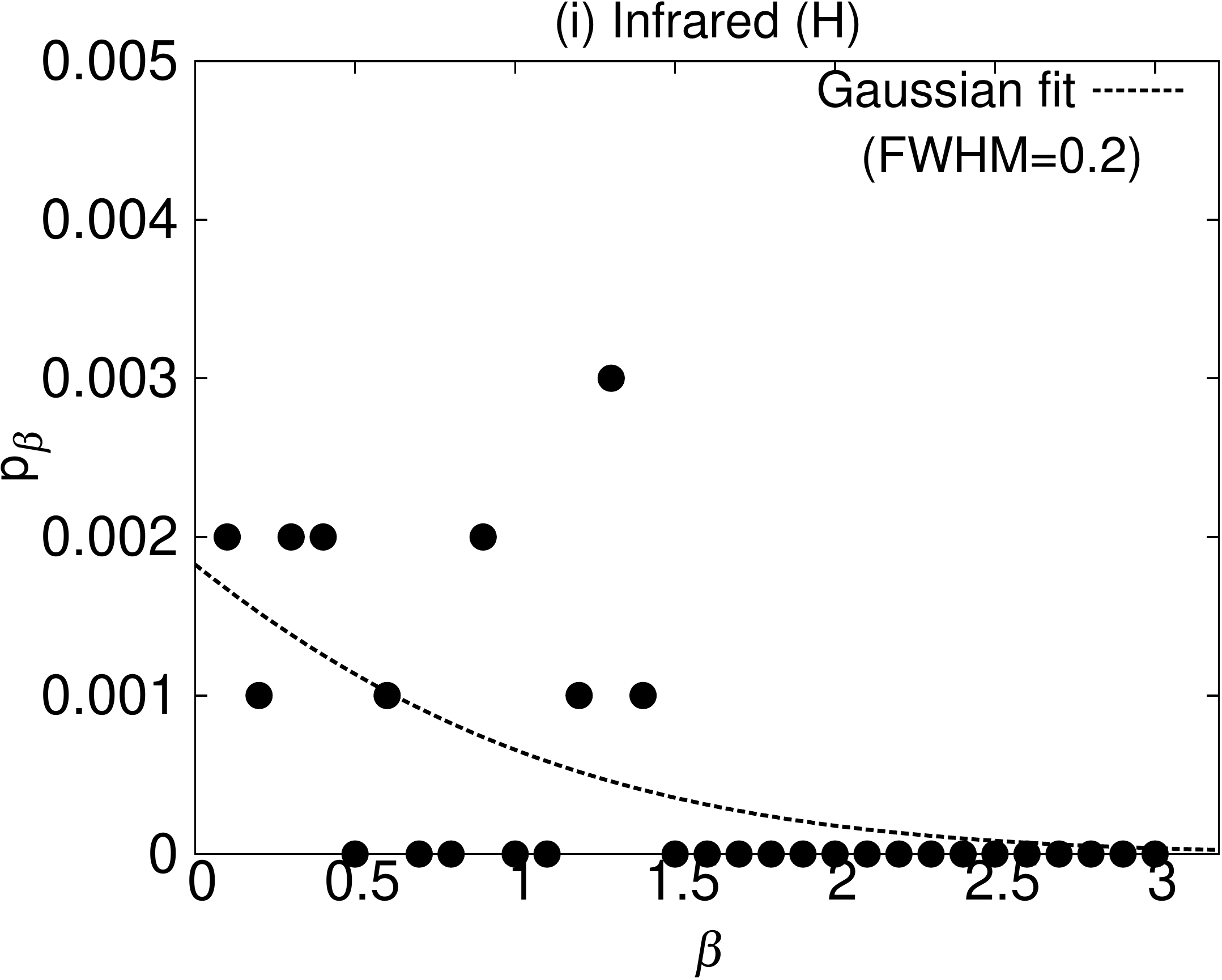}
\includegraphics[width=0.25\textwidth]{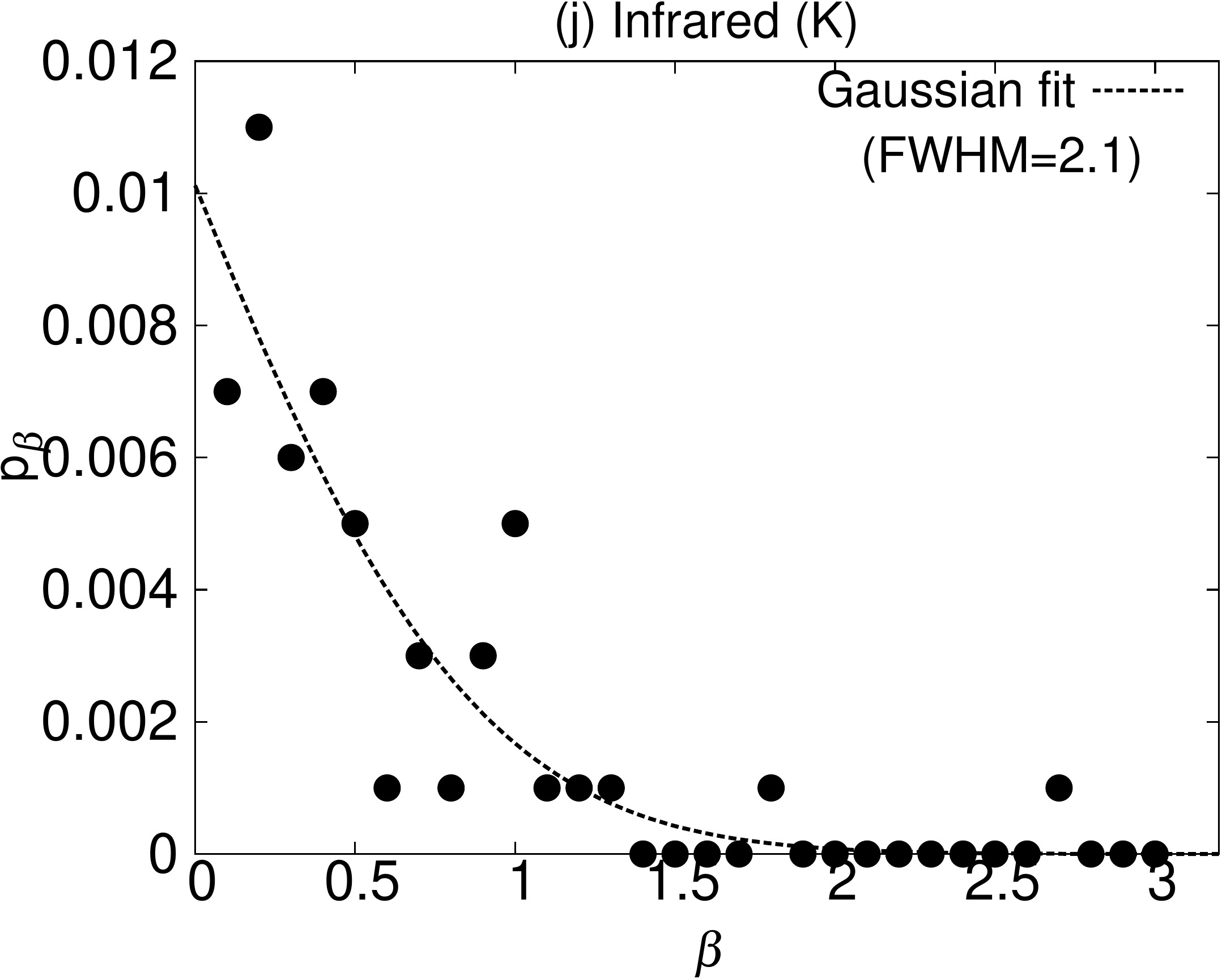}
\includegraphics[width=0.25\textwidth]{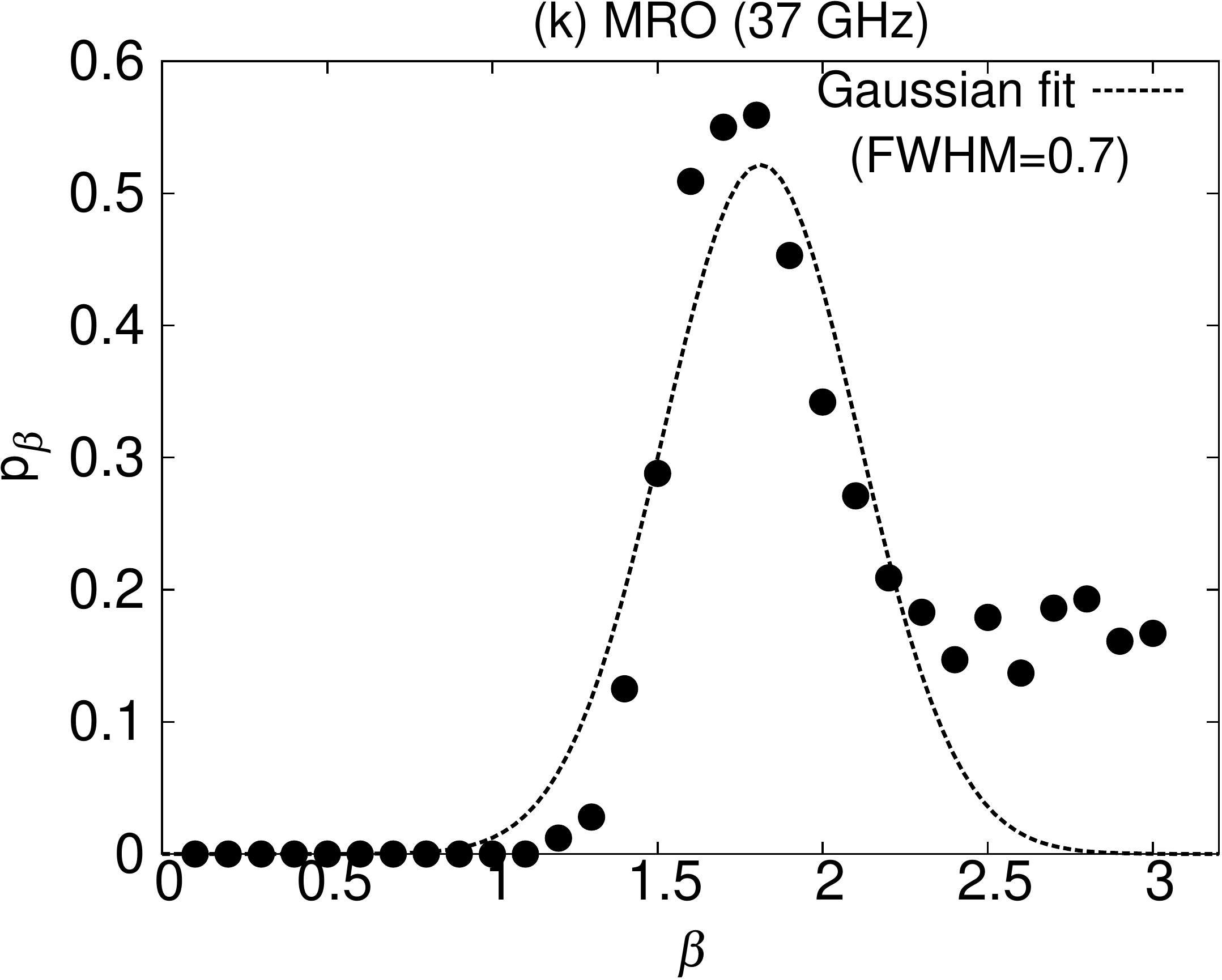}
\includegraphics[width=0.25\textwidth]{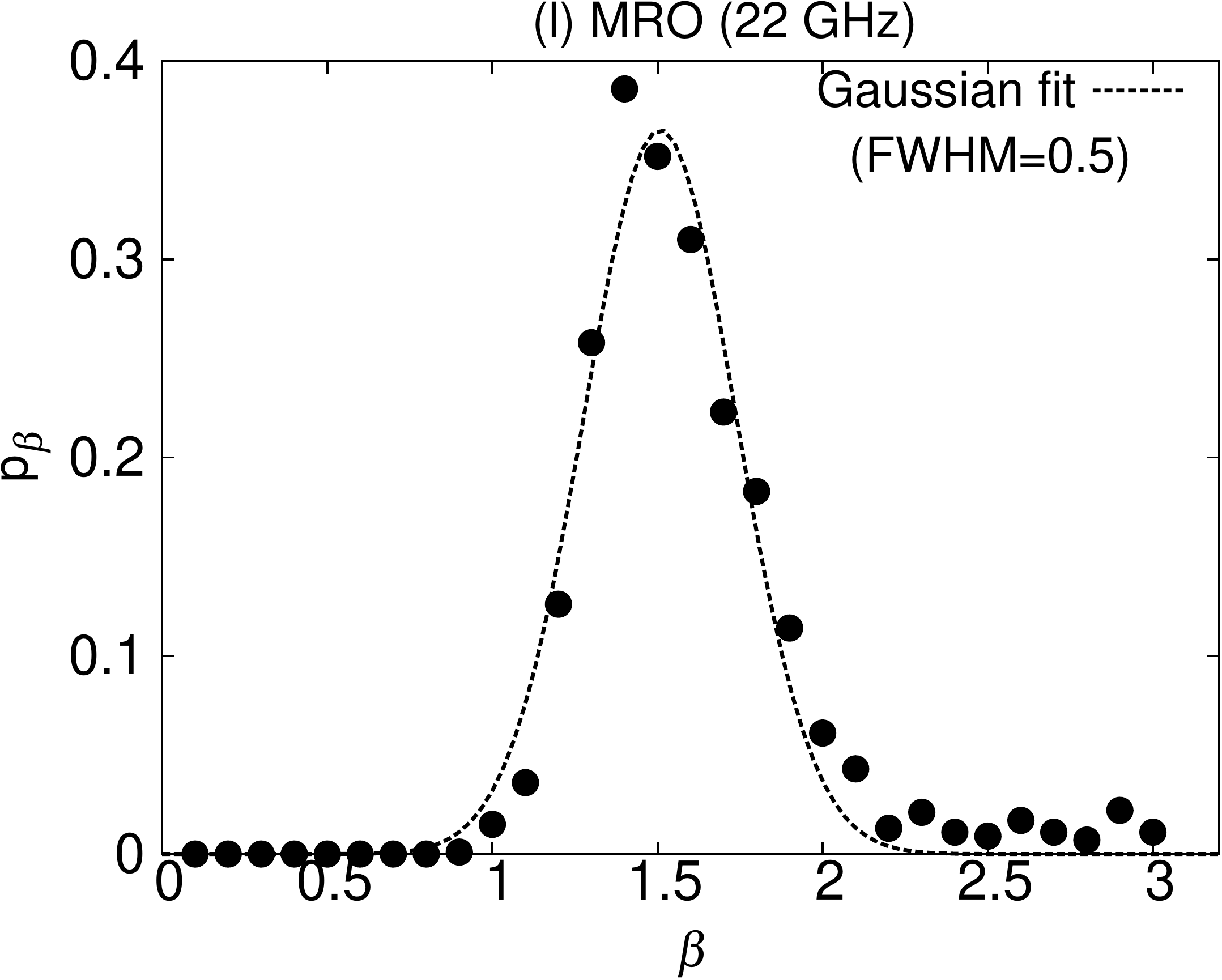}
}
\hbox{
\includegraphics[width=0.25\textwidth]{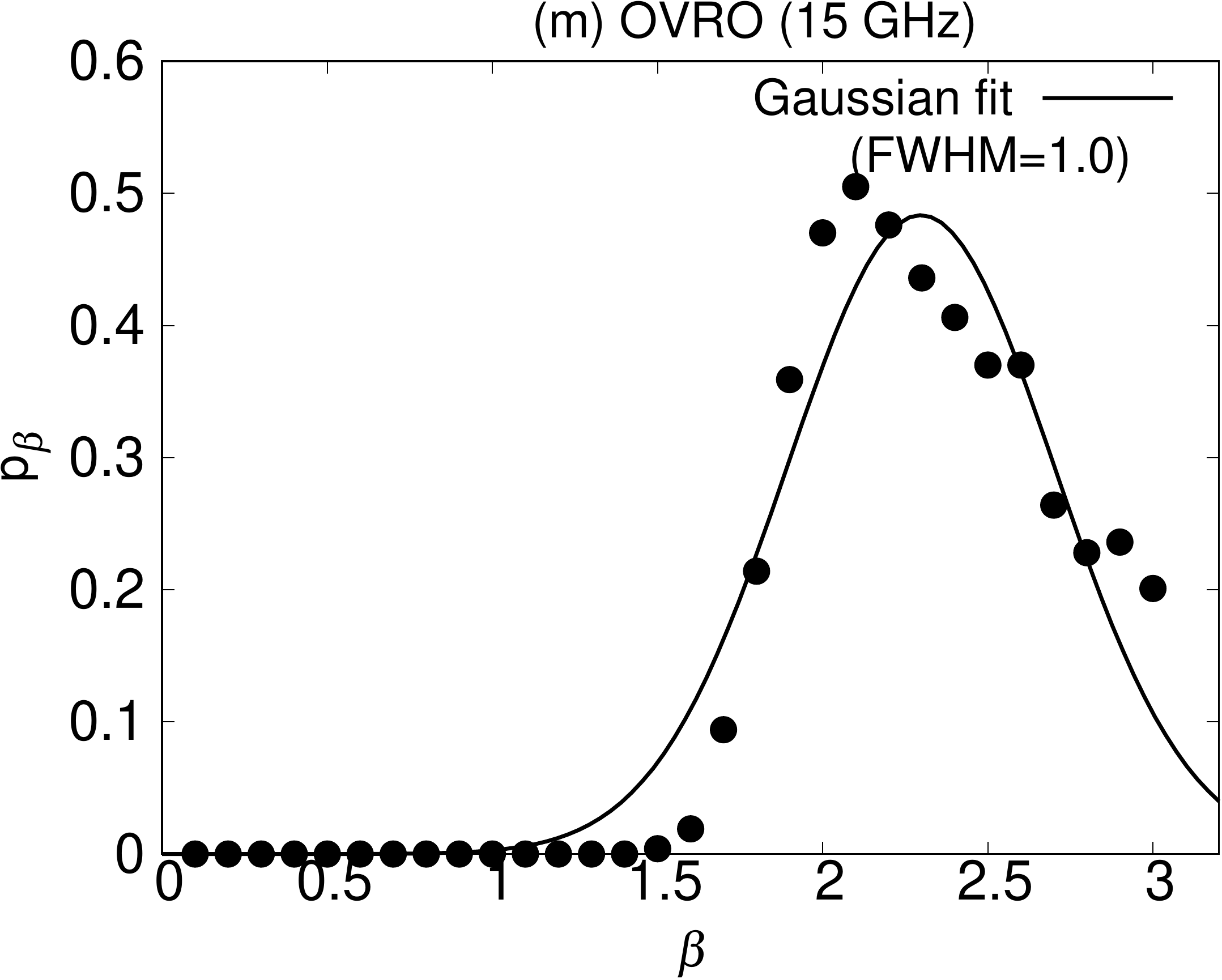}
\includegraphics[width=0.25\textwidth]{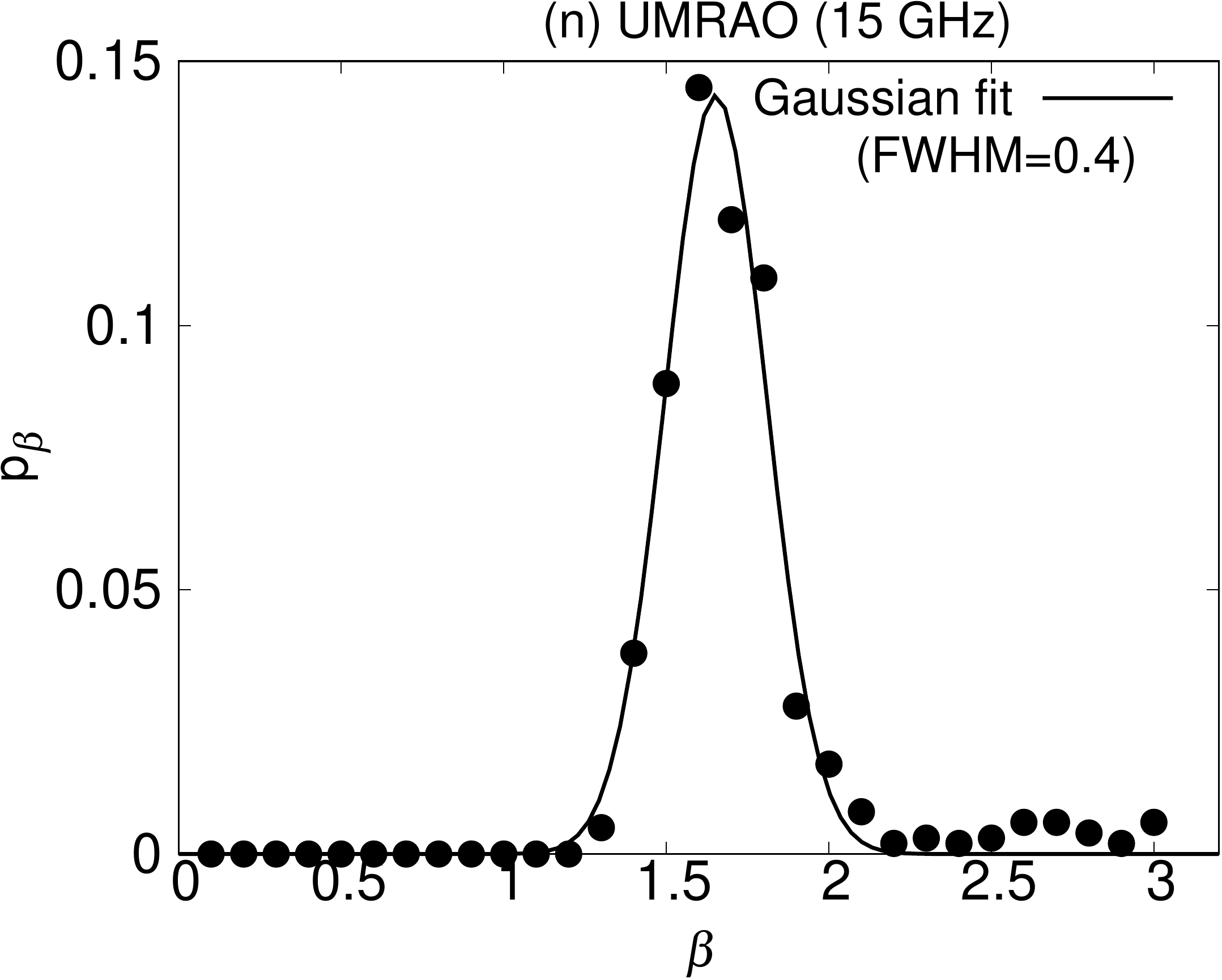}
\includegraphics[width=0.25\textwidth]{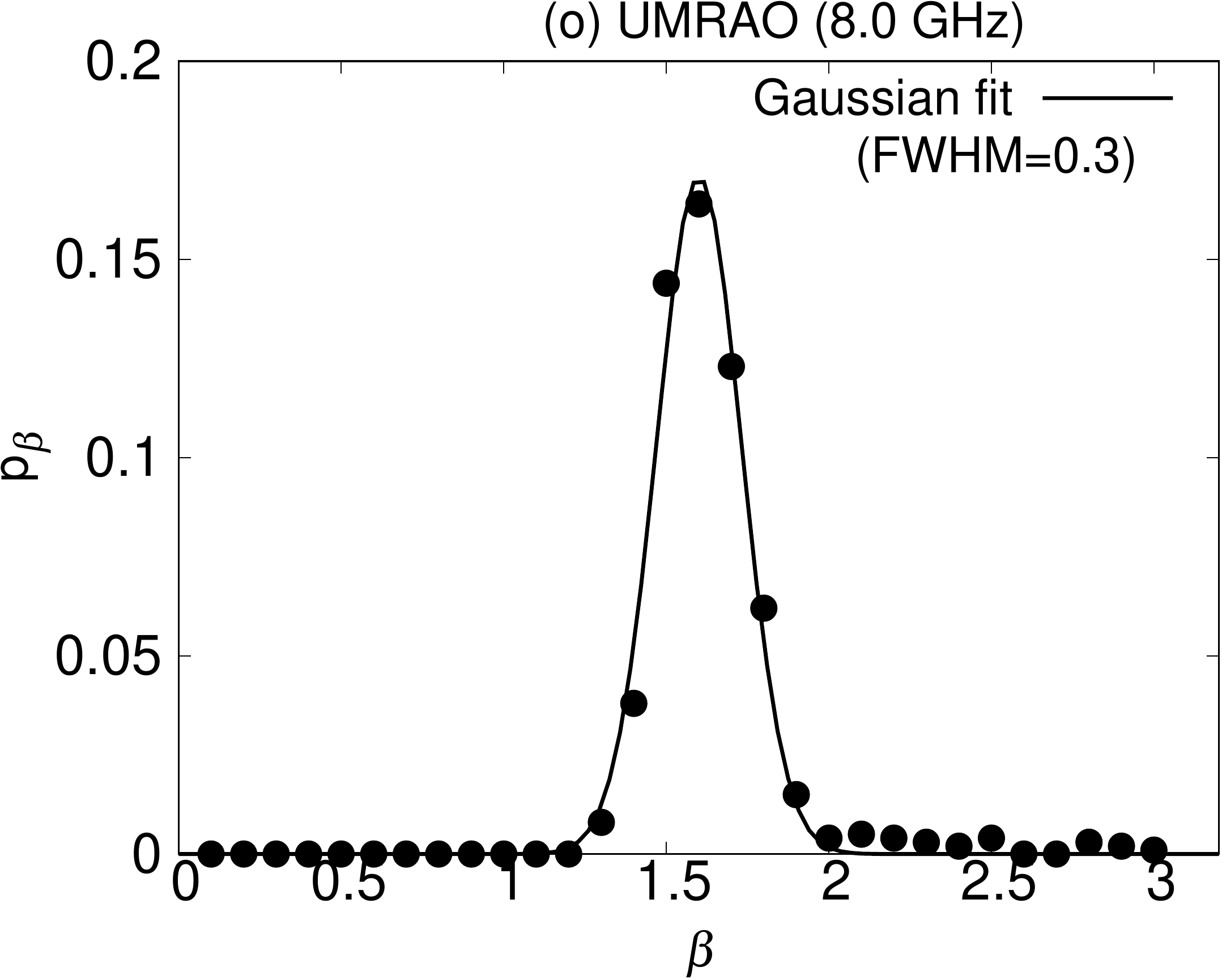}
\includegraphics[width=0.25\textwidth]{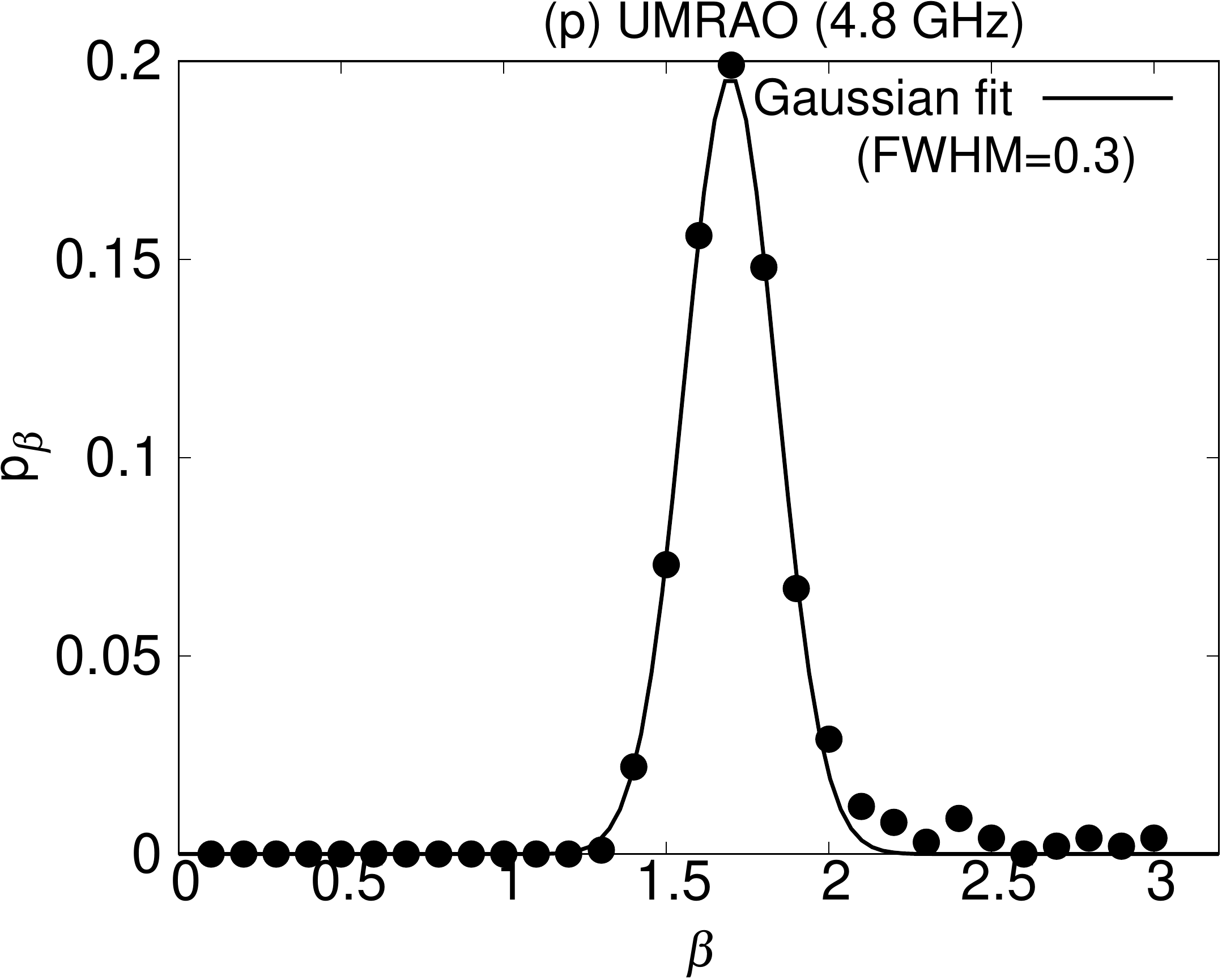}
}
\includegraphics[width=0.25\textwidth]{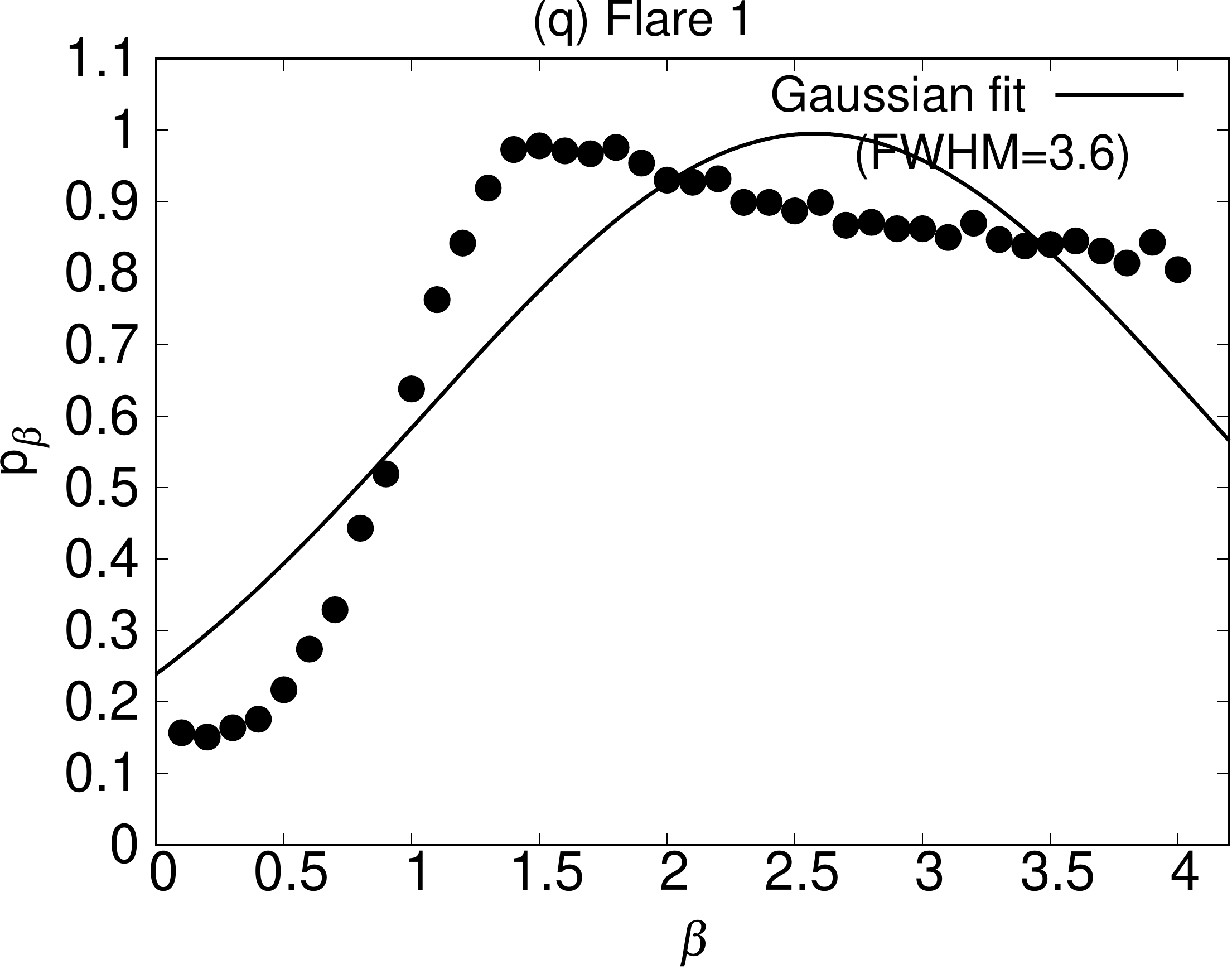}

\caption{As in Figure~\ref{appfig:beta3c}, for the blazar PKS\,1510$-$089.}
\label{appfig:betapks}
\end{figure*}

\clearpage
\newpage

\section{Comparison of PSD shapes using the DFT and LSP methods}\label{app:C}

The LS periodogram (LSP) is famous for naturally handling the data gaps in the time series \citep[][]{Lomb76, Scargle82} as compared to the DFT method, which introduces spurious power in the periodogram for unevenly sampled time series (Appendix~\ref{app:A}). Therefore, to perform DFT of the unevenly sampled time series (as is the case with the observed light curves), a linear interpolation between successive data points is used to obtain an evenly sampled time series. In such cases, the periodograms are obtained down to Nyquist frequencies corresponding to the mean sampling interval of the time series, as linear interpolation introduces a ``lack of variability'' (relative to the intrinsic variability) into the highest frequencies. In the case of LSP, however, the periodograms can be obtained down to the highest Nyquist frequency \citep[determined by the smallest sampling interval;][]{VanderPlas18}. 

\begin{figure*}[!htbp]
	\hbox{
\includegraphics[width=0.5\textwidth]{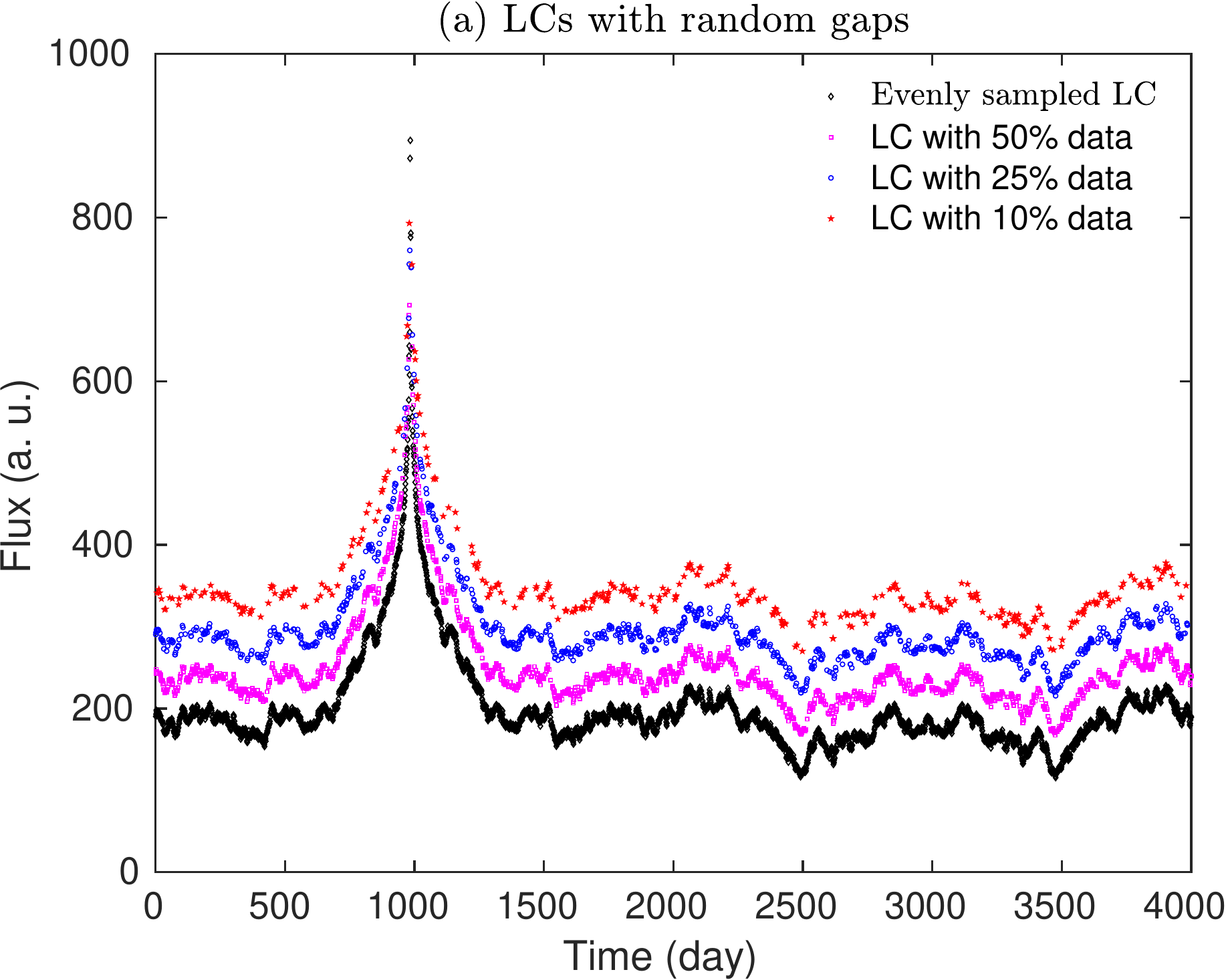}
\includegraphics[width=0.5\textwidth]{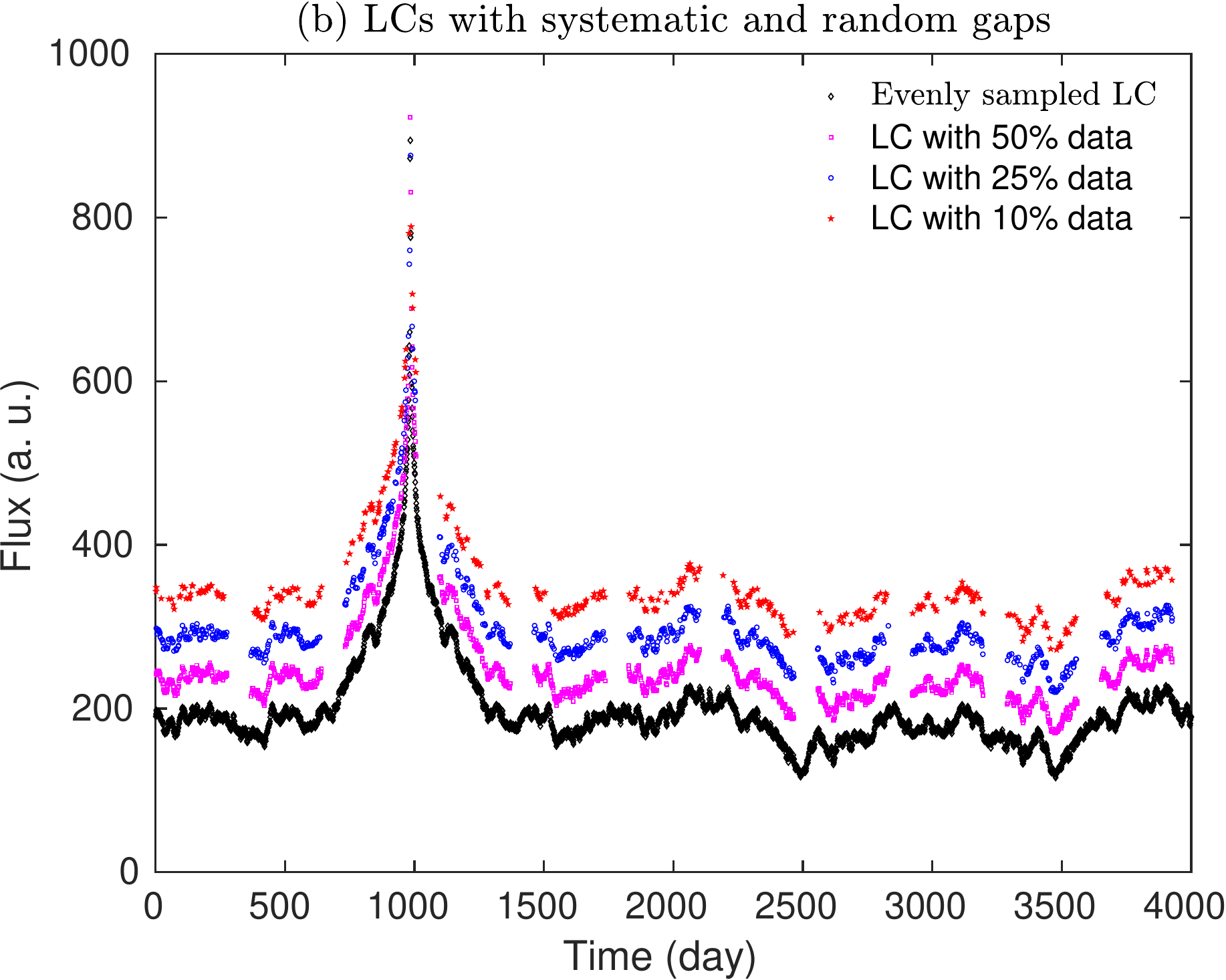}
}
	\caption{Evenly sampled light curve (LC) generated with $\beta$=2 and log-normal flux distribution for a total duration of 4,000\,day with a sampling interval of 1\, day (black cross). Panel (a) shows the unevenly sampled light curves containing 10\% (red star), 25\% (blue diamond), and 50\% (magenta pentagon) of the data, respectively, by introducing random gaps in the time series. Panel (b) shows the unevenly sampled light curves containing 10\% (red star), 25\% (blue diamond), and 50\% (magenta pentagon) of the data, respectively, by introducing seasonal (3 months) and random gaps in the time series. The unevenly sampled LCs are shifted vertically for clarity of visualization.}
\label{fig:simlc}
\end{figure*}

With numerical simulation, however, we demonstrate that contrary to expectations, the LSP PSD does not reproduce the shape of the intrinsic PSD down to the maximum Nyquist frequency  when sampling is relatively sparse. To do so, we simulated a light curve with $\beta = 2$ and a log-normal flux distribution, following the method of \citet{Emmanoulopoulos13}, containing 4,000 data points with a sampling interval of one day. This length is chosen to mimic the nearly decade-long light curves studied in the present study. Using this light curve, we obtained unevenly sampled light curves with mean sampling rates equal to 10\%, 25\%, and 50\%, respectively, in two fashions. (1) We introduced random gaps in the time series such that 90\%, 75\%, and 50\% data were discarded at random times; these sorts of sampling patterns are expected from monitoring instruments such as \textit{Fermi}-LAT when fluxes fall below the TS threshold (shown in panel (a) of Figure~\ref{fig:simlc}). (2) We introduced a 3-month long systematic gap every twelve months in the time series as well as random gaps in the time series; these sorts of sampling patterns are expected at optical and infrared wavelengths from ground-based telescopes (shown in panel (b) of Figure~\ref{fig:simlc}). For simplicity, we did not add Gaussian fluctuations resulting from measurement uncertainties in the simulated LCs. The $T_{\rm mean}$ values correspond to $\sim$10, $\sim$4 and $\sim$2\,days, respectively, for these light curves. 

The LSP \citep[][]{Scargle82} of an unevenly sampled, mean-subtracted light curve $f(t_i)$, with the total duration of $T$, consisting of $N$ data points, is defined as: 

\begin{equation}
P_f(2 \pi \nu_k )  =  \frac{1}{2}\left \{ \frac{ \Bigg\{ \sum_{i=1}^{N} f(t_i)  \cos[2\pi\nu_k (t_i-\tau)]  \Bigg\}^2} {\sum_{i=1}^{N} f(t_i)  \cos^2[2\pi\nu_k (t_i-\tau)]} 
+  \frac { \Bigg\{ \sum_{i=1}^{N} f(t_i)  \sin[2\pi\nu_k (t_i-\tau)]  \Bigg\}^2} {\sum_{i=1}^{N} f(t_i)  \sin^2[2\pi\nu_k (t_i-\tau)]} \right \} \, ,
\label{eq:lsp}
\end{equation}
where a factor $\tau$, which accounts for the uneven sampling, is defined by
\begin{equation}
\tau = \arctan \Bigg\{\frac{1}{2 (2 \pi \nu_k) } \, \frac{ \sum_{i=1}^{N} \sin [2 (2 \pi \nu_k) t_i ]} {  \sum_{i=1}^{N} \cos [ 2 (2 \pi \nu_k) t_i ] }\Bigg\} \, ,  
\label{eq:tau}
\end{equation}

\begin{figure*}[!htbp]
	\hbox{
		\includegraphics[width=0.30\textwidth]{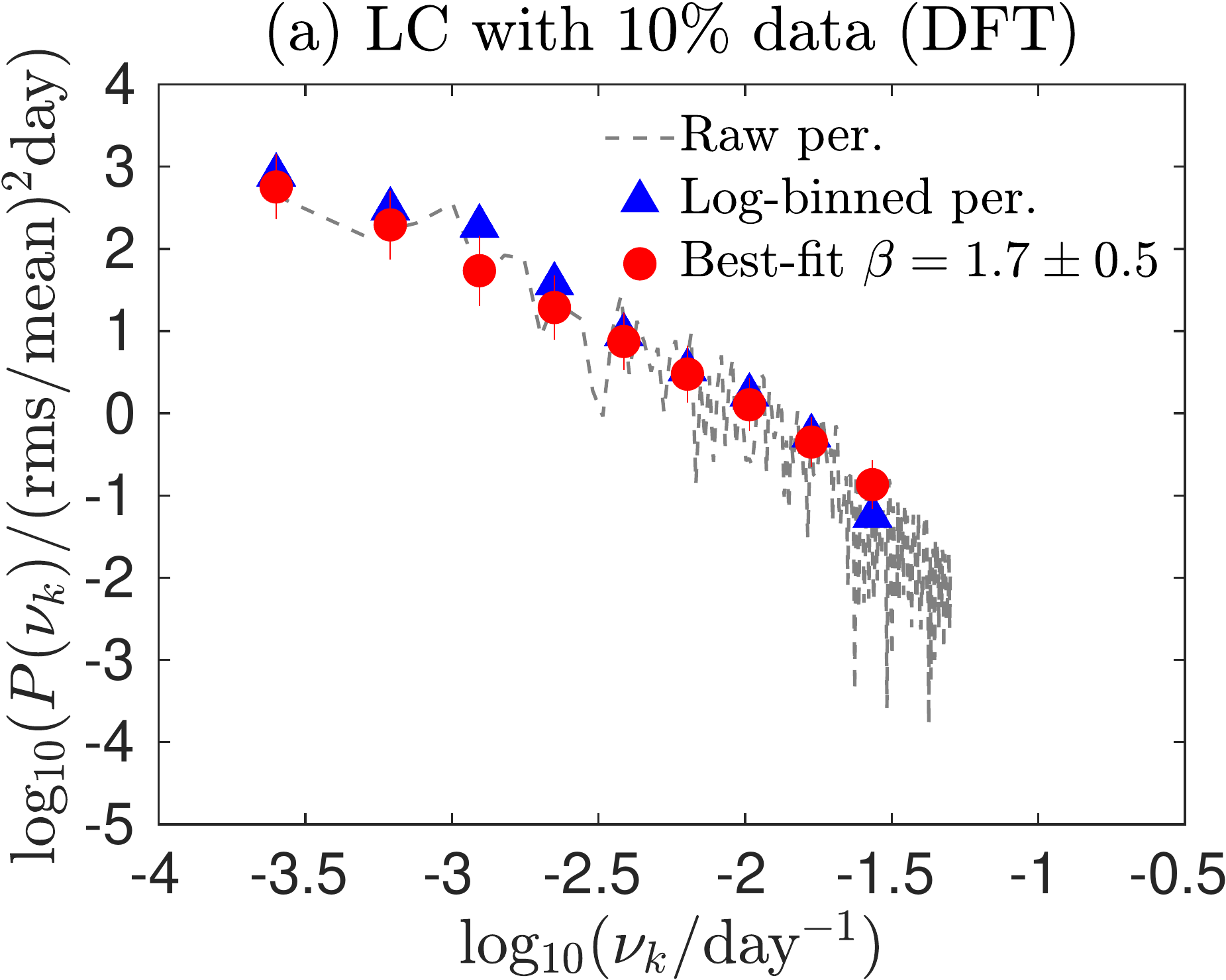}
                \includegraphics[width=0.30\textwidth]{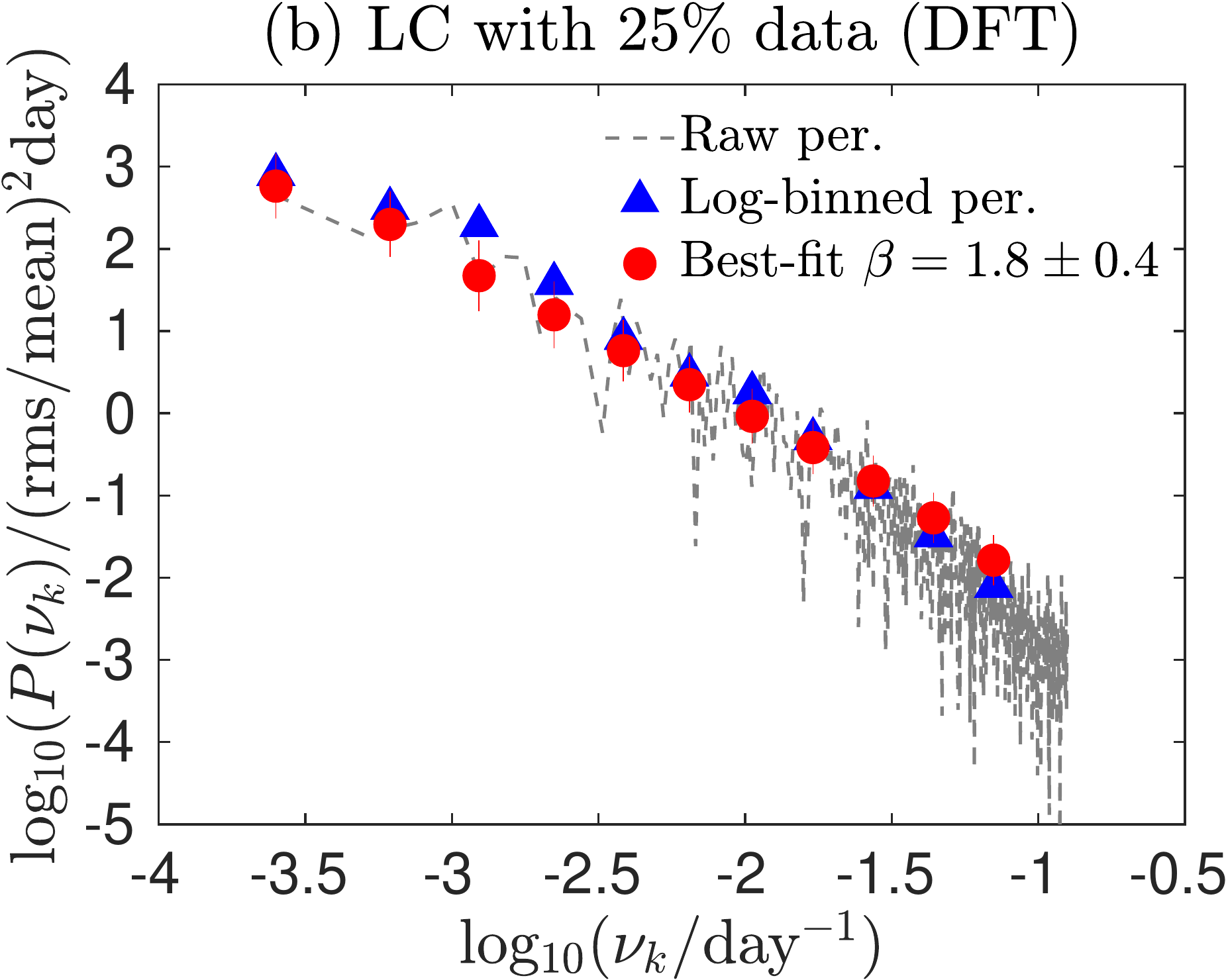}
                \includegraphics[width=0.30\textwidth]{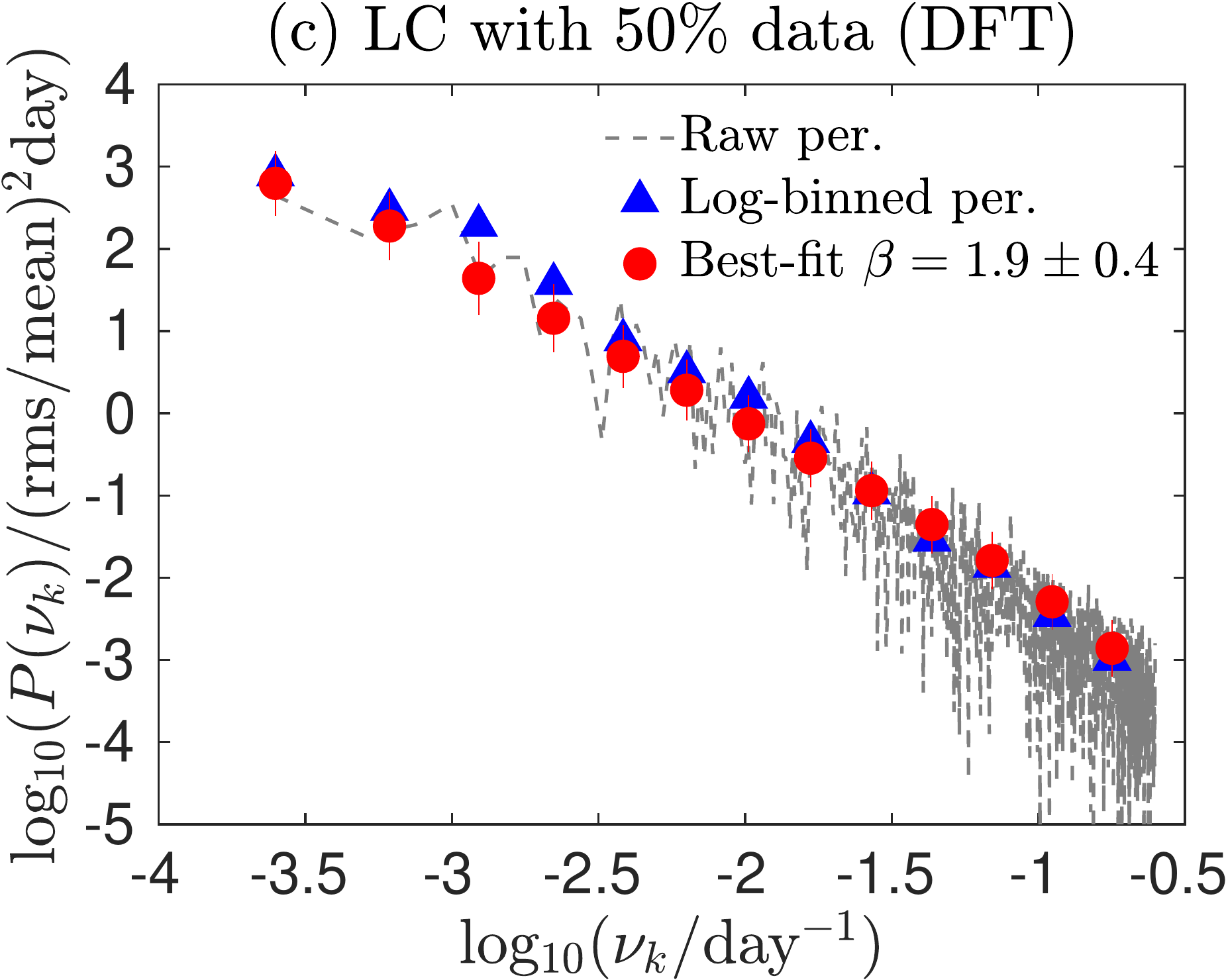}
         }
         	\hbox{
		\includegraphics[width=0.30\textwidth]{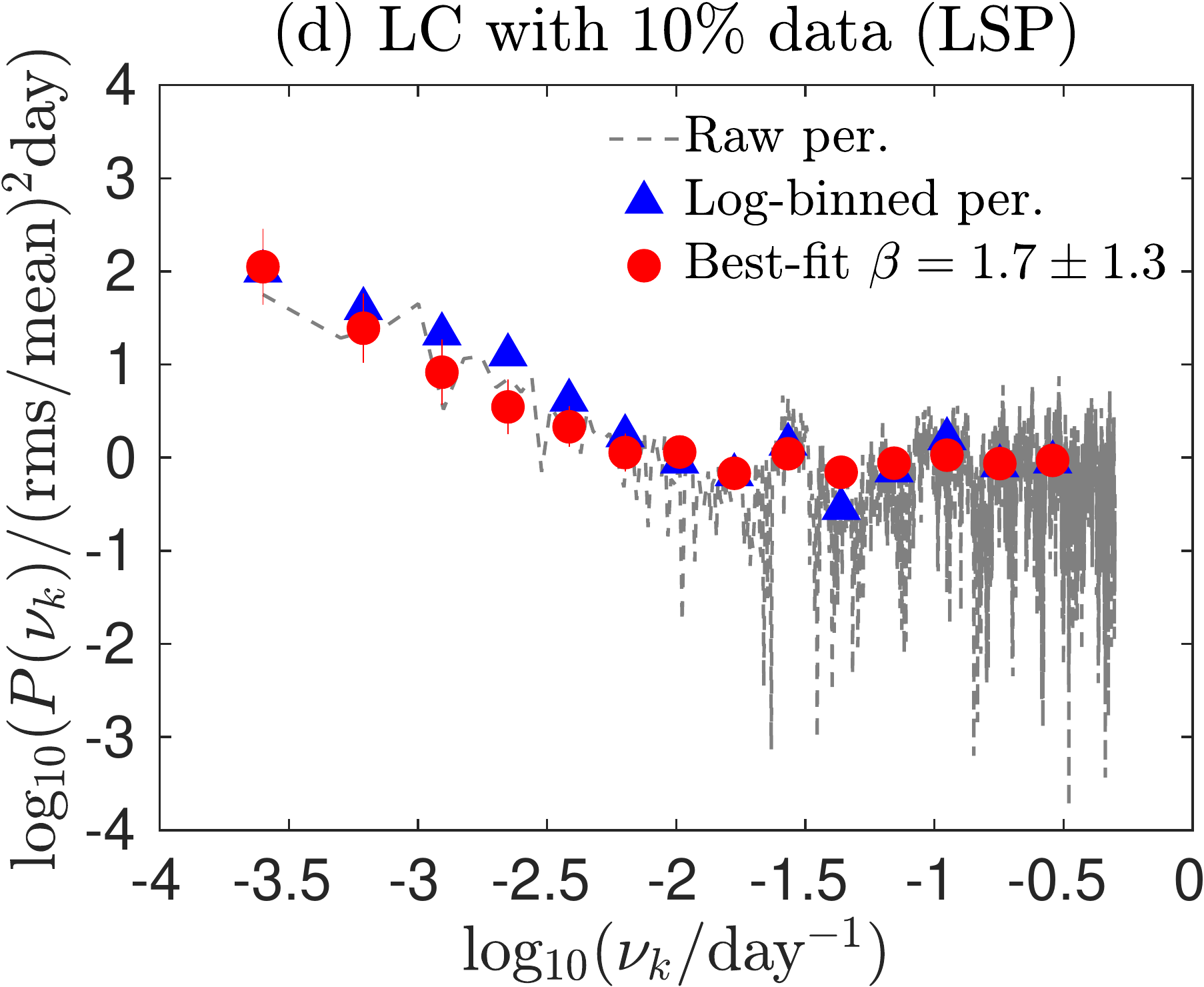}
                \includegraphics[width=0.30\textwidth]{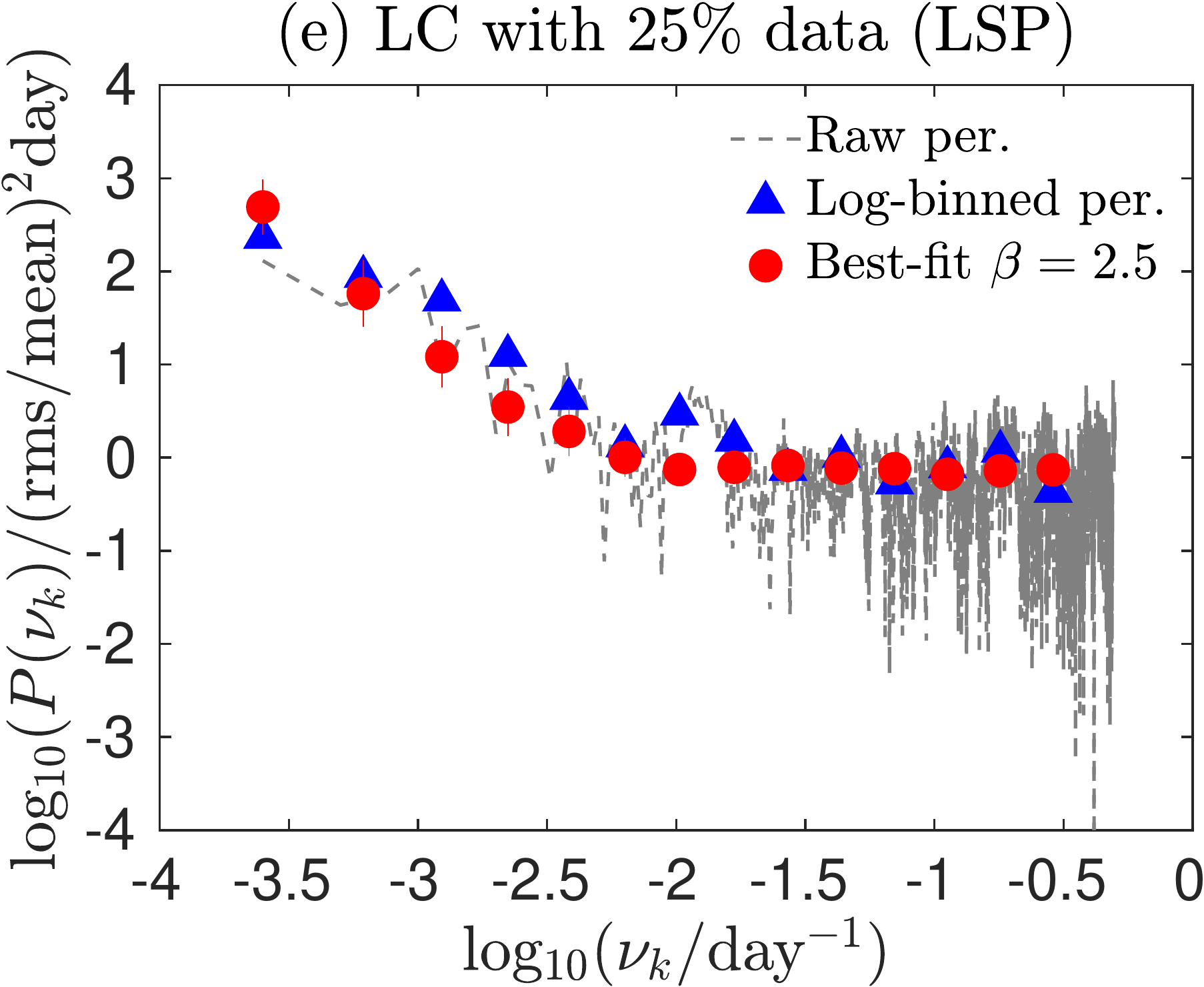}
                \includegraphics[width=0.30\textwidth]{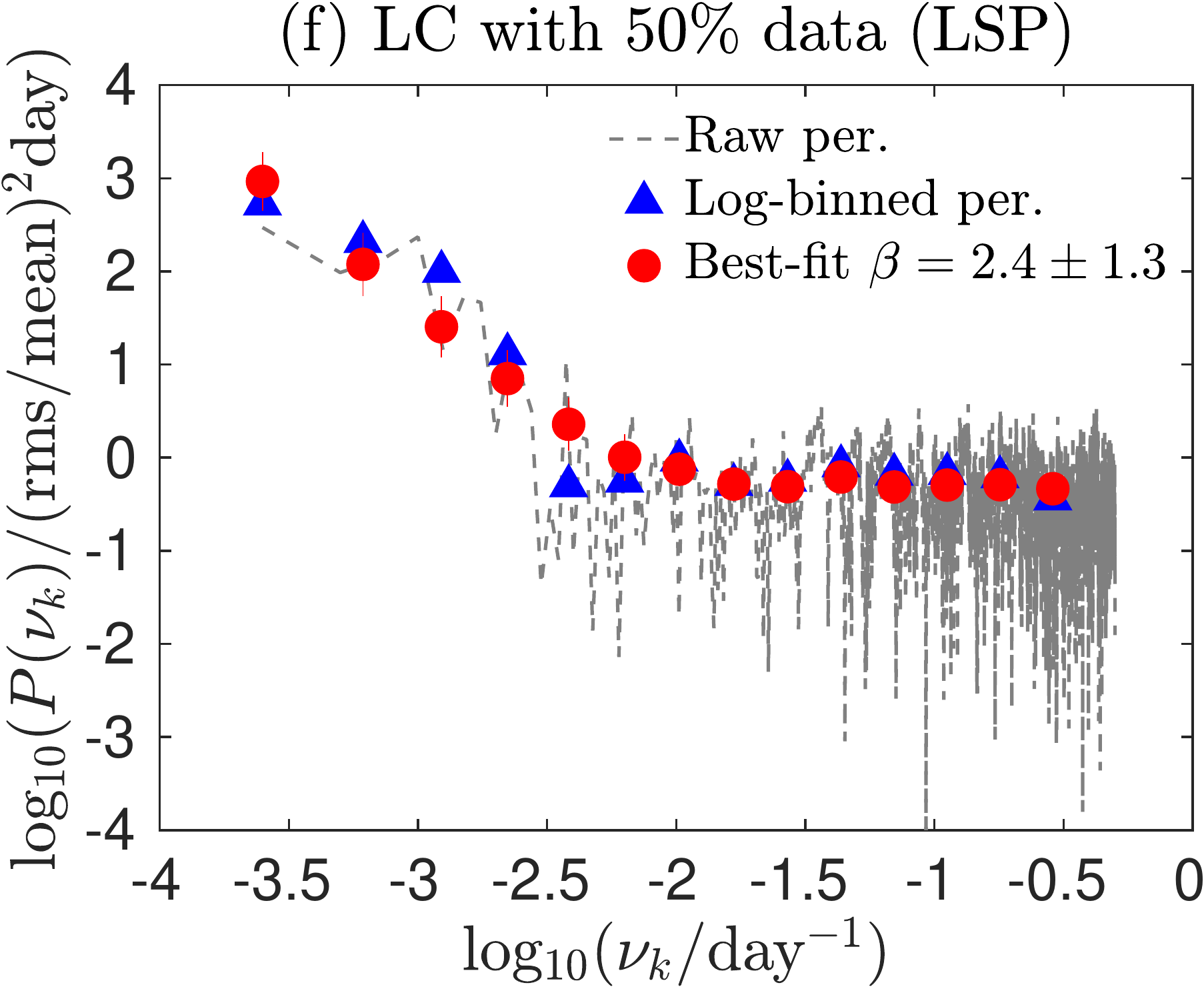}
         }

	\caption{The best-fit PSDs obtained for the unevenly sampled LCs shown in Figure~\ref{fig:simlc}(a) using the PSRESP formalism for the DFT (panels a, b, and c) and LSP (panels d, e, and f) methods (see Section~\ref{sec:psresp} for details). The maximum Nyquist frequencies are 1/2*$T_{\rm mean}$ (day$^{-1}$) and 0.5\,(day$^{-1}$), respectively, for the DFT and the LSP methods. Note that the simulated LCs are free of noise; therefore, the flattening in the LSP PSDs on frequencies $\geq$ 10$^{-1.5}$(day$^{-1}$) arises from the small number of data points available to characterize the variability at the highest frequencies (up to the maximum Nyquist frequency).}
\label{fig:dftlspran}
\end{figure*}

\begin{figure*}[!htbp]
	\hbox{
 		\includegraphics[width=0.30\textwidth]{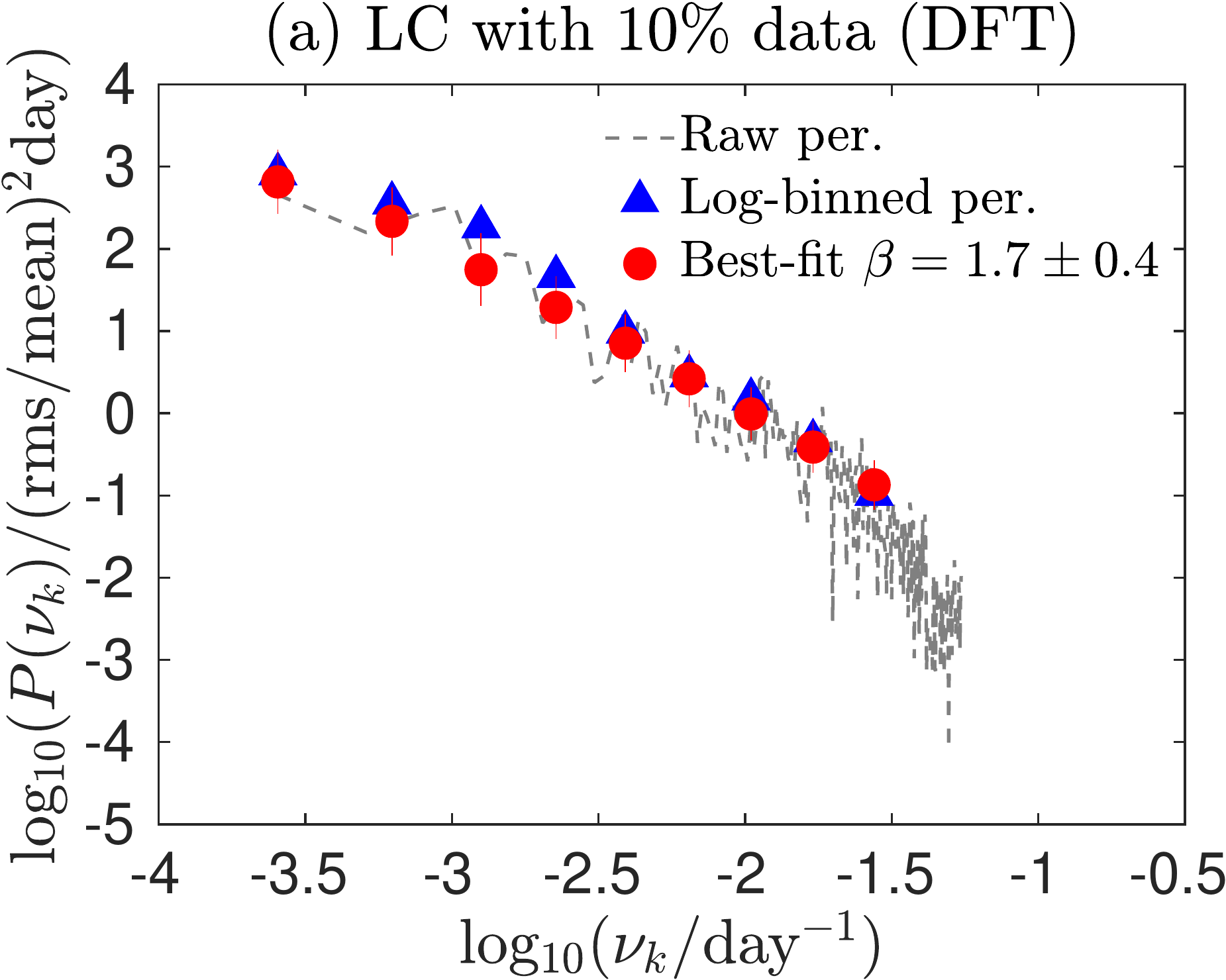}
                \includegraphics[width=0.30\textwidth]{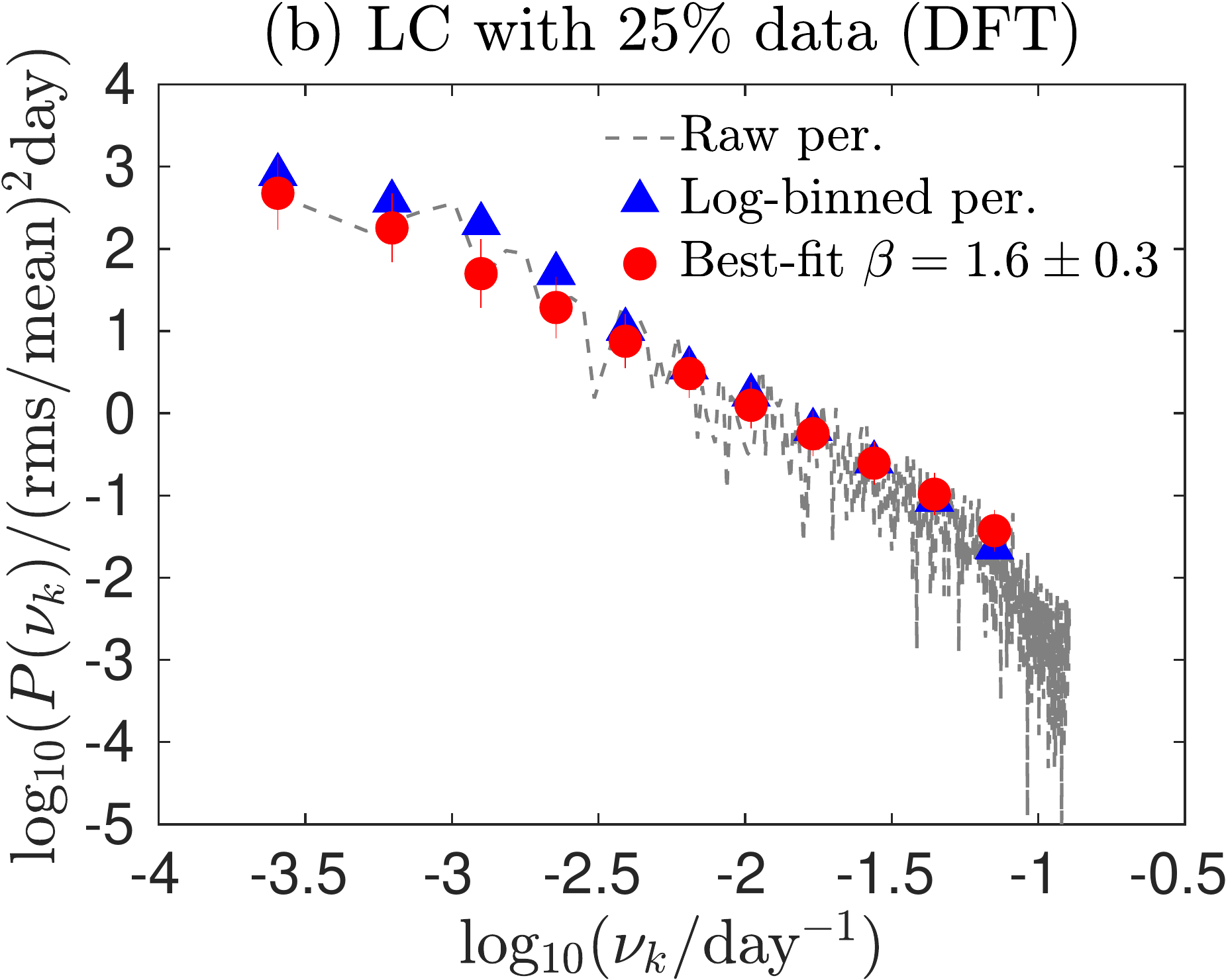}
                \includegraphics[width=0.30\textwidth]{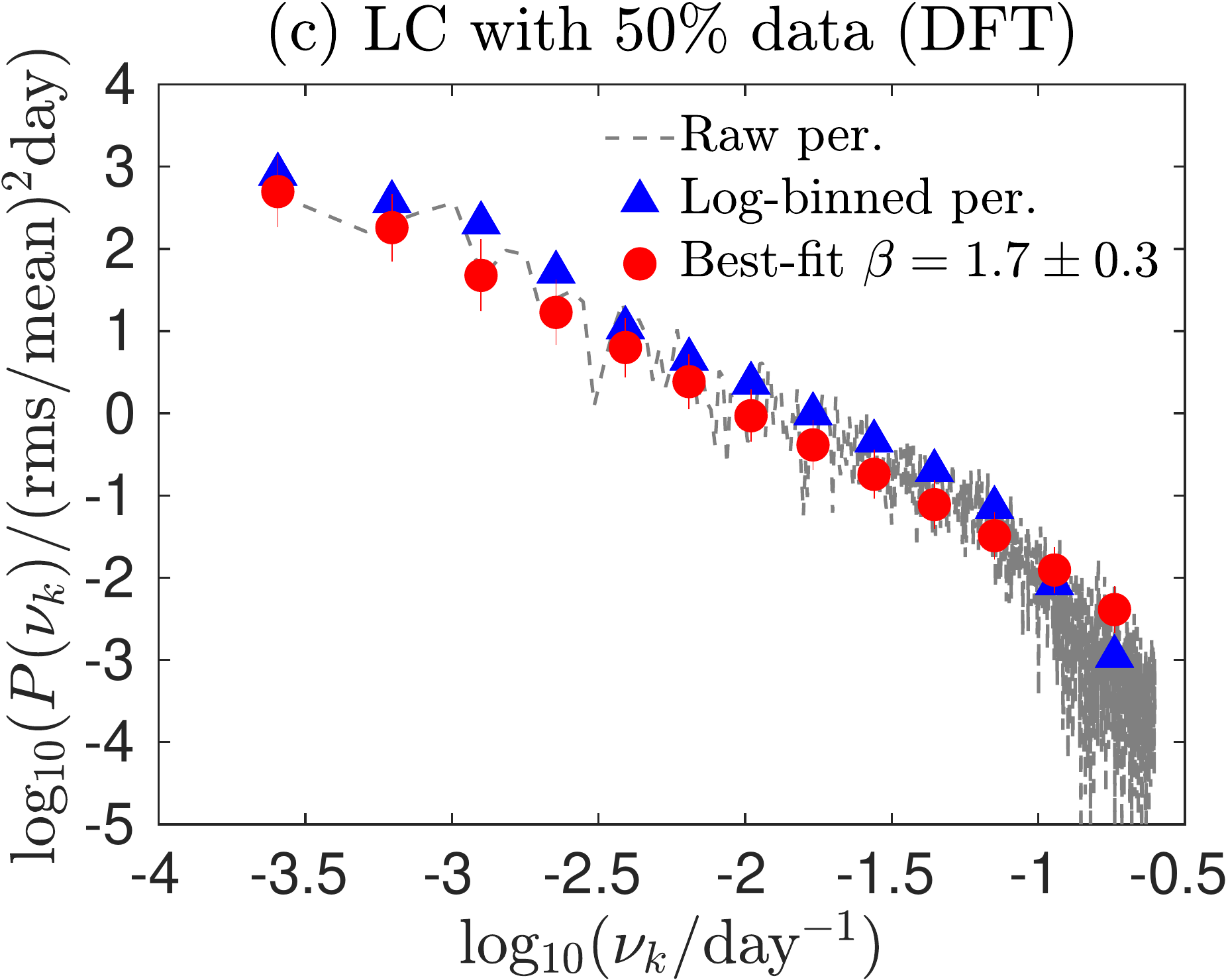}
         }
         	\hbox{
		\includegraphics[width=0.30\textwidth]{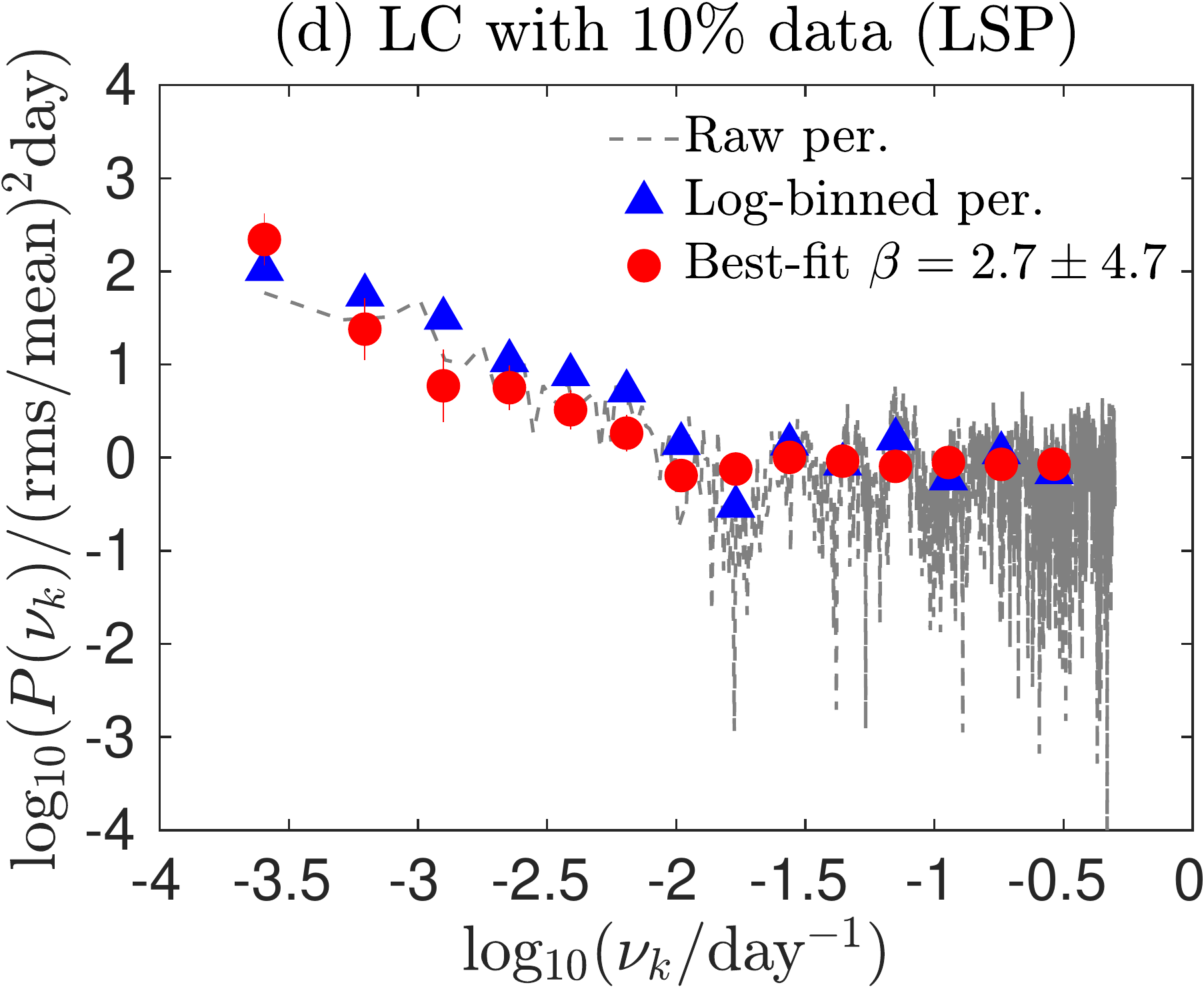}
                \includegraphics[width=0.30\textwidth]{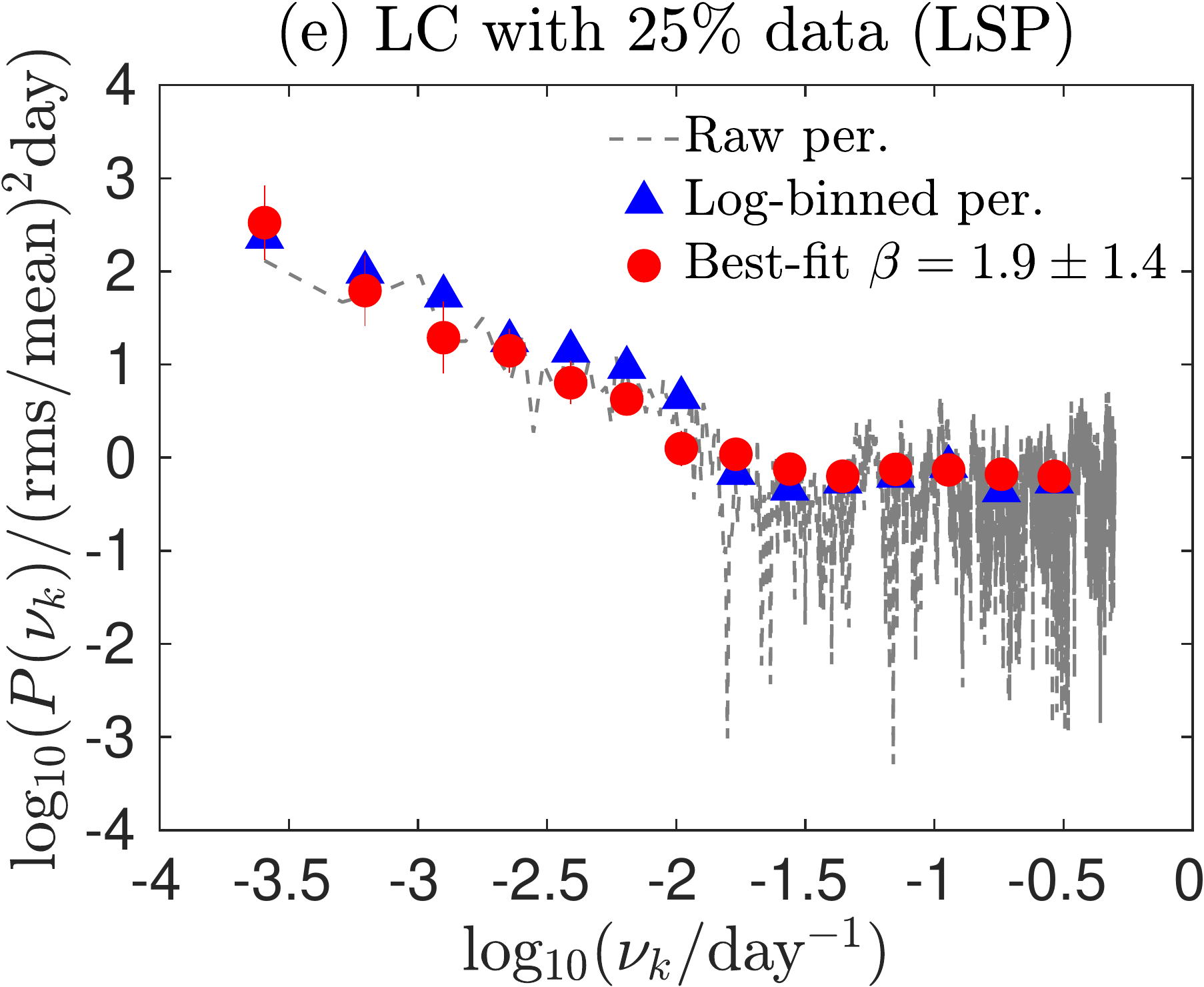}
                \includegraphics[width=0.30\textwidth]{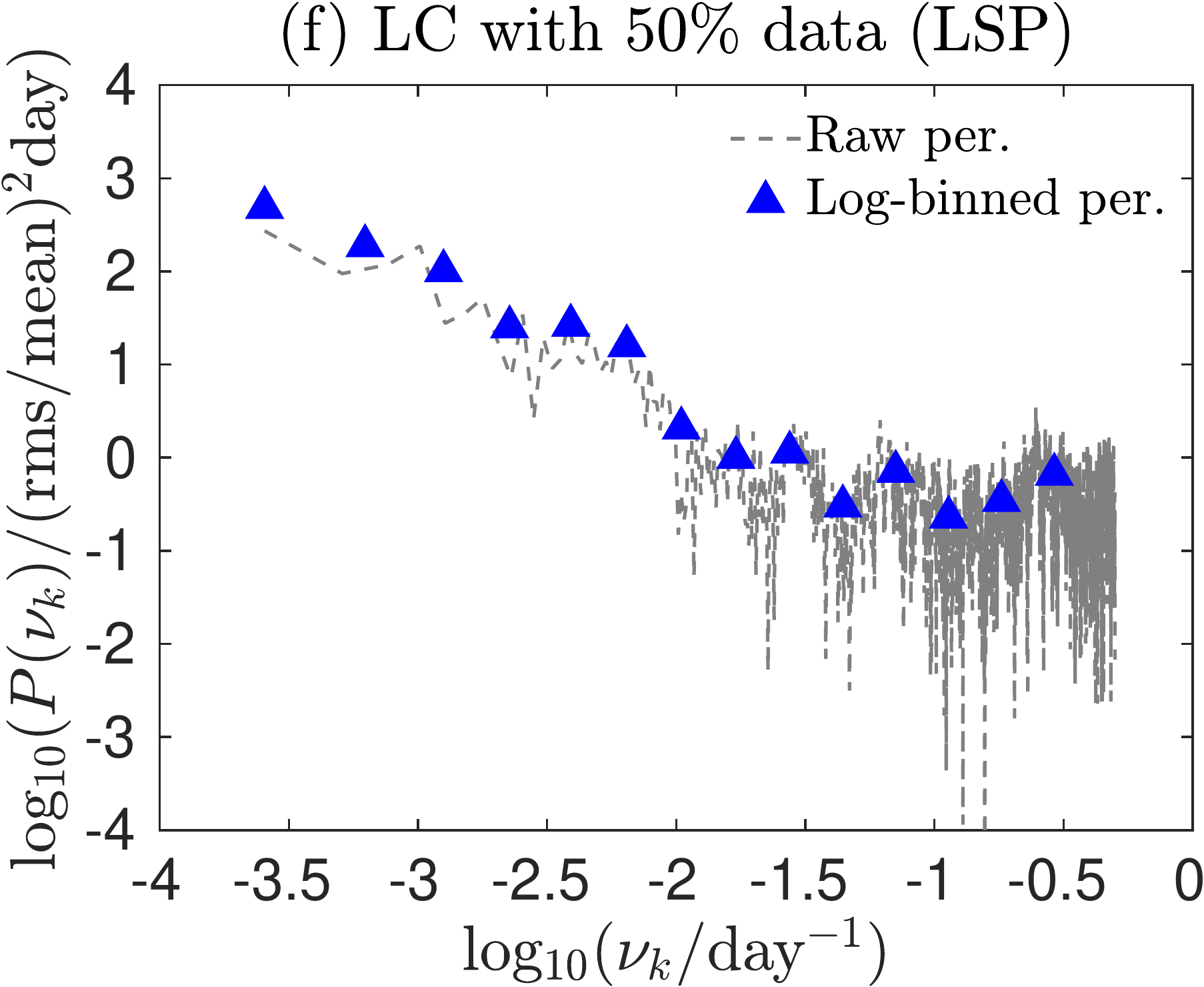}        
		}
	\caption{The best-fit PSDs obtained for the unevenly sampled LCs shown in Figure~\ref{fig:simlc}(b) using the PSRESP formalism for the DFT (panels a, b, and c) and LSP (panels d, e, and f) methods (see Section~\ref{sec:psresp} for details). The maximum Nyquist frequencies are 1/2*$T_{\rm mean}$ (day$^{-1}$) and 0.5\,(day$^{-1}$), respectively, for the DFT and the LSP methods. Note that the simulated LCs are free of noise; therefore, the flattening in the LSP PSDs on frequencies $\geq$ 10$^{-1.5}$(day$^{-1}$) arises from the small number of data points available to characterize the variability at the highest frequencies (up to the maximum Nyquist frequency).}
\label{fig:dftlspsys}
\end{figure*}

\begin{figure*}[!htbp]
	\hbox{
 		\includegraphics[width=0.30\textwidth]{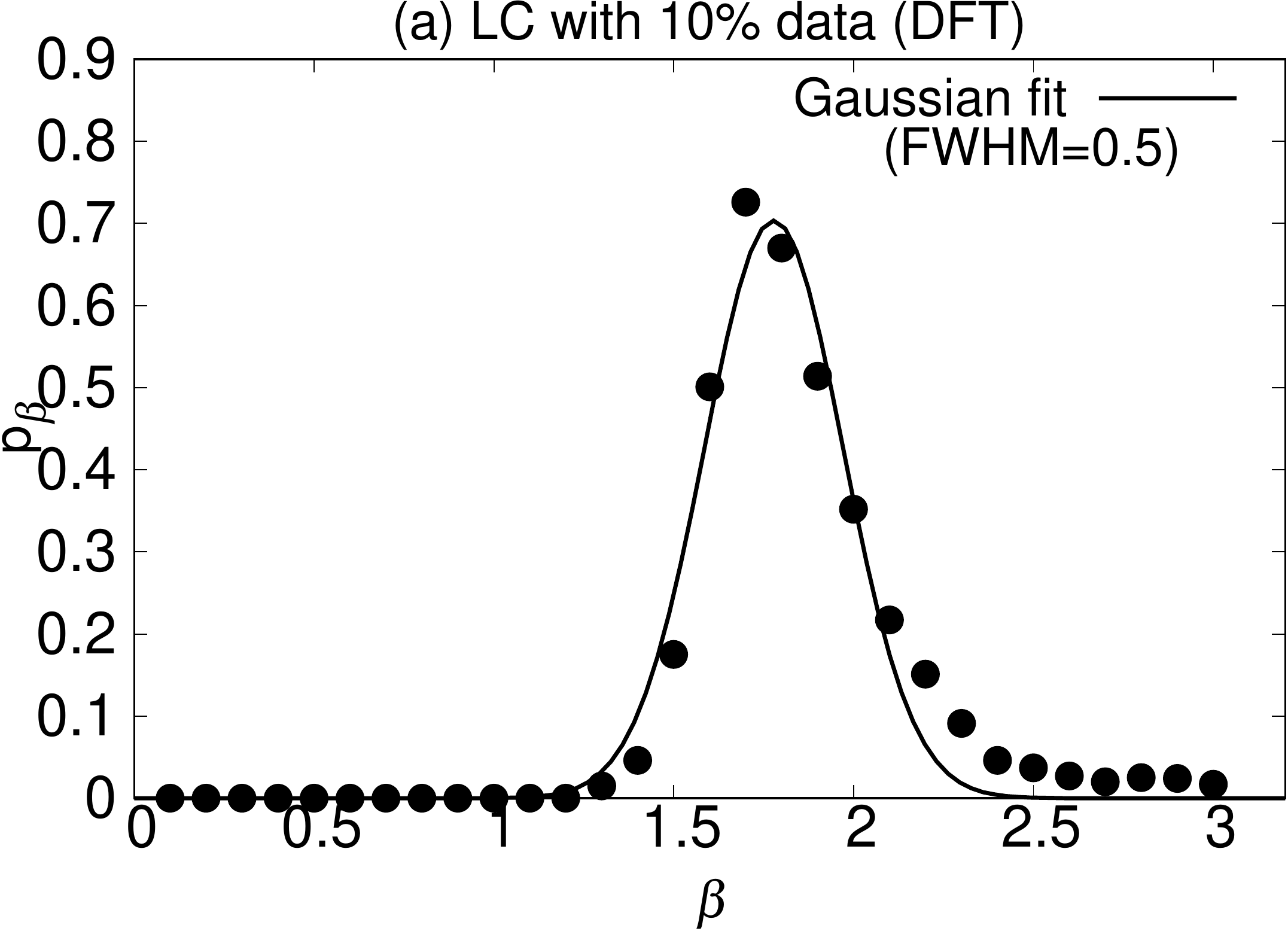}
 		\includegraphics[width=0.30\textwidth]{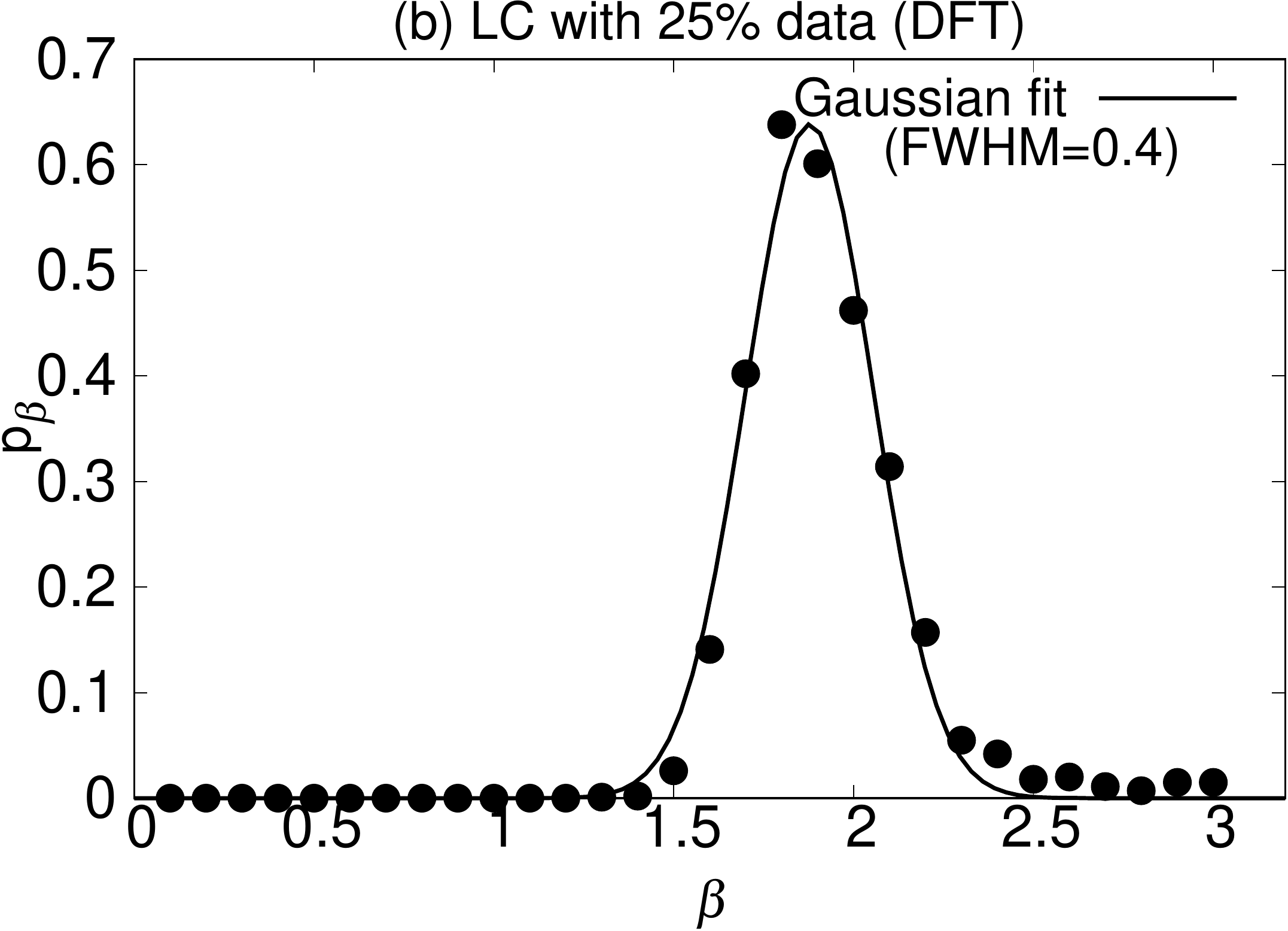}
 		\includegraphics[width=0.30\textwidth]{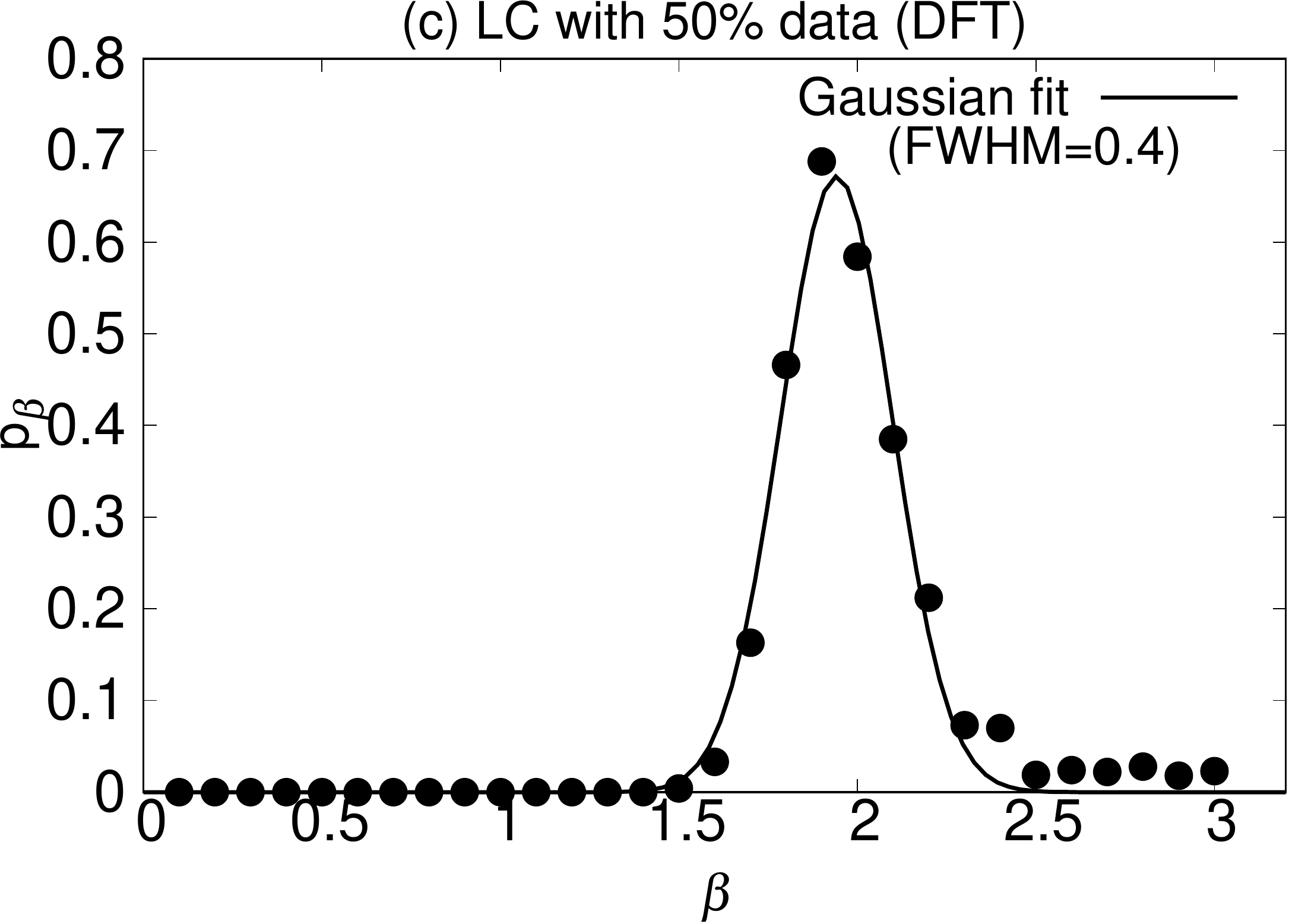}
         }
         	\hbox{
 		\includegraphics[width=0.30\textwidth]{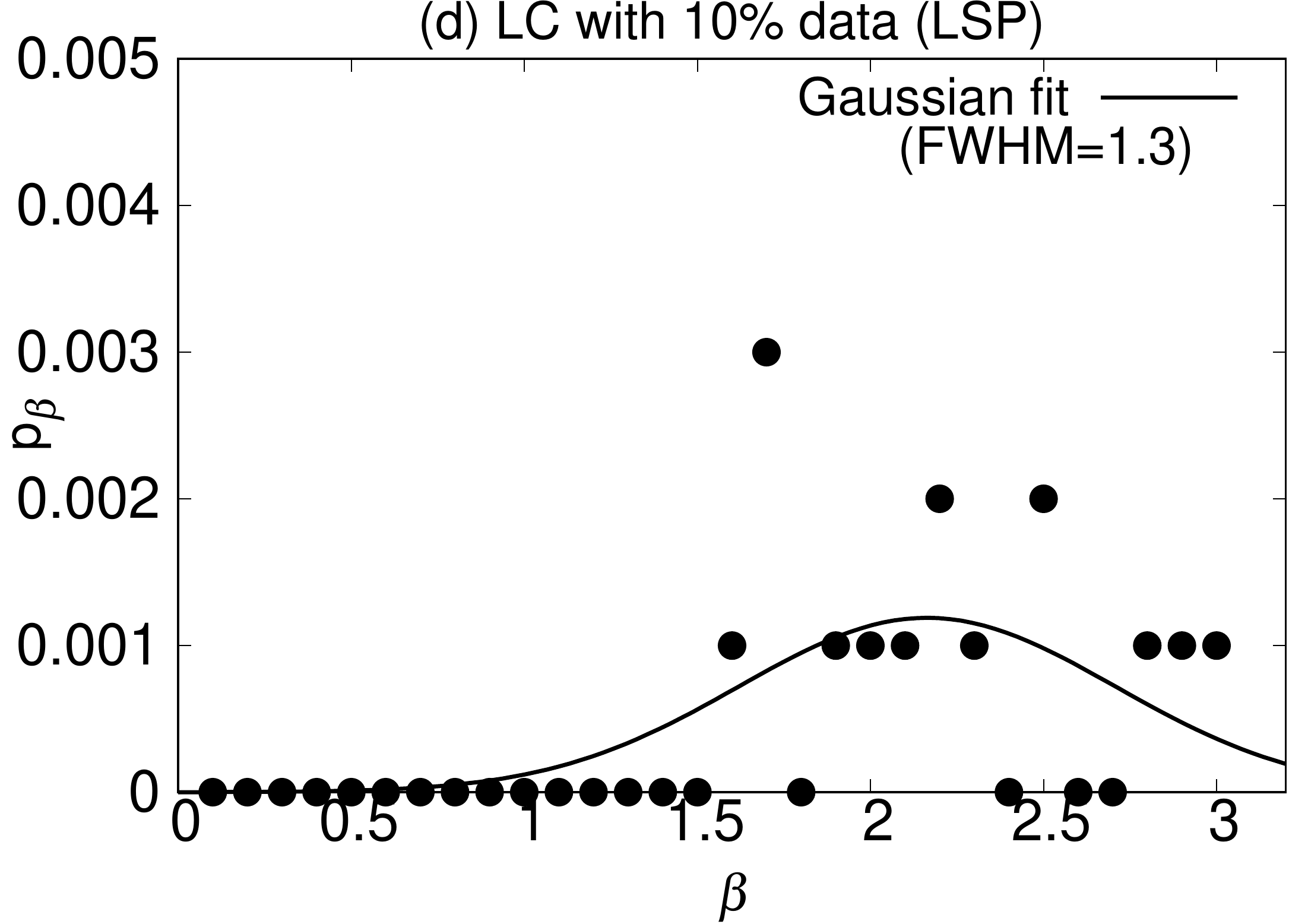}
 		\includegraphics[width=0.30\textwidth]{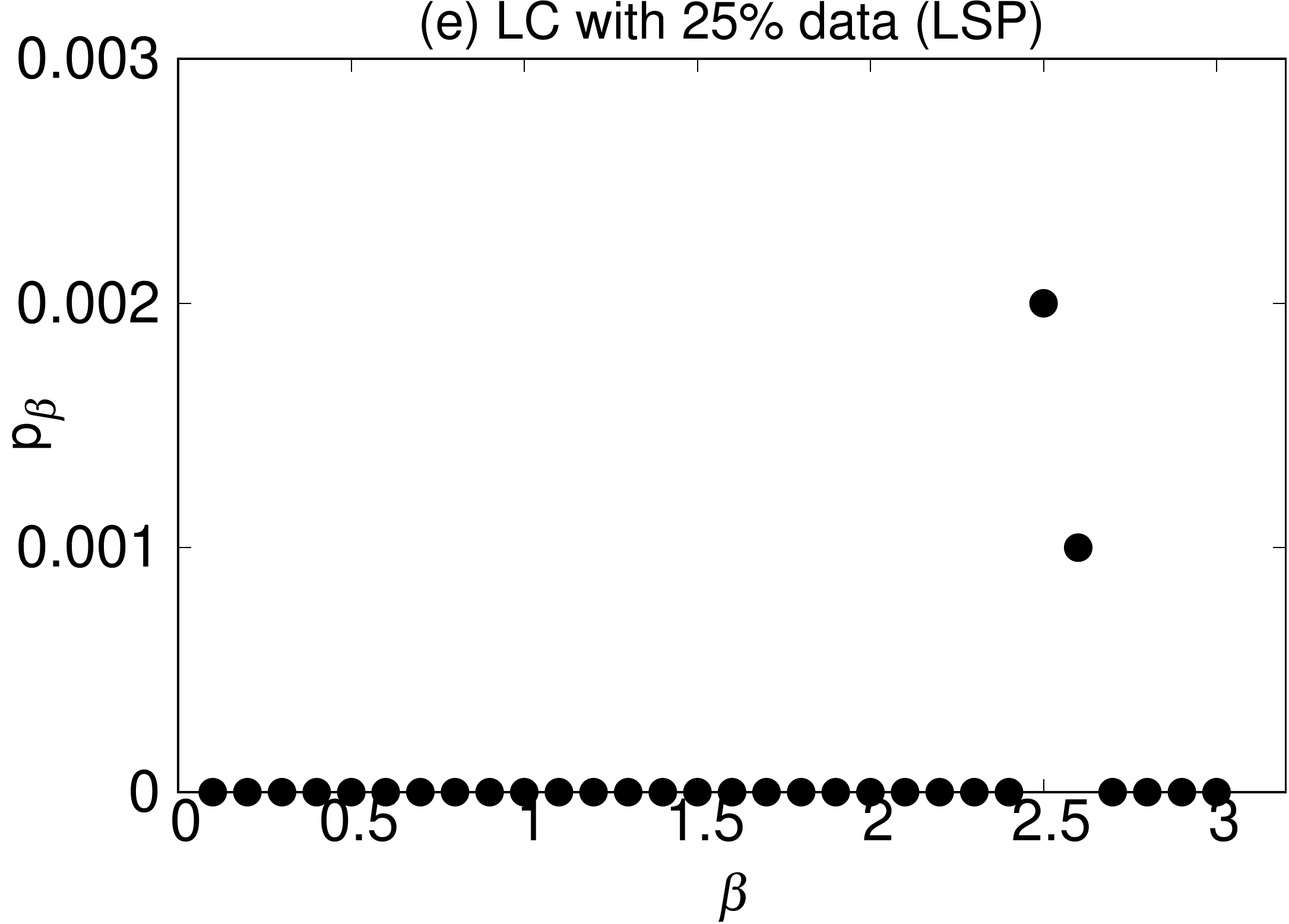}
 		\includegraphics[width=0.30\textwidth]{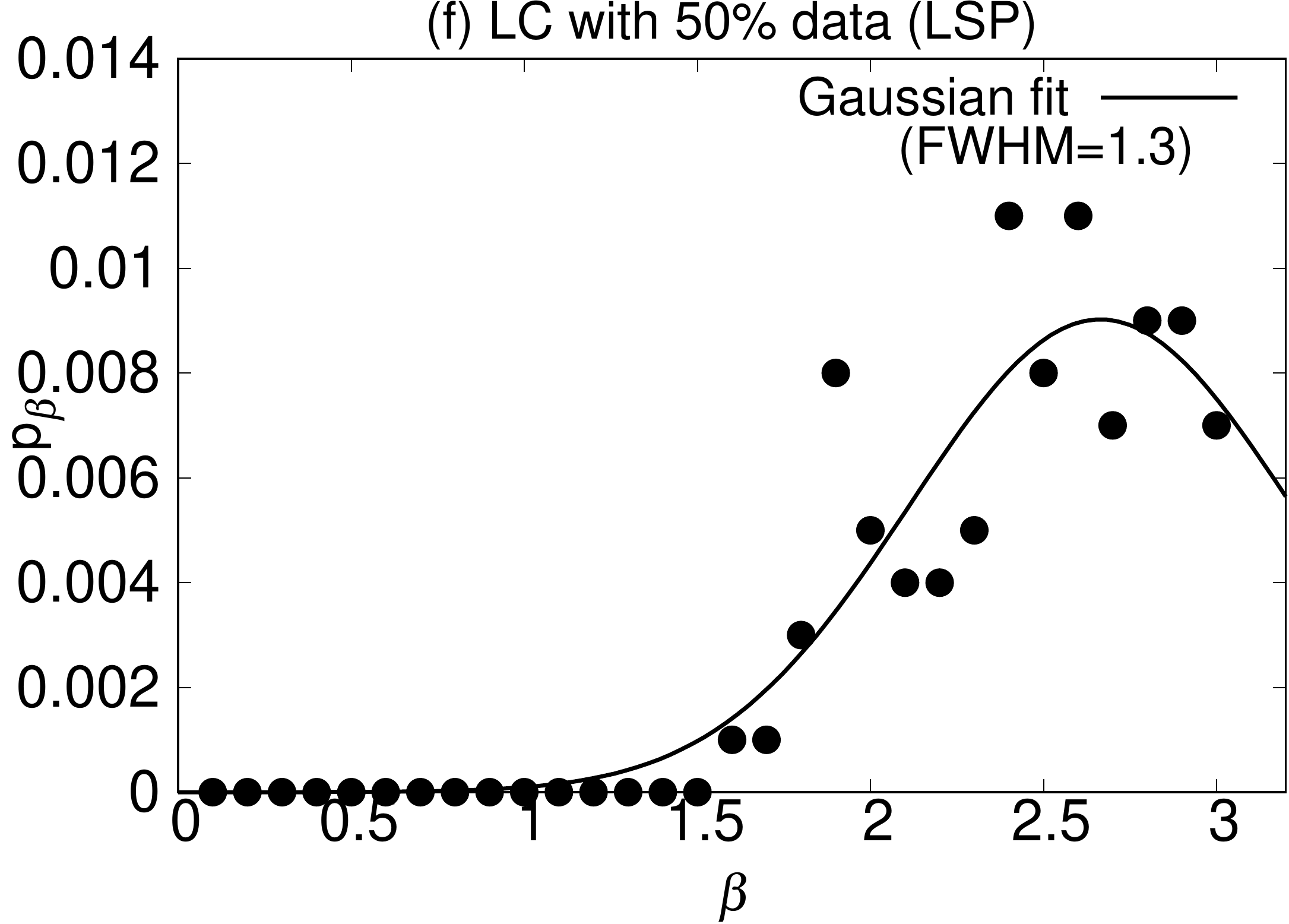}
		}
	\caption{probability curves for DFT (panels a, b, and c) and LSP (d, e, and f) methods for light curves shown in Figure~\ref{fig:simlc}a.}
\label{fig:betasimran}
\end{figure*}

\begin{figure*}[!htbp]
	\hbox{
 		\includegraphics[width=0.30\textwidth]{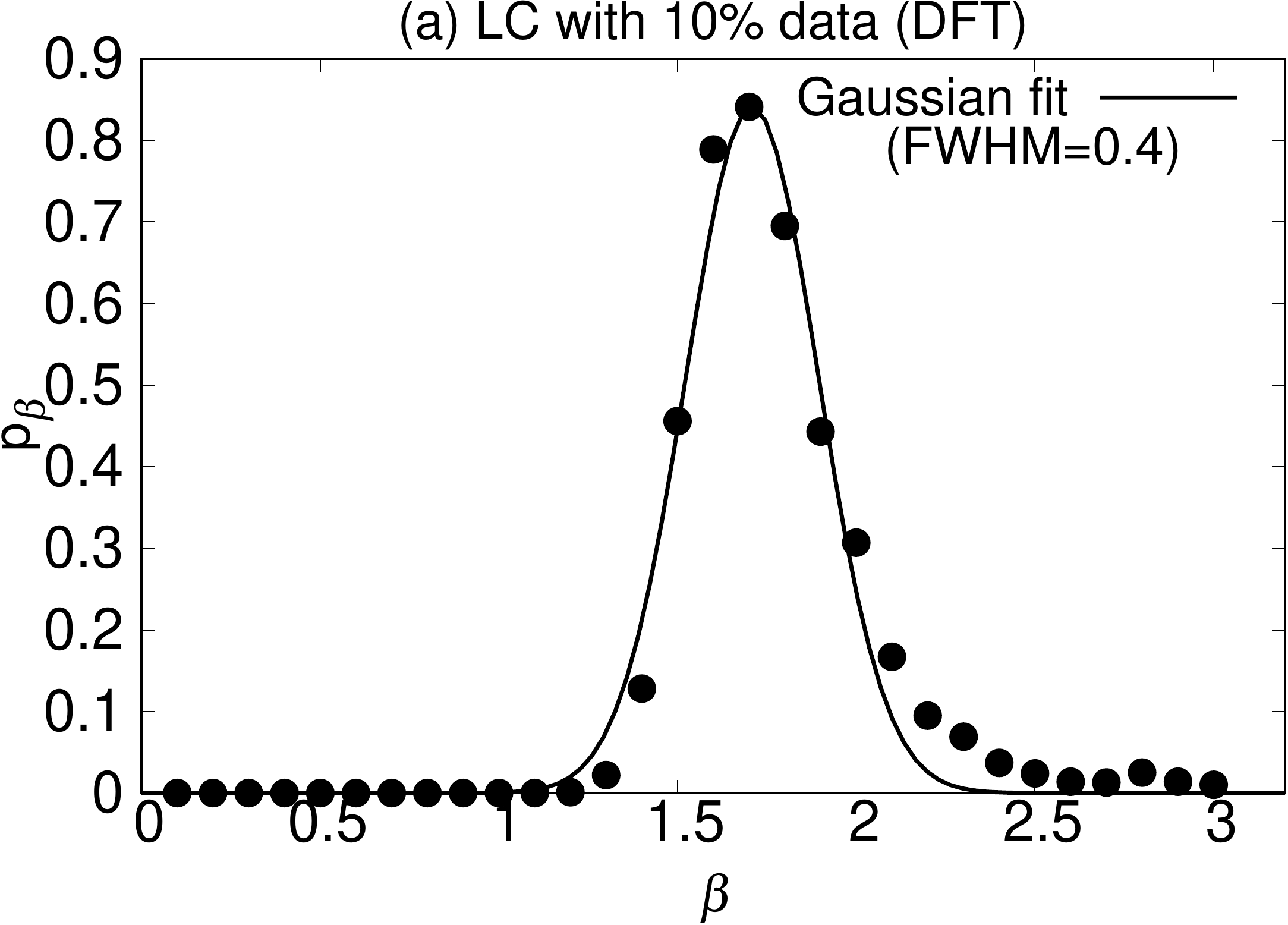}
 		\includegraphics[width=0.30\textwidth]{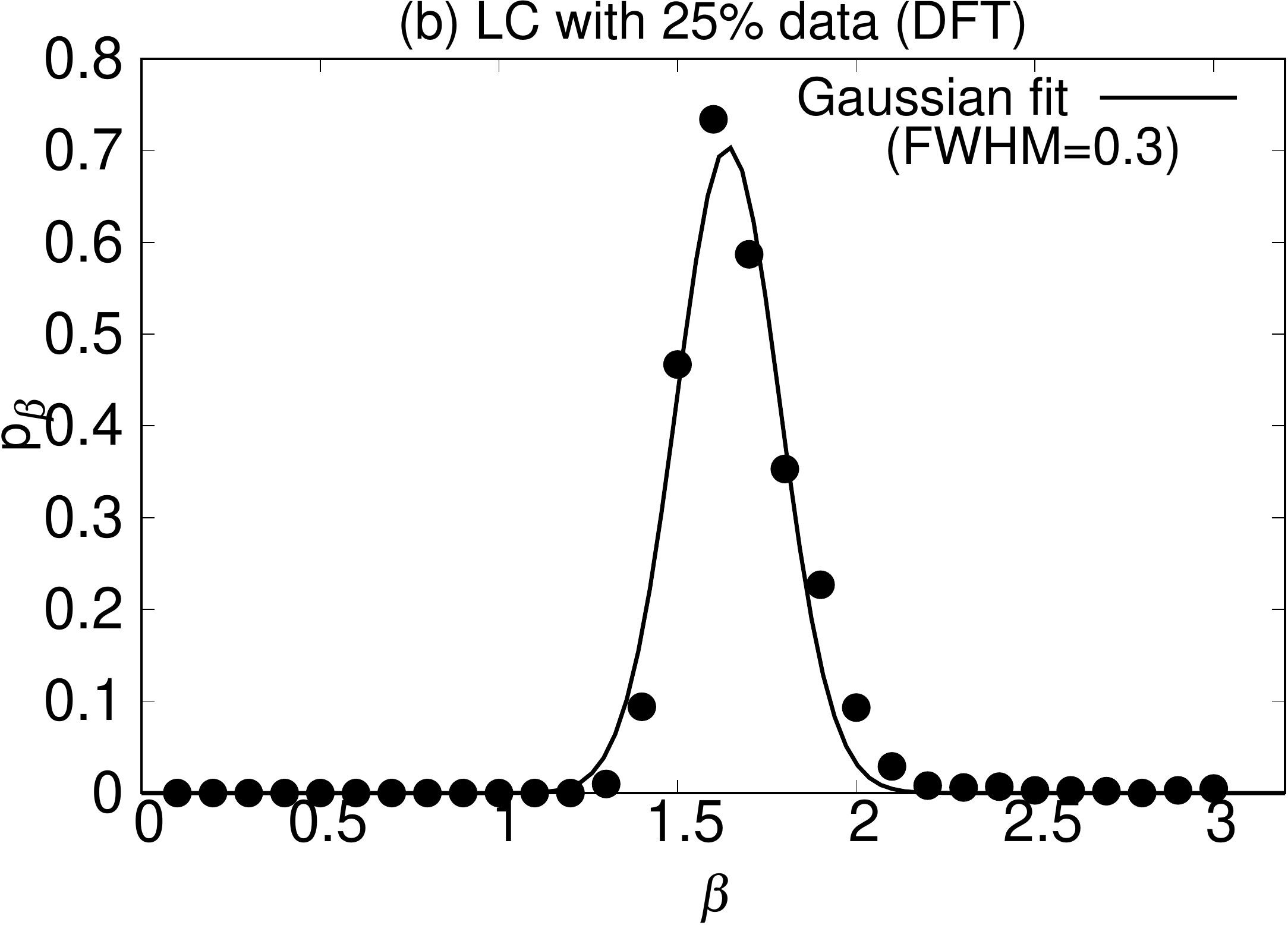}
 		\includegraphics[width=0.30\textwidth]{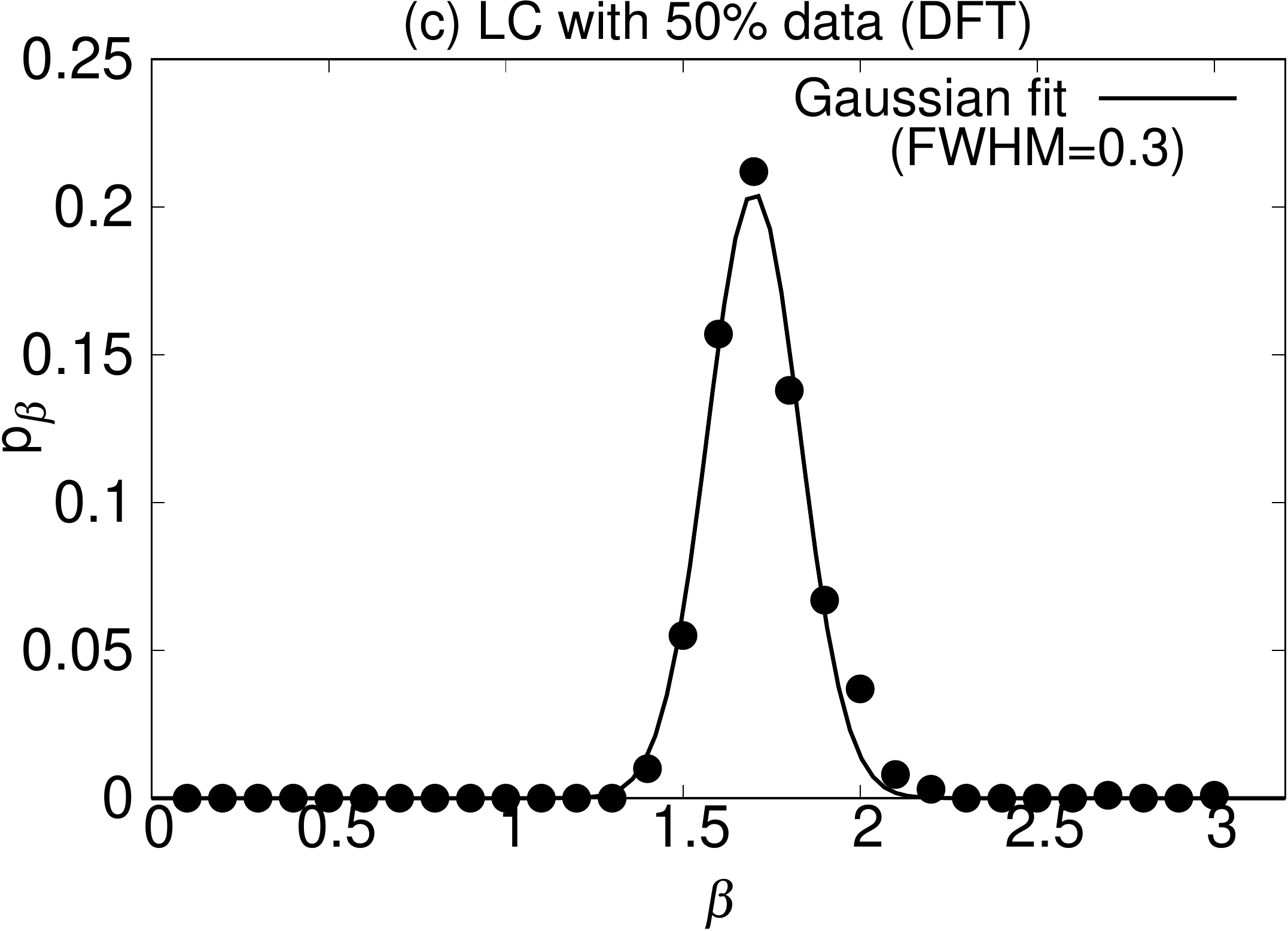}
         }
         	\hbox{
 		\includegraphics[width=0.30\textwidth]{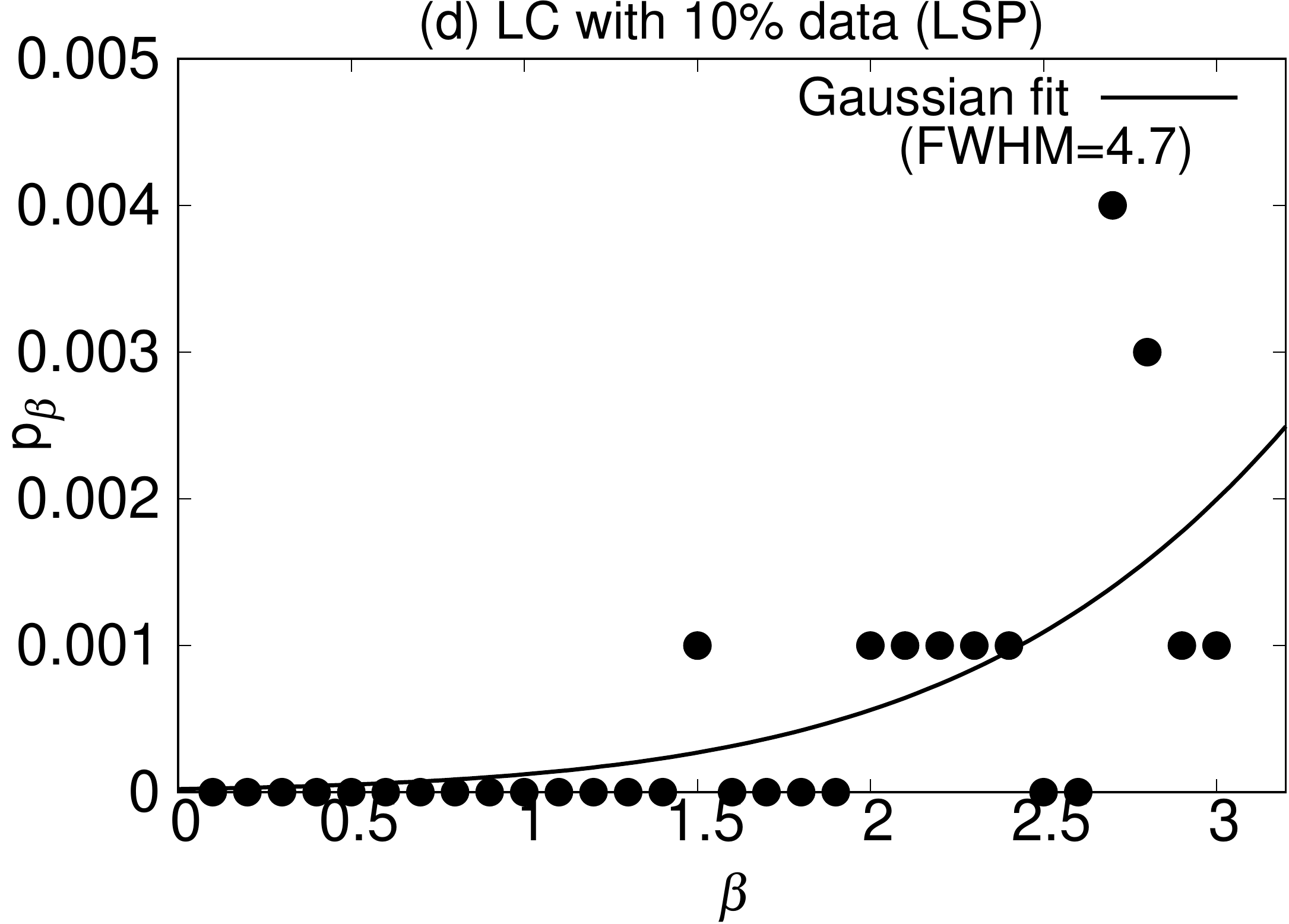}
 		\includegraphics[width=0.30\textwidth]{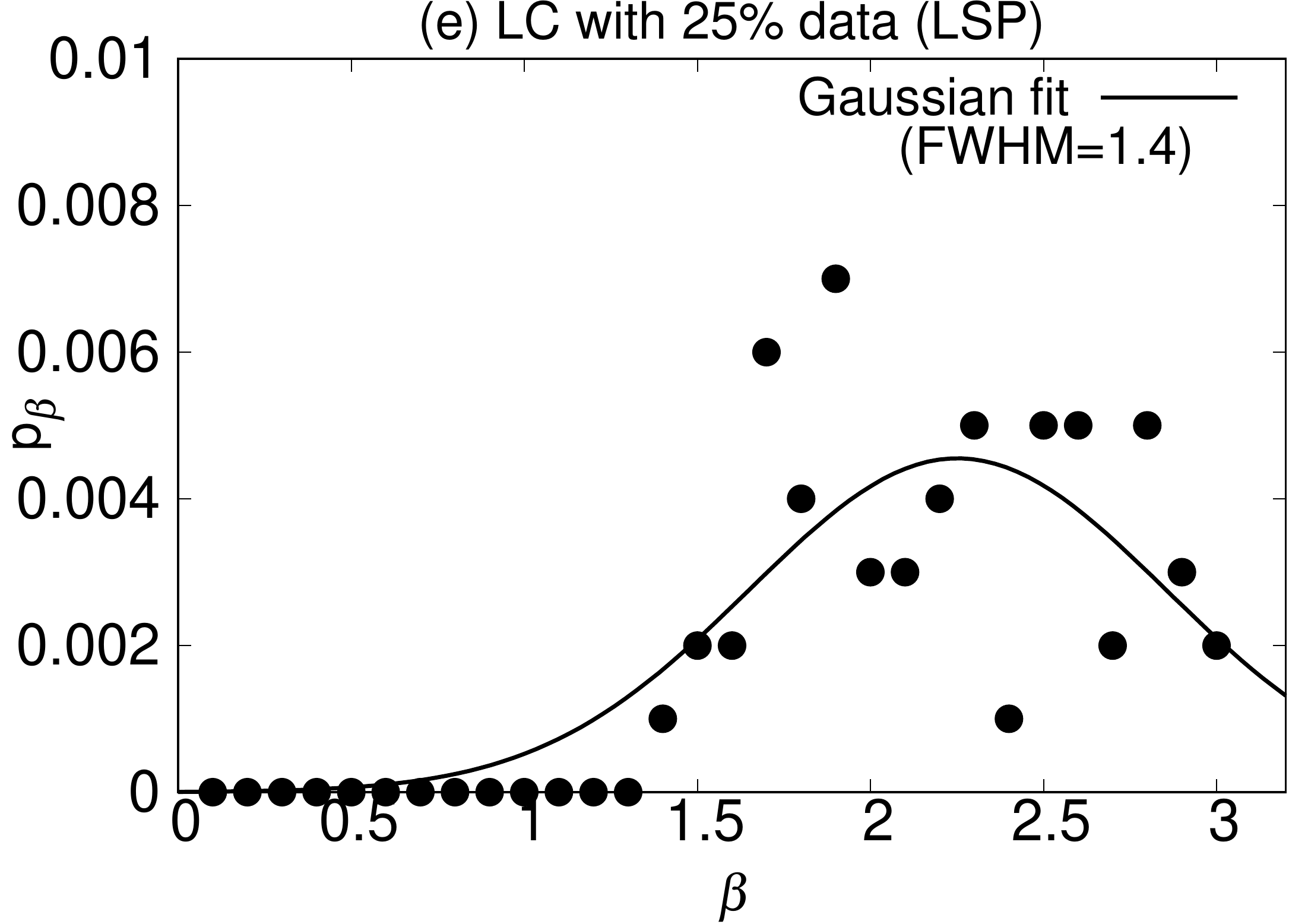}
		}
	\caption{probability curves for DFT (panels a, b, and c) and LSP (d and e) methods for light curves shown in Figure~\ref{fig:simlc}b.}
\label{fig:betasimsys}
\end{figure*}

The LS program was derived for evenly spaced frequencies ranging from the total duration of the light curve, $T$, down to the maximum Nyquist sampling frequency (0.5\,day${^{-1}}$). Figures~\ref{fig:dftlspran} and ~\ref{fig:dftlspsys} show the best-fit PSDs obtained using the PSRESP method for the DFT and LSP methods on the unevenly sampled light curves with 10\%, 25\%, and 50\% sampling rates shown in panels (a) and (b) of Figure~\ref{fig:simlc}. Moreover, to obtain the PSD using the DFT method, the unevenly sampled time series is rendered evenly sampled through linear interpolation with an interval of 0.5\,day. The success fractions for the two methods are given in Figures~\ref{fig:betasimran} (LCs with random gaps only) and ~\ref{fig:betasimsys} (LCs with systematic and random gaps), respectively. The best-fit PSD slope obtained using the LSP method for the 25\% sampling rate data with the random gaps (Figure~\ref{fig:dftlspran}e) does not have error estimation as the probability curve does not allow for fitting the Gaussian function because the success fractions turned out to be non-zero for only two $\beta$ values of the scanned range. On the other hand, we could not derive the best-fit LSP PSD using the PSRESP method for the 50\% sampling rate data with systematic and random gaps because the success fraction turned out to be zero at all $\beta$ values (Figure~\ref{fig:dftlspsys}f). 
 
Figures~\ref{fig:dftlspran} and ~\ref{fig:dftlspsys} demonstrate that the LS periodograms do not reproduce steep power spectra in the case of unevenly sampled data. Generally speaking, in the case of the LSP method, that is because of a small number of degrees of freedom (DOF) available to characterize the variability at the highest frequencies up to the maximum Nyquist frequency. Although the derived PSD slopes using the PSRESP method ($\beta=$1.6--2.7) can be considered reasonably close to the intrinsic PSD slope ($\beta = 2$), a flattening of the PSDs at higher frequencies ($\geq$10$^{-1.5}$ day$^{-1}$) is the result of the DOF issue mentioned above, which introduces artificial power in the high-frequency range of the spectrum and not because of measurement uncertainties, since the simulated light curves are free of scatter that would come from these uncertainties.

The DFT method (panels a, b, and c of Figures~\ref{fig:dftlspran} and ~\ref{fig:dftlspsys}), on the other hand, recovers the actual slope ($\beta$=\,2) with high probability (panels a, b, and c of Figures~\ref{fig:betasimran} and ~\ref{fig:betasimsys}), without introducing additional effects in the derived power spectrum; therefore, the DFT method is the better choice between the two methods \citep[see, also, Appendix of ][ for a relevant discussion]{Goyal17}. Furthermore, we note that the success fractions are low for the LSP PSRESP formalism, most likely due to the contribution of ``extra'' white noise ($\beta=$\,0) at higher frequencies, relative to the input spectrum, due to the undersampling. Finally, we reiterate that the differences in the derived power spectra using the two methods diminish gradually with a progressively increasing sampling of the light curves, with the output becoming the same for the evenly sampled time series \citep[][]{VanderPlas18}.

\end{document}